\documentclass[
reprint,
superscriptaddress,
amsmath,amssymb,
aps,
prb,
twocolumn
]{revtex4-2}
\usepackage[T1]{fontenc}
\usepackage[colorlinks=true, allcolors=blue]{hyperref}
\usepackage{verbatim}
\usepackage{enumitem}
\usepackage{stmaryrd}
\usepackage{xspace}

\usepackage{bm}
\usepackage{amsthm}
\usepackage{array}
\usepackage{mathtools}

\usepackage{physics}
\usepackage{siunitx}

\usepackage[table,svgnames,dvipsnames]{xcolor}
\usepackage{graphicx}
\graphicspath{{figs/sec5}{figs/sec6}{figs/sec9}{figs/sec10}}
\usepackage{tikz}
\usepackage{quantikz}

\usepackage{dcolumn}
\usepackage{tcolorbox}
\usepackage{booktabs}
\usepackage{multirow}
\usepackage{makecell}

\usepackage{subfigure}
\usepackage{epsfig}

\usepackage{orcidlink}

\usepackage{cleveref}
\crefname{equation}{Eq.}{Eqs.}
\crefname{section}{Sec.}{Secs.}
\crefname{appendix}{Appendix}{Appendices}
\crefname{figure}{Fig.}{Figs.}
\crefname{table}{Table}{Tables}
\crefname{theorem}{Theorem}{Theorems}

\begin{document}

\title{Practical Error Suppression and Mitigation for Reliable Quantum Computing}
\author{Han-Ze Li}
\affiliation{Department of Physics, \href{https://ror.org/01tgyzw49}{National University of Singapore}, Singapore 117551}
\affiliation{Institute for Quantum Science and Technology, \href{https://ror.org/006teas31}{Shanghai University}, Shanghai 200444, China}
\author{Mengjie Yang}
\affiliation{Department of Physics, \href{https://ror.org/01tgyzw49}{National University of Singapore}, Singapore 117551}
\author{Xianquan Yan}
\affiliation{Department of Physics, \href{https://ror.org/01tgyzw49}{National University of Singapore}, Singapore 117551}
\affiliation{Department of Computer Science, \href{https://ror.org/01tgyzw49}{National University of Singapore}, Singapore 117417}
\author{Dax Enshan Koh\,\orcidlink{0000-0002-8968-591X}}
\email{dax.koh@singaporetech.edu.sg}
\affiliation{Engineering Cluster, \href{https://ror.org/01v2c2791}{Singapore Institute of Technology}, 1 Punggol Coast Road, Singapore 828608, Republic of Singapore \looseness=-1}
\author{Ching Hua Lee\,\orcidlink{0000-0003-0690-3238}}
\email{phylch@nus.edu.sg}
\affiliation{Department of Physics, \href{https://ror.org/01tgyzw49}{National University of Singapore}, Singapore 117551}
\author{Ruizhe Shen\,\orcidlink{0000-0002-6992-9219}}
\email{e0554228@u.nus.edu}
\affiliation{Department of Physics, \href{https://ror.org/01tgyzw49}{National University of Singapore}, Singapore 117551}
\date{\today}
\begin{abstract}
Quantum computing is entering a transitional regime between noisy intermediate-scale quantum (NISQ) processing and early fault-tolerant quantum computation (FTQC), in which increasingly capable hardware is beginning to support repeated syndrome measurements, partial error correction, and logical-qubit operations, while residual physical and logical errors remain non-negligible. In this regime, error suppression, error mitigation, and quantum error correction are increasingly better viewed as complementary layers of a unified error-reduction strategy rather than as separate approaches, with each acting at a different stage of the quantum computation to improve simulation reliability. Thus, in this review, we provide a practical and forward-looking overview of the principal hardware error sources and the corresponding error suppression and mitigation methods for reducing their impact across the current NISQ--FTQC transition. We discuss hardware-aware circuit design, coherent-error suppression, readout mitigation, noise extrapolation, classical inference, and software-supported workflows, with particular emphasis on their implementation on actual quantum processors. We further examine how error mitigation techniques can be adapted to encoded and logical-qubit settings, so that they can operate alongside quantum error correction to suppress residual logical errors and improve the accuracy of computation in the early fault-tolerant regime.
\end{abstract}
\maketitle

\tableofcontents
\newpage
\onecolumngrid

\section{Introduction}\label{sec0}

Quantum processors have advanced rapidly in scale, controllability, and programmability.
Current examples include IBM's 156-qubit Heron and 120-qubit Nighthawk superconducting processors, Google's 105-qubit Willow processor, Quantinuum's 98-qubit Helios trapped-ion system, and QuEra's reconfigurable 256-atom Aquila processor.
Neutral-atom experiments have also operated programmable logical circuits using arrays of up to 280 physical atoms
\cite{preskill2018quantum,kjaergaard2020superconducting,bruzewicz2019trapped,henriet2020quantum,kimEvidenceUtilityQuantum2023c,bluvstein2024logical}.
These examples show that superconducting, trapped-ion, neutral-atom, and related programmable platforms can now support nontrivial quantum circuits and simulations involving tens to hundreds of physical qubits. Quantum simulation has emerged as one of the principal scientific applications of this hardware.
Analog quantum simulators exploit native interactions to realize effective many-body Hamiltonians directly, whereas digital simulators approximate the target evolution through sequences of programmable quantum gates.
These approaches have been applied to equilibrium and nonequilibrium quantum
magnetism and quantum critical dynamics across trapped-ion, neutral-atom, and
superconducting platforms
\cite{friedenauer2008simulating,bernien2017probing,
zhang2017manybody,ebadi2021quantumphases,
kimEvidenceUtilityQuantum2023c,chen2026robust,
shen2026resolving,koh2022stabilizing,koh2022simulation,
koh2024realization,shen2024enhanced,shen2025observation,
koh2026interacting,
shen2026observation,chen2023high,chen2024direct,richieherm2014nonlocal,jurcevic2014quasiparticle,
smith2016manybody,neill2016ergodic}.
For example, they have enabled the preparation and manipulation of topological matter
\cite{satzinger2021realizing,semeghini2021probing,xu2023digital,
iqbal2024nonabelian,xu2026fractional},
the simulation of quantum dynamics and lattice gauge theories
\cite{martinez2016realtime,kokail2019selfverifying,nguyen2022schwinger,
farrellQuantumSimulationsHadron2024},
and calculations of molecular electronic structure
\cite{peruzzo2014variational,omalley2016scalable,kandala2017hardware,
hempel2018quantum,arute2020hartreefock}.

The field of quantum computing is now entering an intermediate regime: experiments are moving beyond proof-of-principle NISQ demonstrations, while large-scale universal fault-tolerant quantum computation remains out of reach.
Recent experiments have demonstrated repeated quantum-error-correction cycles, logical-qubit operations, logical gates, and elementary fault-tolerant primitives
\cite{ryananderson2021realtime,krinner2022repeated,postler2022faulttolerant,zhao2022surfacecode,bluvstein2024logical,google2025quantum,mayer2024benchmarking}.
    For example, a programmable neutral-atom processor used up to 280 physical qubits to realize circuits with as many as 48 logical qubits, including 228 logical two-qubit gates and 48 logical three-qubit CCZ gates
\cite{bluvstein2024logical}.
Nevertheless, present-day computations are still limited by finite physical error rates, modest code distances, substantial encoding and decoding overheads, restricted logical circuit depth, and workflows in which not all operations are protected in a fault-tolerant manner.
Even the most advanced demonstrations currently involve only a limited set of encoded operations and code families, demonstrated only on carefully selected circuits.
Consequently, for most currently accessible applications, useful information must still be inferred from noisy hardware outputs, making error mitigation an essential bridge between raw device performance and reliable computational estimates.

It is useful at the outset to clarify how the error-suppression and error-mitigation techniques considered in this review relate to quantum error correction (QEC) and fault-tolerant quantum computing. All of these approaches seek to reduce the impact of hardware imperfections, but they intervene at different stages of the computation and provide different levels of protection. QEC acts at the level of encoded quantum information: physical qubits are redundantly encoded into logical qubits, errors are diagnosed through repeated syndrome measurements, and recovery operations are used to prevent physical faults from propagating into logical errors. When the physical error rates are below the relevant fault-tolerance threshold, increasing the code distance can systematically reduce the logical error rate and, in principle, support computations whose reliable depth greatly exceeds the coherence scale of the underlying physical qubits.

The approaches emphasized in this review instead operate mainly in the unencoded or lightly encoded regime, where such full fault-tolerant protection is not yet available or would require prohibitive overhead. Within this regime, errors can be addressed in two complementary ways. One can reduce the errors that enter the computation in the first place, or one can compensate for their effect on the quantities extracted from the noisy computation. We refer to the former as \emph{error suppression}: these techniques modify the physical control, scheduling, compilation, or circuit structure so that the implemented computation accumulates less noise or a less harmful form of noise. Examples include improved calibration and control, hardware-aware compilation, circuit-depth reduction, randomized compiling, twirling, and dynamical decoupling. We use \emph{quantum error mitigation} in the narrower sense for the latter class of approaches, which do not prevent all physical errors during the circuit but instead use additional measurements, calibration information, repeated circuit executions, or classical inference to estimate observables and output statistics closer to their ideal values.

Error suppression and error mitigation should therefore not be viewed as competing alternatives to QEC, but as complementary layers of error management with different objectives and resource requirements. Suppression improves the effective computation that is executed on the hardware, while mitigation improves the information extracted from the resulting noisy data; both can also be used together, and increasingly can complement partially error-corrected computations as hardware moves toward fault tolerance. In the present intermediate regime, quantum error mitigation therefore remains a central practical strategy for extracting useful information from imperfect quantum hardware
\cite{temmeErrorMitigationShortDepth2017a,liEfficientVariationalQuantum2017a,endo2018practical,endo2021hybrid,cai2023quantum,bultrini2023unifying,shen2026simulating,tsubouchi2023universalcost,PhysRevLett.129.020502}.
For unencoded or lightly encoded devices, it can improve estimates of physically relevant quantities without requiring the substantial qubit, syndrome-extraction, and real-time control overhead associated with full QEC. Representative approaches include zero-noise extrapolation
\cite{kandalaErrorMitigationExtends2019,giurgica-tironDigitalZeroNoise2020,caiMultiexponentialErrorExtrapolation2021,kimScalableErrorMitigation2023a,harrisReducingQuantumError2026},
probabilistic error cancellation and related quasi-probability or inverse-channel methods
\cite{temmeErrorMitigationShortDepth2017a,endo2018practical,piveteau2021error,vandenberg2023sparsepec,chen2025fasterpec},
readout-error mitigation
\cite{maciejewski2020mitigation,bravyi2021mitigating,nation2021scalable,smith2021qubit,funcke2022measurement,aasen2024readout},
symmetry verification and symmetry expansion
\cite{bonetmonroig2018lowcost,sagastizabal2019experimental,cai2021symmetry},
Clifford-data-regression-type and learning-assisted methods
\cite{czarnik2021error},
and tensor-network-assisted mitigation
\cite{guo2022mpo,filippov2023tem,filippov2024scalability,ibm2026tem}.

Beyond these widely used families, several additional mitigation paradigms have become increasingly important. One class is based on purification-inspired post-processing, including virtual distillation, derangement-based error suppression, dual-state purification, and related purification-based protocols, where noisy quantum states are processed to estimate observables with respect to an effectively purified density matrix
\cite{hugginsVirtualDistillationQuantum2021,koczorExponentialErrorSuppression2021,huoDualStatePurification2022,obrienPurificationBasedQuantum2023}.
A second class uses subspace or verification ideas, in which information from error-detecting codes, quantum subspace expansions, or phase-estimation circuits is used to project noisy data back toward a physically meaningful sector
\cite{mccleanDecodingQuantumErrors2020,urbanekChemistryQuantumComputers2020,obrienErrorMitigationVerified2021}.
A third rapidly developing direction is data-driven and machine-learning-based mitigation, where corrected estimators are learned from classically simulable circuits, variable-noise data, augmented noisy datasets, or related training ensembles, rather than from a complete microscopic noise model
\cite{loweUnifiedApproachDataDriven2021,strikisLearningBasedQuantum2021,bennewitzNeuralErrorMitigation2022,liaoMachineLearningPractical2024,liaoNoiseAgnosticQuantum2025}.
Recent extensions to mid-circuit measurements and the characterization or mitigation of measurement backaction, and preprocessing-stage mitigation further show that error mitigation is becoming a platform- and workflow-level methodology rather than only a final classical post-processing step
\cite{suErrorMitigationPhotonic2021,giortamis2026mcmit,chuLearningMidCircuit2026,martinQuantumErrorMitigationPreprocessing2026}.

At the same time, quantum error mitigation is increasingly being explored as a bridge technology for early logical processors and partially error-corrected workflows, where some operations may be protected by encoding while other factors, such as non-Clifford gates, finite code distances, circuit-level logical errors, or decoder-level uncertainty, still require mitigation
\cite{piveteau2021error,suzuki2022universal,wahlZeroNoiseExtrapolation2023,smith2024logical,dutkiewicz2025error,zhangDemonstratingQuantumError2026,jeon2026qecmitigated}. 
Dynamic-circuit and feedforward settings further broaden this perspective, because mid-circuit measurements and classically conditioned operations introduce new noise and calibration structures that can also be incorporated into mitigation protocols
\cite{gupta2024dynamicpec,bar2026layered}.
Rather than eliminating errors through complete logical encoding and fully fault-tolerant control, error mitigation reduces, models, or compensates for noise at the level of experimentally accessible quantities, including expectation values, probability distributions, logical observables, and estimator averages. This output-level character also exposes intrinsic limitations: mitigation accuracy can depend strongly on sampling overhead, noise-model validity, circuit depth, system size, temporal drift, non-Markovian effects, and the quality of benchmark or calibration data
\cite{takagi2022fundamental,takagi2023samplinglowerbounds,quek2024tighter,govia2025modelviolation,white2025nonmarkovian,harrisReducingQuantumError2026}.

A substantial literature has now developed around quantum error mitigation, including reviews, benchmarks, conceptual perspectives, and rigorous analyses of its capabilities and limitations
\cite{endo2021hybrid,qin2022overview,cai2023quantum,bultrini2023unifying,aharonov2025importance,quek2024tighter}.
This development has been accelerated by the rapid growth of quantum processors themselves. As devices have increased in qubit number, circuit depth, connectivity complexity, calibration availability, and support for dynamic-circuit primitives, error mitigation has evolved from a set of relatively abstract estimator-correction ideas into a hardware-facing methodology shaped by real device constraints. The purpose of the present review is therefore different. Rather than providing only a method-by-method catalog, we also focus on the \emph{practical logic of error mitigation on contemporary quantum hardware in the transition toward fault tolerance}. In particular, we emphasize that the usefulness of a mitigation protocol depends not only on its formal correctness, but also on its compatibility with hardware connectivity, native gate sets, calibration access, sampling overhead, numerical stability, dynamic-circuit primitives, provider-native runtime support, and software implementation
\cite{maciejewski2020mitigation,bravyi2021mitigating,nation2021scalable,smith2021qubit,aasen2024readout,laroseMitiqSoftwarePackage2022a,hashim2025benchmarking,layden2026theory,bar2026layered,camilo2025compilation}.

A quantum error mitigation method that is attractive at the level of abstract formalism may become ineffective or impractical if its calibration cost scales too quickly, if its inference problem is ill-conditioned, if the required sampling overhead overwhelms the gain in estimator accuracy, or if its assumptions do not match the dominant physical noise processes of the device
\cite{takagi2022fundamental,takagi2023samplinglowerbounds,quek2024tighter,govia2025modelviolation,filippov2024scalability,white2025nonmarkovian,harrisReducingQuantumError2026}.
This hardware-facing perspective is especially important as mitigation begins to interface with early logical processors, partial error-correction workflows, and dynamic circuits involving mid-circuit measurement and feedback
\cite{wahlZeroNoiseExtrapolation2023,dutkiewicz2025error,zhangDemonstratingQuantumError2026,smith2024logical,jeon2026qecmitigated,bar2026layered}.
It is also increasingly relevant for provider-native runtimes and software ecosystems, where mitigation can serve not only as a post-processing routine but also as an integral component of compilation, calibration management, twirling, readout correction, dynamical decoupling, and runtime estimator primitives
\cite{laroseMitiqSoftwarePackage2022a,hashim2025benchmarking,ibm2026tem}.

This practical perspective on hardware implementation is important because error mitigation is rarely deployed as a single isolated protocol. In experimental workflows, mitigation is typically assembled as a layered procedure, combining hardware-aware circuit choices, readout calibration, noise scaling or extrapolation, quasi-probability reconstruction, symmetry filtering, coherent-error tailoring, and data-driven classical inference
\cite{endo2021hybrid,bultrini2023unifying,endo2018practical,mari2021extending,giurgica-tironDigitalZeroNoise2020,bonetmonroig2018lowcost,sagastizabal2019experimental,czarnik2021error}.
For example, measurement calibration and readout-error mitigation provide a particularly common first layer, since they can be incorporated into many experimental pipelines before or together with higher-level mitigation methods
\cite{nation2021scalable,funcke2022measurement,aasen2024readout}.
Similarly, randomization, twirling, and dynamical decoupling can reshape coherent or temporally structured errors into forms that are more compatible with downstream extrapolation, inverse-channel mitigation, or symmetry filtering
\cite{hashim2025benchmarking,layden2026theory,bar2026layered}.
The central question is therefore not simply whether a mitigation method works in principle, but under what physical, statistical, and computational conditions it produces a meaningful improvement at acceptable cost
\cite{takagi2022fundamental,quek2024tighter}.

This review aims to organize error-mitigation and error-suppression strategies according to their operational roles in realistic pre-fault-tolerant, early-logical, and partially fault-tolerant workflows. Although QEC provides the central route toward fault-tolerant quantum computation, it remains a rapidly developing field with evolving codes, decoding methods, and hardware implementations, and a comprehensive review lies beyond the scope of this work. We therefore discuss QEC only where it helps distinguish mitigation and suppression from logical error correction, or where these approaches interface with early-logical and partially fault-tolerant computation. We begin with the hardware landscape and the dominant physical error sources, since the effectiveness of any mitigation or suppression strategy must be understood in relation to concrete device capabilities, noise mechanisms, and control constraints. We then review observable- and circuit-level suppression strategies, measurement-error mitigation, noise scaling and extrapolation, inverse-channel and quasi-probability methods, randomization and twirling protocols, symmetry-based mitigation, sample-efficient estimation strategies, and software ecosystems. Throughout, we emphasize the interplay among error structure, mitigation overhead, statistical stability, benchmarking, and deployment conditions on real hardware. In this sense, the present review is intended not only as a summary of existing methods, but also as a practical guide to when, why, and how different mitigation and suppression strategies can be expected to work, and how their roles evolve as quantum processors progress from NISQ-scale operation toward early logical and fault-tolerant quantum computation.

\section{Overview}

To understand the practical role of error suppression and mitigation, it is useful first to distinguish between analog and digital quantum computation. In analog quantum simulation, the target many-body dynamics is generated directly by a programmed Hamiltonian, as in neutral-atom arrays and other analog simulators
\cite{ebadi2021quantumphases,semeghini2021probing,henriet2020quantum}. Error reduction therefore acts mainly through improved calibration and control of the implemented Hamiltonian, suppression of decoherence and unwanted couplings, symmetry or conservation-law constraints, and post-processing of measured observables. These approaches can substantially improve the reliability of analog simulations, but the available mitigation operations are closely tied to the specific simulator and generally provide less freedom to modify the computation without also modifying the physical dynamics.

Digital quantum processors provide a considerably broader setting for error suppression and mitigation. A target evolution is decomposed into a programmable sequence of gates, measurements, resets, and classical feedforward operations, so errors can be addressed at several distinct stages of the computation. Hardware-aware compilation, pulse optimization, dynamical decoupling, and circuit randomization can suppress or reshape errors before and during execution; readout mitigation corrects the measurement stage; and methods such as zero-noise extrapolation, probabilistic error cancellation, symmetry-based mitigation, and learning-based inference use repeated circuit executions and classical processing to reduce residual bias in the final observables
\cite{endo2021hybrid,cai2023quantum,bultrini2023unifying}. The digital setting is particularly favorable because logically equivalent circuit realizations, controlled noise scaling, calibration circuits, and repeated sampling can be generated systematically without changing the intended computation. Error mitigation has therefore developed most extensively in gate-based quantum computing, where it can be integrated directly with the compiler, runtime, calibration system, and classical analysis stack.

This role is becoming increasingly important as quantum hardware moves from the conventional NISQ regime toward early fault-tolerant operation. Current processors already support nontrivial computations on tens to hundreds of physical qubits, while finite gate errors, decoherence, readout imperfections, crosstalk, leakage, and compilation overhead remain relevant
\cite{preskill2018quantum,kjaergaard2020superconducting,blais2021circuit,
arute2019quantum,wu2021strong,kimEvidenceUtilityQuantum2023c,
bruzewicz2019trapped,pino2021demonstration,moses2023race}.
At the same time, advances in mid-circuit measurement, reset, feedforward, dynamic circuits, and logical-qubit control are beginning to connect physical-qubit mitigation with quantum error correction
\cite{edmunds2020dynamically,botelho2022midcircuit,ryananderson2021realtime,
krinner2022repeated,postler2022faulttolerant,zhao2022surfacecode,
iqbal2024topological,bluvstein2024logical,google2025quantum,
zhangDemonstratingQuantumError2026}.
In this transitional regime, suppression, mitigation, and error correction naturally operate as complementary layers rather than as separate alternatives. The hardware discussion below therefore focuses primarily on the leading gate-based platforms on which this layered digital workflow can be implemented.

\subsection{Superconducting-circuit platforms}

Superconducting-circuit processors are among the most widely used and technologically mature platforms for gate-based quantum computing
\cite{kjaergaard2020superconducting,blais2021circuit,abughanem2025ibm}.
Their practical importance stems from fast gate operations, broad cloud accessibility, and mature software and control stacks for calibration, compilation, benchmarking, runtime execution, error mitigation, and dynamic-circuit operation
\cite{jurcevic2021demonstration,edmunds2020dynamically,botelho2022midcircuit,
laroseMitiqSoftwarePackage2022a,hashim2025benchmarking,koh2026readout}.
Superconducting hardware now spans a regime extending from conventional NISQ computation to early logical-qubit operation, including repeated error-correction cycles, below-threshold logical memories, and error mitigation applied to encoded circuits
\cite{krinner2022repeated,zhao2022surfacecode,google2025quantum,
zhangDemonstratingQuantumError2026}.

\subsubsection{Architecture, connectivity, and native gate operations}

Superconducting qubits are fabricated on-chip and connected through an architecture-dependent coupling graph
\cite{kjaergaard2020superconducting,blais2021circuit,place2021new,sung2021realization}.
Because native two-qubit gates are generally restricted to connected qubit pairs, an abstract circuit must be mapped, routed, and optimized for the topology and calibration state of the target processor
\cite{sivarajah2021t,javadi2024quantum,li2019tackling,murali2019noise}.
Nonlocal interactions therefore introduce additional SWAP operations or equivalent circuit transformations, increasing two-qubit-gate count, circuit duration, and exposure to decoherence and control errors. Connectivity must consequently be considered together with coherence, gate fidelity, readout error, leakage, crosstalk, and calibration stability when designing mitigation strategies
\cite{burnett2019decoherence,carroll2022dynamics,chen2016measuring,
sarovar2020detecting,maciejewski2021modeling,tripathi2022suppression,
perrin2024crosstalk}.

Representative IBM processors illustrate this hardware dependence [Fig.~\ref{fig:device}]. The \emph{Heron} family uses fixed-frequency superconducting qubits with tunable couplers and retains a heavy-hex connectivity architecture, while the newer \emph{Nighthawk} processor adopts a square-lattice topology with increased local connectivity
\cite{abughanem2025ibm}.
Heron r1 contains 133 qubits, whereas later Heron generations operate at the 156-qubit scale; Nighthawk contains 120 qubits
\cite{abughanem2025ibm}.
Heavy-hex connectivity reduces coordination and helps alleviate wiring, frequency-collision, and unwanted-coupling constraints, but can increase routing overhead for nonlocal circuits
\cite{chamberland2020topological,jurcevic2021demonstration,murali2019noise}.
Higher-connectivity square-lattice architectures can reduce this routing burden, although they place greater demands on crosstalk management and scheduling
\cite{sarovar2020detecting,perrin2024crosstalk,hashim2025benchmarking}.

At the circuit level, an ideal computation may be written as
\begin{equation}
U=U_L\cdots U_2U_1,
\end{equation}
but the layers $U_\ell$ must ultimately be translated into the calibrated native instruction set of the backend. Single-qubit gates are implemented using resonant microwave pulses,
\begin{equation}
R_{\hat{\bm n}}(\theta)
=
\exp\left[
-i\frac{\theta}{2}\hat{\bm n}\cdot\bm{\sigma}
\right],
\qquad
\hat{\bm n}=(\cos\varphi,\sin\varphi,0),
\end{equation}
while $Z$ rotations are typically realized through virtual frame updates,
\begin{equation}
R_Z(\phi)
=
\exp\left(
-i\frac{\phi}{2}Z
\right).
\end{equation}
Because virtual $Z$ rotations require no additional microwave pulse, they are particularly useful for compilation, Pauli-frame tracking, and randomized compiling.

The native entangling operation depends on the processor family
\cite{abughanem2025ibm,wei2024native}.
Recent Heron- and Nighthawk-class systems use CZ-type interactions, whereas earlier Eagle-class devices commonly expose the echoed cross-resonance (ECR) gate. An ideal CZ gate is
\begin{equation}
\mathrm{CZ}
=
\mathrm{diag}(1,1,1,-1)
=
\exp\!\left[
-i\frac{\pi}{4}(I-Z)\otimes(I-Z)
\right].
\label{eq:cz_def}
\end{equation}
The ECR gate, by contrast, is a microwave-driven entangling primitive locally equivalent to a CNOT-class interaction. These different native entanglers lead to different decompositions, calibration procedures, coherent-error signatures, and opportunities for frame tracking or echoed suppression
\cite{ganzhorn2020benchmarking,wei2024native,qiskit_transpiler}.

\subsubsection{Control, software, and dynamic-circuit capabilities}

Superconducting processors are increasingly operated through integrated control and software stacks rather than through isolated gate execution. IBM Quantum, for example, exposes a toolbox-style workflow through Qiskit and Qiskit Runtime
[see Sec.~\ref{provider}], where suppression and mitigation can be composed at execution time
\cite{javadi2024quantum,qiskitdocs_mitigation_2026,qiskitdocs_techniques_2026}.
Error-suppression layers include dynamical decoupling, randomized compiling, and Pauli-twirling-based methods, whereas mitigation layers include readout correction, ZNE, and related estimator-level procedures
\cite{viola1999dynamical,hashim2021randomized,ezzell2023dynamical,
temmeErrorMitigationShortDepth2017a,endo2018practical,cai2023quantum}.
This reflects a broader hardware--software co-design strategy in which circuit mapping, pulse control, calibration, execution, and classical inference are optimized together.

Dynamic circuits provide an important extension of this control model. Mid-circuit measurement, reset, and classically conditioned feedforward allow later operations to depend on information obtained during the same execution
\cite{botelho2022midcircuit,gupta2024dynamicpec,bar2026layered,shirizly2025dynamicrb}.
Superconducting platforms are particularly well suited to such workflows because dispersive readout and electronic feedback can provide comparatively low-latency measurement and control. These capabilities enable adaptive state preparation, measurement-based operations, long-range entanglement generation, and repeated quantum-error-correction cycles
\cite{baumer2024efficient,carrera2024combining,krinner2022repeated,
zhao2022surfacecode,google2025quantum}.
For mitigation, however, they also introduce additional internal error sources associated with measurement, reset, feedforward latency, and branch-dependent evolution, which must be treated as part of the complete circuit noise model.

\subsubsection{From NISQ hardware to early logical-qubit operation}

\begin{figure}[h]
    \centering
    \includegraphics[width=0.9\linewidth]{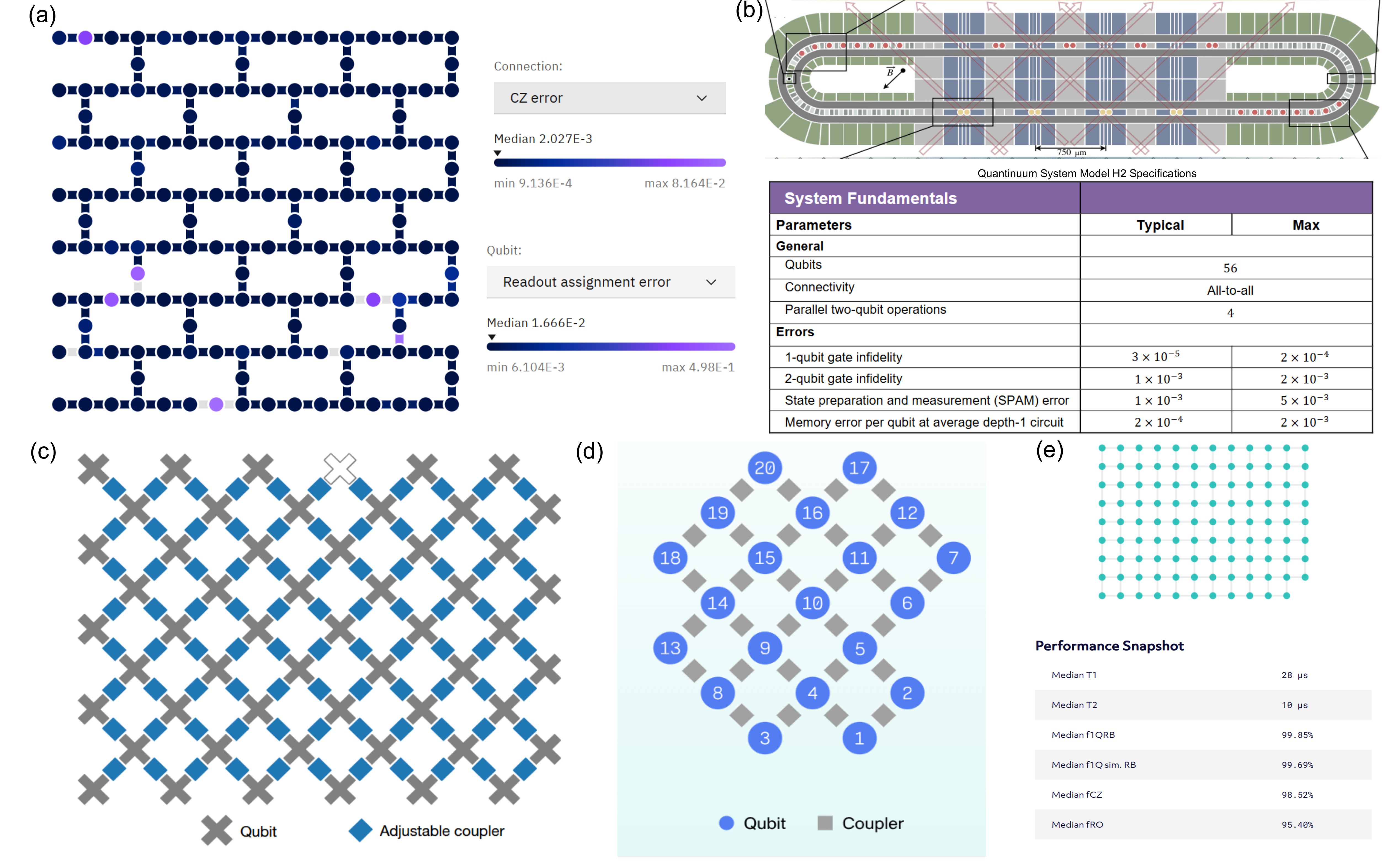}
\caption{{Platform topology and representative error metrics for quantum hardware.}
(a) Schematic of the IBM Quantum device connectivity together with representative calibration data.
(b) Conceptual layout of the Quantinuum System Model~H2 architecture and a summary of its system fundamentals.
(c) Schematic of Google's Sycamore superconducting processor, showing the two-dimensional qubit array and tunable nearest-neighbor couplers.
(d) Representative connectivity map of an IQM superconducting processor, showing qubits and two-qubit couplers.
(e) Representative Rigetti superconducting QPU, showing its two-dimensional qubit-connectivity layout together with a performance snapshot of median coherence times, single- and two-qubit gate fidelities, and readout fidelity.
Panels in (b) are adapted from Refs.~\cite{moses2023race,quantinuum_h2_system}, while panels (c,d,e) are adapted from Refs.~\cite{arute2019quantum}, \cite{iqmacademyNISQ}, \cite{rigetti_qpus} respectively.}
    \label{fig:device}
\end{figure}

The evolution of superconducting hardware is increasingly shaped by requirements associated with quantum error correction rather than by physical-qubit count alone. Google Quantum AI provides a prominent example [Fig.~\ref{fig:device}(c)]. Surface-code experiments have progressed from demonstrating improved logical performance with increasing code distance to below-threshold logical-memory operation on the Willow processor
\cite{acharya2023suppressing,google2025quantum}.
These experiments emphasize the need to combine processor topology, repeated syndrome measurement, control electronics, decoding, and calibration into a single scalable architecture.

Other commercial platforms illustrate related design choices. IQM processors combine transmon qubits, tunable couplers, and square-lattice connectivity [Fig.~\ref{fig:device}(d)], with native single-qubit rotations and CZ entangling gates. Their square-lattice layout is naturally compatible with surface-code geometries and reflects increasing attention to future fault-tolerant requirements
\cite{iqmacademyNISQ}.
Rigetti provides another full-stack superconducting platform [Fig.~\ref{fig:device}(e)], combining tunable superconducting qubits, parametric entangling operations, cloud execution, and hybrid quantum--classical workflows
\cite{azure_rigetti_provider_2026}.

These developments indicate that superconducting processors now occupy an important transition between physical-qubit NISQ computation and early fault-tolerant operation. Progress increasingly depends on the co-design of connectivity, tunable coupling, measurement and reset, runtime control, compilation, error suppression, mitigation, and quantum error correction
\cite{sung2021realization,google2025quantum,zhangDemonstratingQuantumError2026}.
Superconducting hardware therefore provides a central platform for studying not only mitigation on noisy physical qubits, but also how suppression and mitigation can operate alongside emerging logical-qubit and error-corrected computation.

\subsection{Trapped-ion platforms}

Trapped-ion quantum computers encode qubits in long-lived internal electronic
states of atomic ions confined by electromagnetic traps and manipulated using
laser- or microwave-driven control
\cite{cirac1995quantum,harty2014high,bruzewicz2019trapped,
monroe2021programmable,foss2025progress}.
A distinctive feature is that entangling interactions are mediated by collective
motional modes, allowing gates between ions that are not restricted by a fixed
nearest-neighbor coupling graph
\cite{molmer1999multiparticle,blatt2012quantum,ballance2016high,
gaebler2016high,pino2021demonstration,moses2023race}.
This flexible connectivity substantially reduces routing overhead and makes
trapped-ion processors particularly attractive for circuits containing
long-range interactions or nonlocal entangling patterns
\cite{linke2017comparison,grzesiak2020efficient,monroe2021programmable}.

\subsubsection{Connectivity and native gate operations}

Within a single ion chain or reconfigurable trap module, trapped-ion processors
can realize effective near-all-to-all connectivity
\cite{kielpinski2002architecture,debnath2016demonstration,
wright2019benchmarking,pino2021demonstration,moses2023race,
chen2024ionqforte}.
Consequently, many two-qubit interactions can be implemented directly without
the SWAP networks required on sparse-connectivity architectures. This can
reduce both circuit depth and the accumulation of routing-induced errors,
particularly for quantum simulation, fermionic mappings, and other workloads
with intrinsically nonlocal interaction structure
\cite{kim2010quantum,islam2013emergence,richieherm2014nonlocal,
smith2016manybody,grzesiak2020efficient}.

The native gate set is typically built from arbitrary single-qubit rotations
and a M{\o}lmer--S{\o}rensen (MS) entangling interaction
\cite{molmer1999multiparticle,blatt2012quantum,bruzewicz2019trapped}.
For ions $i$ and $j$, a commonly used form is
\begin{equation}
R^{XX}_{ij}(\theta)
=
\exp\!\left(
-i\frac{\theta}{2}X_iX_j
\right),
\label{eq:xx_gate}
\end{equation}
or, more generally,
\begin{equation}
MS_{ij}(\theta,\varphi)
=
\exp\!\left[
-i\frac{\theta}{2}
\left(
\cos\varphi\,X_i+\sin\varphi\,Y_i
\right)
\left(
\cos\varphi\,X_j+\sin\varphi\,Y_j
\right)
\right].
\label{eq:ms_general}
\end{equation}
The optical phase $\varphi$ controls the interaction axis in the $XY$ plane,
while $\theta$ sets the entangling strength. Together with arbitrary
single-qubit rotations, these interactions provide a universal gate set.

IonQ provides a representative example of this rotation-based native gate
language. Its single-qubit primitives include $\mathrm{GPI}$ and
$\mathrm{GPI2}$ operations,
\begin{equation}
\mathrm{GPI}(\phi)
=
\exp\!\left[
-i\frac{\pi}{2}
(\cos\phi\,X+\sin\phi\,Y)
\right],
\qquad
\mathrm{GPI2}(\phi)
=
\exp\!\left[
-i\frac{\pi}{4}
(\cos\phi\,X+\sin\phi\,Y)
\right],
\label{eq:gpi_gpi2}
\end{equation}
together with an MS-type two-qubit entangler
\cite{niroula2024phase,romero2025protein}.
The continuous phase parameters of these native operations provide substantial
freedom for hardware-aware circuit compilation.

\subsubsection{Dynamic control and reconfigurable execution}

Trapped-ion processors also support increasingly sophisticated measurement and
control primitives. Quantinuum's Model~H2, for example, combines arbitrary
single-qubit rotations and MS-type entangling gates with mid-circuit
measurement, reset, and classically conditioned feedforward
\cite{moses2023race,mayer2024benchmarking,iqbal2024topological,
quantinuum_h2_system}.
These capabilities allow circuits to include adaptive state preparation,
conditional operations, qubit reuse, and repeated measurement structures
rather than being restricted to a fixed sequence of unitary gates.

Such capabilities are closely connected to the quantum-charge-coupled-device
(QCCD) architecture, in which ions can be transported between storage,
interaction, and measurement zones
\cite{kielpinski2002architecture,pino2021demonstration,decross2023qubitreuse}.
Physical qubits can therefore be reconfigured during computation instead of
remaining fixed on a static coupling graph. This flexibility can reduce
routing costs and enables measurement, reset, and reuse to become integral
parts of the execution model.

From the perspective of error mitigation, this architecture introduces a
different balance of error sources from superconducting hardware. Routing
through SWAP gates is less important, whereas motional-mode errors, imperfect
laser control, heating, measurement and reset errors, ion transport, and
time-dependent calibration become more prominent. Dynamic-circuit workflows
also require measurement and feedforward errors to be treated as internal
components of the circuit noise rather than solely as terminal readout errors.

\subsubsection{Toward early fault-tolerant operation}

Trapped-ion systems are increasingly relevant to the transition from
high-fidelity physical-qubit computation to early fault-tolerant operation.
Their long coherence times, flexible connectivity, mid-circuit measurement,
reset, feedforward, and qubit reuse are naturally suited to repeated syndrome
extraction and logical-state manipulation
\cite{egan2021fault,ryananderson2021realtime,postler2022faulttolerant,
paetznick2024demonstration,iqbal2024topological}.

This direction is already visible experimentally. Trapped-ion platforms have
demonstrated fault-tolerant control of encoded qubits, real-time quantum error
correction, logical gate operations, and measurement-and-feedforward protocols
for preparing and manipulating encoded many-body states
\cite{egan2021fault,ryananderson2021realtime,postler2022faulttolerant,
iqbal2024topological,paetznick2024demonstration}.
These results show that the relevant hardware challenge is no longer only the
fidelity of individual physical gates, but the integration of coherent
operations, ion transport, measurement, reset, decoding, and classical
feedforward into a repeated logical workflow.

Trapped-ion hardware should therefore be viewed as spanning the regime from
advanced pre-fault-tolerant computation to early logical-qubit operation.
Its combination of flexible interaction graphs and measurement-enabled control
makes it an important platform for studying how error suppression, mitigation,
and quantum error correction can be combined as processors move toward
fault-tolerant computation
\cite{monroe2014large,moses2023race,foss2025progress,
ryananderson2024teleportation}.

\subsection{Physical error sources}\label{sources}

To understand the practical role of error mitigation, we next identify the main physical error sources that limit circuit performance on contemporary quantum hardware.
Present-day processors are affected by several partly interdependent noise mechanisms, including imperfect control pulses, decoherence, leakage out of the computational subspace, measurement and readout errors, crosstalk between qubits, temporal drift, and extra gates introduced during compilation
\cite{kjaergaard2020superconducting,blais2021circuit,bruzewicz2019trapped,foss2025progress,sheldon2016characterizing,chen2016measuring,carroll2022dynamics,hashim2025benchmarking,hazra2025benchmarking,quantinuum_leakage_2026}.
These errors do not merely lower the overall fidelity of a circuit. They also determine which observables can be measured reliably, which circuit structures can be executed at useful depth, and which mitigation or suppression strategies are well matched to a given device
\cite{endo2021hybrid,cai2023quantum,bultrini2023unifying,javadi2024quantum,govia2025modelviolation}.
In particular, correlated and context-dependent noise mechanisms, such as readout correlations, spectator-qubit effects, crosstalk, measurement-induced backaction, and time-dependent calibration drift, often violate the simplified independent-error models assumed by many mitigation protocols
\cite{sarovar2020detecting,maciejewski2021modeling,geller2021toward,fang2022crosstalk,tripathi2022suppression,perrin2024crosstalk,li2025crosstalkqec,hothem2025measuring,hawley2026crossbench,marciniak2026millisecond,magann2025fastfeedback}.
This point becomes even more important as quantum hardware moves from conventional NISQ operation toward early logical and partially error-corrected workflows. The same physical error channels that affect bare circuits also constrain mid-circuit measurement, active reset, feedforward control, syndrome extraction, and logical operations
\cite{ryananderson2021realtime,postler2022faulttolerant,google2025quantum,bar2026layered,koh2026readout,giortamis2026mcmit,shirizly2025dynamicrb,sramek2026observable}.
We therefore begin with the simplest control-level imperfections, namely single-qubit gate errors, before turning to multi-qubit control errors, decoherence and leakage, measurement and readout errors, crosstalk and correlated noise, and compilation-induced errors.

\subsubsection{Single-qubit control errors}

Single-qubit control errors arise when the physically implemented one-qubit gate deviates from the intended ideal operation because of imperfect calibration, pulse distortions, frequency drift, residual detuning, leakage, or other hardware instabilities
\cite{motzoi2009simple,gambetta2011analytic,sheldon2016characterizing,chen2016measuring,cerfontaine2020self,hyyppa2024reducing,hashim2025benchmarking,wesdorp2026mitigating}.
Although these errors are usually smaller than two-qubit gate errors on present-day hardware, they remain important for several reasons.
First, single-qubit gates appear throughout essentially all compiled circuits and therefore contribute systematically to the total control-error budget, especially in deep circuits, dynamically controlled circuits, circuits with extensive basis changes, and highly parallel workloads where simultaneous microwave or laser pulses can introduce additional crosstalk
\cite{sheldon2016characterizing,li2023single,brown2011single,harty2014high,ballance2016high,gaebler2016high,wesdorp2026mitigating}.
Second, single-qubit errors are not always simple stochastic over-rotations. They can include coherent amplitude and phase miscalibrations, drive-induced dephasing, microwave-pulse distortion, AC-Stark shifts, leakage to noncomputational levels, and context-dependent errors that depend on whether other qubits are driven simultaneously
\cite{motzoi2009simple,gambetta2011analytic,sheldon2016characterizing,hyyppa2024reducing,wesdorp2026mitigating}.
Third, in the transition from pre-fault-tolerant processors toward early logical or partially error-corrected hardware, accurate single-qubit control is required not only for bare physical-gate performance, but also for state preparation, syndrome extraction, dynamical decoupling, logical gate implementation, calibration-sensitive mitigation workflows, and high-fidelity feedforward-conditioned operations
\cite{ezzell2023dynamical,ryananderson2021realtime,egan2021fault,postler2022faulttolerant,acharya2023suppressing,google2025quantum,ryananderson2024teleportation}.

For an ideal rotation by angle $\theta$ about the Bloch-sphere axis $\hat{\bm n}$, the target operation is
\begin{equation}
U(\theta,\hat{\bm n})
=
\exp\left(
-i\frac{\theta}{2}\hat{\bm n}\cdot\bm{\sigma}
\right),
\end{equation}
where $\bm{\sigma}=(X,Y,Z)$ denotes the vector of Pauli operators. In practice, however, the realized gate $\widetilde U$ is only approximately equal to $U(\theta,\hat{\bm n})$.

A convenient way to describe coherent single-qubit control errors is to model the implemented gate as a slightly miscalibrated rotation,
\begin{equation}
\widetilde U
=
\exp\left[
-i\frac{\theta+\delta\theta}{2}
(\hat{\bm n}+\delta\hat{\bm n})\cdot\bm{\sigma}
\right],
\label{eq:1q_coherent_param}
\end{equation}
where $\delta\theta$ represents an over-rotation or under-rotation error, and $\delta\hat{\bm n}$ describes a small tilt of the rotation axis away from its intended direction. Physically, $\delta\theta$ may arise from an inaccurate pulse area or duration, while $\delta\hat{\bm n}$ can result from phase miscalibration, crosstalk, quadrature imbalance, or unwanted detuning. More precisely, the perturbed axis should be normalized; Eq.~\eqref{eq:1q_coherent_param} is therefore best understood as a leading-order parametrization for small coherent errors.

As a simple example, consider an intended $X$-axis rotation $R_X(\theta)$ implemented in the presence of a small detuning or phase mismatch, so that the actual rotation axis acquires an unwanted $Z$ component. A minimal perturbative model is
\begin{equation}
\widetilde R_X(\theta)
=
\exp\left[
-i\frac{\theta}{2}\left(X+\eta Z\right)
\right],
\label{eq:rx_detuning}
\end{equation}
where $\eta$ quantifies the strength of the coherent axis error. Physically, such a term may arise from quadrature imbalance, AC Stark shifts, or a residual frequency mismatch in the rotating frame. When $\eta\neq 0$, the gate is no longer a pure rotation about the $X$ axis, but instead rotates about a slightly tilted axis in the $XZ$ plane. By contrast, an amplitude miscalibration preserves the intended axis but changes the rotation angle. In that case one may write
\begin{equation}
\widetilde R_{\hat{\bm n}}(\theta)
=
R_{\hat{\bm n}}(\theta+\delta\theta),
\end{equation}
where $\delta\theta$ denotes the over-rotation or under-rotation error. To leading order, amplitude errors change the rotation angle, while detuning or phase errors change the rotation axis itself.

These errors can be addressed at several levels. The first layer is calibration and characterization: randomized benchmarking, gate-set calibration, and related diagnostic protocols estimate the size and structure of the control error, after which pulse amplitudes, pulse phases, pulse durations, and qubit frequencies can be retuned
\cite{sheldon2016characterizing,cerfontaine2020self,hashim2025benchmarking}.
A second layer is pulse-level error suppression. Pulse shaping, dynamically corrected gates, and leakage-suppression techniques aim to reduce the physical error before the circuit is executed
\cite{chen2016measuring,edmunds2020dynamically}.
A third layer is sequence-level suppression. Dynamical decoupling can reduce the effect of slow dephasing, residual detuning, and idle-time errors by inserting carefully chosen single-qubit pulses into otherwise idle periods
\cite{viola1998dynamical,ezzell2023dynamical}.
Finally, circuit-level noise tailoring can reduce the coherent accumulation of residual single-qubit errors. Randomized compiling, Pauli-frame randomization, and Pauli-conjugation methods insert classically tracked Pauli or Clifford transformations so that coherent control errors are converted, after averaging over random instances, into a more stochastic and easier-to-model effective noise channel
\cite{wallman2016noise,hashim2021randomized,ware2021pauliframe,cai2020pauliconjugation}.

\subsubsection{Two-qubit gate errors}

Two-qubit gate errors are often the dominant control-level error source on present-day quantum hardware, because entangling operations are physically more demanding and more sensitive to calibration imperfections than single-qubit gates
\cite{barends2014superconducting,kjaergaard2020superconducting,ganzhorn2020benchmarking,sung2021realization,ballance2016high,gaebler2016high,schafer2018fast,wei2024native,tannu2019not,marxer2026above}.
This point remains central even as the field moves beyond the earliest NISQ demonstrations and toward early logical or partially error-corrected processors. In current devices, two-qubit performance is still a primary bottleneck not only for deep physical-qubit circuits, but also for stabilizer measurement, syndrome extraction, logical-gate synthesis, lattice surgery, and other building blocks of fault-tolerant computation
\cite{barends2014superconducting,ryananderson2021realtime,postler2022faulttolerant,acharya2023suppressing,google2025quantum,paetznick2024demonstration,ryananderson2024teleportation}.
Understanding the structure of two-qubit errors is therefore essential both for near-term error mitigation and for the broader transition toward scalable fault-tolerant architectures. In superconducting devices, such errors include entangling-angle miscalibration, residual $ZZ$ coupling, flux-pulse distortion, leakage to noncomputational levels, parasitic coupler excitation, microwave or flux crosstalk, and context-dependent errors during simultaneous operations
\cite{wood2018quantification,chen2016measuring,yan2018tunable,sung2021realization,li2025highprecision,zhang2023tunablefluxonium,smirnov2025bipolar,echoCrossResonance2026}.
In trapped-ion processors, the corresponding limitations arise from motional-mode heating, optical phase and intensity noise, off-resonant excitation, spectator-mode coupling, residual spin--motion entanglement, and the calibration of multi-ion entangling pulses
\cite{molmer1999multiparticle,blatt2012quantum,ballance2016high,gaebler2016high,schafer2018fast,mehdi2025fastmixedspecies,liu2025paralleltrappedion}.
Because these error mechanisms are gate-, context-, and architecture-dependent, two-qubit errors are also a central target for noise-aware compilation, randomized compiling, Pauli-twirling-based noise tailoring, dynamical decoupling, and calibration-aware mitigation workflows
\cite{murali2019noise,hashim2021randomized,ezzell2023dynamical,tripathi2022suppression,perrin2024crosstalk,hashim2025benchmarking,li2025highprecision,marxer2026above}.

At the most general level, an intended two-qubit gate on qubits $(i,j)$ can be described by the unitary channel
\begin{equation}
\mathcal{U}_{ij}(\rho)
=
U_{ij}\rho U_{ij}^{\dagger},
\end{equation}
where $U_{ij}$ may be a native $CZ$, $ECR$, $XX(\phi)$, or other platform-dependent entangling primitive. The actually implemented operation is more accurately described as a noisy channel,
\begin{equation}
\widetilde{\mathcal{U}}_{ij}
=
\mathcal{N}_{ij}\circ\mathcal{U}_{ij},
\end{equation}
where $\mathcal{N}_{ij}$ represents the residual noise associated with the physical implementation of the gate. This noise can include coherent miscalibration, stochastic Pauli errors, leakage, amplitude damping, dephasing, spectator-qubit effects, and crosstalk. A useful starting point for analyzing the coherent component is to write the implemented gate as
\begin{equation}
\widetilde{U}_{ij}
=
U_{ij}E_{ij},
\qquad
E_{ij}
=
\exp(-iG_{ij}),
\qquad
\|G_{ij}\|\ll 1,
\label{eq:2q_unitary_error}
\end{equation}
where $E_{ij}$ represents a small residual unitary error. This convention places the coherent error after the ideal gate; equivalently, the error may be moved before the gate by conjugating the error generator.

A convenient parametrization expands $G_{ij}$ in the two-qubit Pauli basis,
\begin{equation}
G_{ij}
=
\frac{1}{2}
\sum_{P\in\mathcal{P}_2\setminus\{II\}}
\epsilon_P P,
\qquad
\mathcal{P}_2=\{I,X,Y,Z\}^{\otimes 2}.
\end{equation}
The identity component is omitted because it contributes only a global phase. This expansion is useful because it decomposes a complicated coherent control error into physically interpretable operator components. The local terms
\begin{equation}
XI,\quad YI,\quad ZI,\quad IX,\quad IY,\quad IZ
\end{equation}
describe single-qubit imperfections that occur during the entangling pulse, such as Stark shifts, phase miscalibration, quadrature imbalance, residual detuning, or imperfect frame tracking. Other terms describe errors in the entangling interaction itself. For instance, an intended $ZZ$ interaction may have the wrong strength, or it may acquire unwanted components such as $ZX$ or $XZ$. Additional two-body terms,
\begin{equation}
XY,\quad YX,\quad XZ,\quad ZX,\quad YZ,\quad ZY
\end{equation}
can arise from coherent crosstalk, residual drive terms, spectator-qubit effects, or imperfect cancellation of parasitic couplings. This operator-level view is valuable because different physical imperfections leave different signatures in the Pauli expansion and therefore call for different calibration or mitigation strategies.

As a concrete example, we can consider an intended Ising-type entangler,
\begin{equation}
U_{ZZ}(\phi)
=
\exp\!\left(
-i\,\frac{\phi}{2}\,Z\otimes Z
\right).
\end{equation}
A simple model for its imperfect implementation is
\begin{equation}
\widetilde{U}_{ZZ}
=
U_{ZZ}(\phi)
\exp\!\left[
-i\,\frac{\delta\phi}{2}\,Z\otimes Z
-i\sum_{P\neq ZZ}\frac{\epsilon_P}{2}\,P
\right],
\label{eq:zz_coherent_noH}
\end{equation}
where $\delta\phi$ represents an entangling-angle miscalibration and the remaining coefficients $\epsilon_P$ quantify parasitic coherent terms. The first correction preserves the intended interaction axis but changes its strength, while the other terms represent qualitatively different control imperfections. The same logic applies to other native entangling gates. Even when the ideal gate family differs across platforms, coherent over-rotation, axis misalignment, spectator-qubit coupling, leakage, and parasitic interactions remain among the most relevant physical error channels.

For small coherent errors, Eq.~\eqref{eq:2q_unitary_error} may be linearized as
\begin{equation}
E_{ij}
\approx
I-iG_{ij}.
\end{equation}
This expression makes clear that coherent two-qubit errors act as unwanted generators appended to the target entangling gate. Such errors are particularly harmful because they can accumulate systematically across a circuit rather than averaging away like purely stochastic noise. In a deep circuit, a small over-rotation or residual coupling may therefore produce a large bias in the measured observable. In early logical or partially error-corrected workflows, the same issue appears in a different form: repeated entangling operations are required for ancilla-data interactions, parity checks, syndrome extraction, and logical control routines. Thus, two-qubit coherent miscalibration can degrade not only raw physical-gate fidelity, but also the reliability of error-correction primitives.

Randomization-based methods provide a route when residual coherent errors cannot be fully removed by calibration [see Sec.~\ref{twirling}]. Pauli twirling, Pauli-frame randomization, Pauli-conjugation methods, and randomized compiling insert classically tracked Pauli or Clifford frame changes around entangling operations
\cite{wallman2016noise,hashim2021randomized,cai2020pauliconjugation,jain2023improved}.
After averaging over randomized circuit instances, coherent and structured miscalibrations can be transformed into a more stochastic effective noise channel. This transformation does not necessarily reduce the underlying physical error rate, but it can prevent coherent errors from adding constructively across the circuit and can make the residual noise easier to characterize, model, and mitigate.

Estimator-level mitigation may then be applied to the remaining effective noise when the required assumptions and overheads are acceptable. Zero-noise extrapolation estimates the zero-noise value of an observable by comparing related circuits executed at different effective noise strengths, while probabilistic error cancellation attempts to reconstruct the ideal operation from calibrated noisy operations using quasi-probability weights
\cite{temmeErrorMitigationShortDepth2017a,liEfficientVariationalQuantum2017a,endo2018practical,cai2023quantum}.
For two-qubit gates, however, these methods face an important practical limitation: noise amplification, quasi-probability overhead, and calibration cost typically grow rapidly with the number of entangling operations. This is why two-qubit errors remain a central concern throughout error mitigation. Improving, diagnosing, suppressing, and reshaping these errors is essential not only for better physical-qubit performance, but also for the reliable logical operations required in scalable fault-tolerant quantum computing.

\subsubsection{Time scales and decoherence}

Beyond coherent control imperfections, the performance of present-day quantum processors is ultimately limited by decoherence and by the total duration of the compiled experiment
\cite{preskill2018quantum,ithier2005decoherence,bylander2011noise,kjaergaard2020superconducting,krantz2019quantum,blais2021circuit,bruzewicz2019trapped,smith2022timestitch,das2021adapt}.
Even if each gate were perfectly calibrated, the quantum state remains exposed to environmental noise throughout the execution time, including idle periods, measurement windows, reset operations, and classically controlled feedforward.
This time-scale constraint remains central not only for conventional NISQ circuits, but also for the transition toward early fault-tolerant hardware: repeated stabilizer measurements, ancilla reuse, real-time feedforward, and syndrome-extraction cycles all make the accumulated exposure to relaxation, dephasing, leakage, and measurement-induced disturbance a central performance bottleneck
\cite{riste2015detecting,kelly2015state,krinner2022repeated,ryananderson2021realtime,postler2022faulttolerant,google2025quantum,hothem2025measuring,bar2026layered}.
It is therefore useful to describe the full computation as a noisy continuous-time evolution over a total exposure time $\tau$. If the ideal circuit implements a unitary $U$, then the realized output state may be modeled as
\begin{equation}
\widetilde{\rho}
=
\mathcal{E}_{\tau}\circ \mathcal{U}(\rho_{0}),
\qquad
\mathcal{U}(\rho)=U\rho U^{\dagger},
\label{eq:effective_noise_map}
\end{equation}
where $\mathcal{E}_{\tau}$ is an effective noise channel that collects the nonunitary effects accumulated during the circuit runtime
\cite{temmeErrorMitigationShortDepth2017a,endo2018practical,coote2025graphdd}.
In general, this map depends not only on the total duration $\tau$, but also on how the circuit is scheduled in time, including idle periods, pulse overlaps, measurement windows, reset steps, feedforward latency, qubit-reuse operations, dynamical-decoupling blocks, and the durations of individual gates
\cite{murali2019noise,ezzell2023dynamical,tripathi2022suppression,marciniak2026millisecond,magann2025fastfeedback,shirizly2025dynamicrb}.

In many platforms, especially superconducting processors, the dominant Markovian decoherence mechanisms are well approximated by energy relaxation and dephasing
\cite{ithier2005decoherence,burnett2019decoherence,carroll2022dynamics,
brand2024markovian,matityahu2019dynamical}.
These processes are conventionally characterized by the relaxation time $T_1$, the coherence time $T_2$, and the pure-dephasing time $T_{\varphi}$.

\begin{itemize}
\item \textbf{Energy relaxation and $T_1$.}
The time $T_1$ characterizes irreversible decay from the excited state $\ket{1}$ to the ground state $\ket{0}$. If the qubit is initialized in $\ket{1}$, the excited-state population decays approximately as
\begin{equation}
    \rho_{11}(t)
    =
    \rho_{11}(0)e^{-t/T_1}.
\end{equation}
Thus, $T_1$ is the characteristic time over which the qubit loses excitation energy to its environment. Relaxation also reduces phase coherence, because decay of the excited-state amplitude causes the off-diagonal density-matrix elements to decrease at a rate $1/(2T_1)$.
\item \textbf{Dephasing and $T_2$.}
The time $T_2$ characterizes the decay of coherence between $\ket{0}$ and $\ket{1}$. For an initial superposition state, the off-diagonal density-matrix element behaves approximately as
\begin{equation}
    \rho_{01}(t)
    =
    \rho_{01}(0)e^{-t/T_2}.
\end{equation}
This decay has two contributions. Energy relaxation already causes coherence loss at the rate $1/(2T_1)$, while fluctuations that randomize the relative phase without changing the populations produce additional pure dephasing characterized by $T_{\varphi}$.
\end{itemize}
For independent Markovian relaxation and pure-dephasing processes,
\begin{equation}
\frac{1}{T_2}
=
\frac{1}{2T_1}
+
\frac{1}{T_{\varphi}}.
\label{eq:t1_t2_relation}
\end{equation}
This implies $T_2\leq2T_1$, with $T_2=2T_1$ in the absence of additional pure dephasing. On current IBM superconducting processors, both $T_1$ and $T_2$ are typically of order $10^2~\mu{\rm s}$ and vary across qubits and calibration cycles. For example, values around $T_1\sim200~\mu{\rm s}$ and $T_2\sim100~\mu{\rm s}$ are representative of recent Heron-class devices.

For a single qubit undergoing relaxation and pure dephasing in the absence of active control, the reduced density matrix may be modeled by the Lindblad master equation \cite{gorini1976completely,lindblad1976generators}
\begin{equation}
\frac{d\rho}{dt}
=
\mathcal{L}(\rho)
=
\frac{1}{T_{1}}
\left(
\sigma_{-}\rho\sigma_{+}
-\frac{1}{2}\{\sigma_{+}\sigma_{-},\rho\}
\right)
+
\frac{1}{2T_{\varphi}}
\left(
Z\rho Z-\rho
\right),
\label{eq:lindblad_t1_tphi}
\end{equation}
with $\sigma_{-}=\ket{0}\!\bra{1}$ and $\sigma_{+}=\ket{1}\!\bra{0}$ \cite{ithier2005decoherence,burnett2019decoherence}. The corresponding completely positive trace-preserving channel after an exposure time $\tau$ is
\begin{equation}
\mathcal{E}_{\tau}
=
e^{\tau\mathcal{L}}.
\end{equation}
This expression makes explicit why circuit duration is so important: for fixed $T_{1}$ and $T_{2}$, reducing the total exposure time $\tau$ through better compilation, shorter schedules, improved routing, or lower measurement-and-feedback overhead directly weakens the accumulated decoherence. This is particularly important in connectivity-limited architectures, where additional routing operations such as inserted SWAP gates increase both circuit depth and total runtime, and in dynamic-circuit or early fault-tolerant workflows, where repeated measurement and control steps can substantially lengthen the effective execution window.

These time-scale considerations also clarify the limits of extrapolation-based mitigation methods \cite{heZeronoiseExtrapolationQuantumgate2020,krebsbachOptimizationRichardsonExtrapolation2022}. In zero-noise extrapolation (ZNE)  [see Sect.~\ref{sec:zne} below], one measures an observable at several amplified noise levels and fits the resulting dependence $\langle O\rangle(\lambda)$ to estimate the noiseless limit $\lambda\to 0$. The basic assumption is that noise amplification produces a smooth and controlled deformation of the observable, so that
\begin{equation}
\langle O\rangle(\lambda)
\approx
\langle O\rangle(0)
+
a_{1}\lambda
+
a_{2}\lambda^{2}
+\cdots
\end{equation}
over the range of sampled noise strengths. This perturbative picture is most reliable when the underlying circuit remains in a weak-noise regime. If decoherence becomes too strong, for example, because the amplified circuits are too long compared with $T_{1}$ and $T_{2}$, then the dependence on $\lambda$ need not remain smooth or low-order, and the extrapolation can become unstable or biased. For this reason, decoherence not only limits raw circuit fidelity, but also constrains the practical regime in which ZNE and related scaling-based mitigation methods can be expected to perform well.

More broadly, the importance of decoherence has acquired a slightly different interpretation in the latest hardware perspective.
In the earliest NISQ setting, decoherence was often discussed mainly as a limit on shallow physical-qubit circuits
\cite{preskill2018quantum,kjaergaard2020superconducting,blais2021circuit,bruzewicz2019trapped}.
In the current transition from pre-fault-tolerant to early fault-tolerant hardware, however, decoherence must also be understood as a constraint on the temporal structure of logical control itself.
Even when error correction is introduced, finite coherence times, measurement latency, reset time, and repeated syndrome-cycle duration continue to determine whether the effective logical workflow remains below threshold and whether mitigation or suppression techniques can still provide useful improvements
\cite{krinner2022repeated,ryananderson2021realtime,postler2022faulttolerant,google2025quantum,paetznick2024demonstration,hothem2025measuring,bar2026layered}.
Accordingly, time scales and decoherence are not merely background device parameters; they are central quantities governing circuit scheduling, compilation, mitigation, dynamic control, syndrome extraction, and the practical path toward scalable fault-tolerant quantum computation.

\subsubsection{Measurement and readout errors}

Measurement and readout errors, often grouped with state-preparation errors under the label SPAM, arise because the physical preparation-and-measurement layer implemented by the hardware is not identical to the ideal initialization and projective measurement assumed in the circuit model; see Sec.~\ref{sec:readout_models} for a more detailed discussion
\cite{greenbaum2015introduction,magesan2011scalable,chen2019detector,maciejewski2020mitigation,geller2021toward,nation2021scalable,funcke2022measurement,bravyi2021mitigating,van2022model,aasen2024readout}.
Equivalently, one may regard the measurement stage as an additional noise process acting immediately before an otherwise ideal computational-basis measurement, so that the recorded classical outcomes are sampled from a distorted probability distribution.
This detector-level distortion is commonly modeled by an assignment matrix or an effective noisy POVM, but realistic devices can also exhibit state-dependent bias, measurement crosstalk, spectator-dependent correlations, temporal drift, and relaxation during the measurement window
\cite{chen2019detector,geller2021toward,bravyi2021mitigating,maciejewski2021modeling,koh2026readout,hothem2025measuring}.
State-preparation errors introduce a related bias by preparing an input state that differs from the intended one, often $\ket{0}$ or $\ket{1}$, and can contaminate detector calibration if preparation and measurement errors are not carefully separated
\cite{greenbaum2015introduction,magesan2011scalable,geller2021toward}.
In many gate-based experiments, however, the dominant SPAM contribution often comes from readout rather than initialization, which is why readout-error mitigation has become one of the most mature and widely used layers of near-term error mitigation
\cite{maciejewski2020mitigation,nation2021scalable,bravyi2021mitigating,funcke2022measurement,van2022model,aasen2024readout,koh2026readout}.
This distinction remains important not only in conventional NISQ experiments but also in the transition toward early fault-tolerant hardware, because repeated measurement, qubit reset, mid-circuit feedforward, and syndrome extraction make measurement performance a central bottleneck for both physical-qubit mitigation and logical-control workflows
\cite{riste2015detecting,kelly2015state,krinner2022repeated,ryananderson2021realtime,postler2022faulttolerant,google2025quantum,hothem2025measuring,giortamis2026mcmit,chuLearningMidCircuit2026,bar2026layered}.

For most gate-based platforms, readout is implemented in the computational (generally $Z$) basis. The measured outcome statistics, therefore, reflect not only the underlying quantum state, but also imperfections in the measurement apparatus itself. A standard and useful model is to treat readout error as a classical confusion process acting on the ideal outcome distribution. Let $p_{\rm id}(0)$ and $p_{\rm id}(1)$ denote the ideal probabilities of obtaining outcomes $0$ and $1$, and let $p_{\rm exp}(0)$ and $p_{\rm exp}(1)$ be the experimentally observed probabilities. Then one writes
\begin{equation}
\begin{pmatrix}
p_{\rm exp}(0)\\[2pt]
p_{\rm exp}(1)
\end{pmatrix}
=
A
\begin{pmatrix}
p_{\rm id}(0)\\[2pt]
p_{\rm id}(1)
\end{pmatrix},
\qquad
A=
\begin{pmatrix}
1-\epsilon_{0\to 1} & \epsilon_{1\to 0}\\
\epsilon_{0\to 1} & 1-\epsilon_{1\to 0}
\end{pmatrix},
\label{eq:single_qubit_assignment}
\end{equation}
where $\epsilon_{0\to 1}$ is the probability of reporting outcome $1$ when the true outcome is $0$, and $\epsilon_{1\to 0}$ is the probability of reporting outcome $0$ when the true outcome is $1$. The matrix $A$ is often called the assignment matrix or confusion matrix, because it quantifies how ideal outcomes are reassigned by the measurement process \cite{chen2019detector,nation2021scalable}.

This model is especially useful because it shows that readout error acts after the quantum evolution and therefore distorts observed classical statistics without changing the underlying quantum state itself. In the symmetric case,
\begin{equation}
\epsilon_{0\to 1}
=
\epsilon_{1\to 0}
=
\epsilon,
\end{equation}
the assignment matrix reduces to a simple binary symmetric channel, and the measured expectation value of $Z$ is attenuated according to
\begin{equation}
\langle Z\rangle_{\rm exp}
=
(1-2\epsilon)\,\langle Z\rangle_{\rm id}.
\label{eq:z_attenuation_symmetric}
\end{equation}
Thus, even when the quantum state is prepared and evolved perfectly, imperfect discrimination between outcomes $0$ and $1$ suppresses the observed signal.

More generally, for a single qubit, one may invert Eq.~\ref{eq:single_qubit_assignment}, provided the matrix $A$ is well conditioned, to estimate the ideal probabilities from the experimental data:
\begin{equation}
\begin{pmatrix}
p_{\rm id}(0)\\[2pt]
p_{\rm id}(1)
\end{pmatrix}
=
A^{-1}
\begin{pmatrix}
p_{\rm exp}(0)\\[2pt]
p_{\rm exp}(1)
\end{pmatrix}.
\end{equation}
This is the basic idea behind measurement-error mitigation by calibration inversion. In practice, however, the situation becomes more complicated for multi-qubit measurements.

In realistic devices, readout errors need not factorize independently across qubits. Measurement crosstalk can produce correlated assignment errors, so that the probability of misidentifying one qubit may depend on the states of neighboring qubits or on the simultaneous measurement of the full register \cite{sarovar2020detecting,nation2021scalable}. In such cases, a tensor-product model built from independent single-qubit calibration matrices can become inaccurate, especially for nonlocal observables and many-body correlators. In addition, readout errors may drift over time as hardware conditions change across calibration cycles. These correlated and time-dependent effects motivate scalable, adaptive, and dynamically updated calibration procedures for reliable readout mitigation on larger quantum processors.

From the latest fault-tolerant perspective, measurement and readout errors should therefore not be viewed merely as a peripheral SPAM contribution to shallow physical-qubit circuits. They are increasingly tied to the viability of dynamic circuits, repeated syndrome extraction, ancilla reset, and logical-state readout. Even when logical encoding is introduced, imperfect measurement can remain a dominant source of overhead, latency, and residual logical error. For this reason, readout characterization and mitigation remain relevant not only in pre-fault-tolerant experiments, but also in the broader pathway toward scalable fault-tolerant quantum computation.

\subsubsection{Crosstalk and correlated errors}

A central limitation of many error-mitigation protocols is the assumption that the effective noise is approximately local and only weakly correlated across qubits and circuit layers.
Actual quantum devices often violate this assumption.
In practice, operations applied to one qubit can disturb nearby qubits through crosstalk, and the resulting errors can be correlated across space, time, and circuit context rather than acting independently on each gate
\cite{sarovar2020detecting,seif2024suppressing,edmunds2020dynamically,fang2022crosstalk,clader2021impact}.
This issue is important not only for conventional NISQ workloads, but also in the transition toward early logical and partially error-corrected hardware, where correlated errors can directly degrade repeated syndrome extraction, parallel stabilizer measurements, decoder assumptions, and logical-error rates
\cite{krinner2022repeated,google2025quantum,paetznick2024demonstration}.

At the process level, crosstalk may be understood as a dependence of the implemented operation on the surrounding circuit context.
Suppose that a circuit is scheduled into parallel layers labeled by $m$, and let $\mathcal{G}_m$ denote the set of operations executed concurrently in layer $m$.
If $\mathcal{G}_g$ denotes the ideal channel associated with a particular gate $g\in\mathcal{G}_m$, then the corresponding implemented operation may be modeled schematically as
\begin{equation}
\widetilde{\mathcal{G}}_g^{(m)}
=
\mathcal{N}_{g|\mathcal{G}_m}^{(m)}
\circ
\mathcal{G}_g .
\label{eq:crosstalk_schedule_dep}
\end{equation}
Here $\mathcal{N}_{g|\mathcal{G}_m}^{(m)}$ denotes an effective noise channel whose form can depend on the full set of operations executed in the same layer.
This expression emphasizes that the error affecting a given gate is not necessarily an intrinsic property of that gate alone, but can depend on the surrounding schedule, spectator qubits, simultaneous measurements, idle periods, and calibration context.

Spatial correlations can be diagnosed using multi-qubit observables. For example, consider the length-$\ell$ parity string
\begin{equation}
P_{\ell}
=
Z_{1}Z_{2}\cdots Z_{\ell}.
\label{eq:string_parity}
\end{equation}
If readout errors are independent across qubits, and the reported outcome of qubit $j$ is flipped with probability $\epsilon_j$, then the measured expectation value satisfies
\begin{equation}
\mathbb{E}[\widehat{P}_{\ell}]
=
\left[
\prod_{j=1}^{\ell}(1-2\epsilon_j)
\right]
\langle P_{\ell}\rangle_{\rm id}.
\label{eq:string_independent_readout}
\end{equation}
Under this independent-error model, each qubit contributes a separate attenuation factor $(1-2\epsilon_j)$.
Deviations from the product form in Eq.~\eqref{eq:string_independent_readout} therefore provide a simple signature of correlated readout errors or measurement crosstalk.
More generally, if calibration data obtained from single-qubit or low-weight measurements fail to predict the behavior of longer Pauli strings, this often indicates that the underlying noise cannot be treated as a tensor product of independent single-qubit channels.

Temporal correlations arise when errors persist across circuit layers instead of being independently redrawn at each operation.
A simple illustration is provided by a repeated coherent over-rotation.
Prepare the state $|+\rangle$, apply $L$ nominally identical $Z$ rotations, and then measure an equatorial observable:
\begin{equation}
|+\rangle
\;\xrightarrow{\;R_Z(\theta)\;}\;
\cdots
\;\xrightarrow{\;R_Z(\theta)\;}\;
\text{measure }X\text{ or }Y .
\label{eq:repeat_gate_protocol}
\end{equation}
If each intended gate $R_Z(\theta)$ is implemented as $R_Z(\theta+\delta\theta)$ with the same systematic over-rotation error $\delta\theta$ at every layer, then the error accumulates coherently.
After $L$ repetitions, the net angle error is $L\,\delta\theta$.
For an ideal final state with equatorial expectations $\langle X\rangle_{\rm id}$ and $\langle Y\rangle_{\rm id}$, the measured $X$ expectation is shifted to
\begin{equation}
\langle X\rangle_{\rm exp}
=
\cos\!\left(L\delta\theta\right)
\langle X\rangle_{\rm id}
-
\sin\!\left(L\delta\theta\right)
\langle Y\rangle_{\rm id}.
\label{eq:coherent_accumulation}
\end{equation}
This relation shows that even a very small per-layer miscalibration can become significant when repeated many times.
In other words, coherent errors are especially dangerous because they build up linearly in rotation angle, leading to order-one deviations in sufficiently deep circuits.

These examples illustrate why correlated noise is challenging for error mitigation.
Many standard protocols, such as local readout correction, probabilistic error models, and low-order extrapolation methods, rely implicitly on some degree of spatial or temporal independence.
When crosstalk or long-range correlations are strong, these approximations can break down, and the mitigation strategy must include more of the physical context of the experiment.

In practice, correlated errors can be reduced or mitigated by combining hardware characterization, schedule-aware compilation, and correlation-aware inference.
Crosstalk diagnostics and simultaneous benchmarking protocols can identify which qubits, gates, or measurement operations interfere with one another
\cite{sarovar2020detecting,hashim2025benchmarking}.
Once these correlations are known, the compiler can avoid particularly harmful simultaneous operations, insert idle-time protection such as dynamical decoupling, or choose schedules that reduce spectator-qubit and residual-coupling effects
\cite{seif2024suppressing,edmunds2020dynamically}.
For measurement crosstalk, readout mitigation can be extended beyond independent single-qubit calibration by using local correlated assignment matrices or reduced models that include the dominant multi-qubit correlations.
For coherent crosstalk, randomized compiling, Pauli twirling, and related noise-tailoring methods can help convert structured coherent errors into more stochastic effective noise, making the residual error easier to model and less likely to accumulate constructively across layers [see Sec.~\ref{twirling}].
Thus, the main practical lesson is that correlated noise cannot usually be mitigated by applying a fixed single-qubit correction independently to each qubit.
Effective mitigation must account for where the qubits are placed, which gates are executed together, how measurements and resets are scheduled, and how the noise changes over time.
  
\subsubsection{Compilation- and routing-induced errors}

On cloud-accessible quantum hardware, a circuit written at the algorithmic level usually cannot be executed directly. Instead, it must first be compiled into the device's native gate set and mapped onto the hardware connectivity graph
\cite{sivarajah2021t,javadi2024quantum,qiskit_transpiler,yan2024quantum,zhu2025compilerdesign}.
This compilation step can itself become a major source of performance loss: even when the native gates are individually well calibrated, the compiled circuit may contain more gates, deeper entangling layers, and a longer execution time than the original logical circuit
\cite{li2019tackling,murali2019noise,wagner2025optimized,molavi2025generating,escofet2024routeforcing}. This point has become even more important in the latest hardware landscape. In the earlier NISQ setting, compilation overhead was often discussed mainly as a limitation on shallow physical-qubit circuits. In the current transition toward early logical and partially error-corrected processors, however, compilation and scheduling also affect repeated syndrome-extraction circuits and dynamic-circuit control flow
\cite{krinner2022repeated,google2025quantum,ryananderson2024teleportation,bar2026layered,optimalSyndromeExtraction2026}.

A principal source of this overhead is limited qubit connectivity. If two qubits that must interact are not directly connected on the processor, the compiler must introduce additional routing operations, typically through SWAP insertion, qubit remapping, bridge constructions, teleportation-like movement primitives, or equivalent circuit rewritings. Because a SWAP operation is usually decomposed into several native two-qubit entangling gates, routing can substantially increase the number of error-prone operations and amplify the total error budget
\cite{li2019tackling,sivarajah2021t,javadi2024quantum,qiskit_transpiler,zhu2025compilerdesign,molavi2025generating,escofet2024routeforcing}.

Routing and scheduling can also extend the total execution time, leaving some qubits idle for longer intervals and increasing their exposure to relaxation and dephasing. If the total circuit duration is denoted by $\tau$, then the effective noise channel $\mathcal{E}_{\tau}$ generally becomes stronger as $\tau$ increases, even when the target logical unitary is unchanged. Compilation overhead therefore degrades performance through two related mechanisms: gate-count inflation and time-dependent decoherence. This is why schedule-aware methods, such as slack-aware instruction scheduling, adaptive dynamical decoupling, coherence-aware mapping, and calibration-aware circuit rewriting, can improve circuit performance by reducing the effect of idle-time errors without changing the intended computation
\cite{smith2022timestitch,das2021adapt,ezzell2023dynamical,tripathi2022suppression,huang2025tram}.

Moreover, because hardware errors can depend on the detailed schedule of simultaneous operations, two different compiled circuits that realize the same logical unitary need not perform equally well on real devices. Crosstalk, spectator-qubit effects, calibration drift, and schedule-dependent control errors can all depend on which gates are executed concurrently and where they are placed on the chip. Consequently, compilation is not merely a formal translation from abstract gates to hardware instructions, but a hardware-sensitive transformation that can significantly modify the observed error profile; see Sec.~\ref{transpilation} below
\cite{murali2019noise,sarovar2020detecting,fang2022crosstalk,wagner2025optimized,perrin2024crosstalk,hashim2025benchmarking,hawley2026crossbench}.

Within the framework reviewed in this work, compilation-induced errors should be addressed primarily before execution through hardware-aware circuit design and transpilation. Observable- and ansatz-aware circuit construction can reduce the required interaction range, circuit depth, and number of entangling gates before routing is performed [see Sec.~\ref{ansatz}]. Noise- and calibration-aware layout selection can map the most important logical interactions onto well-connected physical qubits with comparatively low gate and readout errors, while routing and scheduling passes can minimize SWAP insertion, entangling depth, idle duration, and harmful parallel operations [see Sec.~\ref{transpilation}]. 

Residual errors introduced by the compiled implementation can then be treated using the other mitigation methods discussed in this review. Dynamical decoupling can suppress dephasing accumulated during unavoidable idle periods, whereas randomized compiling and Pauli or Clifford twirling can reduce the coherent and context-dependent accumulation of native-gate errors [see Sec.~\ref{twirling}]. ZNE may be applied to the compiled circuit by amplifying its native-gate noise through folding or pulse stretching [see Sec.~\ref{sec:zne}], but the scaling procedure should preserve the physical layout and routing pattern; otherwise, recompilation may generate a different noise channel rather than a controlled amplification of the original one. PEC similarly requires characterization and inversion of the effective native-gate channels appearing after compilation, rather than only the abstract logical gates [see Sec.~\ref{pec}]. Symmetry and constraint verification can remove routing- or crosstalk-induced faults that transfer the state outside a known physical or logical sector [see Sec.~\ref{verification}].

\subsubsection{Effective benchmarking of error effects}

Hardware parameters such as $T_1$, $T_2$, readout error, and isolated gate
fidelity characterize individual error mechanisms, but they do not fully
describe how errors accumulate when many operations are executed together.
This distinction is important for large circuits, where simultaneous gates,
idle periods, synchronization, and crosstalk can modify the effective error
seen by the computation. IBM Quantum introduced \emph{layer fidelity} (LF) as
a scalable benchmark designed to probe this circuit-level error
\cite{mckay2023benchmarking}.

\begin{figure}
    \centering
    \includegraphics[width=0.8\linewidth]{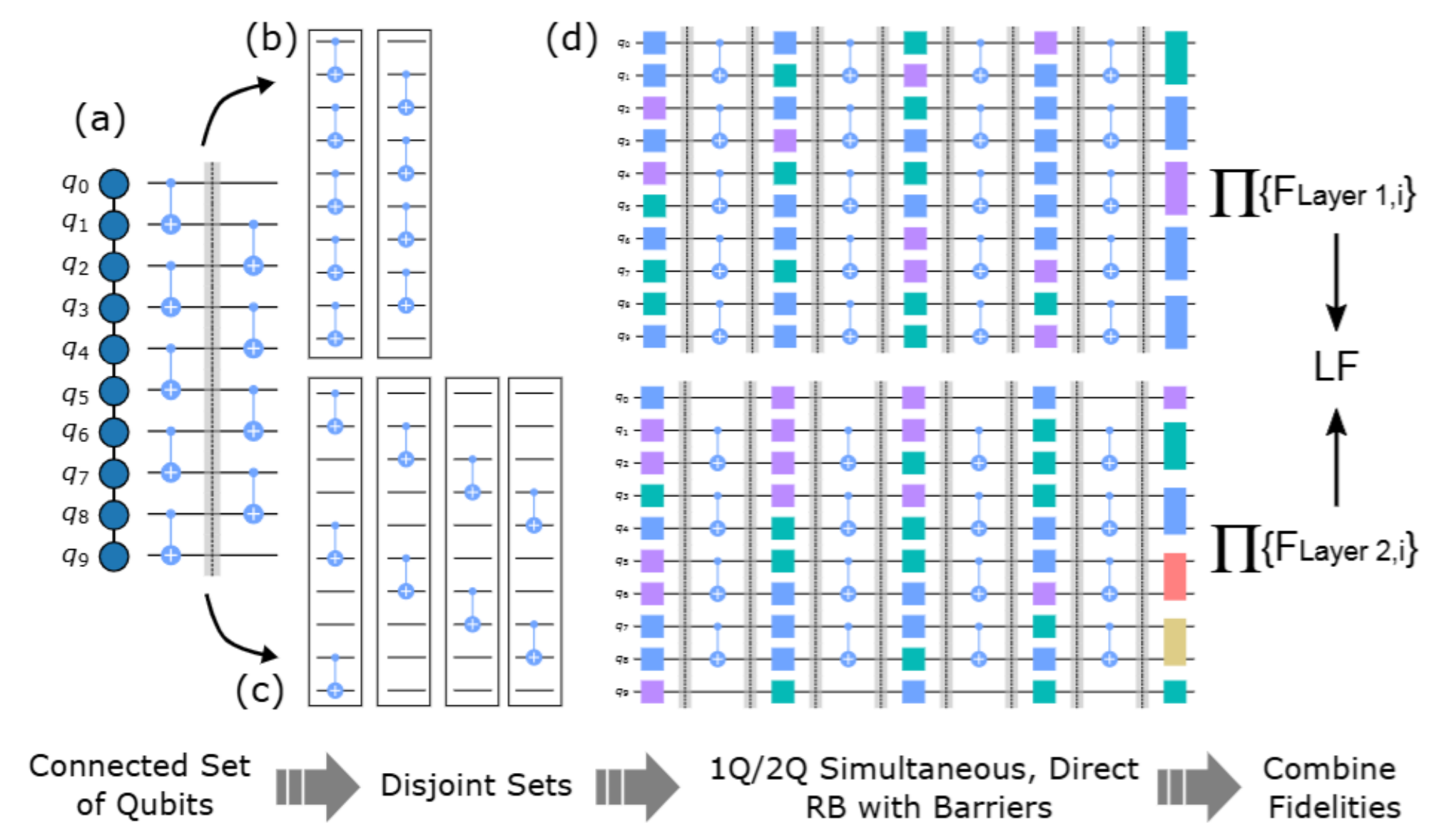}
    \caption{{Layer-fidelity benchmarking for large quantum processors.}
    (a) A connected set of qubits and two-qubit interactions is selected.
    (b) The interactions can be partitioned into maximally parallel disjoint
    layers, or (c) into more sparse layers with fewer simultaneous gates.
    (d) Each disjoint layer is characterized using simultaneous direct
    randomized benchmarking with timing barriers. The process fidelities of
    the individual one- and two-qubit subsystems are combined to obtain the
    fidelity of each disjoint layer, and the resulting layer fidelities are
    multiplied to give the full layer fidelity (LF).
    Adapted from Ref.~\cite{mckay2023benchmarking}.}
    \label{fig:bench}
\end{figure}

Fig.~\ref{fig:bench} illustrates the protocol. As shown in panel~(a), one
first selects a connected set of $N$ qubits and the corresponding set of
two-qubit interactions. Because gates that share a qubit cannot generally be
executed simultaneously, these interactions are divided into disjoint layers.
For a one-dimensional chain, the nearest-neighbor gates can be covered by two
alternating layers. Panel~(b) shows this maximally parallel decomposition,
whereas panel~(c) illustrates a more sparse decomposition into additional
layers. The latter can reduce simultaneous-gate interactions, but increases
the total execution time and hence exposure to decoherence.

As illustrated in Fig.~\ref{fig:bench}(d), each disjoint layer is characterized
using \emph{simultaneous direct randomized benchmarking}. Random Clifford
sequences of increasing length are applied to the active two-qubit pairs and
idle qubits, with barriers enforcing the timing of the complete layer. From
the fitted randomized-benchmarking decay parameter $\alpha_i$, the process
fidelity of subsystem $i$ is obtained as
\begin{equation}
    F_i
    =
    \frac{1+(d_i^2-1)\alpha_i}{d_i^2},
\end{equation}
where $d_i=2$ for a single-qubit subsystem and $d_i=4$ for a two-qubit
subsystem. The fidelity of disjoint layer $m$ is then
\begin{equation}
    \mathrm{LF}_m
    =
    \prod_i F_{i,m},
\end{equation}
and the full layer fidelity is estimated as
\begin{equation}
    \mathrm{LF}
    =
    \prod_m \mathrm{LF}_m .
    \label{eq:layer_fidelity}
\end{equation}

Because $\mathrm{LF}$ decreases as more two-qubit interactions are included, it
is useful to convert it into a size-normalized quantity. If the complete
connected structure contains $n_{\rm 2Q}$ two-qubit gates, the original
definition of the error per layered gate (EPLG) is
\begin{equation}
    \mathrm{EPLG}_{\rm proc}
    =
    1-\mathrm{LF}^{1/n_{\rm 2Q}} .
    \label{eq:eplg_process}
\end{equation}
For a linear chain of $N$ qubits, $n_{\rm 2Q}=N-1$. Current IBM Quantum
backend reporting instead expresses EPLG using the average two-qubit gate-error
convention,
\begin{equation}
    \mathrm{EPLG}
    =
    \frac{4}{5}
    \left(
        1-\mathrm{LF}^{1/n_{\rm 2Q}}
    \right),
    \label{eq:eplg}
\end{equation}
where the factor $4/5$ converts the two-qubit process error to an average gate
error.

The main advantage of LF is that it measures gates under \emph{simultaneous
layered execution}. It is therefore sensitive to crosstalk and timing effects
that may be absent from isolated randomized benchmarking
\cite{mckay2023benchmarking}. Two processors, or two regions of the same
processor, can have similar isolated two-qubit gate errors but substantially
different layer fidelities because their errors behave differently when many
operations are executed in parallel. This makes LF particularly relevant to error mitigation. It provides a compact
measure of the effective noise experienced by the layered circuits. Moreover, for high-fidelity
Pauli-like noise, LF can be related to the sampling overhead of probabilistic
error cancellation, so a reduction in layer fidelity also signals an increase
in the resources required for mitigation
\cite{mckay2023benchmarking,vandenberg2023sparsepec}.

\subsection{Error mitigation and suppression methods}

The methods reviewed here address different stages of the noisy computational workflow and are therefore most naturally organized according to their operational roles. Although these approaches rely on distinct principles, they are often combined in practice because they target different components of the total error. Following the organization of the subsequent sections, we summarize the main classes as follows.

\begin{itemize}

\item \textbf{Observable- and ansatz-level error suppression} [Sec.~\ref{ansatz}].
These approaches reduce sensitivity to noise by adapting what is measured or how the target state and evolution are implemented. Observable selection favors quantities that remain informative under realistic noise, while error-aware transpilation, circuit rewriting, and depth reduction seek shallower and more hardware-compatible implementations. Rather than reconstructing an ideal result from noisy data, these methods reduce the error accumulated during execution.

\item \textbf{Measurement-error mitigation} [Sec.~\ref{read}].
Measurement-error mitigation characterizes the imperfect readout process through calibrated detector models and uses this information to correct measured probability distributions or expectation values. It is among the most mature mitigation techniques on current hardware, although its scalability is limited by calibration cost, correlated readout errors, detector drift, and instability of the associated inverse problem.

\item \textbf{Error mitigation through reduced sampling cost} [Sec.~\ref{sec:sampling}].
Sampling-efficient methods aim to reduce the experimental resources required to estimate observables with a desired precision. Classical-shadow and randomized-measurement protocols, for example, allow information about many observables to be extracted from a common measurement ensemble. These methods do not necessarily reduce the physical noise itself, but can make noisy experiments more practical by improving the efficiency with which useful information is extracted from finite measurement budgets.

\item \textbf{Noise scaling and extrapolation} [Sec.~\ref{sec:zne}].
Zero-noise extrapolation (ZNE) estimates ideal observables from a family of logically equivalent circuits executed at different effective noise strengths. It avoids full microscopic reconstruction of the device noise, but relies on a sufficiently controlled noise-scaling procedure and a stable extrapolation model. Its practical performance is therefore governed by the interplay between extrapolation bias and the increased statistical uncertainty associated with noise amplification.

\item \textbf{Inverse-channel error mitigation} [Sec.~\ref{pec}].
Probabilistic error cancellation (PEC) and related inverse-channel methods seek to reconstruct the inverse of an effective noise process using quasiprobability combinations of experimentally implementable operations. When the noise model is sufficiently accurate, PEC can provide an unbiased estimator of the ideal observable. This stronger guarantee comes at the cost of detailed noise characterization and a sampling overhead that can grow rapidly with circuit depth and noise strength.

\item \textbf{Circuit randomization, twirling, and coherent-error suppression} [Sec.~\ref{twirling}].
Randomized compiling, Pauli or Clifford twirling, and dynamical decoupling act directly on the structure of the effective noise. Randomization and twirling convert coherent, circuit-dependent errors into more stochastic and homogeneous effective channels, whereas dynamical decoupling suppresses slowly varying coherent evolution, particularly during idle periods. These techniques are therefore better viewed as noise-engineering or error-suppression layers that can also improve the performance of downstream mitigation methods.

\item \textbf{Symmetry- and constraint-based mitigation} [Sec.~\ref{verification}].
These approaches exploit prior knowledge that the ideal computation remains within a known physical or logical sector. Measurements can be post-selected, projected, or reweighted according to conserved quantities, parity conditions, stabilizer relations, or gauge constraints. Such methods can efficiently remove symmetry-violating errors without requiring a complete device-noise model, but symmetry-preserving errors remain undetected and the sampling cost increases as the probability of remaining in the target sector decreases.

\end{itemize}

The current experimental landscape spans from few-qubit and intermediate-scale
demonstrations to large-scale circuits. Across the largest reported demonstrations, the relevant scales therefore range from tens to more than one hundred qubits, from several to tens of entangling layers, and from hundreds of active entangling operations to several thousand two-qubit gates.   A representative large-scale example is ZNE combined with Pauli-noise tailoring, which has been applied to a 127-qubit superconducting processor using circuits with up to 60 layers of two-qubit gates and a total of 2,880 CNOT gates \cite{kimEvidenceUtilityQuantum2023c}. Calibration-based measurement-error mitigation has likewise been demonstrated in the preparation and characterization of 127-qubit GHZ states \cite{pokharel2024scalable}. For PEC based on sparse Pauli--Lindblad noise models, the noise-characterization stage was demonstrated for a 20-qubit layer containing ten simultaneously executed CX gates, while the corresponding error-mitigated Ising simulations were performed on systems of up to 10 qubits and seven Trotter steps \cite{vandenberg2023sparsepec}. More recently, a preprint reported unbiased quasiprobabilistic mitigation for a 103-qubit kicked-Ising simulation with seven evolution steps, corresponding to 21 programmed two-qubit layers and an observable-dependent active volume of 301 entangling gates \cite{aharonovReliableHighAccuracy2025}. It is important to note that these examples depend strongly on the mitigation principle, circuit geometry, and target observable.

We emphasize that, in practice, the most reliable strategy is often a layered
workflow rather than the isolated use of a single protocol. Different software
components can target different contributions to the hardware error budget:
hardware-aware compilation reduces routing and mapping overhead, error-suppression
tools reduce selected errors during circuit execution, and error-mitigation
methods reduce residual biases in measured observables or output statistics.
These software stacks increasingly combine these functions with calibration
data and classical post-processing into reproducible workflows
[see Sec.~\ref{software}]:

\begin{itemize}

\item \textbf{Routing overhead and mapping-induced errors --- Qiskit and TKET.}
Qiskit and TKET provide hardware-aware compilation and circuit-optimization
tools that map logical qubits and interactions onto the native device topology
while reducing unnecessary routing operations
\cite{javadi2024quantum,sivarajah2021t}.
Noise-aware placement can additionally favor qubits and couplers with better
calibrated performance, while improved routing reduces additional SWAP gates
and the associated accumulation of two-qubit-gate errors
\cite{murali2019noise}.

\item \textbf{Idle-time coherent errors --- Qiskit Runtime and Quantinuum.}
IBM's Qiskit Runtime stack and Quantinuum System Model H2 provide
dynamical-decoupling capabilities that insert refocusing pulses into idle
periods. These sequences primarily suppress coherent errors accumulated while
qubits are idle, including slowly varying frequency offsets and unwanted
interactions, rather than general irreversible relaxation
\cite{qiskitdocs_techniques_2026,quantinuum_dd_2026}.

\item \textbf{Coherent gate errors --- Qiskit Runtime.}
Qiskit Runtime provides Pauli gate twirling, in which selected gates are
surrounded by randomized Pauli operations while preserving the ideal circuit
action. Averaging over the randomized circuit ensemble converts coherent error
components into a more stochastic Pauli-like effective noise channel, reducing
their systematic accumulation and making the resulting noise more suitable for
subsequent mitigation
\cite{qiskitdocs_techniques_2026,wallman2016noise}.

\item \textbf{Measurement and SPAM errors --- Qiskit Runtime and Qermit.}
Qiskit Runtime implements Twirled Readout Error Extinction (TREX), which
combines randomized measurement twirling with calibration and classical
rescaling to reduce readout-induced bias in Pauli expectation values
\cite{qiskitdocs_techniques_2026}.
Qermit provides complementary correlated and uncorrelated SPAM-mitigation
routines within its modular task-graph framework
\cite{cirstoiu2023qermit,qermitdocs2026}.

\item \textbf{Residual circuit bias --- Qiskit Runtime.}
After error suppression and readout correction, residual bias in expectation
values can be further reduced using ZNE. Qiskit Runtime supports noise
amplification through digital gate folding followed by extrapolation toward the
zero-noise limit. ZNE can be incorporated in the same execution workflow as
dynamical decoupling, gate twirling, and TREX
\cite{qiskitdocs_techniques_2026}.

\item \textbf{Aggregate circuit errors --- Qedma QESEM.}
Qedma's Quantum Error Suppression and Error Mitigation (QESEM) combines
circuit-specific device characterization, noise-aware transpilation, gate
optimization, error suppression, and unbiased quasi-probabilistic error
mitigation. Rather than targeting a single physical error mechanism, QESEM
constructs a characterization-informed workflow for reducing the aggregate
error affecting expectation-value estimation
\cite{aharonovReliableHighAccuracy2025,ibm_qedma_qesem_2026}.

\item \textbf{Structured circuit noise --- Algorithmiq TEM.}
Algorithmiq's Tensor-Network Error Mitigation (TEM) constructs a tensor-network
representation of the inverse effective noise channel and applies it in
classical post-processing to informationally complete measurement data
\cite{filippov2023tem,ibm2026tem}.
In its Qiskit Function implementation, the noise affecting circuit layers is
learned using a sparse Pauli--Lindblad model that can incorporate structured
effects including qubit crosstalk. TEM is particularly effective when the
resulting inverse noise map remains efficiently representable and contractible
as a tensor network
\cite{ibm2026tem}.

\item \textbf{Cross-stack error suppression --- Q-CTRL Fire Opal.}
Q-CTRL's Fire Opal provides an automated error-suppression workflow combining
error-aware compilation, system-wide gate optimization, dynamical-decoupling
embedding, and measurement-error mitigation. It therefore acts across several
layers of the execution stack, targeting mapping and gate errors, coherent
errors accumulated during idle periods, and measurement errors without relying
primarily on probabilistic post-processing
\cite{mundada2023fireopal}.

\item \textbf{Leakage detection and erasure conversion --- Quantinuum Helios.}
Quantinuum Helios provides heralded leakage measurement that distinguishes
population outside the computational subspace from the logical $|0\rangle$ and
$|1\rangle$ outcomes. The resulting leakage flag converts an otherwise
unheralded leakage event into a detectable erasure-type event that can be
discarded or processed explicitly by higher-level protocols
\cite{quantinuum_leakage_2026}.

\item \textbf{Multi-method mitigation workflows --- Mitiq and Qermit.}
Mitiq and Qermit do not target a single hardware error mechanism, but instead
provide modular interfaces for constructing mitigation workflows. Mitiq
supports methods including ZNE, probabilistic error cancellation, and Clifford
data regression across different software frameworks and backends
\cite{laroseMitiqSoftwarePackage2022a}.
Qermit provides a composable graph-based framework containing modules for SPAM
mitigation, frame randomization, ZNE, probabilistic error cancellation,
post-selection, and other mitigation protocols
\cite{cirstoiu2023qermit,qermitdocs2026}.

\end{itemize}

\section{Observable-level and ansatz-level error suppression}
\label{ansatz}

\subsection{Motivation}

Error suppression is commonly discussed in terms of hardware control, pulse engineering, and circuit compilation. However, substantial gains can also be obtained at a higher algorithmic level by exploiting structure in the quantities being measured and in the circuits used to prepare the relevant states. Two complementary strategies are especially important:

\begin{itemize}

\item \textbf{Observable-level error suppression.}
Most quantum algorithms do not require reconstruction of the full quantum state. Instead, the final objective is typically a limited set of quantities such as energies, correlation functions, order parameters, conserved charges, fidelities, or witness operators
\cite{mcclean2016theory,cerezo2021variational,huggins2021efficient,gonthier2022measurements}.
If only these observables are required, it is unnecessary to suppress errors uniformly over the entire Hilbert space. One can instead design the measurement procedure so that the relevant physical information is extracted as efficiently and robustly as possible.

This viewpoint underlies Pauli-grouping strategies, classical shadows, randomized and derandomized measurements, observable-level readout mitigation, and symmetry-aware estimators
\cite{verteletskyi2020measurement,crawford2021efficient,huggins2021efficient,
huang2020predicting,elben2023randomized,funcke2022measurement,
van2022model,oliva2026guess}.
For example, if the target is a small set of Pauli expectation values, one can directly estimate and mitigate those observables rather than reconstructing the complete output distribution. This can reduce both measurement and calibration overhead while concentrating the available sampling budget on the quantities that enter the final physical result.

Observable structure can also provide additional robustness. If the target dynamics preserves a known symmetry $S$, with $[H,S]=0$, the ideal state remains within a fixed symmetry sector. Measurement outcomes inconsistent with that sector can then be identified as error-induced contributions and removed or reweighted. Symmetry verification and symmetry expansion exploit precisely this information
\cite{bonetmonroig2018lowcost,cai2021symmetry,koh2022classical,
nakaji2023measurement,cai2023quantum}.
Observable selection is therefore not merely a diagnostic choice: it determines which components of the noisy state must remain experimentally resolvable and which known physical constraints can be used to improve the reliability of the estimate.

\item \textbf{Ansatz-level error suppression.}
A second opportunity arises from the freedom to choose the circuit used to prepare the target state or approximate the desired dynamics. In variational algorithms, quantum simulation, and problem-inspired state preparation, different ansatz families can represent similar physical states while requiring very different hardware resources
\cite{peruzzo2014variational,kandala2017hardware,mcclean2016theory,
cerezo2021variational}.

A useful ansatz should therefore balance expressibility against implementation cost. In practice, shallower circuits with fewer entangling gates generally accumulate less hardware error, especially when the circuit follows the native connectivity and gate set of the processor
\cite{kandala2017hardware,murali2019noise,javadi2024quantum,
li2019tackling}. Hardware-efficient ans\"atze exploit this principle directly, whereas problem-inspired ans\"atze incorporate known Hamiltonian structure, particle-number conservation, gauge constraints, or other symmetries
\cite{mcclean2017hybrid,barron2021preserving,yao2021adaptive,
huang2022robust,bharti2022noisy}.

When a conserved quantity $S$ is known, a symmetry-preserving ansatz can be designed so that $[U(\boldsymbol{\theta}),S]=0$. The variational evolution then remains within the relevant symmetry sector, reducing exploration of physically irrelevant states and making symmetry-violating errors easier to identify
\cite{barron2021preserving,bonetmonroig2018lowcost,
cai2021symmetry,sagastizabal2019experimental,kakkar2022qaoa}.

Ansatz design is also closely connected to trainability. Highly expressive circuits are not always advantageous, because they often require greater depth and can suffer from barren plateaus, where optimization gradients become increasingly small with system size or circuit depth
\cite{mcclean2017hybrid,arrasmith2021effect,larocca2022diagnosing,
larocca2024review}. Noise can further suppress useful gradients, making deep ans\"atze difficult to optimize even when they are formally more expressive
\cite{holmes2022connecting,katabarwa2022connecting,wang2024trainability}.
The practical goal is therefore not maximal expressibility, but a balance among expressibility, trainability, symmetry structure, native connectivity, and robustness of the target observables.
\end{itemize}

Observable-level and ansatz-level suppression are closely related. The ansatz determines which regions of Hilbert space are explored and which errors are generated during state preparation, while the observable determines which components of the resulting noisy state actually contribute to the final estimate. An effective workflow should therefore co-design the state-preparation circuit and the measurement strategy rather than optimize them independently.

This perspective remains relevant as processors move from conventional NISQ operation toward early logical and partially error-corrected computation. Logical encoding and repeated error-correction cycles reduce physical noise, but logical qubits, syndrome extraction, feedforward, and non-Clifford operations remain expensive
\cite{piveteau2021error,google2025quantum,smith2024logical,bar2026layered}.
Choosing compact symmetry-compatible logical circuits and measuring only the observables required by the application can therefore continue to reduce the effective computational overhead.

Observable-level and ansatz-level methods do not attempt to correct every physical error. Instead, they exploit structure already present in the computational task: which quantities must be measured, which Hilbert-space sectors are relevant, and which circuit realizations are most compatible with the hardware. They therefore provide an important algorithmic layer of error suppression that complements hardware-level control, circuit-level noise tailoring, and post-processing-based error mitigation.

\subsection{Operator and observable selection}

Observable-level error suppression begins from a simple practical fact: on imperfect quantum hardware, not all observables can be estimated with the same useful precision. Even when two observables describe the same underlying physics, their measured estimates can differ substantially in bias, variance, and experimental cost. These differences arise because observables respond differently to gate errors, decoherence, readout noise, measurement-basis changes, and the number of circuit repetitions required for reliable estimation
\cite{verteletskyi2020measurement,gokhale2020minimizing,yen2020measuring,crawford2021efficient,hadfield2022measurements,huggins2021efficient}.
The central aim is therefore to identify an optimal observable that characterizes the physical feature of interest as clearly as possible while remaining robust and experimentally accessible on the target device. Operator and observable selection must consequently balance two requirements: the chosen diagnostic should faithfully capture the scientific property under study, while also retaining sufficient signal under the dominant noise and sampling constraints
\cite{huang2020predicting,huang2021efficient,elben2023randomized}.

This point can be stated formally. For an observable $O$, the experimentally accessible quantity is
\begin{equation}
\langle O\rangle_{\rm exp}
=
\mathrm{Tr}\!\left(O\,\widetilde{\rho}\right),
\qquad
\widetilde{\rho}=\mathcal{E}(\rho),
\end{equation}
where $\rho$ is the ideal state and $\mathcal{E}$ is the effective noise channel. The ideal expectation value is
\begin{equation}
\langle O\rangle_{\rm id}
=
\mathrm{Tr}\!\left(O\rho\right).
\end{equation}
The observable-dependent bias is therefore
\begin{equation}
\Delta_O
=
\langle O\rangle_{\rm exp}
-
\langle O\rangle_{\rm id}.
\end{equation}
Since different observables probe different components of the noisy state, $\Delta_O$ is generally not the same for all $O$. The statistical uncertainty of the estimator also depends on the operator structure, the measurement basis, and the number of measurement shots. Thus, two observables that are equivalent or closely related in the ideal theory may differ greatly in practical measurability on hardware.

A simple example arises in dynamical experiments designed to detect transport, operator spreading, or scrambling. In such settings, highly nonlocal observables can degrade rapidly under realistic noise. For instance, a global quantity such as the full-chain parity, $\prod_{j=1}^{L}Z_j$, is especially sensitive to readout errors, since every measured qubit contributes an additional opportunity for a bit-flip error. By contrast, local densities $\langle Z_j\rangle$ or short-range correlators,
\begin{equation}
\langle Z_i Z_{i+r}\rangle,
\end{equation}
often retain a useful signal at larger circuit depth. More generally, if measurement errors are approximately independent across qubits, then the signal associated with a weight-$w$ Pauli string typically decays with a factor that worsens as $w$ increases. This makes low-weight observables especially attractive whenever they capture the desired physical information.

When the ideal dynamics respects a symmetry or conservation law, it is often advantageous to build the measurement strategy around that structure
\cite{bonetmonroig2018lowcost,cai2021symmetry,sagastizabal2019experimental}.
Examples include particle number, total magnetization, fermion parity, lattice gauge constraints, or other conserved charges generated by an operator $Q$
\cite{mcardle2020quantum,bonetmonroig2018lowcost,sagastizabal2019experimental,gonzales2023paulicheck}.
If the dynamics preserves $Q$, so that
\begin{equation}
[U,Q]=0,
\end{equation}
then observables satisfying $[O,Q]=0$
are naturally adapted to the corresponding symmetry sector. Such observables are often more stable diagnostics because they do not rely on coherences between sectors that should ideally remain decoupled. In practice, noise can cause leakage out of the intended sector, and symmetry-compatible observables, symmetry verification, or symmetry expansion can help identify, filter, or reweight these unwanted contributions (see Sect.~\ref{verification} below).

A closely related idea appears in stabilizer-based settings. Suppose the target manifold is defined by stabilizers $\{S_p\}$ satisfying
\begin{equation}
S_p|\psi\rangle=|\psi\rangle .
\end{equation}
Then measuring local stabilizer expectation values provides a direct test of whether the prepared state remains in the intended code space or constrained subspace. Because these operators are tied directly to the defining structure of the target manifold, they can be more informative than generic observables that do not respect the same constraints. Pauli-check and stabilizer-based mitigation methods build on this idea by using additional check operators to detect or suppress error components that move the state away from the desired subspace
\cite{gonzales2023paulicheck}.

The broader lesson is that observable choice is itself a form of error-aware experimental design. Rather than attempting to recover every feature of a noisy quantum state, one selects diagnostics whose algebraic structure, spatial support, and symmetry properties make them comparatively insensitive to the dominant noise channels. On present-day quantum devices, where noise remains structured and measurement resources are finite, this task-aware choice of observables can substantially improve the reliability of the extracted physical information.

\subsection{Error-aware transpilation and noise-adaptive compilation}
\label{transpilation}
\subsubsection{Error-aware transpilation}
In practice, a circuit $U$ is executed only after it has been transformed into a hardware-native representation.
Error-aware transpilation seeks a mapping, gate decomposition, routing strategy, and schedule that reduce the impact of noise, while noise-adaptive compilation uses calibration data to tailor the compiled circuit to the current state of the device
\cite{murali2019noise,li2019tackling,sivarajah2021t,cowtan2019qubitrouting,javadi2024quantum,yale2025noise,kurniawan2024calibration,zhong2025cycle,wagner2025optimized}.
This perspective is important because the realized error profile is determined not only by the abstract logical circuit, but also by how that circuit is embedded into a specific processor, routed through its connectivity graph, decomposed into native gates, and scheduled under time-dependent calibration constraints
\cite{murali2019noise,sarovar2020detecting,fang2022crosstalk,hashim2025benchmarking,kurniawan2024calibration,yale2025noise,wagner2025optimized}.
Recent calibration-aware studies further show that the benefit of noise-aware compilation depends on the freshness, stability, and processing of backend calibration data, and that minimizing depth or two-qubit gate count alone is not always the best proxy for maximizing circuit fidelity
\cite{javadi2024quantum,kurniawan2024calibration,yale2025noise}.
Back in the NISQ setting, this issue arose mainly as a question of improving physical-circuit fidelity.
In the current transition toward early fault-tolerant hardware, however, compilation quality also affects ancilla placement, measurement scheduling, routing of syndrome-extraction circuits, dynamic-circuit control flow, and logical-qubit layouts
\cite{krinner2022repeated,google2025quantum,bar2026layered,salva2026cphase}.
Thus, transpilation should be viewed not as a preliminary preprocessing step, but as an integral component of practical error suppression and early fault-tolerant hardware--software co-design.

A simple way to formalize this idea is to treat layout selection as an optimization over a mapping from logical qubits to physical qubits.
Let $\pi$ denote such a mapping.
A topology-driven compiler primarily optimizes whether the required two-qubit interactions can be implemented with a small number of routing operations.
By contrast, a noise-adaptive compiler assigns a hardware-dependent cost to the compiled circuit, using quantities such as calibrated gate errors, readout errors, gate durations, coherence times, and possibly crosstalk information.
Schematically, we can use
\begin{equation}
\pi_{\rm opt}
=
\arg\min_{\pi}
C_{\rm hw}\!\left[\mathcal{C}(U,\pi)\right],
\end{equation}
where $\mathcal{C}(U,\pi)$ denotes the circuit obtained by compiling $U$ under the layout $\pi$, and $C_{\rm hw}$ is a hardware-aware cost function estimated from backend noise data \cite{murali2019noise}.
This expression is not meant to prescribe a unique cost function.
Rather, it emphasizes the operational principle: the preferred layout is the one expected to give the most reliable hardware execution, not necessarily the one with the smallest formal circuit depth.

\begin{figure}
    \centering
    \includegraphics[width=0.8\linewidth]{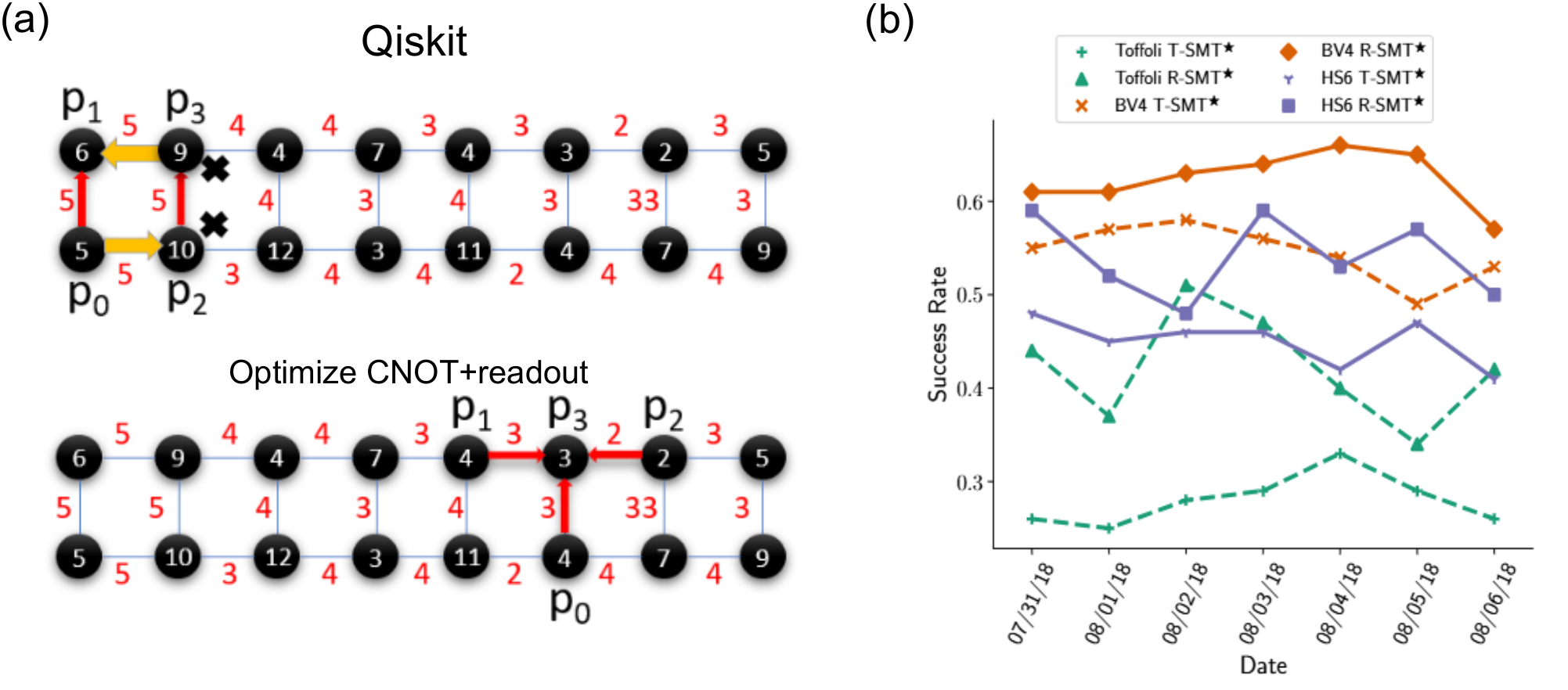}
    \caption{Calibration-aware layout and compilation improve reliability.
(a) Illustration of qubit-layout selection on a fixed hardware coupling graph. (b) Experimental success rate of Toffoli-circuit instances compiled under different layout choices, demonstrating that noise-adaptive compilation
can yield systematically higher success probabilities. Figures are adapted from Ref.~\cite{murali2019noise}. }
    \label{fig:compile}
\end{figure}

Murali \textit{et al.} used this calibration-aware viewpoint to show that exploiting spatial, and in practice also temporal, variations in device parameters can substantially improve empirical success rates compared with purely topology-driven layouts
\cite{murali2019noise} [see Fig.~\ref{fig:compile}].
Their approach used backend calibration information, including gate errors, readout errors, and gate durations, to select physical qubits and couplers that better matched the structure of the input circuit.
This result illustrates a basic lesson for error mitigation: the same logical circuit can have different effective noise depending on where and how it is executed on the device.

The IBM compilation stack has historically exposed this idea through noise-aware layout passes that use backend calibration data to choose a high-quality hardware subgraph for the circuit
\cite{li2019tackling}.
In Qiskit
\cite{javadi2024quantum},
the \texttt{NoiseAdaptiveLayout} pass explicitly targeted this type of calibration-aware placement. Although this pass has been deprecated in later releases and replaced by newer transpiler workflows, the underlying principle remains the same: one should choose the physical qubits and couplers that minimize the expected error of the compiled circuit, not merely the logical depth; see Sec.~\ref{software} for more details.

\subsubsection{Noise-adaptive circuit rewriting}

Noise-adaptive circuit rewriting refers to circuit transformations that preserve the intended ideal computation while changing how the computation is represented at the circuit level.
Unlike noise-adaptive compilation, which primarily chooses a hardware layout, routing path, native-gate decomposition, or schedule, circuit rewriting focuses on replacing a given implementation with an ensemble or family of logically equivalent implementations whose effective noise is more benign.
The target unitary is unchanged in the noiseless limit, but the physical error process can be reshaped by inserting additional gates, changing Pauli or Clifford frames, or exploiting gate identities that are inexpensive on the hardware
\cite{wallman2016noise,hashim2021randomized,ware2021pauliframe,cai2023quantum,santos2024pseudotwirling,tsubouchi2025symmetricclifford}.
This distinction is important: the goal is not simply to find a shorter circuit or a better hardware subgraph, but to engineer the effective noise model seen by the computation.

A central example is randomized compiling, introduced in Ref.~\cite{wallman2016noise} and demonstrated experimentally on a superconducting processor in Ref.~\cite{hashim2021randomized}.
In randomized compiling, the original circuit is rewritten as an ensemble of logically equivalent circuits.
Each circuit in the ensemble implements the same ideal computation, but contains randomly chosen local frame changes together with corresponding correction gates that preserve the noiseless action of each circuit cycle.
After averaging measurement results over the randomized ensemble, coherent and gate-dependent errors can be transformed into an effective stochastic Pauli noise model.
This makes the noise less prone to coherent accumulation and often easier to characterize using benchmarking protocols.

This idea is especially important because coherent control errors can be more damaging than stochastic errors with comparable average infidelity.
A small coherent over-rotation, for example, can add constructively over many circuit layers and produce a large observable bias.
Randomization-based rewriting suppresses this worst-case coherent buildup by changing the representation of the circuit from run to run, while keeping the target computation fixed.
In this sense, randomized compiling is not merely a compilation convenience, but a circuit-level form of error suppression.

Pauli-frame randomization and Pauli twirling are closely related to this principle
\cite{wallman2016noise,hashim2021randomized,ware2021pauliframe,cai2023quantum}.
They insert Pauli operations, often tracked classically rather than physically applied, so that off-diagonal or coherent components of the noise are averaged into a Pauli channel.
The resulting channel can still contain many different Pauli-error rates, but it is usually easier to model than a coherent, circuit-dependent error process.
Clifford twirling uses a larger randomizing group and therefore imposes a stronger symmetry on the effective noise channel.
Depending on whether the twirl is global, local, or symmetry restricted, the effective noise can be pushed closer to a depolarizing or otherwise more homogeneous form
\cite{cai2023quantum,tsubouchi2025symmetricclifford}.

Recent variants extend this logic beyond standard Clifford-gate randomized compiling.
Pseudo-twirling was introduced to mitigate coherent errors in non-Clifford gates, where ordinary randomized compiling may not apply directly
\cite{santos2024pseudotwirling}.
Symmetric Clifford twirling similarly exploits restricted Clifford randomizations compatible with gate symmetries, and has been proposed as a cost-efficient route for mitigating structured noise in early fault-tolerant regimes
\cite{tsubouchi2025symmetricclifford}.
These developments show that circuit rewriting is not limited to conventional NISQ circuits; it can also be adapted to more structured gate sets and to workflows in which some operations are already encoded or partially protected.

From the perspective of error mitigation, the main value of noise-adaptive rewriting is that it can make downstream mitigation assumptions more realistic.
Many mitigation methods, including zero-noise extrapolation, inverse-channel correction, probabilistic error cancellation, and data-driven inference, become more stable when the effective noise is closer to a stochastic or weakly gate-dependent model
\cite{endo2021hybrid,cai2023quantum,bultrini2023unifying,kimScalableErrorMitigation2023a,gupta2024dynamicpec,harrisReducingQuantumError2026}.
Randomization-based rewriting can therefore be used before these estimator-level methods, not to remove all physical errors, but to make the residual error channel more uniform, more benchmarkable, and less sensitive to coherent circuit structure.

The same logic remains relevant in the transition toward early fault-tolerant hardware.
Even when some operations are protected by encoding, coherent physical-level errors can still enter syndrome-extraction cycles, ancilla--data interactions, dynamic-circuit branches, or logical control routines.
Several recent works therefore explore how mitigation and noise tailoring can be integrated with early logical or partially error-corrected workflows
\cite{piveteau2021error,suzuki2022universal,wahlZeroNoiseExtrapolation2023,dutkiewicz2025error,jeon2026qecmitigated,bar2026layered}.
Thus, noise-adaptive circuit rewriting should not be viewed only as a near-term heuristic.
More broadly, it is a way of engineering an effective noise model that is better suited to both error mitigation and the gradual transition toward fault-tolerant quantum computation.

\subsection{Circuit depth reduction}
\label{reduction}

Reducing circuit depth is one of the most direct ways to improve performance on noisy quantum hardware, because both control errors and decoherence accumulate during circuit execution
\cite{preskill2018quantum,murali2019noise,li2019tackling,javadi2024quantum}.
Shallower circuits generally yield higher output fidelity and more reliable estimates of observables, provided that the depth reduction preserves the target unitary or introduces only controlled approximation error.
At the compiler level, circuit depth can be reduced through several strategies, including gate cancellation, gate merging, commutation-based rewriting, peephole optimization, routing-aware synthesis, ZX-calculus simplification, and hardware-aware resynthesis
\cite{maslov2008quantum,nam2018automated,kissinger2020pyzx,sivarajah2021t,iten2022exact,liu2021relaxed,martiel2022lazy,qiskit_transpiler}.
Among these, gate cancellation is the most local and conceptually simplest optimization, yet it often provides substantial practical benefit, especially after basis decomposition, qubit routing, and SWAP insertion have generated redundant adjacent gates or locally simplifiable circuit fragments
\cite{maslov2008quantum,li2019tackling,cowtan2019qubitrouting,sivarajah2021t,iten2022exact,liu2021relaxed}.

\subsubsection{Gate cancellation}

Gate cancellation removes subsequences of operations whose combined action is trivial.
The simplest case arises when two consecutive gates are mutual inverses:
\begin{equation}
U U^{\dagger} = \mathbb{I},
\end{equation}
so the pair can be deleted without changing the implemented unitary.
A particularly important subclass is formed by self-inverse gates, for which
\begin{equation}
H^2 = X^2 = Y^2 = Z^2 = \mathbb{I},
\qquad
\mathrm{CNOT}^2 = \mathbb{I}.
\label{eq43}
\end{equation}
As a result, adjacent patterns such as $HH$, $XX$, or two identical consecutive $\mathrm{CNOT}$ gates can be removed directly.
Although this rule is elementary, such redundancies occur frequently in compiled circuits because gate decomposition, basis conversion, and routing often generate pairs of operations whose net action is identity
\cite{maslov2008quantum,qiskit_transpiler}.

More generally, useful cancellations are not always visible in the original circuit layout.
Two inverse gates may be separated by intermediate operations that commute with them, so that algebraic rewriting can first be used to bring the inverse pair together.
In this sense, gate cancellation is closely connected to commutation analysis, template matching, and peephole optimization, where a compiler identifies short-circuit fragments and replaces them with equivalent but shallower implementations
\cite{maslov2008quantum,iten2022exact,liu2021relaxed,qiskit_transpiler}.
If
\begin{equation}
[U,V]=0,
\end{equation}
then a pattern of the form
\begin{equation}
U V U^{\dagger}
=
V
\end{equation}
can be simplified by commuting $U^{\dagger}$ past $V$ and cancelling the resulting inverse pair.

Gate cancellation is especially valuable on current hardware because it can remove operations that would otherwise contribute directly to the physical error budget.
This is particularly important when the cancelled operations include two-qubit gates, which typically have higher error rates and longer durations than single-qubit gates
\cite{preskill2018quantum,murali2019noise,li2019tackling,javadi2024quantum}.
Even removing a small number of redundant entangling operations can therefore noticeably improve the final output quality.
In addition, every cancelled gate can shorten the total execution time of the computation, thereby reducing exposure to relaxation, dephasing, idle errors, and schedule-dependent crosstalk
\cite{kjaergaard2020superconducting,blais2021circuit,sarovar2020detecting,fang2022crosstalk}.
Thus, gate cancellation does more than reduce the formal gate count: it converts algebraic redundancy into a direct reduction of physical noise.

This perspective remains important in the emerging early fault-tolerant regime.
Even when logical encoding is introduced, unnecessary physical operations can increase the number of fault locations, lengthen syndrome-extraction cycles, and complicate the maintenance of locality and timing constraints
\cite{krinner2022repeated,google2025quantum,salva2026cphase}.
Consequently, gate cancellation helps not only by improving bare physical-circuit fidelity, but also by reducing the effective control overhead in architectures moving toward fault-tolerant operation.

For this reason, current transpilers routinely include cancellation, commutation analysis, peephole optimization, and local resynthesis passes as standard stages of circuit optimization
\cite{maslov2008quantum,nam2018automated,kissinger2020pyzx,sivarajah2021t,iten2022exact,liu2021relaxed,martiel2022lazy,qiskit_transpiler}.
Despite its simplicity, gate cancellation often yields a favorable tradeoff between compilation cost and hardware benefit, making it one of the most broadly useful techniques for circuit-depth reduction in near-term and pre-fault-tolerant quantum computing.
More sophisticated noise-aware routing and circuit-optimization methods build on the same operational principle: reducing unnecessary operations and choosing lower-noise implementations can suppress accumulated circuit errors on real hardware
\cite{cowtan2019qubitrouting,nation2023suppressing,wagner2025optimized}.

\begin{figure}
    \centering
    \includegraphics[width=0.5\linewidth]{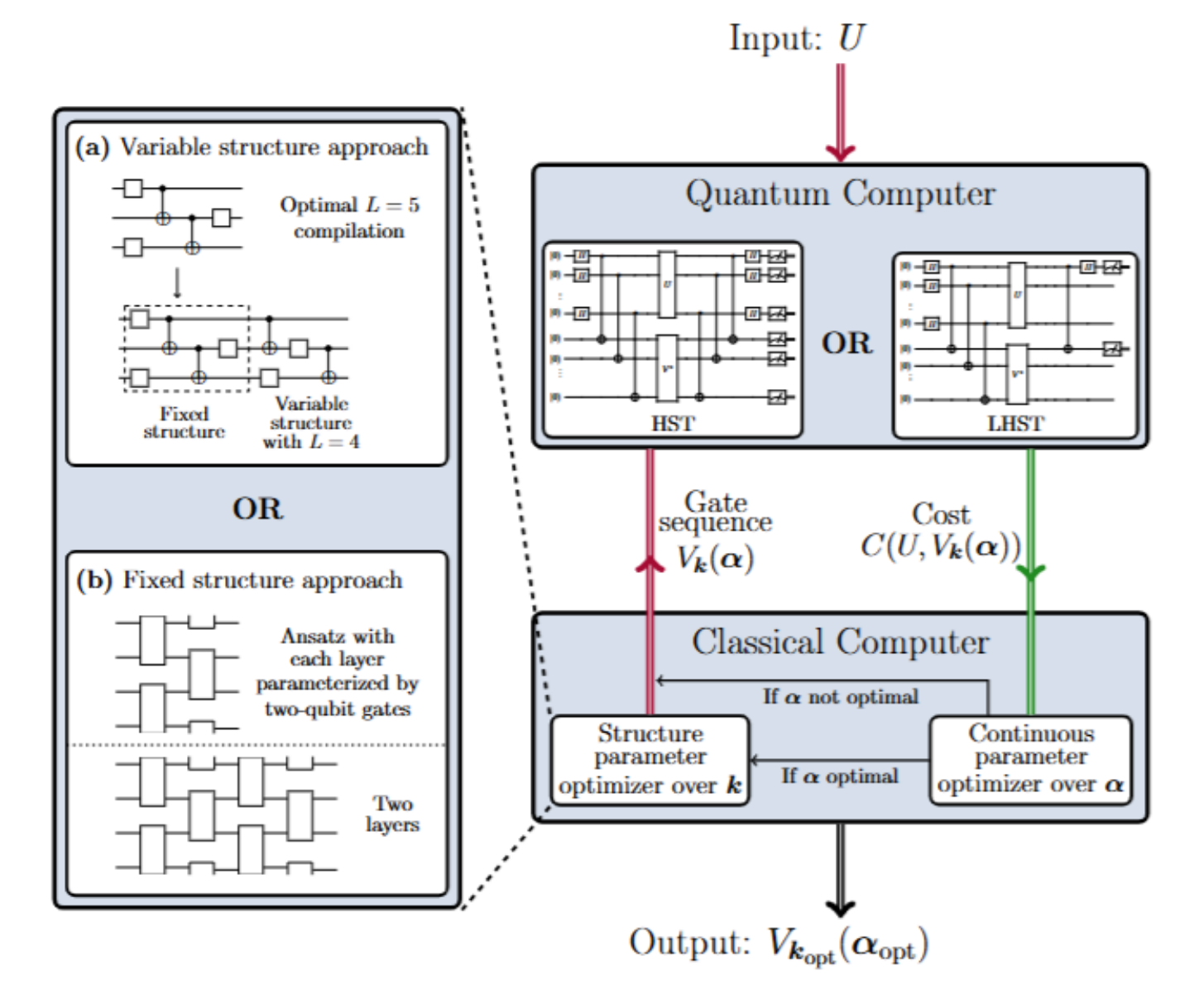}
    \caption{Outline of variational quantum compilation, adapted from Ref.~\cite{khatri2019qaqc}, where gate structures and parameters are optimized for a given input unitary.}
    \label{fig:var_med}
\end{figure}

\subsubsection{Variational methods for circuit compression}

Variational methods provide a flexible route to circuit-depth reduction by replacing an exact but deep target circuit with a shallower parameterized ansatz whose parameters are optimized in a hybrid quantum--classical loop
\cite{liEfficientVariationalQuantum2017a,khatri2019qaqc,yuan2019theory,endo2020general,cerezo2021variational,shen2025circuit}.
Rather than insisting on an exact gate-by-gate synthesis of the target unitary, one instead searches for a restricted-depth circuit $V(\bm{\theta})$ that approximates the desired operation $U_{\mathrm{tar}}$.
Depending on the application, this approximation may be required at the full-process level, at the state-preparation level, or only on the subset of input states and observables relevant to the task at hand
\cite{khatri2019qaqc,endo2020general,cirstoiu2020vff,kanasugi2025subspace,robertson2025approximate}.
In this way, variational methods replace exact compilation by an accuracy--depth tradeoff.
This tradeoff is often better aligned with realistic hardware constraints, because a shallow approximate circuit can outperform an exact but deeper implementation once gate errors, decoherence, routing overhead, and finite coherence time are taken into account
\cite{fontana2021evaluating,wang2024trainability,cerezo2021variational,fauseweh2024quantum}.

The basic idea is to choose a parameterized ansatz $V(\bm{\theta})$, typically constructed to be compatible with the native gate set and connectivity of the target device, and then optimize its parameters by minimizing a cost function that quantifies the discrepancy from the desired evolution.
For full-unitary compilation, a natural choice is a process-level cost,
\begin{equation}
C_{\mathrm{proc}}(\bm{\theta})
=
1-\frac{1}{d^{2}}
\left|
\mathrm{Tr}\!\left[
V^{\dagger}(\bm{\theta})\,U_{\mathrm{tar}}
\right]
\right|^{2},
\end{equation}
where $d=2^{n}$ for an $n$-qubit system.
This cost is closely related to the Hilbert--Schmidt overlap between the compiled ansatz and the target unitary, and it vanishes when the two agree up to a global phase.

In many applications, however, one does not need to reproduce the entire unitary on the full Hilbert space
\cite{khatri2019qaqc,endo2020general,cirstoiu2020vff,heya2023svqs,kanasugi2025subspace,robertson2025approximate}.
If the goal is to prepare the correct output only for a specific initial state $|\psi_{0}\rangle$, or for a restricted manifold of physically relevant states, then a state-level objective is often more appropriate:
\begin{equation}
C_{\mathrm{state}}(\bm{\theta})
=
1-
\left|
\langle\psi_{\mathrm{tar}}|
V(\bm{\theta})
|\psi_{0}\rangle
\right|^{2},
\qquad
|\psi_{\mathrm{tar}}\rangle
=
U_{\mathrm{tar}}|\psi_{0}\rangle .
\end{equation}
Here $C_{\mathrm{state}}(\bm{\theta})$ measures the infidelity between the target output state and the state prepared by the variationally compiled circuit.
This formulation is especially useful in near-term quantum simulation, because many tasks depend only on a limited dynamical sector, a restricted set of input states, or a small collection of target observables rather than on universal compilation of the full unitary
\cite{khatri2019qaqc,cirstoiu2020vff,heya2023svqs,kanasugi2025subspace,robertson2025approximate,fauseweh2024quantum}.

The practical benefit of this approach is that the ansatz depth can be treated as a controllable resource.
Instead of implementing a long Trotter circuit or an exact synthesis with many entangling gates, one restricts $V(\bm{\theta})$ to a hardware-efficient form with fewer layers,
\begin{equation}
V(\bm{\theta})
=
\prod_{\ell=1}^{L_{\mathrm{var}}}
V_{\ell}(\bm{\theta}_{\ell}),
\end{equation}
where $L_{\mathrm{var}}$ is kept small enough to remain compatible with the coherence window and error rates of the device.
If the optimized ansatz achieves sufficiently low cost, then one obtains an approximate implementation of the target evolution with substantially reduced circuit depth and therefore reduced accumulated noise.

From the viewpoint of error suppression, the key advantage of variational compilation is that part of the hardware constraint is absorbed directly into the circuit design.
Because the ansatz is chosen in the native gate language of the device, one can reduce routing overhead, avoid unnecessary basis changes, and tailor the circuit to the connectivity and control primitives of the processor
\cite{khatri2019qaqc,endo2020general,cirstoiu2020vff,heya2023svqs,kanasugi2025subspace,robertson2025approximate}.
In this sense, variational methods do not merely compress circuits algebraically; they seek shallow implementations that are naturally adapted to the hardware
\cite{liEfficientVariationalQuantum2017a,khatri2019qaqc,cerezo2021variational,fontana2021evaluating,fauseweh2024quantum}.

This flexibility is particularly valuable on noisy hardware, where an approximate but shallow circuit can outperform an exact but deep one once realistic noise is taken into account
\cite{fontana2021evaluating,wang2024trainability,cirstoiu2020vff,kanasugi2025subspace}.
Even if the optimized ansatz does not reproduce $U_{\mathrm{tar}}$ perfectly in the noiseless limit, it may still yield a more accurate experimental result because the reduction in depth lowers the total gate error, routing overhead, and decoherence exposure.
Variational methods therefore provide a natural framework for trading a controlled approximation error against a potentially much larger hardware-induced error, and this tradeoff is often favorable in practical near-term computations
\cite{khatri2019qaqc,endo2020general,cerezo2021variational,robertson2025approximate,sharma2020noise,jones2022robust}.

More recent work has pushed this idea toward large-scale many-body simulation.
Local-subspace variational compilation exploits locality and physically motivated subspaces to compile dynamics on small subsystems, and then assembles these local results into a much shallower global simulation circuit
\cite{kanasugi2025subspace}.
In reported benchmarks, this strategy can reduce circuit depth by roughly one order of magnitude relative to straightforward Trotterization while maintaining comparable accuracy for the target dynamics
\cite{kanasugi2025subspace}.
Likewise, approximate quantum compiling methods based on tensor-network or matrix-product-state representations aim to generate shallow circuits that reproduce the desired evolution more efficiently than standard product-formula constructions at the same target depth
\cite{robertson2025approximate}.
More broadly, approximate quantum compiling can be formulated as an optimization problem in which one seeks the best shallow circuit compatible with hardware constraints such as gate alphabet, connectivity, and allowed entangling structure
\cite{khatri2019qaqc,madden2022best,madden2022sketching}.
These developments highlight that, for simulation tasks, the relevant compilation target is often not an exact symbolic gate decomposition, but rather the best shallow circuit within a restricted, hardware-compatible ansatz family.

Variational depth reduction can also be combined with structure learning.
Instead of optimizing only continuous gate parameters, one may variationally optimize the circuit architecture itself, for example, by reinforcement-learning-based gate selection and policy-gradient approaches that divide a deep target circuit into shorter pieces and compile them hierarchically
\cite{herrera2022policy,bilek2022rvqc,ding2022evolutionary,sun2023differentiable}.
Such strategies are useful when the main obstacle is not merely parameter tuning, but the combinatorial search over expressive yet shallow circuit layouts.

The main limitation of variational methods is the classical optimization overhead, including ansatz dependence, local minima, measurement cost, and barren-plateau behavior.
Nevertheless, as variational compiling, fast-forwarding, and local-subspace methods continue to improve, they have become one of the central approaches for reducing effective circuit depth in quantum simulation and compilation
\cite{yuan2019theory,cirstoiu2020vff,heya2023svqs,kanasugi2025subspace,robertson2025approximate}. At the same time, this logic is not restricted to the original NISQ setting.
In the transition toward early fault-tolerant quantum computing, logical resources remain limited, encoded operations still carry substantial spacetime overhead, and one does not always need to spend logical depth, magic-state budget, or ancilla resources on an exact implementation when a shallower task-adapted approximation already suffices
\cite{piveteau2021error,suzuki2022universal,wahlZeroNoiseExtrapolation2023,dutkiewicz2025error,jeon2026qecmitigated,salva2026cphase}.

\begin{figure*}
  \centering
  \includegraphics[width=0.8\linewidth]{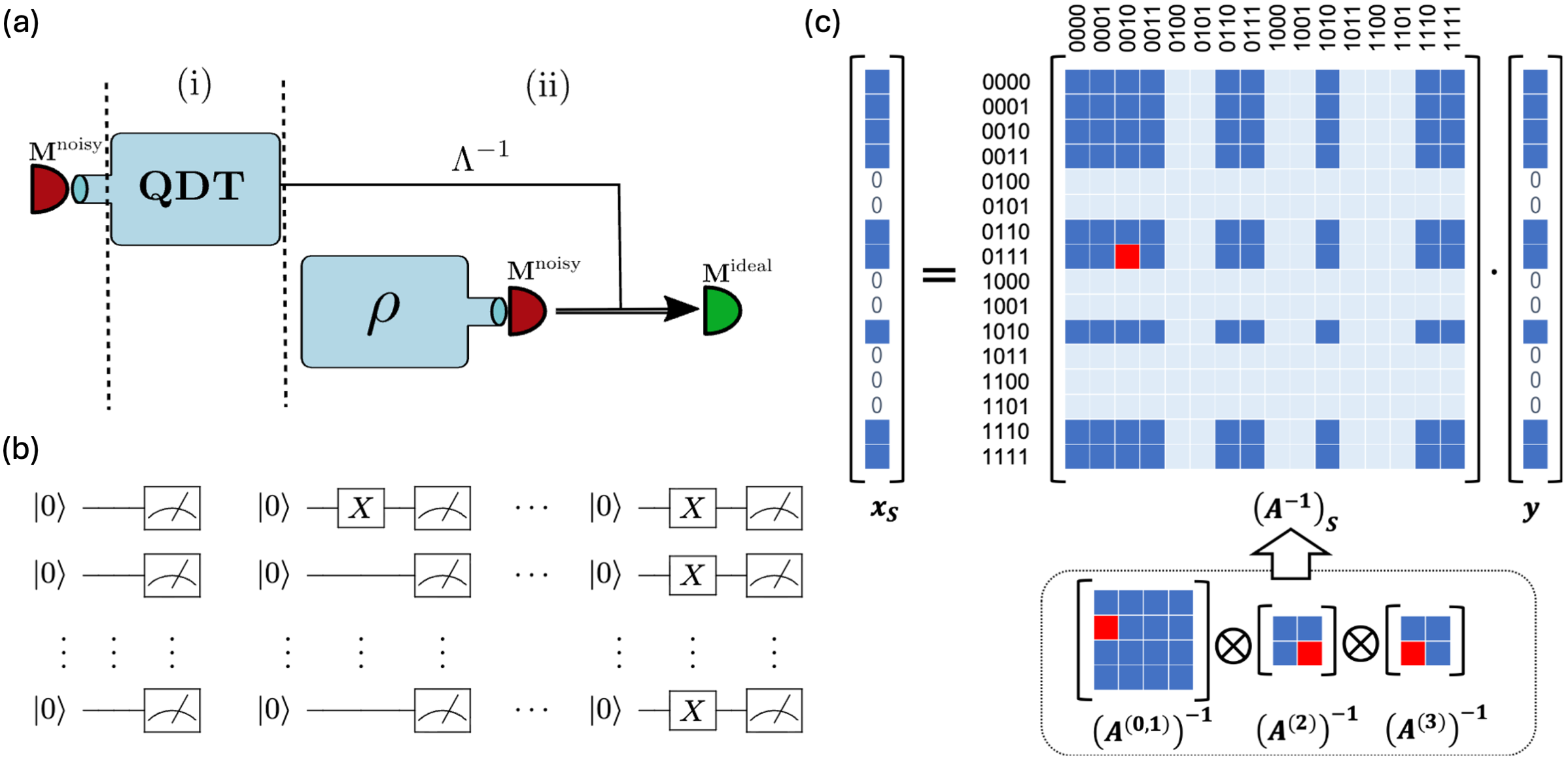}
  \caption{Readout-error mitigation workflow, calibration primitives, and support-restricted mitigation. (a) QDT-based view of readout mitigation, where detector characterization is followed by classical correction to recover ideal measurement statistics from a noisy detector model. (b) Standard basis-state calibration primitive for constructing a full response (assignment) matrix: the $2^n$ computational-basis input states generated by appropriate combinations of $X$ gates before measurement. (c) Reduced-support mitigation: the effective inverse response matrix is restricted to the observed outcome support, enabling scalable correction without explicit construction of the full $2^n\times2^n$ assignment matrix. (a) adapted from Fig.~1 of Ref.~\cite{maciejewski2020mitigation}. (b) adapted from Fig.~3 of Ref.~\cite{nachman2020unfolding}. (c) adapted from Fig.~1(b) of Ref.~\cite{nation2021scalable}.}
  \label{fig:readoutmi_fig1}
\end{figure*}

\begin{figure*}
  \centering
  \includegraphics[width=1\linewidth]{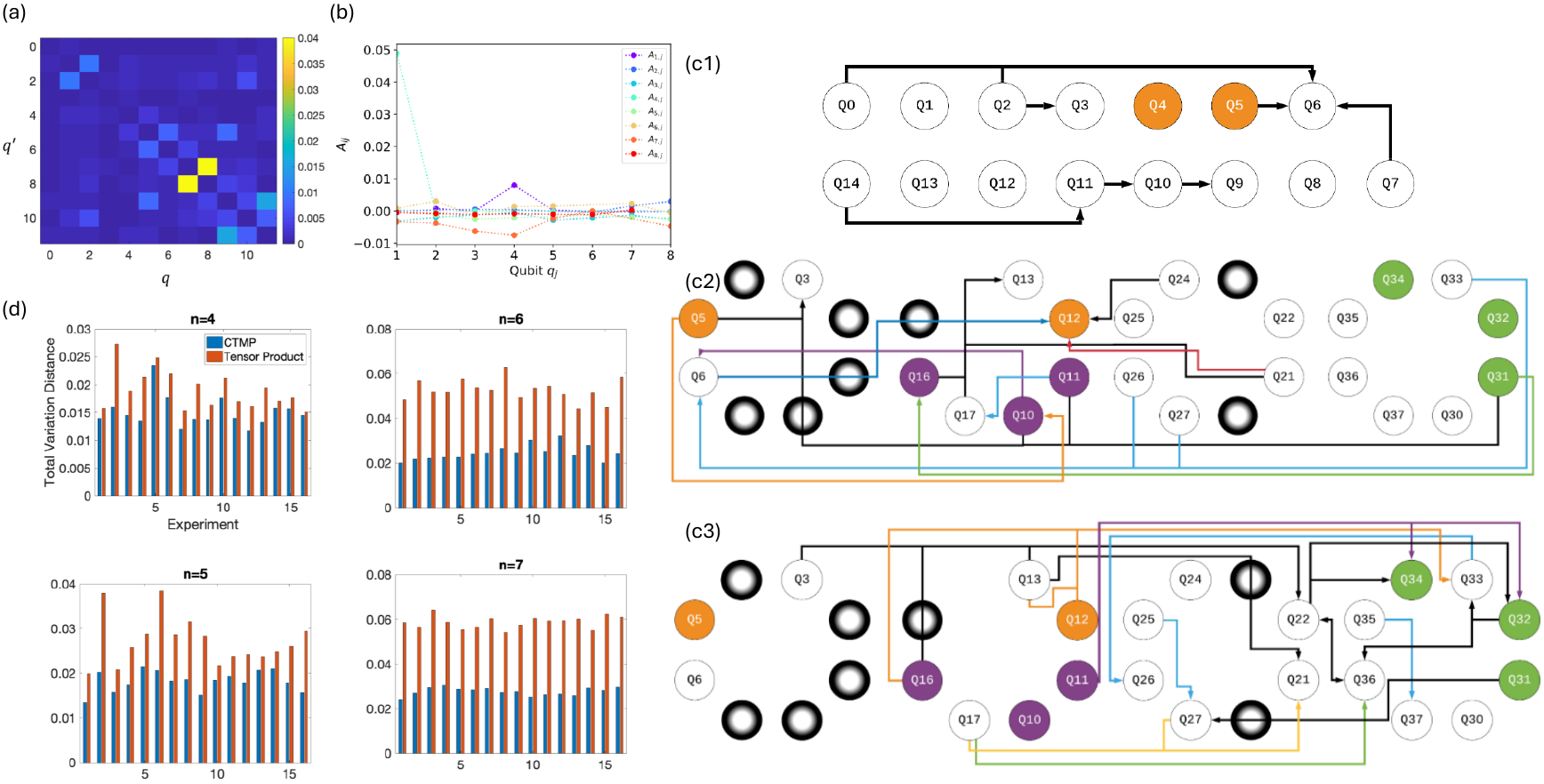}
  \caption{Empirical diagnostics and models for correlated readout errors. (a) Pair-resolved heat map of the combined rate of correlated two-qubit readout errors (e.g., $01\leftrightarrow 10$ and $00\leftrightarrow 11$ processes) inferred within a correlated Markov (CTMP) readout model, showing that specific qubit pairs exhibit few-percent-level correlated readout events. (b) Spectator-response correlator $A_{ij}$ measured on a one-dimensional 8-qubit register, plotted as a function of qubit position $j$ along the chain; the profile directly diagnoses nonlocal (long-range) readout correlations beyond a tensor-product model. (c1--c3) Visualization of a locality-restricted correlated readout-noise model inferred by diagonal detector overlapping tomography (DDOT): (c1) IBM \emph{Melbourne} (15 qubits); (c2) and (c3) a 23-qubit subset of Rigetti \emph{Aspen-8}, with arrows indicating qubits whose states affect the measurement noise on the left and right halves of the device, respectively. (d) Quantitative comparison of structured noise models against the full assignment matrix using total variation distance (TVD), illustrating improved accuracy of correlated models (e.g., CTMP) over tensor-product baselines. (a) adapted from Fig.~4(a) of Ref.~\cite{bravyi2021mitigating}; panel (b) adapted from Fig.~4 of Ref.~\cite{geller2021toward}; panels (c1--c3) adapted from Fig.~2(a--c) of Ref.~\cite{maciejewski2021modeling}; panel (d) adapted from Fig.~2 of Ref.~\cite{bravyi2021mitigating}.}
\label{fig:readoutmi_fig4}
\end{figure*}

\section{Measurement error mitigation}\label{read}

Measurement error mitigation, also known as readout mitigation, addresses errors introduced during the final measurement stage of a quantum circuit.
Having introduced readout errors as a hardware-level noise source, we now discuss the corresponding mitigation framework in more detail.
The basic premise is that the physical detector does not implement the ideal computational-basis projective measurement assumed in the circuit model.
Consequently, because of detector imperfections
[see Fig.~\ref{fig:readoutmi_fig1}(a)], the recorded bit-string frequencies
generally differ from the ideal measurement statistics of the quantum state
\cite{maciejewski2020mitigation,geller2020rigorous,bravyi2021mitigating,nation2021scalable,funcke2022measurement,van2022model,cai2023quantum,aasen2024readout}.
The goal is therefore not to modify the quantum evolution itself, but to infer ideal measurement outcomes, probability distributions, or expectation values from noisy classical data using a calibrated model of the detector response
\cite{maciejewski2020mitigation,geller2021conditionally,nation2021scalable,smith2021qubit,funcke2022measurement,tannu2019mitigating,kwon2020hybrid,barron2020measurement,wang2021measurementTNS,hicks2021readout,hicks2022active,zheng2020bayesian,yu2023efficientSPAM}.

From an implementation viewpoint, readout mitigation is one of the most hardware-facing forms of error mitigation. The calibrated response model is not an abstract property of the circuit alone, but an empirical description of the particular physical qubits, measurement chain, discriminator, calibration time, and readout schedule used in the experiment. On present-day superconducting and trapped-ion devices, a practical readout-mitigation workflow therefore consists of selecting the measured qubit subset and target observable, acquiring calibration data on the same backend and measurement configuration, applying a stable classical inference procedure, and validating the corrected result against drift, finite-shot instability, and model mismatch \cite{chen2019detector,maciejewski2020mitigation,nation2021scalable,aasen2024readout,cai2023quantum}. This hardware dependence becomes even more important in dynamic-circuit settings, where mid-circuit measurement errors can trigger incorrect feedforward branches rather than merely biasing a terminal outcome histogram \cite{koh2026readout,hashim2025quasiprobabilistic,santos2025driftresilient,giortamis2026mcmit}.

To state this process more precisely, consider an $n$-qubit circuit with final pre-measurement state $\rho \in \mathbb{C}^{2^n\times 2^n}$. If the qubits were measured ideally in the computational basis $\{\ket{z}\bra{z}\}_{z\in\{0,1\}^n}$, the resulting probability distribution would be
\begin{equation}
p_{\mathrm{id}}(z)
=
\mathrm{Tr}\!\left(\ket{z}\bra{z}\,\rho\right),
\qquad
z\in\{0,1\}^n,
\label{eq:pid_def}
\end{equation}
where $z=z_1z_2\cdots z_n$ denotes an $n$-bit measurement outcome, and $p_{\mathrm{id}}(z)$ is the ideal probability of observing that outcome. On real hardware, however, the detector can misidentify the bit string. As a result, the experimentally reported outcome $x\in\{0,1\}^n$ is sampled from a noisy distribution $p_{\mathrm{exp}}(x)$ that generally differs from $p_{\mathrm{id}}(x)$. Readout mitigation uses a calibrated classical model of this measurement distortion to estimate the ideal distribution in Eq.~\eqref{eq:pid_def}, or the expectation values derived from it, from the observed data.

This task is conceptually distinct from mitigation methods that target state-preparation or gate errors.
Readout mitigation assumes that the quantum circuit has already been executed and seeks to correct only the noise introduced when the final basis measurement is reported.
It therefore consists of two conceptually separate steps: first, a calibration procedure that characterizes how the device reports computational-basis states; second, a classical post-processing procedure that uses the calibration data to compensate for the resulting distortion
\cite{chen2019detector,geller2020rigorous,geller2021conditionally,bravyi2021mitigating,maciejewski2020mitigation,nation2021scalable}.
Because it operates entirely on classical measurement statistics, without modifying the circuit itself, readout mitigation has become one of the most widely used mitigation layers in noisy quantum experiments
\cite{cai2023quantum,aasen2024readout,funcke2022measurement,van2022model}.
From the current hardware perspective, this role is not limited to conventional NISQ workflows: as platforms move toward dynamic circuits, repeated measurement, reset, feedforward, and early fault-tolerant control, the quality of measurement and the ability to mitigate its imperfections remain central to practical performance \cite{krinner2022repeated,ryananderson2021realtime,zhao2022surfacecode,google2025quantum,koh2026readout,shirizly2025dynamicrb,bar2026layered}. In particular, mid-circuit measurements followed by feedforward introduce a distinct mitigation problem, since faulty per-shot outcomes can trigger the wrong conditional operation rather than merely biasing a terminal histogram \cite{koh2026readout,hashim2025quasiprobabilistic,santos2025driftresilient,giortamis2026mcmit}.

A standard description is obtained by introducing the ideal probability vector
\begin{equation}
\bm{p}_{\mathrm{id}}
=
\bigl(p_{\mathrm{id}}(z)\bigr)_{z\in\{0,1\}^n}
\in \mathbb{R}^{2^n},
\qquad
p_{\mathrm{id}}(z)\ge 0,
\qquad
\sum_{z\in\{0,1\}^n} p_{\mathrm{id}}(z)=1,
\end{equation}
together with the experimentally observed probability vector
\begin{equation}
\bm{p}_{\mathrm{exp}}
=
\bigl(p_{\mathrm{exp}}(x)\bigr)_{x\in\{0,1\}^n}
\in \mathbb{R}^{2^n},
\qquad
p_{\mathrm{exp}}(x)\ge 0,
\qquad
\sum_{x\in\{0,1\}^n} p_{\mathrm{exp}}(x)=1.
\end{equation}
Here, $\bm{p}_{\mathrm{id}}$ describes the ideal computational-basis populations, while $\bm{p}_{\mathrm{exp}}$ is the distribution reconstructed from the measured bit-string frequencies.

The measurement device is then modeled by an assignment matrix
\begin{equation}
A \in \mathbb{R}^{2^n\times 2^n},
\qquad
A_{x,z}
=
\Pr(\text{report }x \mid \text{ideal outcome }z),
\label{eq:assignment_matrix_def}
\end{equation}
where $A_{x,z}$ is the conditional probability that the detector reports the bit string $x$ when the ideal outcome is $z$. Since each column represents a conditional probability distribution, it satisfies
\begin{equation}
A_{x,z}\ge 0,
\qquad
\sum_{x\in\{0,1\}^n} A_{x,z}=1
\quad
\text{for every } z\in\{0,1\}^n.
\end{equation}
With this notation, the observed and ideal distributions are related by the linear map
\begin{equation}
\bm{p}_{\mathrm{exp}}
=
A\,\bm{p}_{\mathrm{id}},
\label{eq:readout_linear_map}
\end{equation}
or equivalently,
\begin{equation}
p_{\mathrm{exp}}(x)
=
\sum_{z\in\{0,1\}^n}
A_{x,z}\,p_{\mathrm{id}}(z).
\label{eq:readout_componentwise}
\end{equation}
In this picture, readout noise acts as a classical stochastic channel on the ideal measurement probabilities.

The central task of readout mitigation is therefore to estimate $\bm{p}_{\mathrm{id}}$ from the experimentally accessible vector $\bm{p}_{\mathrm{exp}}$ together with a calibrated estimate $\widehat A$ of the assignment matrix. In the simplest idealized case, where $\widehat A$ is invertible and sufficiently well conditioned, one may use the linear estimator
\begin{equation}
\widehat{\bm{p}}_{\mathrm{id}}
=
\widehat A^{-1}\bm{p}_{\mathrm{exp}}.
\label{eq:linear_inversion_readout}
\end{equation}

The discussion below is organized accordingly. We first introduce the main readout-error models and the calibration primitives used to estimate their parameters. We then examine practical workflows for implementing mitigation on hardware, including how calibration circuits are acquired and incorporated into classical post-processing. Finally, we summarize the main limitations of these approaches, especially those associated with correlated measurement errors, temporal drift, and the exponential cost of learning the full assignment matrix in large systems \cite{geller2021toward,nation2021scalable,yang2022efficient,peters2023perturbative}.

\subsection{Readout-error models and calibration primitives}
\label{sec:readout_models}

Readout mitigation begins from a simple practical fact: the linear relation $\bm{p}_{\mathrm{exp}}=A\,\bm{p}_{\mathrm{id}}$ [Eq.~\eqref{eq:readout_linear_map}] is only useful if the detector response $A$ can actually be learned from calibration data and its action inverted at acceptable cost. The difficulty is that the most faithful description of $A$ is also the most expensive to calibrate, store, and invert, while the cheapest description may fail to capture the readout correlations that matter for the target observable. Practical readout mitigation is therefore not a single model, but a sequence of controlled approximations, each of which trades descriptive accuracy against calibration and inversion cost \cite{neeley2010generation,dewes2012characterization,chen2019detector,maciejewski2020mitigation,geller2020rigorous,nachman2020unfolding,bravyi2021mitigating,geller2021conditionally,geller2021toward,maciejewski2021modeling,smith2021qubit,nation2021scalable,yang2022efficient,funcke2022measurement,van2022model,peters2023perturbative,pokharel2024scalable,aasen2025correlated,cai2023quantum}.

We organize this subsection according to the tradeoff between model fidelity and calibration cost. We begin with the full assignment-matrix description, which provides the most direct representation of readout errors but scales exponentially with system size. We then introduce the local tensor-product approximation as the simplest scalable baseline, followed by structured correlated models, continuous-time Markov-process (CTMP) descriptions, and locality-restricted overlapping reconstruction, which recover selected correlations without requiring full detector tomography. We subsequently discuss symmetrization and observable-level approaches, which reduce the complexity of the object that must be calibrated, as well as support-restricted representations that exploit sparsity in the measured outcome distribution. Practical software implementations of these readout-mitigation strategies are discussed further in Sec.~\ref{software}. Throughout, the guiding question is the same: what is the simplest detector model that captures the relevant readout structure while remaining practical to calibrate and apply on the target hardware?

\subsubsection{General picture: assignment-matrix method}

The practical problem behind readout mitigation is that the measurement implemented by a quantum device is not an ideal projective measurement. In the most general description, a noisy detector is described by a nonideal positive-operator-valued measure (POVM), namely a set of positive operators whose outcome probabilities are determined by the Born rule. These POVM elements can, in principle, be reconstructed through quantum detector tomography (QDT) [Fig.~\ref{fig:readoutmi_fig1}(a)]
\cite{greenbaum2015introduction,chen2019detector,maciejewski2020mitigation,cai2023quantum}.
Full QDT, however, usually provides more information than is required for readout mitigation. In many NISQ experiments, particularly on superconducting processors, the dominant readout imperfections can be approximated as classical misassignment of computational-basis outcomes. The detector can then be represented by a classical stochastic channel that maps the ideal bit-string distribution to the experimentally observed one
\cite{chen2019detector,maciejewski2020mitigation,geller2020rigorous,bravyi2021mitigating,nation2021scalable}.

Within this approximation, the assignment-matrix framework can be summarized as follows.

\begin{itemize}

\item \textbf{Assignment-matrix model.}
Consider an $n$-qubit measurement in the computational basis, with $x,y\in\{0,1\}^n$ denoting the ideal and observed bit strings, respectively. Let $\bm{p}_{\mathrm{id}}$ and $\bm{p}_{\mathrm{exp}}$ denote the corresponding ideal and experimentally observed probability vectors. The readout process is modeled by a stochastic assignment matrix $A$,
\begin{equation}
    \bm{p}_{\mathrm{exp}} = A \bm{p}_{\mathrm{id}},
\end{equation}
with matrix elements
\begin{equation}
    A_{yx}=P(y|x),
\end{equation}
where $P(y|x)$ is the probability that the device reports $y$ when the ideal outcome is $x$. Each column is normalized as $\sum_y A_{yx}=1$. Readout mitigation is therefore reduced to a classical inference problem: calibrate $A$ and use it to estimate $\bm{p}_{\mathrm{id}}$ from $\bm{p}_{\mathrm{exp}}$.

\item \textbf{Experimental motivation.}
This classical-channel reduction is the standard starting point for transition-matrix readout mitigation
\cite{chen2019detector,maciejewski2020mitigation,geller2020rigorous,bravyi2021mitigating,nachman2020unfolding,cai2023quantum}.
Detector-characterization studies support this approximation while also revealing its limitations. Chen \textit{et al.} used QDT on IBM devices and found few-percent deviations from ideal projective measurements, with correlated two-qubit detector terms required for more accurate modeling
\cite{chen2019detector}.
Maciejewski \textit{et al.} found that readout imperfections on IBM and Rigetti processors are dominated by effectively classical noise, explaining why calibration-based post-processing can substantially improve measured statistics
\cite{maciejewski2020mitigation}.
Related tomography-based studies have incorporated the same principle into state-reconstruction workflows
\cite{aasen2024readout}.

\item \textbf{Calibration and inversion.}
If $A$ were perfectly known and well conditioned, one could formally estimate the ideal distribution through
\begin{equation}
    \bm{p}_{\mathrm{id}} = A^{-1}\bm{p}_{\mathrm{exp}}.
\end{equation}
In practice, however, $A$ is inferred from finite-shot calibration data and may be ill-conditioned, so direct inversion can strongly amplify statistical fluctuations. Practical implementations therefore employ regularized inversion, constrained least-squares fitting, Bayesian unfolding, or related reconstruction procedures that impose physical constraints such as normalization and nonnegative probabilities
\cite{geller2020rigorous,nachman2020unfolding,bravyi2021mitigating}.

\item \textbf{Scalability.}
The principal limitation of the full assignment-matrix approach is its exponential cost. For $n$ measured qubits, exhaustive calibration requires $2^n$ computational-basis input states, while the full response matrix has dimension $2^n\times2^n$ and therefore contains $4^n$ entries
\cite{bravyi2021mitigating,nation2021scalable,yang2022efficient}.
This scaling motivates the tensor-product, correlated-cluster, locality-restricted, and support-restricted models discussed below, which retain the same basic framework while introducing structure that avoids reconstructing the full response matrix.

\item \textbf{Hardware calibration workflow.}
On gate-model hardware, the columns of $A$ are typically calibrated by preparing computational-basis states using appropriate $X$ gates, measuring them on the same physical qubits and with the same readout configuration as the target experiment, and estimating $P(y|x)$ from the observed counts
\cite{maciejewski2020mitigation,nachman2020unfolding,bravyi2021mitigating}.
The resulting matrix is therefore tied to a particular hardware layout and calibration window. Qubit remapping, changes to the readout discriminator, or temporal drift in the measurement chain can invalidate a previously calibrated response matrix
\cite{chen2019detector,maciejewski2020mitigation,aasen2024readout,santos2025driftresilient,lee2025personalizedReadout}.
Assignment-matrix mitigation is thus not only an inverse problem, but also a calibration-scheduling problem on real hardware.

\item \textbf{Interpretation of the calibrated matrix.}
A final caveat concerns what $A$ physically represents. Standard transition-matrix mitigation usually assumes that the calibration states $\ket{x}$ are prepared ideally, so that the measured transition probabilities can be attributed entirely to readout noise
\cite{geller2020rigorous,geller2021conditionally}.
In practice, preparation errors can be absorbed into the inferred matrix. Unless these effects are independently controlled, $A$ should therefore be interpreted as an effective preparation-and-measurement response rather than as a uniquely defined detector-only object
\cite{greenbaum2015introduction,maciejewski2020mitigation,geller2020rigorous,geller2021conditionally}.
For particular biased-measurement models, Geller showed that noisy multiqubit readout can be rigorously represented as an ideal projective measurement followed by a classical Markov process
\cite{geller2020rigorous}.
More generally, however, the interpretation of $A$ depends on how preparation and measurement errors are separated during calibration. This issue motivated the conditionally rigorous transition-matrix framework, which reduces sensitivity to preparation errors and clarifies when the transition-matrix description is justified
\cite{geller2021conditionally}.

\end{itemize}

The assignment-matrix framework converts readout correction into a classical calibration-and-inference problem, but its limitations are both computational and operational: the unrestricted response matrix scales exponentially with system size, while its accuracy depends on calibration stability, model validity, and the interpretation of the experimentally inferred response.

\subsubsection{Local tensor-product approximation}

The full assignment-matrix description is general but not scalable: for an $n$-qubit register, $A$ has dimension $2^n\times 2^n$, and exhaustive calibration requires all $2^n$ computational-basis inputs. The simplest scalable approximation is therefore to assume that readout errors act independently on different qubits. Under this assumption, the global response factorizes into single-qubit assignment matrices
\cite{bravyi2021mitigating,nation2021scalable,funcke2022measurement,cai2023quantum},
\begin{equation}
    A^{(\mathrm{loc})}
    \approx
    \bigotimes_{i=1}^{n} A_i ,
    \label{eq:local_tensor_product}
\end{equation}
where $A_i\in\mathbb{R}^{2\times 2}$ describes the readout response of qubit $i$.

The local tensor-product approximation can be summarized as follows.

\begin{itemize}

\item \textbf{Single-qubit response.}
A convenient parametrization is
\begin{equation}
    A_i
    =
    \begin{pmatrix}
        1-\epsilon_i^{0\to 1} & \epsilon_i^{1\to 0}\\
        \epsilon_i^{0\to 1} & 1-\epsilon_i^{1\to 0}
    \end{pmatrix},
    \label{eq:single_qubit_assignment_matrix}
\end{equation}
where the columns label the ideal outcomes $0$ and $1$, and the rows label the measured outcomes. Here, $\epsilon_i^{0\to 1}$ is the probability of reporting $1$ when the ideal outcome is $0$, while $\epsilon_i^{1\to 0}$ is the probability of reporting $0$ when the ideal outcome is $1$. Each column is normalized, so $A_i$ defines a classical stochastic channel.

\item \textbf{Physical interpretation.}
The model treats readout noise as a collection of independent biased bit-flip processes. The two error probabilities need not be equal. On superconducting transmon devices, for example, $1\to 0$ errors can be enhanced by relaxation during the measurement window, while amplifier noise, imperfect state discrimination, and calibration drift can further contribute to asymmetric readout
\cite{dewes2012characterization,chen2019detector,maciejewski2020mitigation,aasen2024readout}.
The measured distribution is then approximated by
\begin{equation}
    \bm{p}_{\mathrm{exp}}
    \approx
    A^{(\mathrm{loc})}\bm{p}_{\mathrm{id}},
    \label{eq:local_readout_channel}
\end{equation}
and mitigation uses the inverse, or a regularized inverse, of this local response model to estimate $\bm{p}_{\mathrm{id}}$.

\item \textbf{Scalability.}
The main advantage of the tensor-product approximation is that only the $n$ single-qubit response matrices must be calibrated, so the number of model parameters grows linearly rather than exponentially with system size. When the local matrices are invertible,
\begin{equation}
    \left(A^{(\mathrm{loc})}\right)^{-1}
    \approx
    \bigotimes_{i=1}^{n} A_i^{-1}.
    \label{eq:local_inverse}
\end{equation}
This factorized structure avoids explicitly learning a dense $2^n\times2^n$ detector model and is particularly attractive when readout correlations are weak or when only low-weight observables are required
\cite{bravyi2021mitigating,funcke2022measurement,nation2021scalable}.

\item \textbf{Breakdown under correlated readout noise.}
The efficiency of Eq.~\eqref{eq:local_tensor_product} comes from the independence assumption,
\begin{equation}
    P(y|x)
    \approx
    \prod_{i=1}^{n} P_i(y_i|x_i).
    \label{eq:factorized_readout}
\end{equation}
Real devices can violate this relation because measurement crosstalk, resonator-frequency crowding, shared amplification chains, spectator-dependent shifts, or correlated discrimination errors can make the response of one qubit depend on the state of another. In this case, the individual matrices $A_i$ no longer contain enough information to describe the full detector response.

\item \textbf{Diagnosing when the local model is insufficient.}
Rather than assuming independent readout noise a priori, correlations can be characterized experimentally. Geller and Sun developed finite-correlation-volume protocols that identify pairwise and spectator-dependent readout effects without reconstructing the full detector response
\cite{geller2021toward}.
Bravyi \textit{et al.} introduced a structured correlated Markov model that similarly extends the tensor-product approximation while retaining a tractable representation
\cite{bravyi2021mitigating}.
The spectator-response correlator in Fig.~\ref{fig:readoutmi_fig4}(b), for example, directly reveals when one qubit's readout depends on the state of another. Such diagnostics provide a practical criterion for deciding when the local model should be replaced by the correlated models discussed below.

\end{itemize}

The local tensor-product approximation therefore provides a scalable baseline for readout mitigation: it replaces an exponentially large detector response by independently calibrated single-qubit channels. Its validity, however, is controlled by the strength of readout correlations, motivating structured models that retain selected multiqubit dependencies without returning to the full $2^n\times2^n$ assignment matrix.

\subsubsection{Continuous-time Markov-process models}

The local tensor-product model is scalable, but it cannot represent correlations between readout errors on different qubits. At the opposite extreme, a full $2^n\times 2^n$ assignment matrix can capture arbitrary correlations but again requires exponentially many parameters. The practical objective is therefore to incorporate the dominant correlated readout processes while retaining a structured and scalable response model. The continuous-time Markov-process (CTMP) model introduced by Bravyi \textit{et al.} provides such an intermediate description
\cite{bravyi2021mitigating}.

\begin{itemize}

\item \textbf{Markov-generator representation.}
The CTMP model writes the assignment matrix as
\begin{equation}
    A_{\mathrm{CTMP}}
    =
    e^{G},
    \qquad
    G
    =
    \sum_{\alpha} r_{\alpha} G_{\alpha},
    \label{eq:ctmp_generator}
\end{equation}
where $G$ is a continuous-time Markov generator acting on the $2^n$ computational-basis bit strings. Each $G_{\alpha}$ represents an elementary transition process and $r_{\alpha}\geq 0$ is its associated rate. The off-diagonal elements of $G$ are nonnegative transition rates, while each column sums to zero, ensuring that $e^{G}$ defines a normalized stochastic response matrix.

\item \textbf{Sparse few-body structure.}
The central approximation is that only a restricted set of low-weight transitions contributes appreciably. Schematically,
\begin{equation}
    G
    =
    \sum_i G_i
    +
    \sum_{i<j} G_{ij}
    +
    \cdots,
    \label{eq:ctmp_few_body_decomposition}
\end{equation}
where $G_i$ describes single-qubit readout transitions and $G_{ij}$ describes correlated transitions involving a qubit pair. The former recover independent readout-flip processes, whereas the latter capture correlations that cannot be represented by a tensor product of single-qubit assignment matrices. For example, $00\rightarrow11$ and $11\rightarrow00$ on a qubit pair correspond to distinct directed transitions and can therefore carry different fitted rates.

\item \textbf{Mitigation through the calibrated generator.}
Once $G$ has been inferred from calibration data, the measured distribution is modeled as
\begin{equation}
    \bm{p}_{\mathrm{exp}}
    =
    e^{G}\bm{p}_{\mathrm{id}}.
    \label{eq:ctmp_forward_model}
\end{equation}
Formally, the corresponding mitigation map is
\begin{equation}
    \bm{p}_{\mathrm{id}}
    =
    e^{-G}\bm{p}_{\mathrm{exp}}.
    \label{eq:ctmp_inverse_model}
\end{equation}
The advantage of this representation is that the inverse correction is defined directly from the same structured generator used for the forward model. In practice, the sparsity and few-body structure of $G$ can be exploited so that mitigation does not require explicitly constructing or inverting a dense $2^n\times2^n$ assignment matrix.

\item \textbf{Scalability and range of validity.}
The CTMP model replaces the exponentially large set of entries of a full assignment matrix by the rates associated with the selected jump processes. When the dominant detector correlations can be described by one- and two-qubit terms, the number of fitted parameters grows polynomially rather than exponentially with system size. CTMP therefore occupies an intermediate regime between the local tensor-product approximation and an unrestricted response matrix: it captures selected correlated readout processes while remaining substantially more tractable than full detector reconstruction.

Its accuracy, however, depends on whether the dominant readout correlations are indeed sparse and well described by the chosen low-weight transition set. If important higher-order or strongly nonlocal correlations are present, additional generator terms are required, and the calibration cost correspondingly increases.

\end{itemize}

Pair-resolved correlated transition rates inferred within the CTMP framework, together with the improvement over tensor-product baselines, are shown in Fig.~\ref{fig:readoutmi_fig4}(a,d)
\cite{bravyi2021mitigating,cai2023quantum}.

\subsubsection{Locality-restricted overlapping reconstruction}

The CTMP model controls the complexity of correlated readout mitigation by restricting the allowed transition processes. A complementary approach is to restrict the \emph{spatial support} of detector correlations. Rather than assuming either independent single-qubit readout or an unrestricted $2^N\times2^N$ response matrix, one reconstructs correlated detector responses only on small, possibly overlapping subsets of qubits
\cite{geller2021toward,bravyi2021mitigating,maciejewski2021modeling,aasen2025correlated}.
A representative realization is diagonal detector overlapping tomography (DDOT), introduced by Maciejewski \textit{et al.} for characterizing readout crosstalk under a restricted-locality assumption
\cite{maciejewski2021modeling}.

\begin{itemize}

\item \textbf{Local overlapping reconstruction.}
DDOT focuses on the computational-basis response of the detector and reconstructs local detector marginals on overlapping subsets of qubits. The term ``diagonal'' refers to this restriction to computational-basis assignment probabilities, while ``overlapping'' reflects that many local detector responses are inferred simultaneously from shared global calibration circuits. In this way, the method avoids reconstructing the complete assignment matrix while retaining selected multi-qubit correlations.

\item \textbf{Calibration and scaling.}
Let $N$ be the number of measured qubits and $k$ the maximum locality order of the readout correlations to be characterized. Instead of preparing all $2^N$ computational-basis states, DDOT uses random combinations of $X$ and identity gates chosen so that all relevant local bit-string configurations on subsets of size at most $k$ are sampled. For fixed $k$, the number of calibration circuits scales as
\begin{equation}
    O\!\left(k2^k\log N\right),
\end{equation}
up to statistical-sampling and implementation-dependent factors
\cite{maciejewski2021modeling}.
The calibration cost is therefore controlled primarily by the assumed correlation locality $k$, rather than by the full Hilbert-space dimension.

\item \textbf{Inferring the correlated detector structure.}
The calibration data are used to estimate local and conditional readout responses and to identify which qubits significantly influence one another during measurement. This produces an effective detector graph in which edges or local neighborhoods represent statistically relevant readout correlations. The resulting model is more expressive than the tensor-product approximation, while remaining substantially more structured than a full response matrix
\cite{maciejewski2021modeling,nation2021scalable}.

\item \textbf{Detector locality is not hardware connectivity.}
An important result of DDOT is that the inferred detector graph need not coincide with the physical qubit-coupling graph. Readout correlations can arise from shared resonators, multiplexed measurement chains, common amplification hardware, spectator-dependent shifts, or classical discrimination procedures. Consequently, the relevant locality structure should be inferred from readout calibration data rather than imposed directly from geometric qubit connectivity
\cite{maciejewski2021modeling}.
Representative inferred structures are shown in Fig.~\ref{fig:readoutmi_fig4}(c1--c3), including IBM Melbourne and a subset of Rigetti Aspen-8, where the observed correlations are sparse but not restricted to simple nearest-neighbor patterns.

\item \textbf{Range of applicability.}
Locality-restricted reconstruction is most effective when detector correlations are appreciable but remain confined to relatively small neighborhoods. If the chosen neighborhoods are too small, relevant correlations are omitted and residual mitigation bias remains. Increasing the locality order improves model expressiveness but also increases calibration and inference cost. The practical task is therefore to identify the smallest correlated detector model that adequately describes the measured readout response.

\end{itemize}

DDOT thus converts correlated readout mitigation into a locality-aware model-selection problem. Rather than reconstructing every element of the global assignment matrix, it learns only the overlapping local structures supported by calibration data, providing a scalable intermediate description between independent readout models and unrestricted detector reconstruction.

\subsubsection{Symmetrization and observable-level mitigation}

The structured detector models discussed above seek to reconstruct an effective response map for computational-basis readout. Their common challenge is that the model must be sufficiently expressive to capture relevant detector correlations while remaining practical to calibrate and invert. A complementary strategy is therefore to simplify the mitigation target itself. This can be achieved either by symmetrizing the effective readout channel, so that its bias structure becomes easier to characterize, or by correcting only the observable of interest rather than reconstructing the full output distribution
\cite{smith2021qubit,funcke2022measurement,van2022model,aasen2024readout}.

\begin{itemize}

\item \textbf{Readout symmetrization.}
Qubit bit-flip averaging reduces readout asymmetry by randomly applying $X$ gates immediately before measurement and classically relabeling the measured outcomes
\cite{smith2021qubit}.
If the random flip pattern is $s\in\{0,1\}^n$, qubit $i$ is flipped whenever $s_i=1$, and a measured bit string $y$ is relabeled as $y\oplus s$. In the absence of readout noise, this procedure leaves the logical measurement unchanged. In the presence of noise, averaging over the random flip patterns symmetrizes the effective detector response by averaging over different assignments of logical bit values to physical readout outcomes.

The resulting channel has a simpler bias structure and can therefore require less detailed calibration. Importantly, bit-flip averaging is not restricted to independent single-qubit readout errors and can also be combined with correlated readout models
\cite{smith2021qubit}.
It should therefore be viewed as a preprocessing layer that reshapes the effective detector channel before subsequent mitigation.

\item \textbf{Observable-level correction.}
Many NISQ experiments require only a limited set of expectation values rather than the complete corrected bit-string distribution. In this case, reconstructing $\bm{p}_{\mathrm{id}}$ from $\bm{p}_{\mathrm{exp}}$ can be unnecessary and may amplify finite-shot fluctuations. Observable-level mitigation instead constructs a corrected estimator directly for the target observable $O$.

Funcke \textit{et al.} developed operator-level correction rules for classical bit-flip readout models
\cite{funcke2022measurement}.
In this framework, readout noise maps the ideal measured operator to an effective noisy operator, and mitigation constructs a corrected combination of measured outcomes whose expectation value reproduces the ideal result under the calibrated model. This is particularly useful for local observables and local Hamiltonians, where the correction can remain efficient without reconstructing the full probability distribution.

\item \textbf{Observable-specific bias factors.}
Van den Berg \textit{et al.} introduced a related expectation-value mitigation strategy that avoids learning a complete detector model
\cite{van2022model}.
Using additional randomized measurements, the readout bias of the target observable can be reduced to an effective multiplicative factor,
\begin{equation}
    \langle O\rangle_{\mathrm{exp}}
    =
    \gamma_O\,\langle O\rangle_{\mathrm{id}},
    \label{eq:observable_bias_factor}
\end{equation}
so that mitigation reduces to estimating $\gamma_O$ and rescaling the measured expectation value. This directly targets the quantity ultimately reported by the experiment, at the cost of additional sampling. Its effectiveness depends on whether $\gamma_O$ can be estimated with sufficiently small statistical uncertainty.

\item \textbf{Scope and trade-offs.}
These approaches differ from CTMP and DDOT in what they simplify. CTMP and DDOT retain the probability-distribution viewpoint and seek more efficient structured response models. Bit-flip averaging instead simplifies the effective detector channel itself, while observable-level methods avoid full distribution reconstruction when only selected expectation values are required.

Their main advantage is a reduction in calibration, storage, or inversion cost relative to full assignment-matrix mitigation. Their limitation is that they generally provide less information than a fully corrected distribution, and their accuracy depends on whether the symmetrized channel or observable-specific correction adequately captures the relevant readout bias
\cite{smith2021qubit,funcke2022measurement,van2022model,cai2023quantum}.

\end{itemize}

Symmetrization and observable-level mitigation therefore provide an alternative route to scalability: rather than constructing an increasingly detailed global detector model, they reduce the complexity of the effective channel or restrict the correction to the experimentally relevant observable.

\subsubsection{Support-restricted and compressed representations}

Even when the detector response is modeled accurately, a separate scaling problem remains: the full computational-basis probability vector contains $2^n$ components. Structured detector models such as CTMP and DDOT reduce the complexity of the \emph{response model}, but they do not by themselves eliminate the cost of representing and manipulating the full outcome distribution. Support-restricted and compressed methods address this additional bottleneck by exploiting the fact that many near-term circuits produce probability distributions concentrated on a comparatively small number of bit strings
\cite{nation2021scalable,yang2022efficient,peters2023perturbative}.

\begin{itemize}

\item \textbf{Restriction to the relevant support.}
Let $S\subseteq\{0,1\}^n$ denote the subset of bit strings retained in the reduced representation. In a data-driven implementation, $S$ can be chosen from the empirically observed outcomes, for example by retaining bit strings that appear in the samples or exceed a prescribed frequency threshold. The measured distribution is then restricted to $\bm{p}_{\mathrm{exp}}|_S$, and mitigation is performed only within this reduced outcome space [Fig.~\ref{fig:readoutmi_fig1}(c)]
\cite{nation2021scalable,yang2022efficient}.

\item \textbf{Compression of the representation rather than the detector model.}
Support restriction does not require changing the underlying readout model. The response may still be described by a local tensor-product approximation, a correlated model, or another calibrated detector representation. What changes is the space on which that model is evaluated. Instead of constructing and manipulating the full $2^n\times2^n$ response matrix, one forms only the reduced response object associated with the retained outcomes. When $|S|\ll2^n$, this can substantially reduce memory requirements and classical runtime.

\item \textbf{Representative scalable approaches.}
Nation \textit{et al.} developed a scalable mitigation method that avoids constructing the full assignment matrix or its inverse and instead performs the correction in a subspace defined by the observed noisy bit strings
\cite{nation2021scalable}.
Yang \textit{et al.} similarly developed efficient mitigation methods for sparse output distributions, where the computational cost is determined primarily by the number of relevant outcomes rather than by the full Hilbert-space dimension
\cite{yang2022efficient}.
A complementary approach by Peters \textit{et al.} exploits the perturbative suppression of high-weight readout flips, retaining only low-order error processes to obtain a compressed approximation with controlled truncation
\cite{peters2023perturbative}.

\item \textbf{Validity and truncation bias.}
The effectiveness of support restriction depends on whether the retained set $S$ contains the outcomes that are relevant to the quantity being estimated. The approach is particularly favorable for low-entropy or strongly concentrated distributions, such as nearly deterministic outputs, GHZ-like states, or algorithms whose probability mass is localized on a small set of bit strings. It becomes less reliable when the ideal distribution is broad, when readout errors transfer significant weight outside the retained support, or when rare outcomes make important contributions to the target quantity. In these situations, support truncation introduces a systematic bias that cannot be removed simply by improving the inversion within $S$.

\item \textbf{Relation to observable-level mitigation.}
The appropriate compression criterion depends on the final target. For distribution reconstruction, $S$ should capture the bit strings carrying most of the relevant probability weight. For observable estimation, however, the important outcomes are those that dominate the expectation value of the observable, which need not coincide with the most probable outcomes. A low-probability bit string can still contribute strongly to a particular fidelity or nonlocal observable. Support restriction therefore provides a natural intermediate layer between full distribution-level reconstruction and observable-specific mitigation
\cite{nation2021scalable,yang2022efficient,peters2023perturbative}.

\item \textbf{Data-driven correction.}
A conceptually different route is to bypass explicit inversion of a calibrated assignment matrix and instead learn the noisy-to-ideal transformation directly from calibration data. Neural-network-based readout mitigation, for example, trains a classical model on noisy and reference measurement data and then applies the learned map to experimental output statistics
\cite{kim2022deeplearning}.
Such approaches can capture detector structure beyond simple linear inversion models, but require representative training data and generally provide less physical interpretability than assignment-matrix, CTMP, or DDOT descriptions.

\end{itemize}

These methods complete a natural hierarchy of scalable readout mitigation. Full detector tomography provides the most general description, while the assignment-matrix model reduces the problem to a classical response channel. Tensor-product models simplify the detector through independence assumptions; CTMP and DDOT incorporate selected correlations through sparse transition or locality structure; symmetrization and observable-level methods simplify the effective channel or corrected quantity; and support-restricted methods reduce the dimension of the probability space on which the correction is performed
\cite{chen2019detector,maciejewski2020mitigation,geller2020rigorous,geller2021toward,bravyi2021mitigating,maciejewski2021modeling,smith2021qubit,funcke2022measurement,van2022model,cai2023quantum}.

The practical objective is therefore not to construct the most detailed detector model possible, but to choose the simplest combination of detector representation, inference method, and probability-space compression that captures the relevant hardware structure and target quantity at acceptable calibration and computational cost
\cite{cai2023quantum,nation2021scalable,funcke2022measurement,peters2023perturbative}.

\begin{figure*}
  \centering
  \includegraphics[width=0.7\linewidth]{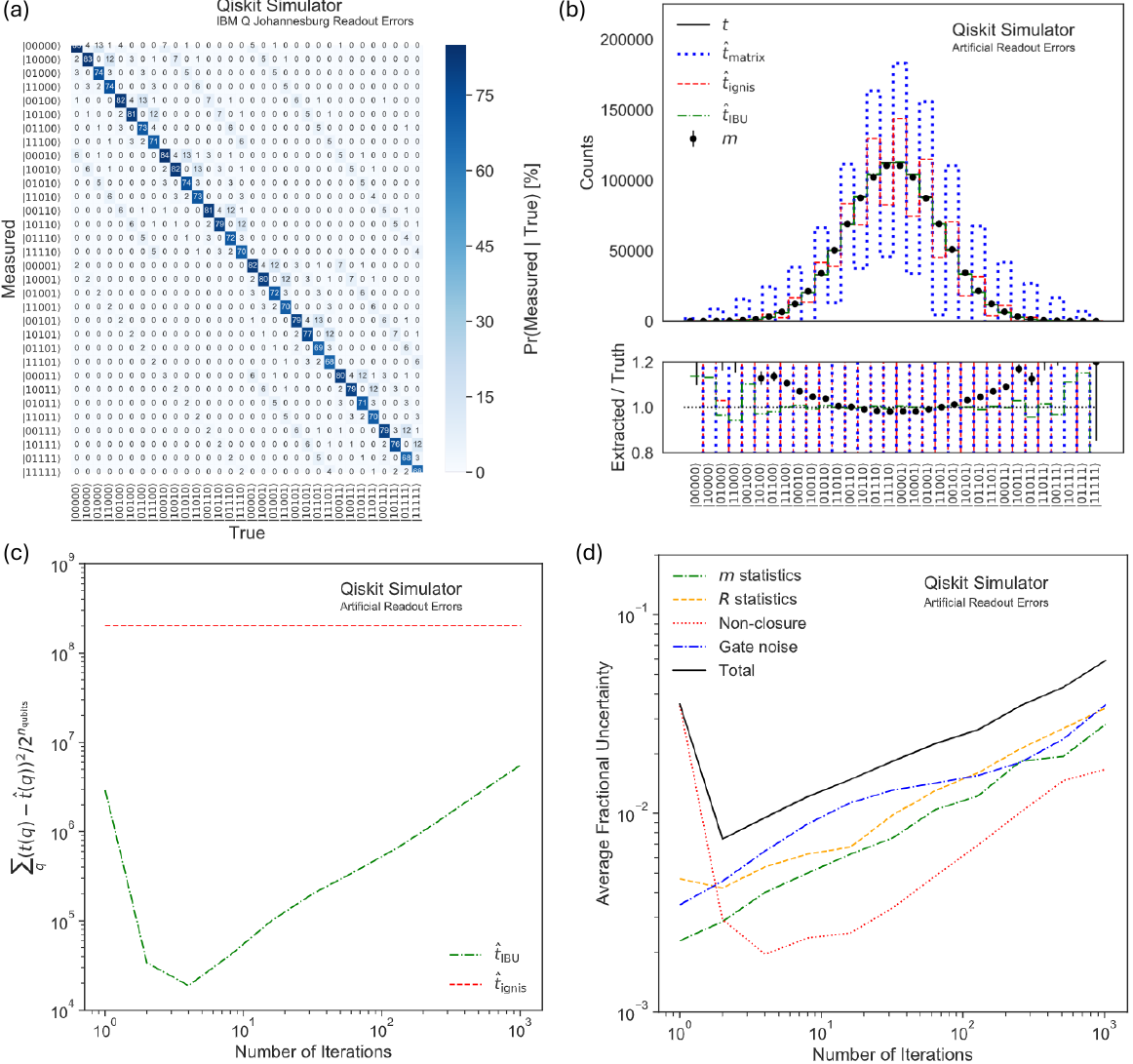}
  \caption{Ill-conditioned unfolding and regularization in readout mitigation. (a) Example empirical response matrix $R$ (assignment/response probabilities) for a small subset of qubits, illustrating off-diagonal migration probabilities that render unfolding an ill-posed inverse problem. (b) A representative failure mode of naive inversion and constrained least-squares approaches: amplified statistical fluctuations and oscillatory artifacts in the unfolded distribution, contrasted with iterative Bayesian unfolding (IBU). (c) Dependence of IBU bias (or residual distortion) on the number of unfolding iterations, highlighting early stopping as an implicit regularizer that controls the bias--variance trade-off. (d) Dependence of uncertainty quantification (e.g., interval width/coverage or related diagnostics) on the number of unfolding iterations, further illustrating iteration count as a practical regularization knob. (a), (b), (c), and (d) are adapted from Fig.~4, Fig.~6, Fig.~9, and Fig.~10 of Ref.~\cite{nachman2020unfolding}, respectively. }
\label{fig:readoutmi_fig2}
\end{figure*}

\subsection{Practical workflows on hardware}
\label{sec:readout_workflows}

A practical implementation of readout mitigation on current hardware can be organized as a calibration-and-inference pipeline. First, one fixes the correction target: a full bit-string distribution, a small set of Pauli-$Z$ expectation values, or a downstream estimator such as an energy. Second, calibration circuits are executed on the same physical qubits and measurement settings as the target circuits. Third, the calibration data are converted into a detector model, ranging from local assignment matrices to correlated, locality-restricted, or support-restricted response models. Fourth, the raw counts are corrected by a stable inference procedure rather than by unrestricted matrix inversion. Finally, the result is validated by checking positivity, normalization, bootstrap stability, calibration freshness, and consistency between local and correlated models \cite{geller2020rigorous,nachman2020unfolding,nation2021scalable,maciejewski2021modeling,cai2023quantum,aasen2024readout}.

\subsubsection{Workflow design}

The practical problem in hardware readout mitigation is that the same formal relation, $\bm{p}_{\mathrm{exp}}=A\bm{p}_{\mathrm{id}}$, can lead to different implementations depending on the experimental target and device constraints. The first step is therefore to decide what quantity must be corrected. In sampling tasks, state tomography, and histogram-based post-processing, the target is often the full computational-basis probability vector $\bm{p}_{\mathrm{id}}=\bigl(p_{\mathrm{id}}(x)\bigr)_{x\in\{0,1\}^n}$, so one calibrates a detector model and solves a distribution-level inverse problem. In many variational, simulation, and benchmarking tasks, however, the experiment only reports a small number of expectation values. In that case, reconstructing the full mitigated histogram may be unnecessary, and observable-level mitigation can be preferable \cite{funcke2022measurement,van2022model,cai2023quantum}.

Once the mitigation target is fixed, the method is to choose the cheapest detector calibration that is still accurate enough for that target. For few-qubit experiments, or for benchmarking small subsystems, one may estimate the full assignment matrix directly by exhaustive basis-state calibration \cite{geller2020rigorous,nachman2020unfolding,bravyi2021mitigating}. For larger systems, this becomes impractical because the full detector model acts on a $2^n$-dimensional outcome space. Practical workflows therefore usually begin with a cheaper baseline, most commonly the local tensor-product approximation, and then escalate to a structured correlated model only if calibration diagnostics show that local readout correction is inadequate \cite{geller2021toward,bravyi2021mitigating,maciejewski2021modeling,cai2023quantum}. The correction is then applied through a numerically stable inference procedure rather than by blindly inverting the calibrated matrix. Finally, the mitigated result must be checked against calibration drift, solver instability, and model mismatch. In this sense, readout mitigation on hardware is a closed loop connecting target selection, detector calibration, stable inference, and validation.

At the implementation level, calibration circuits should be run on the same measured qubits, with the same measurement settings, and as close in time as possible to the target experiment \cite{maciejewski2020mitigation,cai2023quantum,aasen2024readout}. The reason is operationally simple: readout mitigation can only correct the detector that was actually calibrated. If the measurement chain drifts between calibration and application, then the inferred response model may no longer describe the detector seen by the target circuit. On superconducting platforms, detector-characterization studies and readout-mitigated tomography experiments have shown explicitly that the effective measurement model depends on physical effects such as relaxation during the measurement window, amplification and thresholding procedures, and the operating point of the readout chain \cite{chen2019detector,maciejewski2020mitigation,aasen2024readout}. For this reason, practical workflows usually interleave calibration and data acquisition on a timescale short compared with detector drift.

\subsubsection{Distribution-level inference}

Once the calibration data have been collected, the practical problem is to infer the ideal computational-basis distribution from noisy measured counts. Formally, this follows from the forward model $\bm p^{\mathrm{raw}}=A\bm p^{\mathrm{id}}$. However, directly applying $A^{-1}$ is often unstable: finite-shot fluctuations in $\widehat{\bm p}^{\mathrm{raw}}$ and poor conditioning of $A$ can be amplified by the inverse map, producing large statistical oscillations or even unphysical negative probabilities \cite{geller2020rigorous,nachman2020unfolding,nation2021scalable,peters2023perturbative}. This instability is the distribution-level inference problem illustrated in Fig.~\ref{fig:readoutmi_fig2}(a,b), where off-diagonal response probabilities make the unfolding problem ill-conditioned and naive inversion can generate unstable reconstructed distributions.

The standard method is therefore not to invert the response matrix blindly, but to solve a stabilized inverse problem. A common formulation is
\begin{equation}
\widehat{\bm p}^{\mathrm{id}}
=
\arg\min_{\bm p\ge 0,\ \mathbf{1}^{\mathsf T}\bm p=1}
\left\|A\bm p-\widehat{\bm p}^{\mathrm{raw}}\right\|_2^2,
\label{eq:readout_constrained_inference}
\end{equation}
where the constraints $\bm p\ge 0$ and $\mathbf{1}^{\mathsf T}\bm p=1$ enforce the physical interpretation of $\bm p$ as a probability distribution. This procedure works by searching for the physical distribution whose forward image under the calibrated detector model best matches the measured data. Related iterative and uncertainty-aware variants serve the same purpose: they regularize the inverse problem while preserving a probability-level output \cite{nation2021scalable,peters2023perturbative,cai2023quantum}.

A widely used iterative alternative is iterative Bayesian unfolding (IBU). If $p_i^{(n)}$ denotes the current estimate of the ideal probability of outcome $i$ after $n$ iterations, then the update rule is
\begin{equation}
p_i^{(n+1)}
=
\sum_j
\frac{A_{ji}\,p_i^{(n)}}{\sum_k A_{jk}\,p_k^{(n)}}
\,\widehat{p}^{\mathrm{raw}}_j,
\label{eq:readout_ibu}
\end{equation}
where $\widehat{p}^{\mathrm{raw}}_j$ is the measured probability of outcome $j$ \cite{nachman2020unfolding}. This update uses the current estimate $p_i^{(n)}$ as a prior, compares its forward image with the measured data through the response matrix $A$, and then redistributes the observed counts back to the ideal outcomes. The iteration number acts as a regularization parameter: early stopping suppresses variance amplification, while too many iterations can reintroduce noise from the ill-conditioned inverse problem. This bias--variance tradeoff is shown in Fig.~\ref{fig:readoutmi_fig2}(c,d).

For larger systems, the same inference problem also faces a representation-cost problem, because the full $2^n$-dimensional outcome space may be too large to store or manipulate. A scalable method is support-restricted inference. In many NISQ applications, the measured distribution is concentrated on a relatively small set of bit strings. Nation \textit{et al.} showed that one can avoid constructing the full assignment matrix and instead carry out mitigation only on the subspace spanned by the noisy bit strings that are actually observed \cite{nation2021scalable}. Yang \textit{et al.} further developed efficient mitigation strategies for sparse measurement outcomes, where the reduced support lowers memory cost and classical runtime \cite{yang2022efficient}.

\begin{figure*}
  \centering
  \includegraphics[width=1\linewidth]{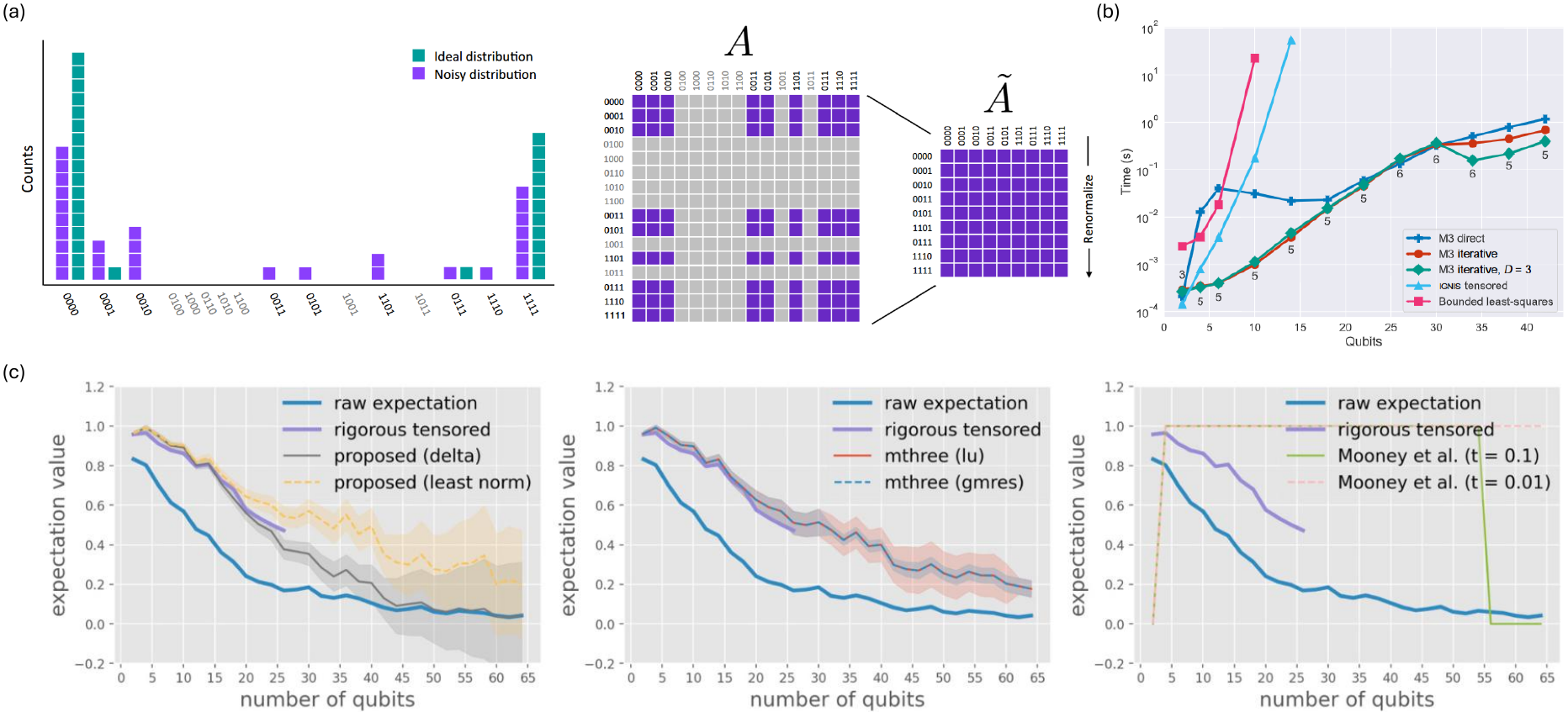}
  \caption{Scalable readout-error mitigation: subspace reduction and large-$n$ benchmarks. (a) Subspace reduction underlying matrix-free measurement mitigation: a sparse/noisy outcome distribution defines an observed support, and the reduced assignment matrix is constructed by selecting the corresponding elements of the full assignment matrix, followed by column renormalization. (b) Classical post-processing time (best of repeated runs) for mitigation methods as a function of qubit number, with annotations indicating the number of iterations required to reach a prescribed solver tolerance. (c) Expectation values versus qubit number for GHZ benchmarks, comparing multiple scalable QREM approaches (including subspace-based mitigation) against raw data and a rigorous baseline where available; shaded regions indicate error bounds. (a) is adapted from Fig.~1(a) and Fig.~1(b) of Ref.~\cite{nation2021scalable}; (b) is adapted from Fig.~3(b) of Ref.~\cite{nation2021scalable}; (c) is adapted from Fig.~2 of Ref.~\cite{yang2022efficient}.} 
  \label{fig:readoutmi_fig3}
\end{figure*}

This support-restricted strategy may be written as
\begin{equation}
\widehat{\bm p}^{\mathrm{id}}_{S}
=
\arg\min_{\bm p_{S}\ge 0,\ \mathbf{1}^{\mathsf T}\bm p_{S}=1}
\left\|A_{S}\bm p_{S}-\widehat{\bm p}^{\mathrm{raw}}_{S}\right\|_2^2,
\label{eq:readout_support_restricted}
\end{equation}
where $S$ denotes the empirically relevant support of the observed outcomes, $\widehat{\bm p}^{\mathrm{raw}}_{S}$ is the measured probability vector restricted to that support, and $A_{S}$ is the corresponding restricted response object \cite{nation2021scalable,yang2022efficient,peters2023perturbative}. The method works by replacing the full inverse problem with a smaller one defined only on the retained support. Fig.~\ref{fig:readoutmi_fig3}(a) illustrates this subspace reduction, Fig.~\ref{fig:readoutmi_fig3}(b) shows the resulting reduction in post-processing cost, and Fig.~\ref{fig:readoutmi_fig3}(c) shows its use in large-$n$ expectation-value benchmarks. Thus, distribution-level inference is not just matrix inversion: it is a stabilized and often support-restricted reconstruction problem whose practical form is chosen according to the conditioning of the detector model and the sparsity of the measured data.

\subsubsection{Observable-level mitigation, symmetrization, and escalation beyond local models}

A different workflow is needed when distribution-level reconstruction is either unnecessary or insufficiently reliable. The first problem is target mismatch: many variational, simulation, and benchmarking experiments report only a small number of expectation values, so reconstructing an entire mitigated bit-string histogram can be excessive. The corresponding method is observable-level mitigation, where one corrects the measured estimator itself rather than first reconstructing $\bm p_{\mathrm{id}}$ \cite{funcke2022measurement,van2022model,cai2023quantum}. The second problem is detector bias: the raw readout channel may be asymmetric or calibration-sensitive, in which case one can first symmetrize the effective readout response by randomized pre-measurement bit flips \cite{smith2021qubit}. The third problem is model inadequacy: if local readout correction fails, the workflow must escalate to correlation diagnostics or structured correlated detector models. Thus, this subsection describes three practical choices beyond direct distribution-level inference: observable-level correction, symmetrization, and escalation beyond local detector models.

A basic example arises when readout noise is modeled as independent symmetric bit-flip errors. Suppose $S$ is a subset of measured qubits and
\begin{equation}
Z_S=\prod_{q\in S} Z_q
\end{equation}
is the corresponding Pauli-$Z$ string. If qubit $q$ is flipped during readout with probability $p_q$, then the raw and ideal expectation values are related by
\begin{equation}
\langle Z_S\rangle_{\mathrm{raw}}
=
\left(
\prod_{q\in S}(1-2p_q)
\right)
\langle Z_S\rangle_{\mathrm{id}},
\end{equation}
so that the ideal value can be estimated as
\begin{equation}
\langle Z_S\rangle_{\mathrm{id}}
=
\frac{\langle Z_S\rangle_{\mathrm{raw}}}
{\prod_{q\in S}(1-2p_q)}.
\label{eq:readout_observable_level}
\end{equation}
This formula shows that, under a simple readout model, one can correct an observable directly without first reconstructing the full probability distribution.

More generally, observable-level mitigation may be written schematically as
\begin{equation}
\langle O\rangle_{\mathrm{raw}}
=
\eta_O\,\langle O\rangle_{\mathrm{id}},
\qquad
\langle O\rangle_{\mathrm{id}}
=
\eta_O^{-1}\langle O\rangle_{\mathrm{raw}},
\label{eq:readout_model_free_obs}
\end{equation}
where $O$ is the observable of interest and $\eta_O$ is a calibration factor determined from auxiliary measurements \cite{van2022model}. In this viewpoint, the central task is not to learn the entire detector response, but only the correction factor relevant for the specific observable being measured. Funcke \textit{et al.} derived such operator-level correction rules for classical bit-flip models, while van den Berg \textit{et al.} proposed a model-free protocol in which the readout bias is converted into a directly measurable multiplicative factor \cite{funcke2022measurement,van2022model}.

When the main issue is systematic detector bias rather than the reconstruction of a full probability distribution, practical workflows can also include a bias-reduction step before the final correction stage. Smith \textit{et al.} showed that bit-flip averaging can symmetrize the effective readout channel by randomizing pre-measurement $X$ gates together with the corresponding classical relabeling of outcomes \cite{smith2021qubit}. Operationally, this preprocessing step can reduce systematic bias and simplify the effective detector seen by the later inference procedure. It is therefore best viewed as an enhancement of the calibration-and-correction pipeline, rather than as a completely separate mitigation family.

A separate problem arises when observable-level correction or local readout mitigation is not accurate enough because the detector contains correlated readout errors. A practical strategy is to begin with correlation diagnostics. Geller and Sun introduced efficient protocols based on a finite correlation volume, which can detect pairwise and spectator-dependent readout structure before one commits to a more expensive correlated reconstruction \cite{geller2021toward}. Similarly, diagonal detector overlapping tomography (DDOT) and related correlated-readout protocols reconstruct overlapping local neighborhoods instead of a fully dense global detector, thereby adapting the calibration cost to the effective locality scale of the measurement apparatus \cite{maciejewski2021modeling,aasen2025correlated}. On real hardware, this staged escalation is often the most practical choice: one starts from the cheapest local baseline and moves to a correlated model only when the calibration data show that the local approximation is inadequate.

Finally, every practical hardware workflow requires validation. At minimum, one should check whether mitigated probabilities remain statistically stable under resampling, whether the corrected result is consistent with the assumed detector model, and whether recalibration changes the inferred answer significantly \cite{geller2020rigorous,nation2021scalable,cai2023quantum}. When possible, one may also validate the workflow on circuits or states with known approximate structure, or compare local and correlated mitigation pipelines on the same dataset to determine whether the additional calibration overhead is justified \cite{geller2021toward,maciejewski2021modeling,aasen2024readout}. The practical lesson is that readout mitigation on hardware is not a single formula applied blindly after measurement. Rather, it is a closed experimental loop in which calibration scheduling, detector modeling, numerical inference, support restriction, and validation must be co-optimized for the specific device, workload, and target observable \cite{cai2023quantum,nation2021scalable,van2022model,peters2023perturbative}.

\subsection{Limitations and scalability}
\label{sec:readout_limitations}

In this subsection, we summarize the main limitations that determine when readout mitigation remains scalable and reliable. The central problem is that readout mitigation must learn, store, and invert an effective detector response, but each of these steps can become costly or unstable as the number of measured qubits increases. The methods used to address this problem include structured detector models, constrained or regularized inference, support-restricted correction, correlated readout models, and observable-level strategies. Each method works by reducing a different part of the mitigation burden: structured models reduce calibration cost, regularized solvers stabilize the inverse problem, sparse-support methods reduce the effective outcome space, and observable-level methods avoid full distribution reconstruction. However, these advantages are conditional. Local models can fail in the presence of correlated readout errors, sparse-support methods can introduce truncation bias, detector drift can make calibration stale, and readout mitigation cannot correct gate or state-preparation errors unless they are explicitly included in a larger effective model. The practical question is therefore not whether readout mitigation scales in an absolute sense, but which detector representation and inference strategy remain accurate enough for the specific hardware, workload, and observable of interest.

\subsubsection{Scalability of detector representations and calibration}

The first scalability problem is representational: readout mitigation requires an effective detector response, but the amount of detector information that can be learned, stored, and inverted decreases rapidly as the measured register grows. Thus, the issue is not whether readout mitigation is useful in principle, but whether the detector model is accurate enough for the target task while remaining calibratable on the available hardware \cite{geller2020rigorous,maciejewski2020mitigation,cai2023quantum,bravyi2021mitigating,nation2021scalable}. This gives the basic problem--method structure of scalable readout mitigation: the full assignment matrix is the most direct representation, while local, correlated, or otherwise compressed models reduce the calibration burden by imposing structure on the detector response.

This issue already appears at the most basic level of detector representation. For an $n$-qubit computational-basis measurement, the full assignment matrix
\begin{equation}
A \in \mathbb{R}^{2^n \times 2^n}
\end{equation}
contains $4^n$ real entries, and exhaustive basis-state calibration requires preparing and measuring all $2^n$ computational-basis input states \cite{geller2020rigorous,bravyi2021mitigating,nation2021scalable}. As a result, the difficulty is not only experimental. It also appears in the classical memory needed to store the model, in the cost of estimating it reliably from finite-shot data, and in the complexity of the later correction step. Full-matrix workflows are therefore best viewed as small-system or subsystem tools, rather than as a default strategy for large devices.

A convenient summary is
\begin{equation}
A \in \mathbb{R}^{2^n\times 2^n},
\qquad
\mathrm{storage}(A)=\mathcal{O}(4^n),
\qquad
N_{\mathrm{circ}}^{\mathrm{full}} = 2^n,
\label{eq:full_matrix_scaling}
\end{equation}
where $N_{\mathrm{circ}}^{\mathrm{full}}$ is the number of circuits needed for exhaustive calibration. If each prepared basis state is sampled to additive precision $\epsilon$, the total shot cost scales as
\begin{equation}
N_{\mathrm{shot}}^{\mathrm{full}}
=
\mathcal{O}(2^n \epsilon^{-2}),
\label{eq:full_shot_scaling}
\end{equation}
up to estimator-dependent constants \cite{geller2020rigorous,bravyi2021mitigating,nation2021scalable}. This is the basic reason why scalable mitigation cannot rely on an unrestricted detector description.

Once this point is recognized, a second limitation follows naturally. To reduce cost, one usually replaces the full detector response by a more structured approximation. The simplest example is the factorized model: $A \approx \bigotimes_{i=1}^n A_i$,
where each $A_i \in \mathbb{R}^{2\times 2}$ is the single-qubit assignment matrix. This is appealing because the number of calibration parameters grows only linearly with system size. However, the gain in efficiency comes from a physical assumption, namely that multiqubit readout correlations are weak enough to neglect \cite{bravyi2021mitigating,geller2021toward,funcke2022measurement}. When this assumption fails, the price of scalability is model mismatch.

More expressive constructions such as CTMP- and DDOT-type models attempt to recover part of the missing correlated structure \cite{bravyi2021mitigating,maciejewski2021modeling}. They often improve accuracy, but they also require more parameters, more calibration data, and stronger assumptions about which correlations are important. There is therefore no universally best detector model. A model that is too simple leaves systematic bias, while a model that is too detailed becomes statistically noisy or operationally expensive \cite{cai2023quantum,maciejewski2021modeling}. In practice, scalable readout mitigation is governed less by formal model richness than by the balance between descriptive accuracy and calibration cost.

A related conceptual issue is that the calibrated response matrix is not always a purely detector-only object in a strict operational sense. Standard transition-matrix mitigation assumes that the calibration states are prepared ideally, so that the measured transition probabilities can be attributed entirely to readout noise \cite{geller2020rigorous}. In reality, state-preparation errors during calibration can contaminate the inferred matrix and blur the separation between detector noise and preparation noise \cite{geller2020rigorous,geller2021conditionally}. For a review article, this point is worth making explicit: the limits of readout mitigation are not only numerical, but also interpretive, because the calibrated model depends on how the calibration experiment itself is defined.

\subsubsection{Inference instability and regularization}

Even if the detector model is accurate enough to describe the dominant readout response, mitigation still faces a second problem: the correction step is an inverse problem, and inverse problems are often unstable \cite{geller2020rigorous,nachman2020unfolding,nation2021scalable,peters2023perturbative}. Here the issue is not how the detector is represented, but how sensitively the reconstructed ideal distribution depends on small fluctuations in the measured data. A standard perturbative estimate is
\begin{equation}
\frac{\|\delta p^{\mathrm{id}}\|}{\|p^{\mathrm{id}}\|}
\lesssim
\kappa(A)
\frac{\|\delta p^{\mathrm{raw}}\|}{\|p^{\mathrm{raw}}\|},
\qquad
\kappa(A)=\|A\|\,\|A^{-1}\|,
\label{eq:readout_condition_number}
\end{equation}
where $\kappa(A)$ is the condition number in the chosen norm \cite{geller2020rigorous,peters2023perturbative}. If $\kappa(A)$ is large, then ordinary shot noise in the raw counts can be amplified strongly during correction.

This explains why bare matrix inversion often performs poorly in practice. Even when the forward detector model is reasonable, direct inversion can produce oscillatory artifacts, negative reconstructed probabilities, or large statistical fluctuations that obscure the intended correction \cite{nachman2020unfolding,nation2021scalable,peters2023perturbative}. Fig.~\ref{fig:readoutmi_fig2}(a,b) illustrates this instability: the empirical response matrix contains off-diagonal migration probabilities, and naive unfolding can amplify statistical fluctuations into unphysical or highly oscillatory corrected distributions.

The corresponding method is to replace direct inversion by a constrained, regularized, or iterative inference procedure \cite{nachman2020unfolding,nation2021scalable,peters2023perturbative,cai2023quantum}. These methods work by suppressing poorly resolved directions of the inverse problem while preserving the physical interpretation of the output distribution. Enforcing nonnegativity and normalization prevents the corrected vector from leaving the probability simplex; truncating unstable directions or stopping an iterative solver early suppresses variance amplification; uncertainty-aware variants quantify how much of the corrected signal is supported by the calibration and shot data. The price is a bias--variance tradeoff: stronger regularization gives more stable estimates, but can also leave residual bias.

This balance becomes more important as the system grows. In larger problems, calibration matrices are themselves noisy, the empirically sampled support may be incomplete, and different solvers can produce materially different corrected distributions \cite{peters2023perturbative,cai2023quantum}. Fig.~\ref{fig:readoutmi_fig2}(c,d) shows this point in the context of iterative Bayesian unfolding, where the iteration count acts as a regularization parameter controlling both residual bias and uncertainty. From this perspective, classical post-processing is not merely an implementation detail after calibration. It is part of the effective mitigation model and should be analyzed as such.

\subsubsection{Sparse-support tradeoffs}

A further scalability problem remains even after the inverse problem has been stabilized: the full $2^n$-dimensional outcome space may still be too large to represent, store, or search efficiently. In many NISQ experiments, however, the observed bit-string distribution is not spread across the entire space $\{0,1\}^n$, but concentrated on a much smaller subset of outcomes. This motivates support-restricted mitigation, where one solves the correction problem only on an empirically relevant support $S\subset\{0,1\}^n$ rather than on the full outcome space \cite{nation2021scalable,yang2022efficient}. The method works by replacing the full response object with a reduced response object acting only on the retained support. Its attraction is immediate: it reduces both memory cost and classical runtime without changing the basic detector-correction picture.

The approximation, however, has a precise price. Schematically, we can write
\begin{equation}
\|p-p_{\mathrm{mit}}\|_1
\le
\|p-p_S\|_1
+
\|p_S-p_{\mathrm{mit}}\|_1,
\label{eq:readout_support_truncation}
\end{equation}
where $p$ is the target distribution, $p_S$ is its restriction to the retained support, the first term is the truncation error from discarded outcomes, and the second term is the residual modeling and inference error within the reduced space \cite{nation2021scalable,yang2022efficient,peters2023perturbative}. In other words, support restriction does not eliminate error; it redistributes it.

This tradeoff is favorable when the true distribution is genuinely sparse, as is often the case for structured circuits or low-entropy outputs. Then the discarded outcomes carry little physical weight, and the reduced problem captures most of the relevant statistics at far lower cost \cite{nation2021scalable,yang2022efficient}. But the same strategy becomes less reliable when the distribution is broad, when rare outcomes matter for the observable of interest, or when the observed support has itself been distorted by readout noise. In such cases, the missing part of the outcome space introduces a systematic truncation bias that cannot be removed simply by solving the reduced problem more accurately.

The broader lesson is that support restriction works best when sparsity is a physical property of the problem rather than an accident of limited sampling. Fig.~\ref{fig:readoutmi_fig3}(a) illustrates how the observed support defines the reduced subspace and the corresponding reduced assignment matrix, Fig.~\ref{fig:readoutmi_fig3}(b) shows the resulting reduction in classical post-processing cost, and Fig.~\ref{fig:readoutmi_fig3}(c) shows large-$n$ benchmark results for scalable readout mitigation \cite{nation2021scalable,yang2022efficient}.

\subsubsection{Correlated models, detector drift, and nonstationarity}

Support restriction addresses the size of the outcome space, but it does not by itself address correlations in the detector response. The problem is that the error probability for one measured qubit can depend on the state or measurement outcome of other qubits, due to measurement crosstalk, spectator effects, shared readout hardware, or thresholding and amplification effects. In this case, a local tensor-product response matrix can give a scalable but biased correction, because the true assignment process is not a product of independent single-qubit channels \cite{geller2021toward,bravyi2021mitigating,maciejewski2021modeling,barron2020measurement,dangwal2023varsaw}. The method is to use structured correlated models, including pair-correlation diagnostics, finite-correlation-volume methods, CTMP generators, and DDOT-style reconstructions, which bridge the gap between a purely local model and a fully dense detector description \cite{geller2021toward,bravyi2021mitigating,maciejewski2021modeling}. These methods work by learning only the dominant low-order or locality-restricted correlations, rather than reconstructing the entire $2^n\times2^n$ assignment matrix.

This is often a reasonable assumption, but it is not guaranteed. If the true detector contains long-range, context-dependent, or otherwise complicated correlations, then even a carefully designed structured model may remain incomplete or expensive to calibrate \cite{geller2021toward,bravyi2021mitigating,maciejewski2021modeling,cai2023quantum}. Moreover, the effective correlation graph inferred from readout data need not coincide in any simple way with the hardware connectivity graph \cite{maciejewski2021modeling}. Detector complexity therefore cannot always be read off from device geometry alone; it must itself be inferred experimentally. That requirement adds a further layer of calibration overhead.

Even when the correlation structure is well captured, another limitation remains: detector nonstationarity. Readout mitigation implicitly assumes that the detector calibrated during the characterization stage is still the detector encountered by the target circuit. In practice, however, this need not be true. The measurement chain can drift because of changes in operating point, amplifier gain, threshold settings, residual excitation, thermal conditions, or relaxation during the readout window \cite{chen2019detector,maciejewski2020mitigation,aasen2024readout}. If calibration and application are separated by too much time, the inferred response matrix can become stale.

This point matters especially for experimental practice. From an asymptotic viewpoint, one might count only the nominal number of calibration circuits. From a hardware viewpoint, however, recurring recalibration may dominate the real cost of mitigation \cite{cai2023quantum,aasen2024readout}. Detector drift therefore ties scalability directly to runtime and scheduling: the longer or more distributed the workload, the more difficult it becomes to guarantee that a single calibration remains valid \cite{cai2023quantum,aasen2024readout,santos2025driftresilient,lee2025personalizedReadout}.

\subsubsection{Scope of applicability and residual limitations}

The preceding discussion shows that readout mitigation is limited not only by calibration cost and numerical stability, but also by scope. The basic problem is that readout mitigation acts on the final measurement layer. It corrects the classical distortion between the ideal computational-basis distribution and the reported bit strings, but it does not by itself undo coherent gate errors, control imperfections, non-Markovian evolution, decoherence during the circuit, or state-preparation errors, unless those effects are deliberately absorbed into a larger effective model \cite{maciejewski2020mitigation,geller2020rigorous,cai2023quantum}. Consequently, even a perfectly calibrated detector response cannot recover the ideal output distribution if the pre-measurement quantum state has already been significantly degraded before readout.

The corresponding practical method is to treat readout mitigation as one layer in a larger mitigation stack rather than as a complete correction of the experiment. When measurement noise is the dominant contribution to the total error budget, readout mitigation can produce substantial improvements. Once gate errors, decoherence, or state-preparation errors dominate, however, the improvement naturally saturates: correcting the detector more accurately does not translate into an equally accurate final observable, because the main error source lies earlier in the computation \cite{cai2023quantum}. In this regime, readout mitigation should be combined with other methods, such as noise-aware compilation, dynamical decoupling, zero-noise extrapolation, symmetry-based filtering, or probabilistic error cancellation, depending on which physical error source limits the target observable.

Observable-level and model-free methods modify this scope by narrowing the target of mitigation. If the experimental goal is only to estimate a few expectation values, then reconstructing a full mitigated probability distribution may be unnecessary \cite{funcke2022measurement,van2022model}. These methods work by correcting the measured estimator itself: operator-level correction rules or directly measurable multiplicative bias factors convert a noisy expectation value into an estimate of the corresponding ideal value. Their advantage is efficiency, because they avoid full distribution reconstruction. Their limitation is specialization, because they typically correct only a restricted class of outputs and rely on assumptions such as a classical bit-flip readout model, weak noise, approximate locality, or a well-defined target observable structure \cite{funcke2022measurement,van2022model}.

A similar qualification applies to symmetrization methods such as bit-flip averaging. These methods work by randomizing pre-measurement $X$ gates and classically relabeling the outcomes, thereby simplifying or symmetrizing the effective readout channel seen by the data \cite{smith2021qubit}. This can reduce systematic readout bias and lower calibration sensitivity, but it does not eliminate the need for calibration or for a residual noise model. It changes the cost structure and bias structure of readout mitigation rather than abolishing the underlying measurement error.

In summary, readout mitigation does not fail to scale in an absolute sense. Its scalability and reliability are conditional and workload-dependent \cite{bravyi2021mitigating,nation2021scalable,yang2022efficient,peters2023perturbative,cai2023quantum}. Full response-matrix methods are limited by exponential representation and calibration cost. Local approximations are limited by unmodeled correlations. Structured correlated models are limited by calibration overhead and modeling assumptions. Sparse-support methods are limited by truncation bias. Observable-level strategies are limited by the restricted class of outputs they correct. The practical objective is therefore not to remove all limitations at once, but to choose the simplest detector model, correction target, and inference strategy that still capture the dominant readout structure relevant to the scientific quantity being measured.

For present-day devices, the practical default is therefore not full detector tomography on the complete register. Rather, one usually mitigates only the measured subsystem or observable with the lowest-complexity model that passes validation: local assignment matrices when readout correlations are weak, CTMP or overlapping correlated models when calibration data reveal crosstalk, support-restricted solvers when the output distribution is sparse, and observable-level correction when only a small set of expectation values is required \cite{bravyi2021mitigating,geller2021toward,maciejewski2021modeling,nation2021scalable,funcke2022measurement,van2022model,yang2022efficient,peters2023perturbative}. This makes readout mitigation a hardware-adaptive layer in the experimental workflow rather than a universal post-processing formula.

\section{Error mitigation through reduced sampling cost}\label{sec:sampling}

In many quantum-computing applications, the limiting factors include not only imperfect state preparation, limited circuit depth, and hardware noise, but also the number of measurements required to extract useful information from repeatedly prepared quantum states. This issue is especially acute in variational quantum algorithms, where the objective function is typically a weighted sum of many Pauli observables and the estimation of expectation values can dominate the total runtime
\cite{preskill2018quantum,peruzzo2014variational,mcclean2016theory,cerezo2021variational,huggins2021efficient,wang2021minimizing,koh2022foundations,gonthier2022measurements}. Similar constraints arise in Hamiltonian learning, device characterization, many-body state diagnostics, and quantum simulation, where one may wish to estimate energies, correlation functions, fidelities, purities, entanglement measures, or topological indicators from the same experimental state
\cite{dasilva2011practical,bairey2019learning,brydges2019probing,elben2020many}.

From this perspective, error mitigation should not be understood only as suppressing, extrapolating, or inverting physical noise channels. It can also mean reducing the measurement burden required to reach a useful statistical precision. A protocol that extracts more observables from the same experimental data effectively increases the information obtained per state preparation, thereby reducing the sampling cost of a target computation or diagnostic task. This form of mitigation is especially relevant when the hardware can prepare the desired state with reasonable fidelity, but the number of observables is large enough that direct term-by-term measurement becomes impractical.

Thus, this section reviews such measurement-efficient mitigation strategies from this sampling-cost perspective. Classical shadows and randomized-measurement protocols provide a natural framework for this goal. In these methods, the quantum state is repeatedly measured in randomly chosen bases, and the resulting bit strings and measurement settings are stored as a reusable classical data set. Many observables can then be estimated in post-processing without rerunning the quantum experiment for each observable separately
\cite{huang2020predicting,elben2023randomized,cieslinski2024analysing}. The key advantage is therefore not that randomized measurements remove physical noise by themselves, but that they reduce the number of experimental repetitions needed to estimate a large family of quantities. In this sense, they mitigate the statistical component of the total error budget. 

On current cloud-accessible quantum processors, this sampling cost has a direct hardware meaning. Each distinct measurement basis usually corresponds to a distinct compiled circuit or circuit batch, and each additional shot consumes state-preparation time, measurement time, queue or runtime resources, and classical post-processing bandwidth. Reduced-sampling methods are therefore practically useful not only because they improve asymptotic sample complexity, but also because they reduce the number of distinct experimental settings that must be scheduled, calibrated, executed, and stored on a real backend \cite{verteletskyi2020measurement,huggins2021efficient,gonthier2022measurements,hadfield2022measurements,javadi2024quantum,laroseMitiqSoftwarePackage2022a}. This gives the sampling problem an explicitly hardware-facing interpretation: the goal is to maximize the useful observable information extracted per prepared noisy state.

\subsection{Motivation}
\label{subsec:reduced_sampling_motivation}

A major bottleneck in current quantum experiments is the sampling cost required to estimate many observables from repeatedly prepared quantum states. This issue is especially visible in variational quantum algorithms and Hamiltonian-estimation tasks, where the objective function is often decomposed into many Pauli terms. If each term, or each mutually incompatible group of terms, is measured separately, the number of circuit repetitions can become comparable to or even larger than the cost of state preparation itself
\cite{verteletskyi2020measurement,huggins2022nearly,hadfield2022measurements}. For near-term hardware, where preparing a useful many-qubit state may already require a deep and noisy circuit, reusing each prepared state as efficiently as possible becomes an important part of the overall mitigation strategy.

This perspective shifts the emphasis from estimating one observable at a time to designing measurement protocols that preserve broadly reusable information about the prepared state. The relevant question is no longer only how to measure a single operator with minimal variance, but how to construct a measurement record from which many observables, diagnostics, or order parameters can be estimated in post-processing. Reduced-sampling protocols are motivated precisely by this change of viewpoint: they mitigate the measurement layer of the resource budget by increasing the amount of useful information extracted per experimental run.

\begin{figure*}
    \centering
    \includegraphics[width=0.95\linewidth]{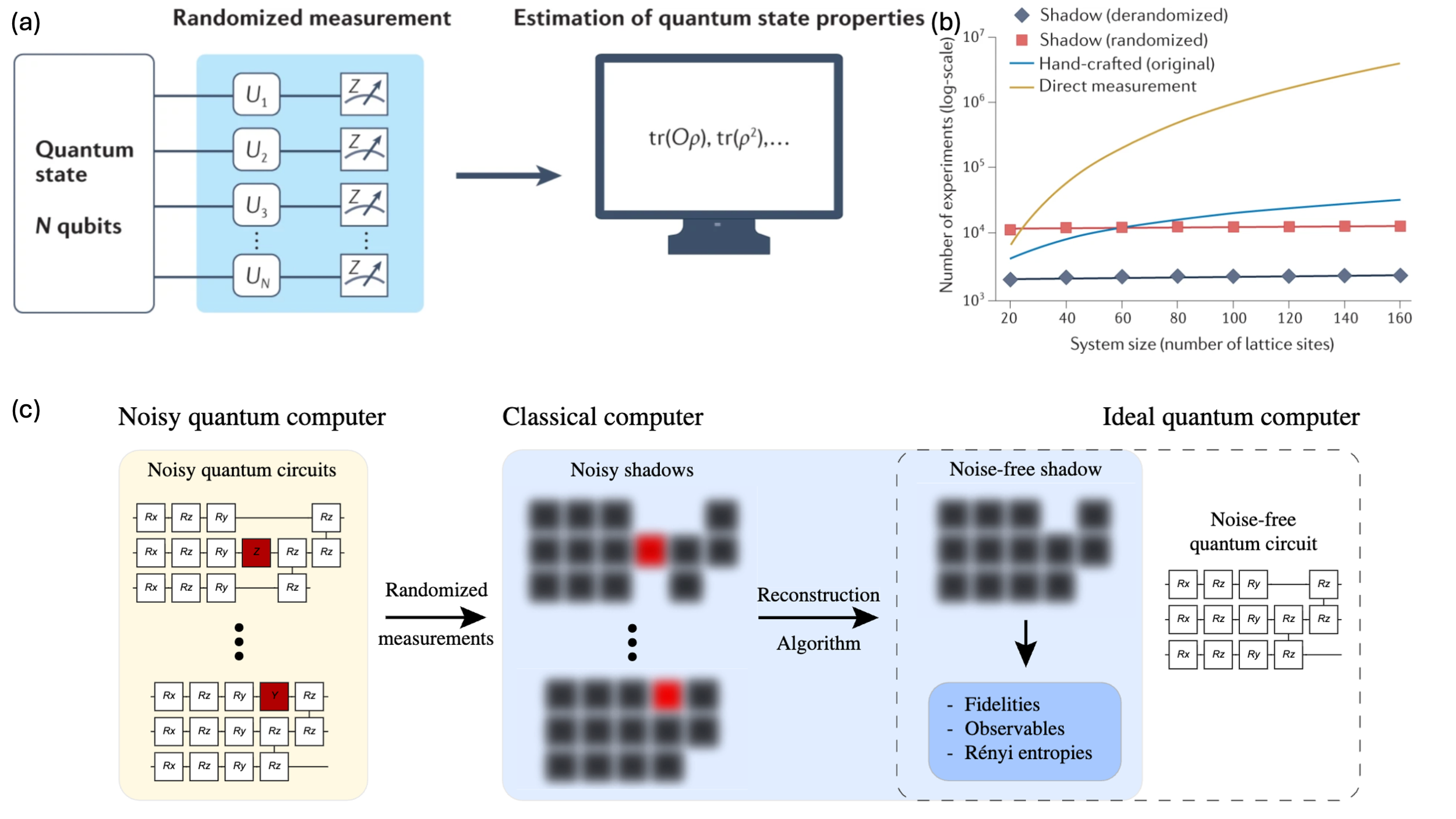}
    \caption{Classical shadows and randomized measurements for reduced-sampling quantum property learning. (a) In a classical-shadow workflow, a quantum state $\rho$ is repeatedly measured after randomly chosen basis rotations, and the resulting outcomes are post-processed into classical snapshots that can later be reused to predict many target properties from the same measurement record. (b) Compared with direct measurement and hand-crafted strategies, shadow-based protocols can require substantially fewer experiments for multi-observable estimation tasks; in particular, for the Schwinger-model Hamiltonian-variance estimation example shown here, the required number of randomized measurements scales only logarithmically with system size, while derandomization can further reduce the cost for target-aware tasks. (c) On noisy hardware, randomized-measurement data can be combined with error-mitigation layers to infer properties of an ideal target state, thereby linking reduced-sampling protocols with the broader quantum-error-mitigation framework. Panel (a) is adapted from Fig.~1 of Ref.~\cite{elben2023randomized}, panel (b) from Fig.~7 of Ref.~\cite{elben2023randomized}, and panel (c) from Fig.~1 of Ref.~\cite{jnane2024quantum}.}
    \label{fig:reduced_sampling}
\end{figure*}

Classical shadows provide the clearest example of this idea. Instead of choosing a separate measurement basis for each target observable, one repeatedly applies randomized measurements and stores the resulting basis choices and bit strings as a classical data set. Observables can then be selected and estimated later, after the quantum experiment has already been completed
\cite{huang2020predicting,aaronson2018shadow,zhang2021experimentalCS}. This ``measure first, predict later'' workflow is illustrated in Fig.~\ref{fig:reduced_sampling}(a), where randomized basis rotations followed by computational-basis readout produce a collection of shadow snapshots that can be queried for many downstream tasks.

It is useful to clarify what is being mitigated in this setting. Classical shadows and related randomized-measurement protocols do not, by themselves, remove gate noise, readout noise, or decoherence. Instead, they reduce the number of distinct experimental settings and state preparations needed to estimate a large family of quantities to a target precision. They therefore mitigate the statistical and measurement-allocation component of the total error budget, and can be combined with more conventional noise-mitigation layers such as readout correction, symmetry verification, or extrapolation
\cite{huang2020predicting,elben2023randomized}.

This sampling-cost viewpoint also connects classical shadows to the broader randomized-measurement toolbox. Randomized measurements have been used to estimate entanglement entropies and many-body correlations, extract topological invariants, and diagnose scrambling from statistical correlations in measurement outcomes
\cite{elben2019statistical,brydges2019probing,elben2020many,garcia2021quantum,elben2023randomized,levy2024shadowQPT,kunjummen2023shadowProcess}. These applications share the same operational advantage: a suitably randomized data set can be reused for many observables or nonlinear diagnostics that would otherwise require more specialized measurement protocols.

The advantage is not universal. Reduced-sampling protocols are most effective when many observables are queried from the same state and when the relevant shadow norms or estimator variances remain moderate. For highly structured Hamiltonians, sparse observable sets, or observables with a large shadow norm under a chosen measurement ensemble, specialized grouping, derandomized measurement design, or locally biased measurement strategies can outperform fully generic shadow protocols
\cite{huang2020predicting,huggins2022nearly,hadfield2022measurements,elben2023randomized,huang2021efficient,wu2023overlapped,hadfield2021adaptivePauli,zhao2021fermionic,hu2023classical,bu2024pauliInvariant,grier2024sampleOptimal,west2025real}. Thus, the role of reduced-sampling mitigation is not to replace observable-specific measurement optimization, but to provide a flexible and reusable measurement layer when the target information is broad.

\subsection{Classical shadows and randomized measurement protocols}

\subsubsection{Basic protocol and origin of the sampling advantage}

The motivation discussed above leads naturally to the classical-shadow workflow: instead of assigning a separate measurement circuit to each observable, one first constructs a reusable randomized measurement record and only later decides which observables to estimate from it. Operationally, the protocol repeatedly prepares the same quantum state $\rho$, applies a randomly chosen measurement basis rotation $U$, and measures in the computational basis \cite{huang2020predicting,elben2023randomized}. Each experimental shot produces a pair $(U,b)$, where $b$ is the observed bit string.

The pair $(U,b)$ is then converted into a classical snapshot by applying the inverse of the average measurement channel,
\begin{equation}
\hat{\rho}(U,b)
=
\mathcal{M}^{-1}
\left(
U^\dagger |b\rangle\langle b| U
\right),
\label{eq:shadow_snapshot}
\end{equation}
where $\mathcal{M}$ is the measurement map induced by the chosen ensemble of random bases. The central property is unbiasedness,
\begin{equation}
\mathbb{E}_{U,b}\!\left[\hat{\rho}(U,b)\right]
=
\rho.
\label{eq:shadow_unbiased}
\end{equation}

Therefore, for any observable $O$, the same collection of snapshots gives the estimator
\begin{equation}
\widehat{\langle O\rangle}
=
\frac{1}{N}
\sum_{s=1}^{N}
\mathrm{Tr}\!\left[
O\,\hat{\rho}_s
\right],
\label{eq:shadow_observable_estimator}
\end{equation}
where $\hat{\rho}_s$ is the snapshot obtained from the $s$-th experimental shot. This is the operational content of Fig.~\ref{fig:reduced_sampling}(a): the quantum device is used to generate a randomized measurement data set, and many observables are subsequently estimated by classical post-processing. The sampling advantage is therefore an amortization advantage. It does not come from making each circuit less noisy, but from reusing the same experimental data for many target quantities.

This advantage can be stated more quantitatively. For a collection of observables $\{O_i\}_{i=1}^{M}$, classical-shadow theory gives a sample complexity of the form
\begin{equation}
N
=
O\!\left[
\frac{\log(M/\delta)}{\epsilon^2}
\max_i \|O_i\|_{\mathrm{sh}}^2
\right],
\label{eq:shadow_sample_complexity}
\end{equation}
where $\epsilon$ is the target additive accuracy, $\delta$ is the failure probability, and $\|O_i\|_{\mathrm{sh}}$ is the shadow norm determined jointly by the observable and the measurement ensemble \cite{huang2020predicting}. Equation~\eqref{eq:shadow_sample_complexity} identifies both the strength and the limitation of the method. The dependence on the number of queried observables is only logarithmic, which is why shadows are powerful for multi-observable prediction. However, the prefactor is controlled by the shadow norm, so the cost can still be high for observables poorly matched to the chosen measurement ensemble.

This distinction is important for interpreting classical shadows as a reduced-sampling mitigation strategy. They are most useful when many observables, Hamiltonian terms, correlation functions, or diagnostic quantities are estimated from the same state. In that regime, a single randomized data set can replace many separate measurement campaigns. For nonlinear quantities, such as purities, entanglement diagnostics, and scrambling measures, the same randomized-measurement logic is used in a slightly different way: the estimator is built from correlations between multiple randomized measurement outcomes or multiple shadow snapshots rather than from the linear estimator in Eq.~\eqref{eq:shadow_observable_estimator} alone \cite{elben2019statistical,brydges2019probing,garcia2021quantum,elben2023randomized}.

\subsubsection{Implementation on present-day hardware}

On gate-based devices, a classical-shadow experiment is implemented by compiling the randomized measurement basis into basis-rotation gates followed by the native computational-basis readout. For local Pauli shadows, each qubit is assigned a basis from $\{X,Y, Z\}$ in each shot or circuit batch. Operationally, an $X$-basis measurement is implemented by applying an appropriate single-qubit rotation before the standard $Z$-basis measurement, a $Y$-basis measurement is implemented by a phase rotation followed by an $X$-basis rotation, and a $Z$-basis measurement requires no additional basis change. Thus, local Pauli shadows do not require a new measurement apparatus; they require shallow pre-measurement rotations, accurate bookkeeping of the chosen basis string, and storage of the measured bit string for each shot \cite{huang2020predicting,elben2023randomized,zhang2021experimentalCS}.

This implementation has a favorable hardware profile because the extra quantum depth is usually only a layer of single-qubit gates. The main overhead is instead organizational and classical: generating randomized or derandomized basis strings, batching circuits with identical or similar measurement settings, submitting the resulting circuit batches to the backend, storing the pairs $(U,b)$, and evaluating many observable estimators in post-processing \cite{huang2020predicting,elben2023randomized,wu2023overlapped,fu2024improvedMoM}. For superconducting and trapped-ion processors, this makes local Pauli shadows comparatively easy to deploy, since both platforms already support high-fidelity single-qubit rotations followed by standard computational-basis readout \cite{kjaergaard2020superconducting,bruzewicz2019trapped,monroe2021programmable,javadi2024quantum}.

The choice of measurement ensemble is therefore hardware-dependent. Global Clifford shadows have strong theoretical guarantees for broad observable families, but implementing a random global Clifford generally requires many entangling gates and substantial compilation overhead. On noisy near-term devices, this extra depth can erase part of the sampling advantage by introducing additional gate errors, decoherence, leakage, and crosstalk. Local Pauli, locally biased, derandomized, or shallow-shadow variants are often more practical because they trade some universality for lower circuit depth and better compatibility with hardware connectivity \cite{huang2021efficient,hadfield2022measurements,wu2023overlapped,helsen2023thrifty,bertoni2024shallow,ippoliti2024classical,brieger2025stability}. The practical optimization is therefore not to minimize the ideal shadow norm alone, but to minimize the total estimation error after including both sampling variance and implementation noise.

\subsubsection{Task-aware measurement design}

The basic shadow protocol is deliberately generic. Once the target observable family is known, however, one can often reduce the sampling cost further by adapting the measurement ensemble to the task. This leads to a second level of the hierarchy: from universal randomized measurements to task-aware randomized or derandomized measurement design.

Locally biased classical shadows use prior information about the target Hamiltonian, observable weights, or a reference state to bias the distribution of local Pauli measurement settings. The purpose is to reduce the variance of the most important observables without increasing the depth of the state-preparation circuit \cite{hadfield2022measurements}. Derandomized classical shadows go further by replacing random local basis choices with deterministic measurement settings chosen for a specified observable family. In this case, the measurement schedule is no longer universal, but it can perform substantially better for the target task, especially when the relevant observables contain high-weight Pauli strings or nonuniform Pauli structure \cite{huang2021efficient}.

Building on this idea, tensor-network-based derandomization schemes optimize shallow measurement circuits gate by gate to reduce the sample complexity for a prescribed Pauli-string target set \cite{vankirk2024derandomized}. A related but distinct strategy is to bias the measurement ensemble using known symmetries or gauge constraints of the target state rather than observable weights alone, which can yield large sample-complexity improvements for symmetry- or gauge-invariant observables at the cost of more complex circuits \cite{sauvage2024symmetries,bringewatt2026gauge}.

Other variants adjust the tradeoff between experimental overhead, circuit depth, and estimator variance. Thrifty shadow estimation reduces overhead by reusing random circuits or measurement data more economically, while shallow-shadow and tensor-network shadow constructions interpolate between strictly local Pauli measurements and deeper Clifford-type ensembles \cite{helsen2023thrifty,akhtar2023scalable,bertoni2024shallow,vankirk2024derandomized}. Locally entangled measurement bases built from few-body Bell- or GHZ-type rotations offer a hardware-efficient intermediate point that can outperform both local Pauli and shallow-Clifford shadows on certain Pauli-estimation tasks \cite{ippoliti2024classical}, while holographic constructions based on tree or hierarchical tensor-network circuits achieve optimal sample-complexity scaling for geometrically local operators at any length scale \cite{zhang2025holographic}.

These developments sharpen the interpretation of Fig.~\ref{fig:reduced_sampling}(b). The observed reduction in sampling cost should not be read as the consequence of a single protocol, but as the manifestation of a broader design principle: once the target observable class is partially known, measurement design can be adapted to trade universality for lower variance. Local Pauli shadows are flexible and shallow; biased or derandomized shadows are more target-specific; shallow Clifford-type shadows can improve access to less local information but introduce additional circuit noise. The optimal choice is therefore determined by the structure of the target observables and by the noise cost of implementing the measurement ensemble.

The same logic also motivates broader shadow ensembles beyond purely local random Pauli measurements. Recent work has explored shallow random Clifford circuits and locally entangled measurement bases, which interpolate between strictly local and more global measurement designs \cite{bertoni2024shallow,ippoliti2024classical,rozon2024optimal}. These ensembles can improve the tradeoff between circuit depth and estimator variance for suitable observable classes, although that improvement introduces new constraints once realistic hardware noise is taken into account \cite{cioli2025approximateinverse}. In other words, a measurement design that lowers the shadow norm may also require additional basis-rotation gates, so the variance gain must be compared with the extra physical noise introduced by the measurement circuit. This observation motivates separating the role of randomized measurements as a sampling-efficient data layer from the separate problem of recovering ideal-state properties on noisy devices.

\subsubsection{Relation to noisy hardware and error mitigation}

The depth--variance tradeoff just described becomes nontrivial on real devices because the snapshots constructed from randomized measurements are snapshots of the state actually sampled by the experiment. In the absence of additional correction, they are unbiased estimators of the noisy output state, not of the ideal target state. Thus, classical shadows and randomized measurements should be viewed primarily as sampling-compression protocols: they reduce the number of experimental settings and state preparations needed to estimate many quantities, but they do not automatically remove gate errors, readout errors, decoherence, or state-preparation errors \cite{huang2020predicting,elben2023randomized}.

This distinction determines how shadows enter an error-mitigation workflow. Randomized measurements provide a reusable data layer, while conventional mitigation procedures act either before the snapshots are formed or during their classical post-processing. Readout mitigation can be incorporated into the measurement outcomes; symmetry verification can filter or reweight snapshots inconsistent with a target symmetry sector; and extrapolation or learning-based procedures can be applied to observables estimated from the shadow data \cite{zhao2024symmetry,hu2022logicalShadow,seif2023shadowDistillation,wu2024fermionicNoisy}. Robust shadow-estimation protocols make this connection more explicit by modifying the calibration and reconstruction map so that, under suitable assumptions on the noise, the resulting shadow estimator can again be made unbiased for ideal-state properties \cite{chen2021robust,koh2022classical,hu2025robustshallow}.

This is the role of Fig.~\ref{fig:reduced_sampling}(c). The figure should not be read as saying that randomized measurements alone constitute full error mitigation. Rather, it illustrates how randomized-measurement data can be embedded inside a larger mitigation stack. In such a stack, the shadow protocol reduces the sampling cost of estimating many properties, while the additional mitigation layer addresses the mismatch between noisy hardware data and ideal target quantities \cite{seif2023shadowDistillation,jnane2024quantum}. The practical message is therefore conceptual: shadows are useful not because they remove physical noise by themselves, but because they create a reusable measurement record to which noise-aware post-processing can be applied. This point leads directly to the quantitative limitations and scalability issues discussed next.

\subsection{Limitations and scalability}

\subsubsection{Observable dependence and limits of the $\log M$ advantage}

The first limitation is intrinsic to the shadow formalism itself. Although Eq.~\eqref{eq:shadow_sample_complexity} shows that the dependence on the number of queried observables $M$ is only logarithmic, the same expression also shows that the cost is controlled by the observable-dependent factor $\max_i\|O_i\|_{\mathrm{sh}}^2$ [Eq.~\ref{eq:shadow_sample_complexity}]. Thus, the $\log M$ dependence should not be interpreted as a uniform guarantee that all large observable families are cheap to estimate \cite{huang2020predicting,elben2023randomized}.

The shadow norm depends on the operator structure, measurement ensemble, support size, and locality of the target observable. For local Pauli observables measured with local Pauli shadows, the prefactor can remain moderate. For high-weight Pauli strings or strongly nonlocal observables, however, the shadow norm can grow rapidly, often exponentially with operator support under local measurement ensembles \cite{huang2020predicting,huang2021efficient,elben2023randomized,ippoliti2024classical}. In that case, the logarithmic dependence on $M$ may be overwhelmed by the observable-dependent prefactor.

This is why task-aware protocols are not merely optional refinements. Locally biased shadows, derandomized measurement schedules, and shallow-shadow constructions should be understood as attempts to reduce the relevant shadow norm for a restricted observable family \cite{hadfield2022measurements,huang2021efficient,bertoni2024shallow}. They trade universality for variance reduction. This tradeoff is favorable when the target observables are known in advance, but less useful when the experiment must remain open-ended and many unrelated observables may be queried later.

\subsubsection{Noise-induced overhead and estimator bias}

A second limitation is that noise changes both the bias and the cost of shadow estimation. In the ideal protocol, the inverse measurement map $\mathcal{M}^{-1}$ converts randomized measurement outcomes into unbiased snapshots of the measured quantum state. On real hardware, however, state-preparation errors, basis-rotation errors, readout errors, and decoherence modify the effective measurement channel. If these effects are not explicitly accounted for, the resulting estimator can remain statistically efficient but become biased relative to the ideal target state \cite{chen2021robust,koh2022classical}.

Robust or noise-aware shadow protocols address this problem by modifying the calibration and reconstruction map. Recent extensions include robust shallow-shadow implementations that learn and mitigate noise in post-processing \cite{hu2025robustshallow}, robust ultra-shallow protocols for low-depth random circuits under noise-invariance assumptions \cite{farias2025robustultrashallow}, and self-calibrating randomized-measurement schemes that infer the noisy measurement channel from the same randomized-measurement data rather than from a separate calibration experiment \cite{onorati2024selfcalibrating}. Under suitable assumptions on the noise, one can replace the ideal inverse map by a corrected inverse map that restores unbiased estimation of the desired ideal quantity \cite{chen2021robust,zhao2024symmetry,hu2025robustshallow}. Related analyses show that, when the relevant noisy channels are sufficiently characterized, classical post-processing can compensate for the noise at the estimator level \cite{koh2022classical,hingane2026real}. These results clarify that classical shadows can be made compatible with error mitigation, but they also show that this compatibility is not automatic.

The price is increased sampling overhead. Inverting a noisy or imperfectly conditioned measurement channel can amplify statistical fluctuations, in close analogy with inverse-channel mitigation and probabilistic error cancellation. Thus, there is a cost--bias tradeoff. Without noise correction, the estimator may have lower variance but can be biased relative to the ideal state. With explicit correction, the estimator can be unbiased under the assumed model, but the variance and required number of shots can increase substantially \cite{chen2021robust,koh2022classical}.

\subsubsection{Depth--noise trade-off in measurement ensembles}

A third limitation concerns the measurement ensemble itself. In noiseless shadow theory, more global or more expressive measurement ensembles can reduce the shadow norm for certain nonlocal observables. For example, shallow random Clifford circuits and locally entangled measurement bases interpolate between strictly local Pauli measurements and fully global Clifford shadows, and can improve the estimation of some extended observables \cite{bertoni2024shallow,ippoliti2024classical,rozon2024optimal,cioli2025approximateinverse}.

On current quantum hardware, however, these measurement ensembles require additional gates before readout. Those gates introduce decoherence, control errors, leakage, and crosstalk. Therefore, a measurement circuit that is information-theoretically favorable in an ideal model need not remain optimal once realistic hardware noise is included \cite{rozon2024optimal,brieger2025stability,hu2025robustshallow}. The relevant optimization is not simply to minimize the ideal shadow norm, but to minimize the total error at a fixed experimental budget, including both statistical variance and noise-induced bias.

This tradeoff is particularly important for nonlocal observables. Local Pauli shadows are shallow and hardware-friendly, but they can have large shadow norms for high-weight operators. More entangling measurement ensembles can reduce this norm, but only by increasing the depth and noise sensitivity of the measurement circuit. The practical optimum therefore depends on the device error rates, the observable family, and the number of quantities to be estimated. In this sense, the measurement ensemble plays a role analogous to an ansatz: it must be chosen not only for formal expressiveness, but also for hardware compatibility.

Taken together, these limitations define the practical scope of mitigation via reduced sampling cost. Classical shadows are most effective when many observables are queried from the same state, when the relevant shadow norms are moderate, and when the measurement ensemble can be implemented without adding excessive hardware noise \cite{huang2020predicting,elben2023randomized}. They are less advantageous when the observable set is small, when specialized grouping already gives a lower-variance estimator, or when the target observables have large shadow norm under the available measurement ensemble.

On noisy devices, this scope narrows further because ideal-state estimation requires additional assumptions, calibration information, and post-processing overhead. The important point is therefore not simply that randomized measurements can be combined with mitigation layers, but that each such combination changes the total resource balance. Sampling compression reduces the number of measurement settings, while noise-aware reconstruction may increase shot cost, estimator variance, or calibration burden. Classical shadows and randomized measurements are therefore best understood as a flexible sampling-compression framework whose practical value depends on matching the measurement ensemble, observable structure, hardware noise, and downstream mitigation target.

In a practical hardware workflow, reduced-sampling protocols should therefore be treated as a measurement-allocation layer rather than as a complete noise-mitigation stack. One first chooses the randomized, locally biased, or derandomized measurement schedule; then executes the corresponding basis-rotation circuits on the device; then applies readout mitigation, symmetry verification, robust-shadow reconstruction, shadow distillation, or extrapolation to the estimators extracted from the same measurement record \cite{smith2021qubit,funcke2022measurement,van2022model,chen2021robust,seif2023shadowDistillation,zhao2024symmetry,hu2022logicalShadow,hu2025robustshallow,wu2024fermionicNoisy}. This ordering separates two roles: the shadow protocol determines which data are collected efficiently, while the downstream mitigation layer determines how noisy hardware data are converted into estimates of ideal target observables.

\section{Noise scaling and extrapolation}
\label{sec:zne}

Noise scaling and extrapolation, more commonly known as \emph{zero-noise extrapolation} (ZNE), is one of the most widely used quantum error-mitigation strategies for NISQ hardware
\cite{liEfficientVariationalQuantum2017a,temmeErrorMitigationShortDepth2017a,endo2018practical,kandalaErrorMitigationExtends2019,giurgica-tironDigitalZeroNoise2020,cai2023quantum,majumdarBestPracticesQuantum2023}.
The central idea is simple: instead of trying to remove noise during a single circuit run, one deliberately runs related versions of the same circuit at different effective noise strengths and then infers what the measured result would have been in the zero-noise limit.
This is attractive because ZNE does not require logical encoding or ancillary qubits, and it directly targets expectation values, which are the main outputs of many near-term quantum algorithms
\cite{endo2018practical,cerezo2021variational,huggins2021efficient,gonthier2022measurements}.

More concretely, ZNE starts from a target circuit and constructs a family of circuits that ideally implement the same computation but experience different amounts of noise on hardware.
The observable of interest is measured for each member of this family, and the resulting data are fitted as a function of the noise strength.
The final estimate is obtained by extrapolating this fitted curve back to the zero-noise point
\cite{temmeErrorMitigationShortDepth2017a,liEfficientVariationalQuantum2017a,kandalaErrorMitigationExtends2019,heZeronoiseExtrapolationQuantumgate2020,giurgica-tironDigitalZeroNoise2020}.
Different implementations realize the noise-scaling axis in different ways, including pulse stretching, global or local unitary folding, identity insertion, probabilistic error amplification after Pauli twirling, hardware-layout variation, noise-aware folding, circuit unoptimization, and layerwise or multivariate scaling
\cite{kandalaErrorMitigationExtends2019,heZeronoiseExtrapolationQuantumgate2020,giurgica-tironDigitalZeroNoise2020,kimScalableErrorMitigation2023a,uvarovMitigatingQuantumGate2024,russoQuantumErrorMitigation2024a,hour2024improving}.
This makes ZNE broadly compatible with Hamiltonian simulation, variational algorithms, observable estimation, and, increasingly, early logical or partially error-corrected workflows
\cite{caiMultiexponentialErrorExtrapolation2021,kimScalableErrorMitigation2023a,zhangDemonstratingQuantumError2026,umbrarescu2026infinitedistance,babukhin2026runtimeefficient}.
At the same time, its performance depends strongly on two practical choices: how faithfully the noise-scaling axis represents the relevant physical error processes, and how stably the extrapolation model can be fitted from finite-shot data
\cite{caiMultiexponentialErrorExtrapolation2021,majumdarBestPracticesQuantum2023,mohammadipourDirectAnalysisZeroNoise2025a,koenigInvertedcircuitZeronoiseExtrapolation2024a,harrisReducingQuantumError2026}.
Recent extensions further show that ZNE is no longer limited to small gate-model demonstrations: it has been applied in utility-scale superconducting experiments, silicon spin-qubit hardware, quantum annealing, and logical-qubit or hybrid physical--logical settings
\cite{kimScalableErrorMitigation2023a,sohnApplicationZeroNoise2025,raymondQuantumErrorMitigation2025,zhangDemonstratingQuantumError2026,umbrarescu2026infinitedistance,babukhin2026runtimeefficient}.

In practice, extrapolation is only the final inference step.
The more consequential issue is whether the constructed noise axis actually tracks the dominant hardware errors over the sampled range.
Existing approaches span several regimes.
In analog noise scaling, gate durations or pulse shapes are modified so that the circuit is exposed to more physical noise while preserving the intended ideal unitary
\cite{temmeErrorMitigationShortDepth2017a,kandalaErrorMitigationExtends2019}.
In digital noise scaling, the ideal circuit is lengthened by inserting gate identities, for example through unitary folding or repeated self-inverse gates
\cite{giurgica-tironDigitalZeroNoise2020,heZeronoiseExtrapolationQuantumgate2020,majumdarBestPracticesQuantum2023}.
In probabilistic error amplification, learned Pauli or Pauli--Lindblad noise models are used to insert calibrated stochastic faults, producing a more controlled scaling of the effective noise channel
\cite{kimEvidenceUtilityQuantum2023c,kimScalableErrorMitigation2023a}. More recent developments have moved beyond nominal scale factors toward hardware-aware or directly measured noise coordinates
\cite{uvarovMitigatingQuantumGate2024,koenigInvertedcircuitZeronoiseExtrapolation2024a,russoQuantumErrorMitigation2024a}.
These extensions reflect a broader shift from treating the noise scale as a simple parameter toward constructing device- and circuit-specific axes that better track the actual implemented noise.
In this section, we review the core principle of ZNE, the main families of noise-scaling protocols, and practical hardware workflows.

\subsection{Core principles, estimator families, and statistical cost}
\label{subsec:zne_core}

At its core, ZNE converts hardware noise into an inference axis. Rather than attempting to identify and correct every microscopic error process, one measures the same target observable along a family of ideally equivalent implementations with different effective noise strengths and then estimates the zero-noise intercept
\cite{temmeErrorMitigationShortDepth2017a,liEfficientVariationalQuantum2017a,endo2018practical,cai2023quantum,shen2025robust}. 
This viewpoint is useful because most NISQ applications report expectation values rather than full quantum states. If the observable of interest is $O$, the experimental data may be viewed schematically as samples of a noise-dependent response curve,
\begin{equation}
E(\lambda)
=
\operatorname{Tr}\!\left[
O\,\mathcal{N}_{\lambda}\circ\mathcal{U}(\rho_0)
\right],
\label{eq:zne_noisy}
\end{equation}
where $\mathcal{U}$ denotes the ideal circuit, $\mathcal{N}_{\lambda}$ is the effective noise channel along the chosen scaling path, and $\lambda=0$ corresponds to the ideal value. ZNE then estimates $E(0)$ from measurements at $\lambda\geq \lambda_0$, where $\lambda_0$ is the native device-noise level.

The essential assumption is not simply that the noise is weak, but that the chosen scaling procedure generates a smooth and physically meaningful trajectory in noise space. In the perturbative regime, one often writes
\begin{equation}
E(\lambda)
=
E(0)+a_1\lambda+a_2\lambda^2+\cdots ,
\label{eq:zne_series}
\end{equation}
which motivates polynomial or Richardson-type extrapolation. However, this expansion should be understood as a local approximation to a particular experimental scaling path, not as a universal property of the device. Different scaling procedures may amplify different combinations of relaxation, dephasing, coherent over-rotation, crosstalk, leakage, and readout-induced bias. If the scaling path changes the character of the noise rather than only its strength, the extrapolated intercept can be biased even when the fitted curve appears smooth
\cite{majumdarBestPracticesQuantum2023,schultzImpactTimecorrelatedNoise2022,mohammadipourDirectAnalysisZeroNoise2025a}.

Several estimator families are used in practice. Richardson extrapolation and low-order polynomial fits are the most common because they make minimal assumptions about the detailed noise model and are straightforward to implement from a small number of noise-scaled data points
\cite{temmeErrorMitigationShortDepth2017a,giurgica-tironDigitalZeroNoise2020,majumdarBestPracticesQuantum2023}. 
For scale factors $s_i=\lambda_i/\lambda_0$, a Richardson estimator combines the measured values linearly,
\begin{equation}
\widehat{E}_{\mathrm{ZNE}}
=
\sum_i \gamma_i \widehat{E}(s_i\lambda_0),
\qquad
\sum_i \gamma_i s_i^m=\delta_{m0},
\label{eq:zne_richardson_1}
\end{equation}
so that the first few powers of the noise strength are cancelled. This construction is appealing because it is transparent and unbiased up to the targeted perturbative order, but it can also strongly amplify shot noise when the coefficients $\gamma_i$ become large.

Exponential and multi-exponential estimators are motivated by situations in which the measured observable behaves more like a decay curve than a low-order polynomial over the accessible noise range
\cite{caiMultiexponentialErrorExtrapolation2021,kimScalableErrorMitigation2023a}. 
Such models can be effective when the dominant noisy evolution is approximately stochastic or Pauli-like, as in probabilistic error amplification after twirling. They are less reliable when coherent errors, leakage, or nonstationary noise produce a response that is not well described by a simple decay law. Thus, the choice between polynomial and exponential extrapolation is not merely a numerical fitting preference; it reflects an assumption about the effective noise model sampled by the scaling protocol.

The statistical cost of ZNE follows from the same linear-combination structure that makes extrapolation possible. For a linear estimator,
\begin{equation}
\widehat{E}_{\mathrm{ZNE}}
=
\sum_i \gamma_i \widehat{E}_i ,
\end{equation}
the variance scales as
\begin{equation}
\operatorname{Var}
\!\left[
\widehat{E}_{\mathrm{ZNE}}
\right]
=
\sum_i \gamma_i^2
\operatorname{Var}\!\left[
\widehat{E}_i
\right],
\end{equation}
assuming statistically independent measurements at the different noise levels. This expression summarizes the main bias--variance tradeoff. Higher-order extrapolation can reduce systematic bias from truncating the noise expansion, but it usually increases the coefficient norm and therefore increases the number of shots required to reach a fixed statistical precision
\cite{majumdarBestPracticesQuantum2023,mohammadipourDirectAnalysisZeroNoise2025a}. Conversely, low-order fits are statistically stable but may leave larger residual bias.

Modern ZNE practice is therefore less about applying a fixed formula and more about balancing three linked choices: the noise-scaling method, the extrapolation model, and the shot-allocation strategy. Reliable applications typically keep the scaled circuits within a regime where the observable has not saturated, sample enough noise levels to diagnose curvature, interleave data acquisition to reduce drift-induced bias, and report the fit model and uncertainty-estimation procedure
\cite{majumdarBestPracticesQuantum2023,kimScalableErrorMitigation2023a,harrisReducingQuantumError2026}. 
In this sense, ZNE is best understood as a family of experimentally calibrated inference protocols. Its success depends not only on the mathematical extrapolator, but also on whether the constructed noise axis remains faithful to the hardware errors that actually limit the target observable.

\begin{figure*}[t]
    \centering
    \includegraphics[width=0.85\textwidth]{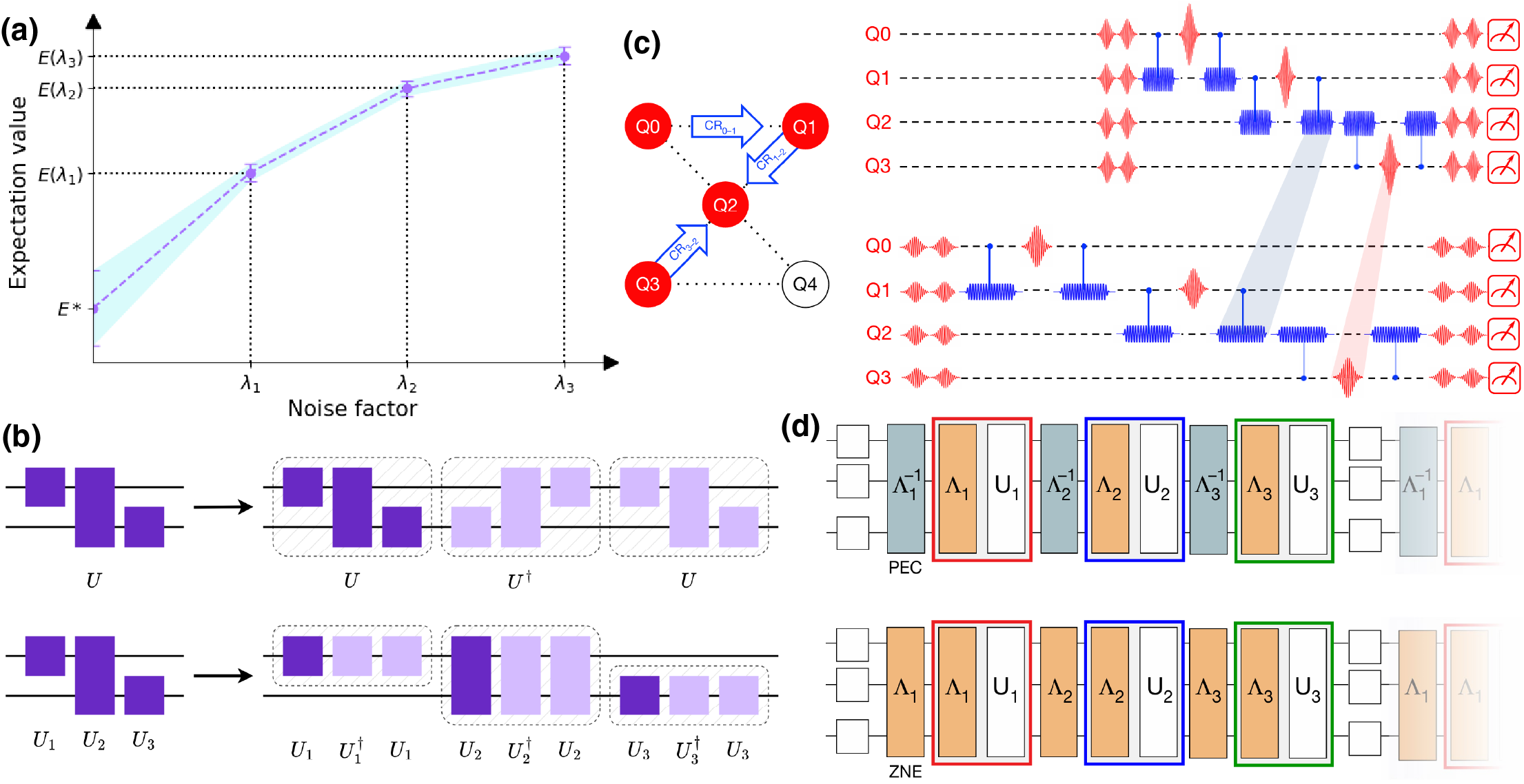}
    \caption{Foundational views of noise scaling and extrapolation.  
    (a) The conceptual ZNE picture: estimate an expectation value at several amplified noise levels and extrapolate to the zero-noise intercept.
    (b) Two standard digital scaling constructions, namely global and local unitary folding.
    (c) Pulse-level noise scaling via calibrated pulse stretching, where the intended ideal gate is restored while the exposure to decoherence is increased.
    (d) Probabilistic error amplification after Pauli twirling, which illustrates a calibrated effective-channel route to noise scaling and enables utility-scale experiments.
    Adapted from Figs.~1 and 3 of Ref.~\cite{majumdarBestPracticesQuantum2023}, Fig.~1(b) of Ref.~\cite{kandalaErrorMitigationExtends2019}, and Fig.~1(d) of Ref.~\cite{kimEvidenceUtilityQuantum2023c}.}
    \label{fig:zne_overview}
\end{figure*}

\subsection{Extrapolation models}
\label{subsec:zne_extrapolant}

Once measurements have been obtained at several effective noise levels, ZNE becomes a classical inference problem. We write the amplified noise strengths as
\begin{equation}
    \lambda_j=s_j\lambda_0,
    \qquad
    s_j\geq1,
\end{equation}
where $\lambda_0$ denotes the native noise level and $s_j$ the nominal relative scale factor
\cite{temmeErrorMitigationShortDepth2017a,liEfficientVariationalQuantum2017a,
endo2018practical,giurgica-tironDigitalZeroNoise2020,
majumdarBestPracticesQuantum2023}.
The qualifier ``nominal'' is important: $s_j$ specifies a point along the chosen experimental scaling path, rather than a complete microscopic rescaling of every hardware error mechanism. The appropriate extrapolation model must therefore be interpreted together with the physical procedure used to generate the scaled circuits.

\begin{itemize}

\item \textbf{Polynomial and Richardson extrapolation.}
In the weak-noise regime, the observable can often be expanded perturbatively as
\begin{equation}
    E(\lambda)
    =
    E^\star
    +
    a_1\lambda
    +
    a_2\lambda^2
    +
    \cdots,
    \label{eq:zne_poly_expansion}
\end{equation}
where $E^\star=E(0)$ is the desired zero-noise value. Evaluating this expression at $\lambda_j=s_j\lambda_0$ gives
\begin{equation}
    E(s_j\lambda_0)
    =
    E^\star
    +
    \sum_{m\geq1}
    a_m s_j^m\lambda_0^m .
\end{equation}
Richardson extrapolation forms the estimator
\begin{equation}
    \widehat{E}_{\mathrm{R}}^\star
    =
    \sum_{j=0}^{K}
    \gamma_j\,\widehat{E}(s_j\lambda_0),
    \label{eq:zne_richardson_2}
\end{equation}
with coefficients satisfying
\begin{equation}
    \sum_{j=0}^{K}\gamma_j s_j^m
    =
    \delta_{m0},
    \qquad
    m=0,1,\ldots,K.
    \label{eq:zne_constraints}
\end{equation}
The $m=0$ condition preserves the zero-noise contribution, while the remaining constraints cancel the first $K$ orders in $\lambda_0$
\cite{temmeErrorMitigationShortDepth2017a,liEfficientVariationalQuantum2017a,
endo2018practical,krebsbachOptimizationRichardsonExtrapolation2022}.

For distinct scale factors, the corresponding weights are
\begin{equation}
    \gamma_j
    =
    \prod_{\substack{m=0\\m\neq j}}^{K}
    \frac{s_m}{s_m-s_j},
    \label{eq:zne_lagrange_weights}
\end{equation}
which are the Lagrange interpolation weights evaluated at the zero-noise point. The simplest case is the two-point linear estimator,
\begin{equation}
    \widehat{E}_{\mathrm{lin}}^\star
    =
    \frac{
        s_1\widehat{E}(\lambda_0)
        -
        \widehat{E}(s_1\lambda_0)
    }{
        s_1-1
    }.
    \label{eq:zne_linear}
\end{equation}
Low-order extrapolation is often preferred experimentally because increasing $K$ reduces perturbative truncation error but generally increases the magnitude of the coefficients $\gamma_j$, thereby amplifying finite-shot fluctuations and sensitivity to deviations from the assumed polynomial response
\cite{krebsbachOptimizationRichardsonExtrapolation2022,
majumdarBestPracticesQuantum2023,
mohammadipourDirectAnalysisZeroNoise2025a}.
When more scale factors are available, lower-degree least-squares fits can provide a more stable alternative to exact high-order interpolation.

\item \textbf{Exponential and multi-exponential models.}
Polynomial fitting is not always the most natural description of the scaled-noise response. For approximately stochastic Pauli, depolarizing, or Lindblad-type noise, an observable may instead exhibit an approximately exponential decay,
\begin{equation}
    E(s)
    \approx
    c_0+c_1e^{-bs},
    \label{eq:zne_exp}
\end{equation}
or, more generally,
\begin{equation}
    E(s)
    \approx
    c_0+\sum_{\ell=1}^{M}c_\ell e^{-b_\ell s}.
    \label{eq:zne_multi_exp}
\end{equation}
The zero-noise estimate is then obtained by evaluating the fitted model at $s=0$.
Multi-exponential forms can capture several decay modes and are particularly natural when the effective noise has been tailored toward a stochastic Pauli description
\cite{caiMultiexponentialErrorExtrapolation2021,
kimEvidenceUtilityQuantum2023c,kimScalableErrorMitigation2023a}.

These models should nevertheless be used only when the scaled data support such a decay structure. If the amplification procedure changes the composition of the noise, introduces leakage, modifies coherent interference, or probes different parts of a time-correlated noise spectrum, a simple exponential model can yield a biased intercept
\cite{majumdarBestPracticesQuantum2023,
schultzImpactTimecorrelatedNoise2022,
mohammadipourDirectAnalysisZeroNoise2025a}.

\item \textbf{Statistical cost and shot allocation.}
For any linear extrapolation estimator, finite-shot fluctuations are amplified by the extrapolation coefficients. Assuming statistically independent estimates at different noise scales,
\begin{equation}
    \operatorname{Var}
    \!\left[
        \widehat{E}_{\mathrm{R}}^\star
    \right]
    =
    \sum_{j=0}^{K}
    \gamma_j^2
    \operatorname{Var}
    \!\left[
        \widehat{E}(s_j\lambda_0)
    \right]
    \approx
    \sum_{j=0}^{K}
    \gamma_j^2
    \frac{\sigma_j^2}{N_j},
    \label{eq:zne_variance}
\end{equation}
where $N_j$ is the number of shots and $\sigma_j^2$ the corresponding per-shot variance
\cite{temmeErrorMitigationShortDepth2017a,
krebsbachOptimizationRichardsonExtrapolation2022,cai2023quantum,
mohammadipourDirectAnalysisZeroNoise2025a}.

For a fixed total shot budget,
\begin{equation}
    N_{\mathrm{tot}}
    =
    \sum_j N_j,
\end{equation}
minimizing Eq.~\eqref{eq:zne_variance} gives
\begin{equation}
    N_j^{\mathrm{opt}}
    =
    N_{\mathrm{tot}}
    \frac{|\gamma_j|\sigma_j}
    {\sum_m|\gamma_m|\sigma_m}.
    \label{eq:zne_opt_shots}
\end{equation}
If the per-shot variances are similar, this reduces to
$N_j^{\mathrm{opt}}\propto|\gamma_j|$
\cite{krebsbachOptimizationRichardsonExtrapolation2022}.
Thus, scale-factor selection, extrapolation order, and shot allocation form a coupled bias--variance optimization problem rather than independent experimental choices.

\item \textbf{Model selection in practice.}
No extrapolation form is uniformly optimal. Polynomial and Richardson estimators are natural when the scaled response remains in a weak-noise perturbative regime, whereas exponential or multi-exponential models are better motivated when the effective channel exhibits identifiable stochastic decay modes. In realistic experiments, lower-order or least-squares fits are often preferable to increasingly high-order interpolation because they trade some residual bias for substantially improved statistical stability
\cite{majumdarBestPracticesQuantum2023,caiMultiexponentialErrorExtrapolation2021,
mohammadipourDirectAnalysisZeroNoise2025a,harrisReducingQuantumError2026}.

\end{itemize}

The extrapolator should therefore not be selected independently of the noise-scaling protocol. Reliable ZNE requires a scaling procedure that produces a smooth and reproducible response over a useful range of noise strengths, together with an inference model whose bias and statistical amplification remain controlled over that range.

\begin{figure}[t]
    \centering
    \includegraphics[width=\textwidth]{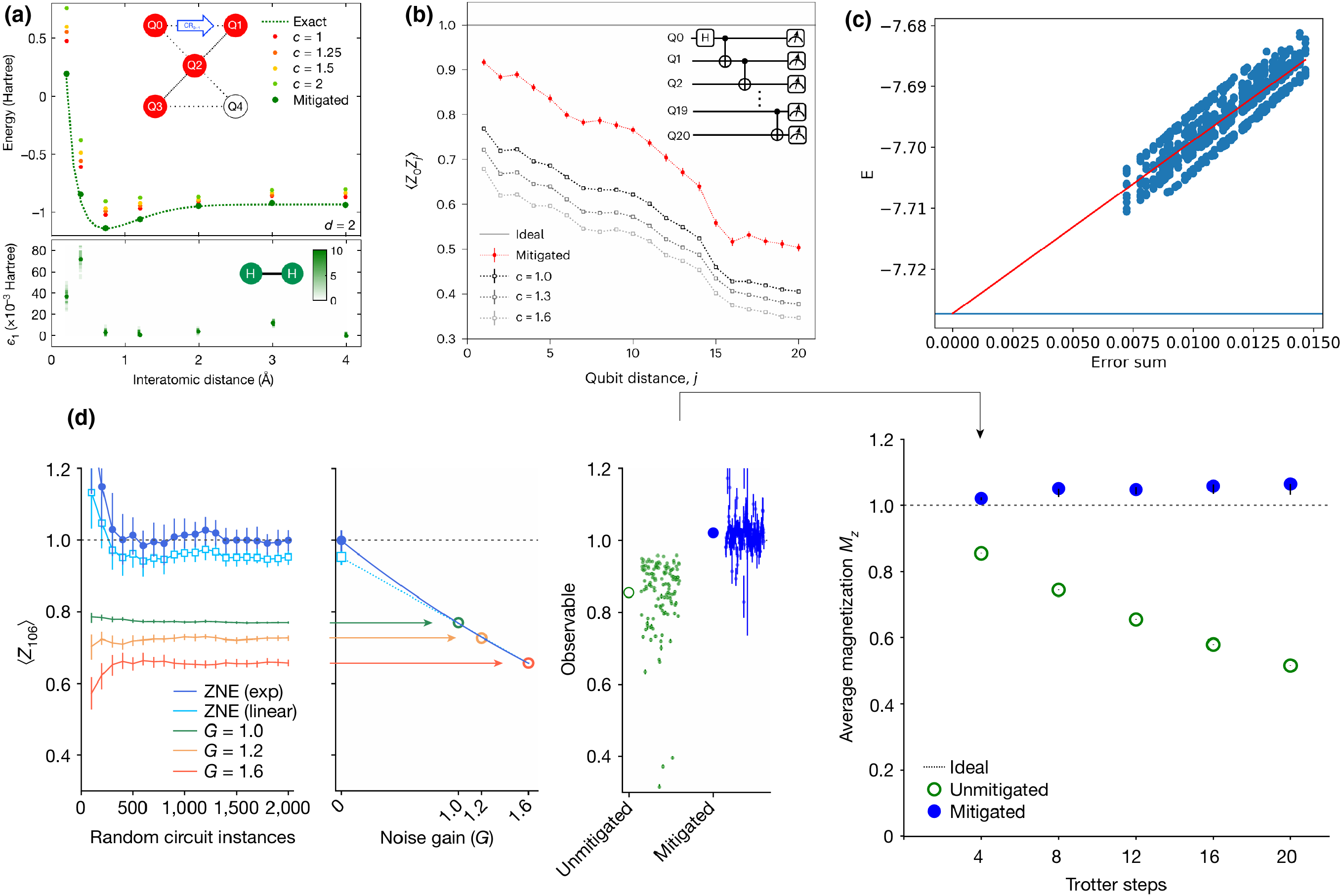}
    \caption{Representative experimental regimes for ZNE. 
    (a) An early pulse-stretching variational experiment (matching \cref{fig:zne_overview}  (c)) that shows error-mitigated molecular energies on superconducting hardware.
    (b) Large-qubit-number pulse-stretch ZNE, where mitigation quality depends strongly on the locality and structure of the measured observable.
    (c) Hardware-inspired ZNE based on mapping-dependent circuit error sums rather than explicit gate folding.
    (d) Utility-scale probabilistic error amplification with extrapolation on 127-qubit Trotterized Ising circuits.
    Adapted from Fig.~4(a) of Ref.~\cite{kandalaErrorMitigationExtends2019}, Fig.~2(d) of Ref.~\cite{kimScalableErrorMitigation2023a}, Fig.~2(a) of Ref.~\cite{uvarovMitigatingQuantumGate2024}, and Fig.~2 of Ref.~\cite{kimEvidenceUtilityQuantum2023c}.}
    \label{fig:zne_benchmarks}
\end{figure}

\subsection{Implementing the noise-scaling axis on quantum hardware}
\label{subsec:zne_scaling_axis}

A central experimental choice in ZNE is how to construct the noise-scaling axis.
A useful scale factor must produce a controlled and interpretable change in the
errors affecting the target observable. On real hardware, however, relaxation,
dephasing, coherent miscalibration, crosstalk, leakage, and temporal drift need
not scale in the same way. The ZNE scale factor should therefore be regarded as
a coordinate along a particular trajectory through a multidimensional noise
space, rather than as a universal multiplier of the complete hardware noise
\cite{majumdarBestPracticesQuantum2023,schultzImpactTimecorrelatedNoise2022,
mohammadipourDirectAnalysisZeroNoise2025a}.
Representative approaches are summarized in \cref{fig:zne_overview}.

\begin{itemize}

\item \textbf{Analog scaling by pulse stretching.}
Pulse stretching amplifies noise by increasing the duration of a physical gate
while rescaling its control amplitude so that the ideal unitary is nominally
unchanged
\cite{temmeErrorMitigationShortDepth2017a,kandalaErrorMitigationExtends2019}.
If an ideal gate is generated by $H_c(t)$ over a duration $T$, an idealized
stretched implementation with scale factor $s>1$ satisfies
\begin{equation}
U
=
\mathcal{T}\exp\!\left[
-i\int_0^T H_c(t)\,dt
\right]
=
\mathcal{T}\exp\!\left[
-i\int_0^{sT}\frac{1}{s}H_c(t/s)\,dt
\right].
\label{eq:zne_pulse_stretch}
\end{equation}
The desired coherent gate action is preserved while the system remains exposed
to decoherence for a longer time. This approach was used in early
superconducting-hardware demonstrations of ZNE
\cite{kandalaErrorMitigationExtends2019} and subsequently in larger experiments
with up to 26 qubits
\cite{kimScalableErrorMitigation2023a}.

Pulse stretching provides a direct physical noise axis but requires pulse-level
control. Moreover, the effective noise need not scale linearly with $s$ once
pulse distortion, drive-dependent decoherence, leakage, or calibration
nonlinearities become important
\cite{majumdarBestPracticesQuantum2023,schultzImpactTimecorrelatedNoise2022}.
The method is therefore most reliable when stretching changes the noise strength
without substantially changing its character.

\item \textbf{Digital scaling by circuit folding.}
Digital ZNE constructs the noise axis through logically redundant gate
operations and therefore does not require pulse-level access
\cite{giurgica-tironDigitalZeroNoise2020,
heZeronoiseExtrapolationQuantumgate2020}.
For a target circuit $U$, global folding replaces
\begin{equation}
U
\longrightarrow
U(U^\dagger U)^m,
\qquad
s=2m+1,
\label{eq:zne_global_fold}
\end{equation}
which preserves the ideal unitary while increasing the number of noisy gate
applications. If
\begin{equation}
U=U_L\cdots U_2U_1,
\end{equation}
local folding instead replaces selected gates or layers according to
\begin{equation}
U_\ell
\longrightarrow
U_\ell(U_\ell^\dagger U_\ell)^{m_\ell}.
\label{eq:zne_local_fold}
\end{equation}
For self-inverse gates such as CNOT, this reduces to
\begin{equation}
\mathrm{CNOT}
\longrightarrow
\mathrm{CNOT}^{\,2m+1}.
\label{eq:zne_cnot_insertion}
\end{equation}
Global and local folding are illustrated in
\cref{fig:zne_overview}(b)
\cite{giurgica-tironDigitalZeroNoise2020,cai2023quantum}.

The main advantage of digital folding is portability across gate-based
platforms. Its limitation is that increasing circuit depth does not necessarily
scale every physical error mechanism uniformly. Global and local folding can
therefore generate different effective noise trajectories, particularly in the
presence of time-correlated or context-dependent noise
\cite{schultzImpactTimecorrelatedNoise2022}.

\item \textbf{Fractional scaling by partial folding.}
Integer scale factors such as $s=3$ or $5$ can amplify the noise too strongly,
causing the observable to lose useful signal before a reliable extrapolation can
be performed. Fractional scale factors provide finer control by folding only a
subset of the circuit
\cite{giurgica-tironDigitalZeroNoise2020,
majumdarBestPracticesQuantum2023}.
If a fraction $f$ of gates is folded once and the gate contributions are
approximately uniform, the nominal scale factor is
\begin{equation}
s\simeq1+2f.
\label{eq:zne_partial_scale}
\end{equation}

Because real processors are spatially and temporally inhomogeneous, two partial
foldings with the same $f$ need not produce the same effective noise strength.
Randomized partial folding therefore averages over several choices of folded
gates at a fixed nominal $s$, reducing sensitivity to any particular circuit
realization
\cite{giurgica-tironDigitalZeroNoise2020,
majumdarBestPracticesQuantum2023}.
This strategy is illustrated in
\cref{fig:zne_practice_extensions}(a).

\item \textbf{Probabilistic error amplification.}
Probabilistic error amplification (PEA) constructs the noise axis at the channel
level rather than by geometrically lengthening the circuit. The effective noise
is first tailored and characterized, after which additional stochastic faults
are sampled so that the learned channel is amplified by a controlled amount.
This approach was central to the 127-qubit superconducting experiments of
Kim \textit{et al.}, where Pauli twirling and sparse noise characterization were
combined with controlled Pauli-error injection
\cite{kimEvidenceUtilityQuantum2023c}.

PEA is more model dependent than circuit folding because the imposed scale
factor relies on the accuracy of the characterized effective noise channel.
Its advantage is that substantial noise amplification can be achieved without
the large circuit-depth increase associated with repeated gate folding, making
it particularly attractive for large-scale circuits
\cite{kimEvidenceUtilityQuantum2023c}.

\item \textbf{Hardware-aware and measured error coordinates.}
More recent approaches replace, or calibrate, the nominal amplification factor
using quantities that more directly reflect the error experienced by the
compiled circuit
\cite{uvarovMitigatingQuantumGate2024,
koenigInvertedcircuitZeronoiseExtrapolation2024a,
russoQuantumErrorMitigation2024a,
sayapinZeroNoiseExtrapolationCyclic2025a,
harrisReducingQuantumError2026}.

One possibility is to exploit hardware inhomogeneity. Uvarov \textit{et al.}
executed the same logical circuit on different physical layouts and used
layout-dependent circuit-error estimates as the extrapolation coordinate
\cite{uvarovMitigatingQuantumGate2024}. Cyclic layout-permutation ZNE similarly
uses different physical embeddings to generate controlled variation in the
effective circuit error
\cite{sayapinZeroNoiseExtrapolationCyclic2025a}.

A second approach is to measure a circuit-specific error proxy. In
inverted-circuit ZNE, the target circuit is followed by its inverse and the
survival probability
\begin{equation}
p_{\mathrm{surv}}
=
\Pr(0^n\,|\,U^{-1}U)
\label{eq:zne_ic_survival}
\end{equation}
is used to estimate the accumulated circuit error
\cite{koenigInvertedcircuitZeronoiseExtrapolation2024a}.
This measured quantity can then calibrate the effective noise coordinate used
for extrapolation.

Finally, layerwise extrapolation allows the noise to vary independently across
different circuit blocks rather than representing the entire computation by one
scalar parameter
\cite{russoQuantumErrorMitigation2024a}. Schematically,
\begin{equation}
E(\boldsymbol{\lambda})
=
E^\star
+
\sum_{b=1}^{B}a_b\lambda_b
+
\sum_{b\leq c}a_{bc}\lambda_b\lambda_c
+\cdots,
\label{eq:zne_layerwise_expansion}
\end{equation}
where $\lambda_b$ denotes the effective noise strength of block $b$. This
multivariate viewpoint is particularly useful when the error distribution is
strongly nonuniform across the circuit.

\end{itemize}

These implementations emphasize that the ZNE scale factor is fundamentally a
\emph{noise coordinate}, not merely a numerical multiplier. Pulse stretching,
global or local folding, partial folding, probabilistic amplification, and
layout-based scaling can probe different mixtures of the underlying hardware
errors. Under correlated or nonstationary noise, these trajectories can lead to
different extrapolated values even for the same circuit and observable
\cite{majumdarBestPracticesQuantum2023,schultzImpactTimecorrelatedNoise2022,
mohammadipourDirectAnalysisZeroNoise2025a,harrisReducingQuantumError2026}.
The practical objective is therefore to construct a scaling axis that changes
the relevant error strength in a controlled manner while preserving, as closely
as possible, the physical structure of the native noise affecting the target
observable.

\subsection{Practical workflow: Transpilation, calibration, composition, and validation}
\label{subsec:zne_workflow}

A recurring lesson from experimental ZNE studies is that the method should be treated as a complete workflow rather than as a curve fit applied only after data collection. The final zero-noise estimate depends jointly on the compiled circuit, the noise-scaling procedure, the accessible scale factors, the extrapolation model, the acquisition schedule, and the validation protocol. Many practical failures arise because the implemented noise axis does not faithfully track the hardware errors that dominate the target observable
\cite{majumdarBestPracticesQuantum2023,schultzImpactTimecorrelatedNoise2022,mohammadipourDirectAnalysisZeroNoise2025a,harrisReducingQuantumError2026}.

\begin{figure}[t]
    \centering
    \includegraphics[width=\textwidth]{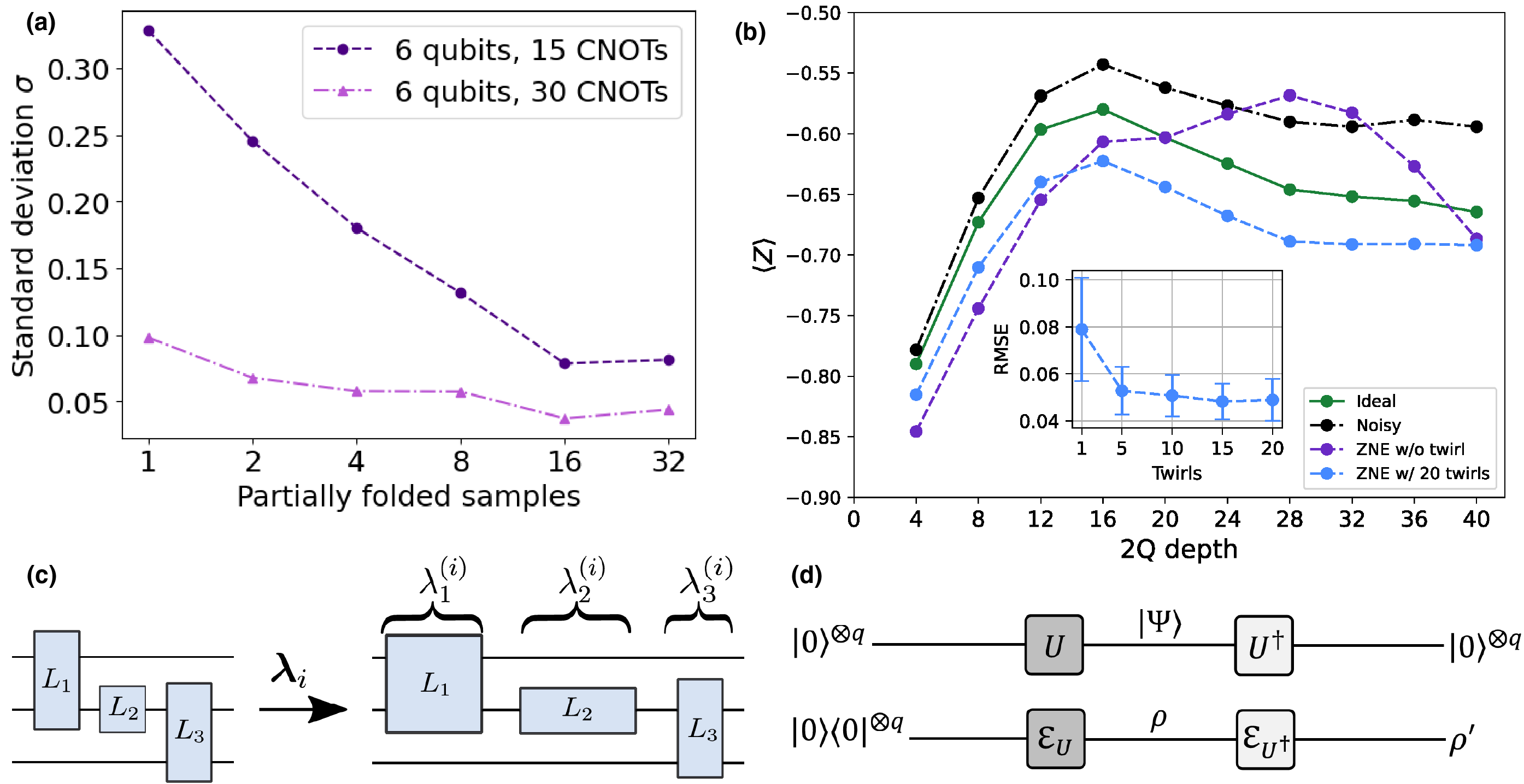}
    \caption{Workflow-level lessons and recent extensions of ZNE. 
    (a) Multiple random partial folds at the same nominal scale reduce the variance of the mitigated estimate.
    (b) Pauli twirling or randomized compiling can reshape coherent errors into a more ZNE-friendly stochastic response.
    (c) Layerwise Richardson extrapolation introduces several independently scaled noise coordinates instead of a single global one.
    (d) Inverted-circuit ZNE replaces the nominal scale factor by a circuit-specific survival-probability proxy for accumulated error.
    Adapted from Figs.~4 and 8 of Ref.~\cite{majumdarBestPracticesQuantum2023}, Fig.~1 of Ref.~\cite{russoQuantumErrorMitigation2024a}, and Fig.~2 of Ref.~\cite{koenigInvertedcircuitZeronoiseExtrapolation2024a}.}
    \label{fig:zne_practice_extensions}
\end{figure}

\begin{itemize}

\item \textbf{Transpile first, then amplify.}
For digital ZNE, the circuit should generally be compiled to the target backend before gate folding or identity insertion is applied. If amplification is introduced before hardware compilation, subsequent optimization passes may cancel or simplify the inserted inverse pairs, so that the executed circuit no longer realizes the intended noise-scaling path
\cite{majumdarBestPracticesQuantum2023}.
This issue is especially relevant on cloud platforms where part of the final routing or optimization may be performed server-side.

Circuit unoptimization provides a complementary approach. Instead of inserting easily recognizable inverse pairs, it constructs longer circuits that remain logically equivalent to the target circuit but are less likely to be simplified back to the original form during compilation
\cite{pelofskeDigitalZeroNoiseExtrapolation2025a}.
In practice, amplified circuits should therefore be generated after backend-aware transpilation and, whenever possible, checked again after the final compilation stage.

\item \textbf{Choose scale factors, extrapolants, and shots jointly.}
The choice of noise scale factors is inseparable from the extrapolation model. Large scale factors increase the extrapolation lever arm but can also amplify statistical uncertainty and push the observable into a saturated or noise-dominated regime. Conversely, scale factors too close to the native point may provide insufficient variation to constrain the extrapolation
\cite{majumdarBestPracticesQuantum2023,mohammadipourDirectAnalysisZeroNoise2025a}.

For shallow circuits and weak native noise, low-order linear or polynomial extrapolation is often more stable because it limits coefficient growth. When the measured response exhibits a clear decay structure, exponential or multi-exponential models may instead be appropriate
\cite{caiMultiexponentialErrorExtrapolation2021,kimEvidenceUtilityQuantum2023c,kimScalableErrorMitigation2023a}.
There is therefore no universally optimal extrapolant: scale-factor selection, fitting model, and shot allocation should be optimized together for the observed noise response and the chosen amplification mechanism.

\item \textbf{Compose ZNE with compatible suppression and mitigation layers.}
ZNE is often more reliable when the effective noise has first been simplified by complementary techniques. Readout-error mitigation is a basic example: because gate folding or pulse stretching primarily amplifies circuit-execution noise rather than terminal measurement noise, readout bias should generally be corrected separately at each scale factor before extrapolation
\cite{majumdarBestPracticesQuantum2023,cai2023quantum}.

Pauli twirling and randomized compiling provide another useful combination. Coherent errors can generate strongly circuit-dependent or oscillatory responses under noise scaling, whereas randomization reshapes them into a more stochastic effective channel and can produce smoother extrapolation curves
\cite{wallman2016noise,kuritaSynergeticQuantumError2023,kimEvidenceUtilityQuantum2023c}.
This synergy is illustrated in \cref{fig:zne_practice_extensions}(b). Any suppression layer used in the target computation should also be applied consistently to the calibration, scaled, and validation circuits so that the extrapolation is performed under the same effective noise conditions.

\item \textbf{Track drift and report uncertainty.}
Temporal drift can distort the apparent ZNE scaling relation. If native and amplified circuits are measured at widely separated times, slow changes in coherence, calibration parameters, or crosstalk may be mistaken for a dependence on the intended noise scale. Interleaving measurements from different scale factors can reduce this effect
\cite{kandalaErrorMitigationExtends2019}.
Recent superconducting experiments further show that mitigation performance can improve when low-frequency hardware fluctuations are actively stabilized
\cite{kimErrorMitigationStabilized2025}.

The uncertainty of the mitigated estimate should be reported together with the central value. Relevant information includes the chosen scale factors, number of randomized foldings or twirled instances, shot allocation, fitting model, acquisition order, and resampling procedure
\cite{majumdarBestPracticesQuantum2023,kimEvidenceUtilityQuantum2023c}.
Bootstrap and related resampling techniques are commonly used to propagate both finite-shot and fitting uncertainty.

\item \textbf{Validate on workload-proximate benchmarks.}
Because ZNE is both biased and dependent on the chosen noise-scaling path, validation is essential whenever a reliable reference is available. Useful benchmarks include exactly simulable Clifford circuits, low-depth instances, conserved quantities, reduced light-cone observables, and verifiable circuits that reproduce the native-gate structure of the target application
\cite{kimEvidenceUtilityQuantum2023c,anandClassicalBenchmarkingZero2023,harrisReducingQuantumError2026}.
The most informative benchmarks reproduce not only the circuit size, but also the compilation pattern, gate composition, and observable structure of the target workload.

Validation itself becomes difficult beyond classically verifiable regimes. Anand \textit{et al.} showed that different classical approximation methods can disagree substantially when used to benchmark ZNE
\cite{anandClassicalBenchmarkingZero2023}.
Benchmarked-noise approaches address this problem by constructing verifiable circuits that closely match the application circuit and using them to estimate the residual mitigation bias
\cite{harrisReducingQuantumError2026}.
Validation should therefore be regarded as an integral part of the ZNE workflow rather than as a separate post-processing check.

\end{itemize}

A reliable ZNE implementation therefore requires co-design of compilation, noise amplification, mitigation composition, data acquisition, and validation. In practice, much of this workflow can be automated through software frameworks. For example, \texttt{Mitiq} provides circuit-folding, noise-scaling, and extrapolation routines across multiple backends, while \texttt{Qiskit Runtime} integrates ZNE directly into estimator-based execution workflows
\cite{laroseMitiqSoftwarePackage2022a,mitiqdocs2026,qiskitdocs_techniques_2026}.
These software implementations are discussed further in Sec.~\ref{provider}. The extrapolation step is therefore only one component of a broader experimental and software workflow.

\subsection{Limitations and current frontiers}
\label{subsec:zne_limits_frontiers}

The central advantage of ZNE is that it converts a complicated hardware-noise
problem into an experimentally accessible inference task. Rather than
reconstructing or inverting the complete noise channel, ZNE measures an
observable along a controlled noise-scaling path and extrapolates the response
toward the zero-noise limit. This simplification, however, also determines its
main limitations: the scaling coordinate may not faithfully represent the
underlying hardware noise, and the extrapolation itself introduces a
bias--variance trade-off.

\begin{itemize}

\item \textbf{Noise scaling in a multidimensional error landscape.}
A real quantum processor is generally not characterized by a single scalar noise parameter. Calibration errors, relaxation, dephasing, crosstalk, leakage, readout bias, temporal drift, and time-correlated fluctuations can respond differently to pulse stretching, gate folding, layout changes, or stochastic fault injection [see Sec.~\ref{sources}]. The nominal ZNE scale factor should therefore be interpreted as a coordinate along a particular path through a multidimensional noise space rather than as a complete description of the hardware noise
\cite{majumdarBestPracticesQuantum2023,schultzImpactTimecorrelatedNoise2022,
koenigInvertedcircuitZeronoiseExtrapolation2024a,
mohammadipourDirectAnalysisZeroNoise2025a,harrisReducingQuantumError2026}.
Different scaling procedures can consequently amplify different mixtures of physical errors and may lead to different zero-noise estimates. Reliable ZNE therefore requires not only an appropriate extrapolation model, but also a scaling procedure that preserves the error structure relevant to the target observable.

\item \textbf{Bias, variance, and sampling complexity.}
ZNE is generally a biased inference procedure because the extrapolated value depends on the assumed functional relation between the observable and the noise scale. This differs from PEC [see Sec.~\ref{pec}], which is unbiased when the noise model and quasiprobability representation are exact. In ZNE, increasing the extrapolation order can reduce model bias but also amplify finite-shot fluctuations, producing a characteristic bias--variance trade-off
\cite{krebsbachOptimizationRichardsonExtrapolation2022,
mohammadipourDirectAnalysisZeroNoise2025a}.
The useful range of scale factors is therefore limited: weak amplification may not constrain the extrapolation, whereas excessive amplification can drive the signal into a noise-dominated or saturated regime.

More generally, worst-case analyses show that the sampling cost required for fixed accuracy can grow rapidly, and in broad settings exponentially, with circuit depth or system size
\cite{takagi2022fundamental,takagi2023samplinglowerbounds,quek2024tighter}.
ZNE is consequently most useful when the native noise is moderate, the observable retains measurable signal, and the chosen scaling procedure produces a smooth response
\cite{kandalaErrorMitigationExtends2019,kimScalableErrorMitigation2023a,
kimEvidenceUtilityQuantum2023c}.
Its relative cost compared with PEC or tensor-network error mitigation also depends strongly on circuit geometry, observable structure, and available classical resources
\cite{filippov2024scalability}.

\item \textbf{Noise-aware and benchmark-assisted extrapolation.}
Recent developments increasingly focus on constructing more informative noise coordinates rather than simply increasing the extrapolation order. Purity-assisted ZNE supplements the measured observable with purity information under Pauli-diagonal or Pauli-twirled noise assumptions
\cite{jinPurityAssistedZeroNoiseExtrapolation2024}. Noise-aware folding uses device calibration information to decide which gates or circuit regions should be amplified
\cite{hour2024improving}. Cyclic layout-permutation ZNE exploits hardware inhomogeneity and circuit symmetries to generate a set of physically distinct but logically equivalent noise realizations
\cite{sayapinZeroNoiseExtrapolationCyclic2025a}. Benchmarked-noise ZNE instead uses verifiable benchmark circuits with native-gate structure close to that of the target computation to estimate and reduce residual extrapolation bias
\cite{harrisReducingQuantumError2026}.
Although operationally different, these approaches share the same goal: to replace a purely nominal scale factor with an experimentally meaningful coordinate that better tracks the errors relevant to the observable.

\item \textbf{Experimental scale and extension to logical qubits.}
The experimental scope of ZNE has expanded substantially. On superconducting hardware, ZNE has been demonstrated on circuits with up to 26 qubits, depth 120, and approximately 1080 CNOT gates
\cite{kimScalableErrorMitigation2023a}. This was extended to a 127-qubit processor, where probabilistic error amplification and extrapolation were applied to Trotterized two-dimensional Ising circuits with up to 60 two-qubit layers and 2880 CNOT operations
\cite{kimEvidenceUtilityQuantum2023c}.
These experiments show that large qubit number alone does not determine feasibility; circuit depth, observable locality, and extrapolation overhead remain equally important.

ZNE has also been demonstrated on other physical platforms. In photonic quantum computing, photon distinguishability was used as the controllable noise coordinate
\cite{borzenkovaErrorMitigatedVariational2024}, while silicon spin-qubit experiments compared global folding, local folding, and pulse stretching
\cite{sohnApplicationZeroNoise2025}.
More recently, ZNE has been applied to error-corrected and logical-qubit circuits
\cite{zhangDemonstratingQuantumError2026}. Repetition codes with distances $d=3,5,7$ used up to 13 physical qubits, while a distance-$3$ rotated surface code used 17 physical qubits. For the distance-$7$ repetition code, the residual bias was reduced to approximately $10^{-4}$ with a sampling overhead of about $5$. This establishes an important extension of ZNE from physical-noise mitigation toward suppression of residual logical errors in the early fault-tolerant regime.

\item \textbf{Beyond physical-noise extrapolation.}
ZNE belongs to a broader family of controlled-extrapolation methods. Mari \emph{et al.} showed that ZNE and PEC can be formulated within a common linear-combination framework, highlighting their different trade-offs between noise knowledge, estimator bias, and sampling variance
\cite{mari2021extending}. The extrapolation coordinate also need not represent only physical hardware noise. In Hamiltonian simulation, one may extrapolate in the Trotter step size or jointly treat physical noise and algorithmic discretization errors
\cite{endoMitigatingAlgorithmicErrors2019,
vazquezEnhancingQuantumLinear2022,
mohammadipourDirectAnalysisZeroNoise2025a}.
This broader viewpoint places ZNE within a general class of controlled-inference methods rather than restricting it to a single physical-noise amplification protocol.

\end{itemize}
Overall, ZNE is best understood as a hardware-, circuit-, and
observable-dependent family of extrapolation protocols. It is most reliable
when the chosen scaling path faithfully tracks the relevant errors, the
observable retains sufficient signal over the sampled scale factors, and the
extrapolation model controls the resulting bias--variance trade-off. The
current frontier is therefore the joint design of noise amplification,
calibration, benchmarking, and statistical inference, together with the
extension of these ideas from physical-qubit NISQ circuits to encoded and
logical-qubit computation.

\section{Inverse-channel error mitigation: quasi-probabilities and channel inversion}\label{pec}

Within quantum error mitigation, inverse-channel methods are conceptually distinct from 
noise scaling and extrapolation (Sec.~\ref{sec:zne}), symmetry- and constraint-based filting (Sec.~\ref{verification}), and readout-only correction (Sec.~\ref{read}). Rather than fitting observables as a function of an amplified noise parameter, filtering a known symmetry sector, or correcting only terminal measurement errors, they seek to compensate for a learned effective noise map itself \cite{temmeErrorMitigationShortDepth2017a,endo2018practical,cai2023quantum}. In their most direct form, one models the noisy implementation of a circuit, gate, or circuit layer as an ideal operation followed by an effective error channel, and then attempts to apply the inverse of that channel at the level of expectation values. This makes inverse-channel mitigation one of the most ambitious forms of quantum error mitigation: when the noise model is sufficiently accurate, it can in principle produce unbiased estimates of ideal observables without requiring additional encoded qubits \cite{temmeErrorMitigationShortDepth2017a,takagi2021optimal}.
This inverse-channel viewpoint is also becoming relevant beyond conventional unencoded NISQ circuits. In the transition toward early fault-tolerant and partially protected computation, some errors may already be suppressed by encoding, syndrome extraction, or restricted logical operations, while other residual errors remain too large to neglect. In this regime, inverse-channel mitigation provides a natural language for correcting the effective noise of encoded operations, dynamic-circuit primitives, or finite-distance logical blocks \cite{suzuki2022universal,piveteau2021error,gupta2024dynamicpec,dutkiewicz2025error,zhangDemonstratingQuantumError2026}. Thus, the same conceptual framework can apply both to physical NISQ circuits and to early logical circuits, although the noise object being inverted changes from a physical gate or layer channel to a residual logical or syndrome-conditioned channel.
Meanwhile, the price of this generality is that the inverse of a noisy physical channel is not, in general, itself a physical channel. It need not be completely positive, trace preserving, or implementable as a deterministic quantum operation. The central technical idea is therefore to simulate this nonphysical inverse indirectly, by expressing it as a quasiprobabilistic linear combination of physical operations that the device can implement \cite{temmeErrorMitigationShortDepth2017a,endo2018practical}. This idea appears under several closely related names, including quasiprobability decomposition, inverse-channel mitigation, and PEC. Among these, PEC has become the standard hardware-oriented formulation and remains the canonical representative of unbiased inverse-channel mitigation \cite{cai2023quantum}. Conceptually, PEC is important because it makes explicit the central tradeoff of inverse-channel methods: one exchanges deterministic physical implementability for a stochastic reconstruction procedure whose cost is controlled by the quasiprobability overhead of simulating the inverse channel.

The discussion  below is organized around this logic. We first introduce the general idea of inverting an effective noise channel. We then discuss PEC as the leading practical instantiation of this idea, including its relation to quasiprobability decompositions, physical implementability, and Monte Carlo sampling. Next, we review experimental implementations on superconducting and trapped-ion platforms, together with more recent extensions to sparse noise models, dynamic circuits, and partially fault-tolerant settings. Finally, we discuss the central limitation of the whole approach: the sampling overhead required to simulate a nonphysical inverse channel can scale unfavorably with circuit size, noise strength, or model mismatch \cite{takagi2022fundamental,tsubouchi2023universalcost,takagi2023samplinglowerbounds}. For this reason, inverse-channel mitigation is best understood not as a universally scalable cure for noise, but as a powerful and principled framework whose practical usefulness depends sensitively on how accurately the effective channel can be learned and how much sampling overhead can be tolerated.

\subsection{Mitigating errors by inverting an effective noise channel}
Inverse-channel mitigation starts from an explicit model of how noise modifies the implemented computation. Let $\mathcal{H}$ be the system Hilbert space, let $\mathcal{D}(\mathcal{H})$ denote the set of density operators on $\mathcal{H}$, let $\rho_{0}\in\mathcal{D}(\mathcal{H})$ be the input state, and let $O$ be the Hermitian observable whose expectation value is to be estimated. If $\mathcal{C}:\mathcal{D}(\mathcal{H})\rightarrow\mathcal{D}(\mathcal{H})$ denotes the ideal circuit channel and $\widetilde{\mathcal{C}}$ its noisy implementation, one may model the noisy circuit as
\begin{equation}
    \widetilde{\mathcal{C}}
    =
    \mathcal{E}_{\mathrm{eff}}\circ \mathcal{C},
    \label{eq:inverse_channel_eff}
\end{equation}
where $\mathcal{E}_{\mathrm{eff}}$ is an effective noise channel. Depending on the chosen description, $\mathcal{E}_{\mathrm{eff}}$ may represent the noise of the full circuit, a circuit layer, an idle period, or an individual gate \cite{temmeErrorMitigationShortDepth2017a,endo2018practical,cai2023quantum}. The ideal expectation value is
\begin{equation}
    \langle O\rangle_{\mathrm{id}}
    =
    \Tr\!\left[
        O\,\mathcal{C}(\rho_{0})
    \right].
    \label{eq:ideal_expectation_inverse_channel}
\end{equation}
If $\mathcal{E}_{\mathrm{eff}}$ were known exactly and an inverse could be defined on the relevant support, the same quantity could be written in the Heisenberg picture as
\begin{equation}
    \langle O\rangle_{\mathrm{id}}
    =
    \Tr\!\left[
        (\mathcal{E}_{\mathrm{eff}}^{-1})^{\dagger}(O)\,
        \widetilde{\mathcal{C}}(\rho_{0})
    \right].
    \label{eq:formal_inverse_estimator}
\end{equation}
Here, $\mathcal{E}_{\mathrm{eff}}^{-1}$ denotes a formal inverse within the assumed noise model, or an inverse defined only on the relevant image or support of $\mathcal{E}_{\mathrm{eff}}$. Eq.~\eqref{eq:formal_inverse_estimator} should not be interpreted as applying a deterministic physical inverse channel to the quantum state. Instead, it defines the observable-level linear functional that inverse-channel mitigation aims to estimate from noisy hardware data.
This formulation highlights both the method and its limitations. The method is to learn or assume an effective noise channel and then represent the inverse action of that channel through classical post-processing or quasiprobabilistic sampling. The first approximation is the noise model itself: gate-level, layer-level, and circuit-level descriptions can differ in calibration cost and accuracy \cite{endo2018practical,cai2023quantum}. The second approximation is the representation of the inverse. Even when $\mathcal{E}_{\mathrm{eff}}$ is completely positive and trace-preserving, its inverse need not be completely positive, trace-preserving, or implementable as a deterministic quantum operation. Moreover, leakage, non-Markovian memory, temporal drift, and gate-context dependence can invalidate the assumption of a single stable inverse channel \cite{endo2018practical,govia2025modelviolation}. Thus, inverse-channel mitigation is controlled by both the accuracy of the effective noise model and the cost of representing its formal inverse.

This framework is broader than mitigation strategies that correct only a restricted part of the noise \cite{kandalaErrorMitigationExtends2019,bonetmonroig2018lowcost,cai2021symmetry,maciejewski2020mitigation,nation2021scalable}. Readout mitigation, for example, inverts or regularizes a classical assignment matrix acting only at the measurement stage \cite{chen2019detector,maciejewski2020mitigation,geller2020rigorous,bravyi2021mitigating,nation2021scalable,smith2021qubit,funcke2022measurement,van2022model}. By contrast, inverse-channel mitigation can in principle target gate, idle, layer, or circuit noise, provided that the noisy operations can be characterized and the inverse map can be represented by implementable operations with quasiprobability weights \cite{temmeErrorMitigationShortDepth2017a,endo2018practical,cai2023quantum}. PEC is the canonical example: it gives an unbiased estimator under the assumed noise model, but the sampling overhead grows with the quasiprobability cost of the inverse.
This cost has a useful operational and resource-theoretic interpretation. Takagi showed that the optimal cost of inverse-channel mitigation can be formulated as the minimal quasiprobability overhead required to represent the desired inverse using the noisy operations available on the device \cite{takagi2021optimal}. The difficulty of inverse-channel mitigation is therefore not only a technical feature of a particular decomposition scheme; it reflects how far the desired inverse operation lies outside the physically implementable operation set. Exact inverse-channel cancellation is also not the only possible use of this idea. Mari \textit{et al.} introduced a linear-combination framework that connects PEC and noise-scaling methods by representing ideal operations using noisy operations at different virtual noise strengths \cite{mari2021extending}. In this broader view, PEC is the limiting case that targets a fully unbiased estimator, whereas partially mitigated linear-combination or noise-scaling strategies can trade residual bias for reduced sampling overhead.

\subsection{Probabilistic error cancellation}
PEC is the canonical example of inverse-channel error mitigation \cite{temmeErrorMitigationShortDepth2017a,endo2018practical,takagi2021optimal,cai2023quantum}. The problem it addresses is that the inverse of a noisy quantum channel is generally not itself a physical quantum operation. PEC solves this problem by representing the formal inverse as a quasiprobabilistic linear combination of experimentally implementable operations. In this way, the deterministic inversion of noise is replaced by a statistical reconstruction procedure. When the effective noise model is exact and the sampled basis operations match their calibrated descriptions, PEC gives an unbiased estimator of the ideal expectation value \cite{temmeErrorMitigationShortDepth2017a,endo2018practical,cai2023quantum}.
Let $\mathcal{E}$ denote the effective noise channel associated with a noisy gate, layer, or circuit fragment. PEC seeks a decomposition of the formal inverse map $\mathcal{E}^{-1}$ as
\begin{equation}
    \mathcal{E}^{-1}
    =
    \sum_{\alpha}\eta_{\alpha}\,\mathcal{B}_{\alpha},
    \label{eq:pec_decomposition}
\end{equation}
where $\eta_{\alpha}\in\mathbb{R}$ are quasiprobability coefficients and $\{\mathcal{B}_{\alpha}\}$ are experimentally implementable noisy basis channels \cite{temmeErrorMitigationShortDepth2017a,endo2018practical}. Depending on the implementation, the basis elements may correspond to noisy gates, state-preparation primitives, measurements, or short circuit fragments. Because some $\eta_{\alpha}$ can be negative, Eq.~\eqref{eq:pec_decomposition} is not a physical stochastic mixture. It instead defines a signed Monte Carlo estimator whose unbiasedness is conditional on the assumed noise model and basis decomposition \cite{temmeErrorMitigationShortDepth2017a,endo2018practical,takagi2021optimal,cai2023quantum}.
The standard sampling rule is to choose a basis channel $\mathcal{B}_{\alpha}$ with probability
\begin{equation}
    p_{\alpha}
    =
    \frac{|\eta_{\alpha}|}{\gamma},
\end{equation}
where
\begin{equation}
    \gamma
    =
    \sum_{\alpha}|\eta_{\alpha}|
    \label{eq:pec_gamma}
\end{equation}
is the quasiprobability overhead, also called the sampling overhead or negativity factor. Each sampled circuit is assigned the sign and weight associated with $\eta_{\alpha}$. Averaging these weighted samples reconstructs the ideal expectation value under the calibrated model. For gatewise or layerwise decompositions, the total overhead multiplies across circuit locations,
\begin{equation}
    \Gamma
    =
    \prod_{k}\gamma_{k},
    \label{eq:pec_total_gamma}
\end{equation}
where $\gamma_k$ is the local overhead at circuit location $k$. Consequently, for fixed target precision, the required number of samples scales roughly as $\Gamma^{2}$ \cite{cai2023quantum,takagi2022fundamental}. This is the central tradeoff of PEC: it can remove bias in principle, but the sampling cost grows with how nonphysical the inverse channel is. The cost is manageable for shallow circuits or weak noise, but can become prohibitive for deeper circuits or stronger noise \cite{takagi2021optimal,takagi2022fundamental}.

\begin{figure*}[t]
    \centering
    \includegraphics[width=0.85\textwidth]{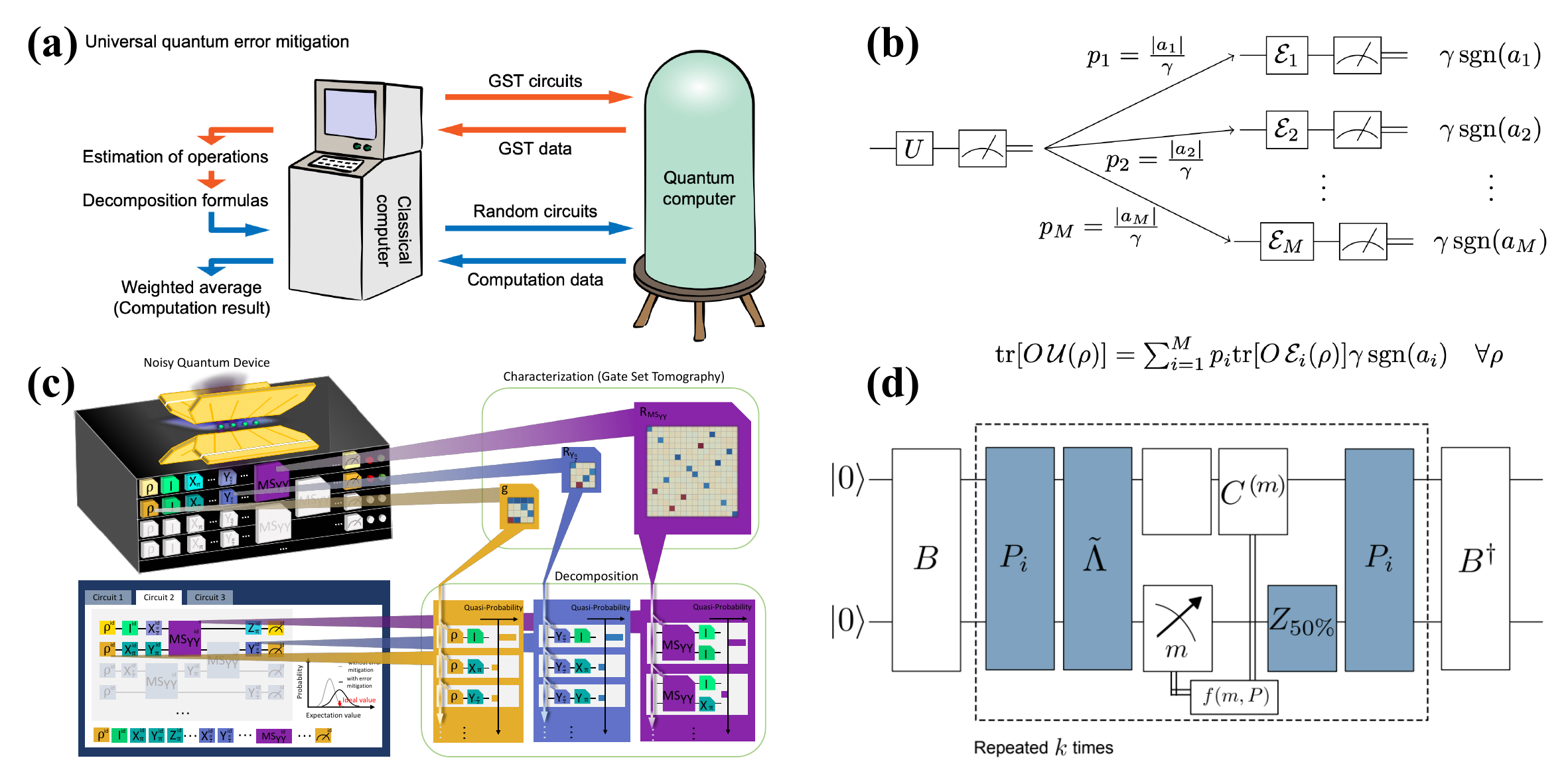}
    \caption{
    Inverse-channel mitigation and PEC.
    (a) A generic workflow for universal inverse-channel mitigation, where experimentally characterized noisy operations are decomposed quasiprobabilistically and recombined through weighted sampling to estimate ideal observables.
    (b) Schematic of quasiprobability decomposition: a target ideal operation is represented as a signed linear combination of implementable noisy basis operations, with sampling probabilities set by the absolute values of the quasiprobability coefficients.
    (c) Hardware-oriented realization of this idea on a superconducting platform, combining gate-set characterization with circuit decomposition into noisy basis operations.
    (d) Extension of PEC to dynamic circuits, illustrating that the same inverse-channel logic can be generalized beyond static unitary circuits to measurement-based and feedforward channels.
    Panels (a,c) are adapted from Ref.~\cite{song2019universal}, panel (b) from Ref.~\cite{zhang2020trappedionpec}, and panel (d) from Ref.~\cite{gupta2024dynamicpec}.}
    \label{figS9_1}
\end{figure*}

The practical performance of PEC depends on both the accuracy of the learned noise model and the choice of experimentally implementable basis operations $\{\mathcal{B}_{\alpha}\}$. The general workflow is illustrated in Fig.~\ref{figS9_1}(a): noisy gates are first characterized, for example through gate-set tomography; their inverse action is expressed as a quasiprobability decomposition, and randomized circuit instances are then sampled and recombined to estimate the ideal observable
\cite{temmeErrorMitigationShortDepth2017a,endo2018practical,song2019universal}.
For an ideal operation $\mathcal{U}$, one writes schematically $\mathcal{U}
    \simeq
    \sum_{\alpha}\eta_{\alpha}\mathcal{B}_{\alpha}$,
where the coefficients $\eta_{\alpha}$ can be negative. As shown in Fig.~\ref{figS9_1}(b), the implementable operations are sampled with probabilities proportional to $|\eta_{\alpha}|$, while the measured outcomes are multiplied by the corresponding sign and normalization factor
\cite{zhang2020trappedionpec}. A representative hardware realization is shown in Fig.~\ref{figS9_1}(c), where experimentally characterized noisy gates on a superconducting processor are used directly as the basis operations for constructing the PEC estimator
\cite{song2019universal}.

The choice of decomposition strongly affects the sampling overhead. Optimized quasiprobability representations can substantially reduce the cost
\cite{piveteau2022reducedsampling}, while locality can be exploited when only local observables are required
\cite{tran2023locality}. More recent approaches based on sparse noise models, efficient Pauli-error propagation, and improved sampling representations further reduce the practical cost of implementing the same formal inverse channel
\cite{koczor2024sparse,scheiber2024reduced,chen2025fasterpec}. Thus, PEC performance is determined not only by whether an inverse decomposition exists, but also by how efficiently that inverse can be represented and sampled on the target hardware.

PEC is unbiased only with respect to the assumed effective noise model. If the calibrated model differs from the actual hardware noise, or if the cancellation operations themselves contain unmodeled errors, a residual bias remains. Jin \textit{et al.} generalized PEC to settings with noisy or restricted cancellation resources and showed explicitly how imperfect implementability modifies both the mitigation overhead and the remaining error
\cite{jin2025noisypec}. In practice, PEC therefore requires both sufficiently accurate characterization and a basis of accessible operations with manageable quasiprobability cost.

The same inverse-channel principle has also been extended beyond static gate-based circuits. Mari \textit{et al.} placed PEC and ZNE within a common linear-combination framework based on virtual noise scaling
\cite{mari2021extending}. More recently, Gupta \textit{et al.} extended PEC to dynamic circuits containing mid-circuit measurements, conditional operations, and feedforward Clifford gates, as illustrated in Fig.~\ref{figS9_1}(d)
\cite{gupta2024dynamicpec}. Related work has emphasized that probabilistic cancellation is part of a broader class of linear error-reduction methods that can remain relevant as quantum processors move from the NISQ regime toward early fault-tolerant operation
\cite{suzuki2022universal}. In this sense, PEC is not restricted to unencoded near-term circuits, but provides a general framework for compensating residual effective noise whenever accurate models and sufficient sampling resources are available.

\subsection{Implementation on quantum hardware}
\begin{figure*}[t]
    \centering
    \includegraphics[width=0.65\textwidth]{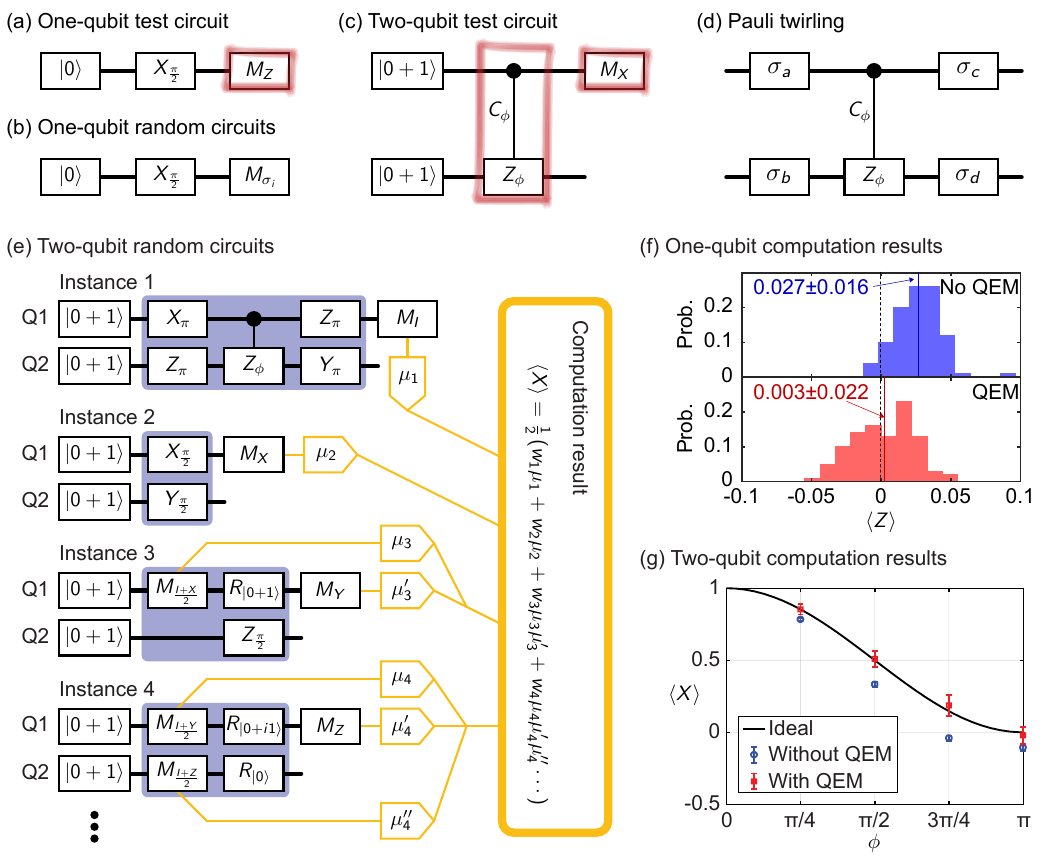}
    \caption{First superconducting proof-of-principle implementation of PEC. Characterized one- and two-qubit noisy operations are used to generate quasiprobability-sampled circuit instances. The signed, weighted average of their outcomes reconstructs the target expectation value. The displayed one-qubit distributions and two-qubit benchmark compare unmitigated and PEC-corrected results, showing reduced systematic deviation from the ideal prediction. Adapted from Ref.~\cite{song2019universal}.}
    \label{fig:pec_superconducting_demo}
\end{figure*}
Hardware implementations of inverse-channel mitigation began at small scale because PEC requires both an accurate effective-noise model and a large number of samples set by the quasiprobability overhead \cite{temmeErrorMitigationShortDepth2017a,endo2018practical,song2019universal,zhang2020trappedionpec}. A first superconducting demonstration was reported by Song \textit{et al.}, who combined gate-set tomography with quasiprobability decomposition for one- and two-qubit circuits on a superconducting processor \cite{song2019universal}. This experiment established that PEC is not only a formal inversion procedure: a characterized noisy gate set can be sampled and recombined to emulate an effectively noiseless one at the level of expectation values.
A closely related trapped-ion milestone was reported by Zhang \textit{et al.}, who benchmarked PEC at the gate level and showed that the effective mitigated gate fidelities can exceed the raw physical gate fidelities \cite{zhang2020trappedionpec}. This result provides a transparent demonstration of the central PEC tradeoff. Given a sufficiently accurate noise model, one can exchange additional sampling cost for improved effective gate performance. At the same time, the experiment also illustrates why PEC is demanding: the improvement relies on accurate characterization and on keeping the quasiprobability overhead under control.
A major step toward larger superconducting experiments was provided by van den Berg \textit{et al.}, who introduced sparse Pauli--Lindblad noise models for PEC \cite{vandenberg2023sparsepec}. Instead of performing full process tomography, they learned a sparse effective description of layer noise that can include correlated multiqubit errors and then inverted this model quasiprobabilistically. This work clarified that the main bottleneck in scalable PEC is often not the formal inverse itself, but the ability to learn a sufficiently accurate and compact noise model.
Trapped-ion implementations also moved from gate-level benchmarks toward many-body simulation. Chen \textit{et al.} applied PEC to digital simulations of interacting fermionic dynamics using up to four trapped-ion qubits, combining tomographically reconstructed error channels with positive-probability filtering and symmetry constraints \cite{chen2023fermions}. This example shows how PEC is often used in practice: not as an isolated correction layer, but as part of a broader mitigation stack that also exploits structure in the target problem.

More recent work has extended inverse-channel ideas to dynamic-circuit settings. Gupta \textit{et al.} developed PEC for circuits containing mid-circuit measurements and classically conditioned Clifford feedforward, extending sparse Pauli--Lindblad characterization from unitary gate layers to measurement-based channels \cite{gupta2024dynamicpec}. Related work on dynamic-circuit benchmarking, readout mitigation for mid-circuit measurements and feedforward, and layered dynamic-circuit mitigation further reflects the need to characterize and mitigate errors beyond static unitary circuits \cite{shirizly2025dynamicrb,koh2026readout,bar2026layered}. These developments are important because adaptive algorithms and early error-correction routines rely precisely on measurement, feedback, and real-time control.


\begin{figure*}[t]
    \centering
    \includegraphics[width=0.99\textwidth]{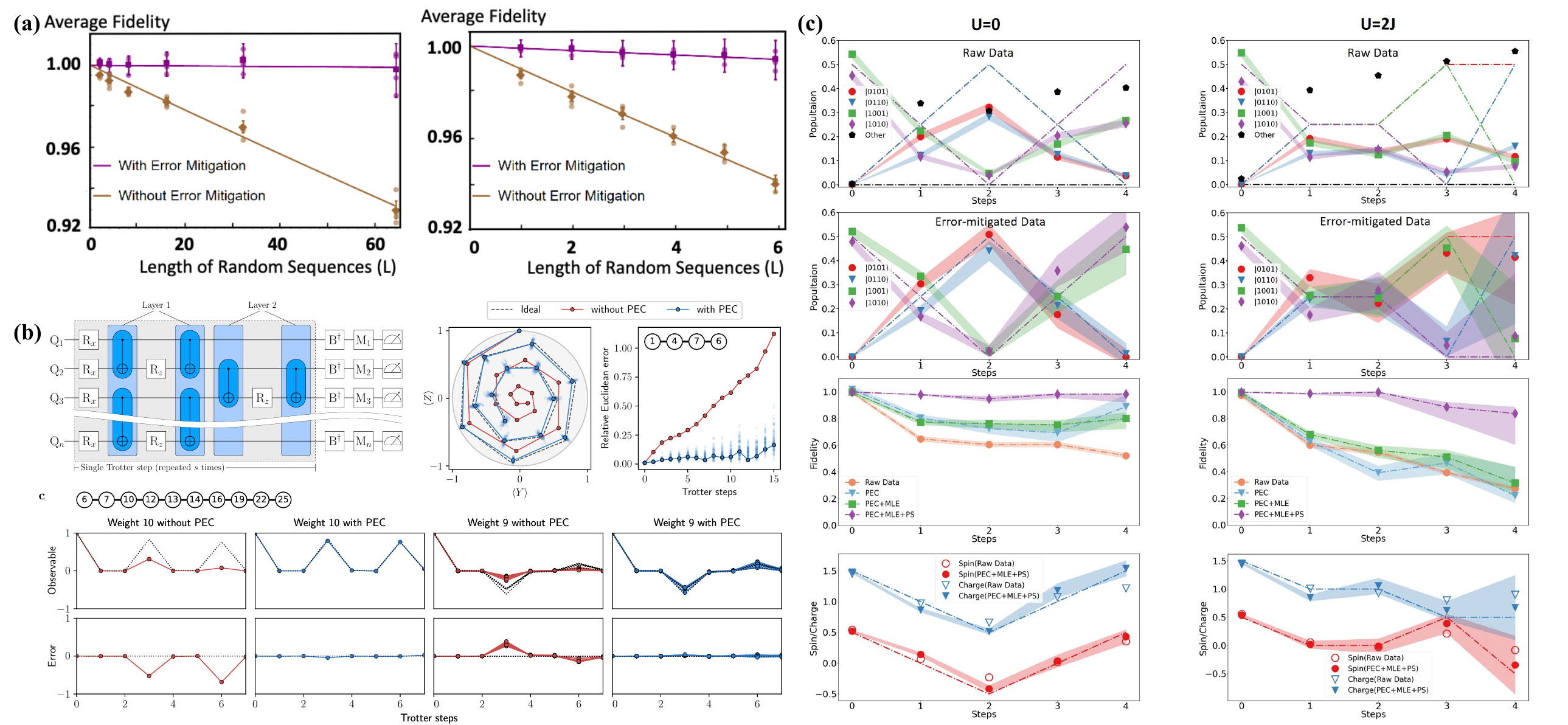}
    \caption{Later hardware implementations of inverse-channel mitigation and PEC. (a) Trapped-ion gate-level benchmarks: PEC suppresses the sequence-length-dependent loss of average fidelity for one- and two-qubit random circuits. (b) Sparse Pauli--Lindblad PEC on a superconducting processor: a learned layer-noise model is inverted quasiprobabilistically, yielding observables and Trotter-step errors closer to the ideal dynamics. (c) Error-mitigated trapped-ion simulation of interacting fermions: PEC and structural filtering improve populations, state fidelities, and spin/charge observables. Panels (a)--(c) are adapted from Refs.~\cite{zhang2020trappedionpec}, \cite{vandenberg2023sparsepec}, and \cite{chen2023fermions}, respectively.}
    \label{fig:pec_hardware_implementations}
\end{figure*}
Early experiments established that PEC could be implemented directly with
hardware-characterized operations. The superconducting experiment of
Song \textit{et al.}, summarized in Fig.~\ref{fig:pec_superconducting_demo},
combined gate-set tomography with quasiprobability decomposition to mitigate
both measurement and entangling-gate errors in one- and two-qubit circuits
\cite{song2019universal}. An important feature of this experiment was that the
basis operations entering the quasiprobability representation were themselves
experimentally characterized rather than assumed to be ideal. The figure
therefore illustrates the basic experimental structure of PEC: characterize
the available noisy operations, construct an inverse representation, and
estimate observables from a signed ensemble of executable circuits.

The later experiments collected in Fig.~\ref{fig:pec_hardware_implementations}
illustrate how this basic idea evolved toward more scalable characterization
and application-level simulations. In trapped ions, Zhang \textit{et al.}
applied PEC to elementary one- and two-qubit operations and showed that the
\emph{effective} gate fidelities inferred from mitigated expectation values
could substantially exceed the corresponding unmitigated values
\cite{zhang2020trappedionpec}. These effective fidelities characterize the
accuracy of the PEC estimator and should not be interpreted as an improvement
of the underlying physical gates.

A complementary development was to reduce the cost of noise characterization.
Van den Berg \textit{et al.} introduced a sparse Pauli--Lindblad model capable
of describing local and correlated errors while remaining efficiently
learnable
\cite{vandenberg2023sparsepec}. As represented in
Fig.~\ref{fig:pec_hardware_implementations}(b), the same characterization
framework was demonstrated for a 20-qubit layer containing ten simultaneous
CX gates, while PEC was subsequently applied to transverse-field Ising
simulations on four and ten qubits. This work is important because it shifts
the practical bottleneck from generic process tomography toward learning a
structured noise model whose inverse can be sampled efficiently. PEC has also progressed from gate benchmarks to many-body applications.
Fig.~\ref{fig:pec_hardware_implementations}(c) summarizes the trapped-ion
simulation of interacting fermions by Chen \textit{et al.}, where PEC was
applied primarily to the noisier two-qubit gates and combined with
positivity and normalization constraints as well as post-selection based on
fermion-number conservation
\cite{chen2023fermions}. The resulting improvement in state populations,
fidelities, and spin- and charge-related observables illustrates an important
practical trend: PEC need not operate in isolation, but can be combined with
problem-specific physical constraints and other mitigation layers.

These experiments highlight three ingredients that govern the
practical development of PEC: an experimentally accessible quasiprobability
basis, a noise model that can be characterized at the relevant circuit scale,
and a sampling overhead that remains manageable as circuit depth increases.
The progression from the early superconducting demonstration to sparse
many-qubit noise models and interacting-fermion simulations therefore reflects
not simply an increase in system size, but a gradual improvement in how the
inverse noise channel is represented, learned, and integrated with the target
application.

PEC and related linear-mitigation methods are increasingly being discussed beyond the original NISQ setting. Suzuki \textit{et al.} formulated probabilistic cancellation and linear mitigation as error-reduction tools that can also be relevant to partially error-corrected or early fault-tolerant architectures \cite{suzuki2022universal}. Related work on encoded gates, logical-qubit mitigation, and early fault-tolerant phase-estimation workflows develops this direction further \cite{piveteau2021error,dutkiewicz2025error,smith2024logical,zhangDemonstratingQuantumError2026,mayer2024benchmarking}. Thus, the hardware literature shows a clear progression: PEC has been demonstrated on superconducting and trapped-ion processors, scaled using sparse noise models, applied to small many-body simulations, and generalized toward dynamic and logical-circuit settings \cite{song2019universal,zhang2020trappedionpec,vandenberg2023sparsepec,chen2023fermions,gupta2024dynamicpec}.

\subsection{Cost and scalability}

\subsubsection{Sampling overhead and a simple bit-flip example}

The central limitation of probabilistic error cancellation is its sampling
overhead. In PEC, the inverse noise channel is represented as a signed
quasiprobability combination of experimentally implementable operations,
\begin{equation}
    \mathcal{E}^{-1}
    =
    \sum_{\alpha}\eta_{\alpha}\mathcal{B}_{\alpha},
\end{equation}
with local quasiprobability norm
\begin{equation}
    \gamma
    =
    \sum_{\alpha}|\eta_{\alpha}|.
\end{equation}
For a circuit containing $M$ independently mitigated gates or noisy layers, the
total normalization factor is
\begin{equation}
    \Gamma
    =
    \prod_{m=1}^{M}\gamma_m .
\end{equation}
If the measured observable is bounded by unity, the variance of the PEC
estimator scales as
\begin{equation}
    \operatorname{Var}
    \!\left(
        \widehat O_{\rm PEC}
    \right)
    \lesssim
    \frac{\Gamma^2}{N},
\end{equation}
so that reaching statistical uncertainty $\epsilon_{\rm stat}$ requires
approximately
\begin{equation}
    N_{\rm PEC}
    \gtrsim
    \frac{\Gamma^2}{\epsilon_{\rm stat}^2}.
    \label{eq:pec_general_sampling_cost}
\end{equation}
Thus, even modest local quasiprobability overheads can become large because
they multiply across the mitigated circuit
\cite{takagi2021optimal,takagi2022fundamental}.

A one-qubit bit-flip channel makes this scaling explicit. Consider
\begin{equation}
    \mathcal{E}_p
    =
    (1-p)\mathcal{I}
    +
    p\mathcal{X},
    \qquad
    \mathcal{X}(\rho)=X\rho X ,
    \label{eq:pec_bitflip_channel}
\end{equation}
with $0\leq p<1/2$. Because $X^2=I$, applying the conjugation channel twice
returns the original state,
\begin{equation}
    \mathcal{X}\circ\mathcal{X}
    =
    \mathcal{I}.
\end{equation}
The inverse channel is therefore
\begin{equation}
    \mathcal{E}_p^{-1}
    =
    \frac{1-p}{1-2p}\mathcal{I}
    -
    \frac{p}{1-2p}\mathcal{X},
    \label{eq:pec_bitflip_inverse}
\end{equation}
with
\begin{equation}
    \gamma
    =
    \frac{1}{1-2p}.
    \label{eq:pec_bitflip_gamma}
\end{equation}
For $M$ identical noisy locations,
\begin{equation}
    \Gamma^2
    =
    (1-2p)^{-2M},
    \qquad
    N_{\rm PEC}
    \gtrsim
    \frac{(1-2p)^{-2M}}
    {\epsilon_{\rm stat}^2}.
    \label{eq:pec_exact_bitflip_cost}
\end{equation}
In the weak-noise regime,
\begin{equation}
    (1-2p)^{-2M}
    \simeq
    e^{4pM},
    \qquad
    p\ll1,
    \label{eq:pec_sampling_estimate}
\end{equation}
which makes the exponential growth with circuit depth or volume explicit.

\begin{table}[t]
\caption{
Selected PEC variance-amplification factors for the bit-flip model, calculated
directly from Eq.~\eqref{eq:pec_exact_bitflip_cost}. The values illustrate how
small local error rates can generate large sampling overheads when many noisy
locations are mitigated
\cite{takagi2021optimal,takagi2022fundamental}.}
\label{tab:pec_bitflip_overhead}
\centering
\small
\begin{tabular}{ccc}
\toprule
Bit-flip rate $p$ &
Mitigated locations $M$ &
Variance factor $\Gamma^2$ \\
\midrule
$0.5\%$ & $100$  & $7.46$ \\
$0.5\%$ & $500$  & $2.32\times10^{4}$ \\
$0.5\%$ & $1000$ & $5.37\times10^{8}$ \\
$1\%$   & $100$  & $56.9$ \\
\bottomrule
\end{tabular}
\end{table}

These numbers are not universal predictions for real hardware, since practical
PEC decompositions involve different noise channels and operation bases. They
simply illustrate the basic mechanism: the signed norm is accumulated before
the variance is evaluated, so small local overheads can become expensive over
many circuit locations.

\subsubsection{Locality and structured reductions of PEC cost}

The global circuit size does not always determine the relevant PEC overhead.
If the target observable $O$ is local, only noisy operations inside its
\emph{backward light cone} can influence its expectation value. If that light
cone contains $M_O$ noisy locations, the bit-flip example gives
\begin{equation}
    \Gamma_O^2
    =
    (1-2p)^{-2M_O},
\end{equation}
rather than the corresponding factor based on all $M$ circuit locations.
For geometrically local circuits and low-weight observables,
$M_O\ll M$ can therefore substantially reduce the sampling cost
\cite{tran2023locality}.

More generally, recent work has explored several ways of reducing the
quasiprobability overhead. Optimized decompositions minimize the signed norm of
the inverse representation
\cite{piveteau2022reducedsampling}, while sparse noise models avoid learning and
inverting a completely generic channel
\cite{vandenberg2023sparsepec,koczor2024sparse}.
Pauli-error propagation and improved sampling representations can further reduce
the effective computational cost
\cite{scheiber2024reduced,chen2025fasterpec}.
Noise bias can also be exploited in architectures where one error channel is
strongly suppressed relative to others
\cite{rennela2025lowbitflip}.

\begin{figure*}[t]
    \centering
    \includegraphics[width=0.82\textwidth]{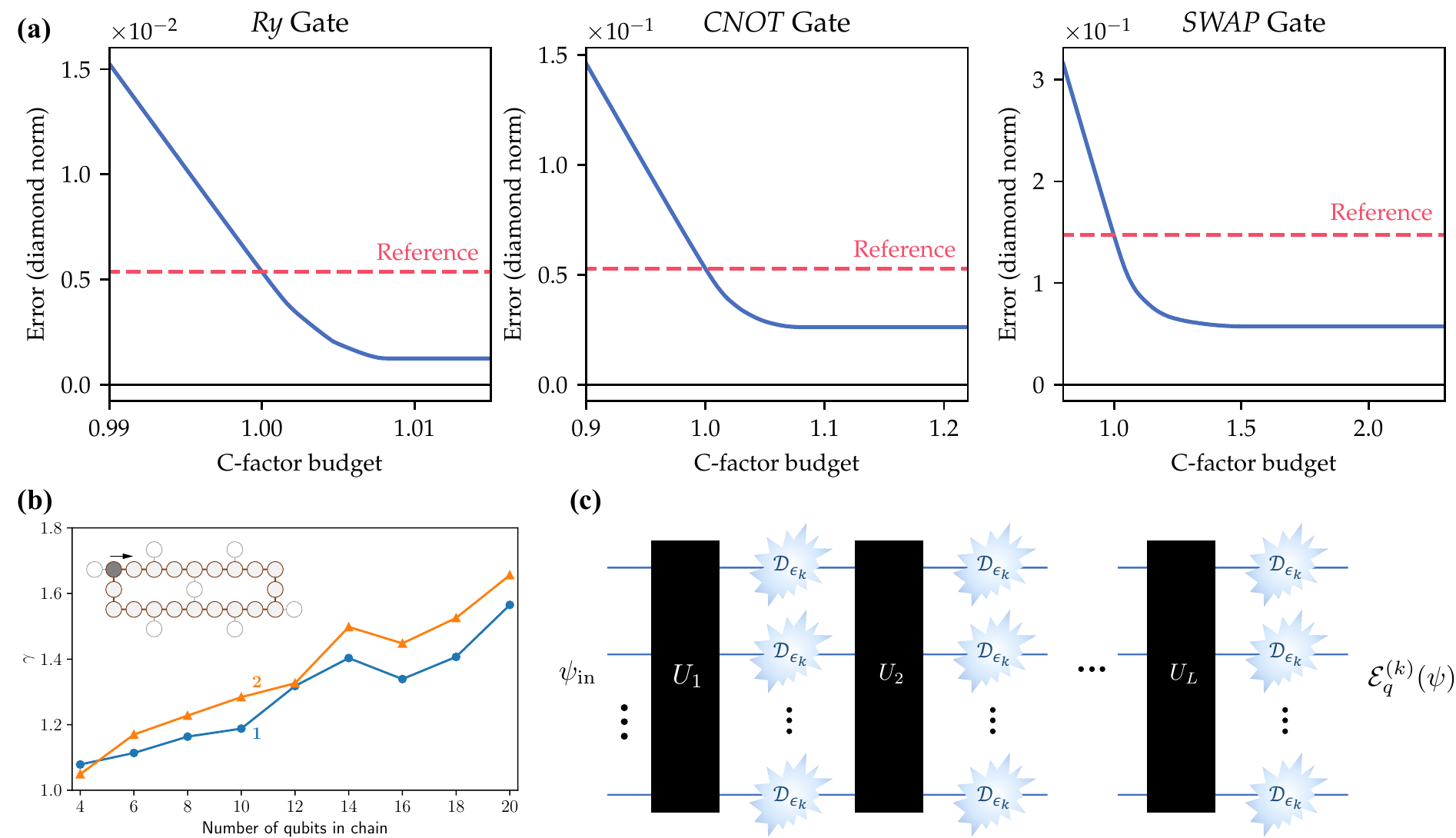}
    \caption{Operational sources of PEC sampling cost. (a) Accuracy--overhead trade-off for reduced-cost quasiprobability decompositions of representative gates: a larger budget $\gamma$ permits a smaller decomposition error. (b) Empirical growth of the sparse-model PEC overhead with system size, showing that efficient characterization and efficient sampling are distinct requirements. (c) Layered noisy-circuit model used to analyze mitigation lower bounds, in which the cancellation cost accumulates over noisy locations and circuit depth. Panels (a)--(c) are adapted from Refs.~\cite{piveteau2022reducedsampling}, \cite{vandenberg2023sparsepec}, and \cite{takagi2022fundamental}, respectively.}
    \label{fig:pec_cost_mechanisms}
\end{figure*}

These mechanisms are summarized in Fig.~\ref{fig:pec_cost_mechanisms}.
Panel~(a) illustrates how improved quasiprobability decompositions trade
approximation accuracy against signed norm
\cite{piveteau2022reducedsampling}.
Panel~(b) shows that sparse Pauli--Lindblad characterization can make the noise
model scalable while the resulting PEC overhead can still increase with system
size
\cite{vandenberg2023sparsepec}.
Panel~(c) emphasizes that, in layered circuits, local inverse costs accumulate
across the relevant circuit depth
\cite{takagi2022fundamental}.

\begin{figure*}[t]
    \centering
    \includegraphics[width=0.52\textwidth]{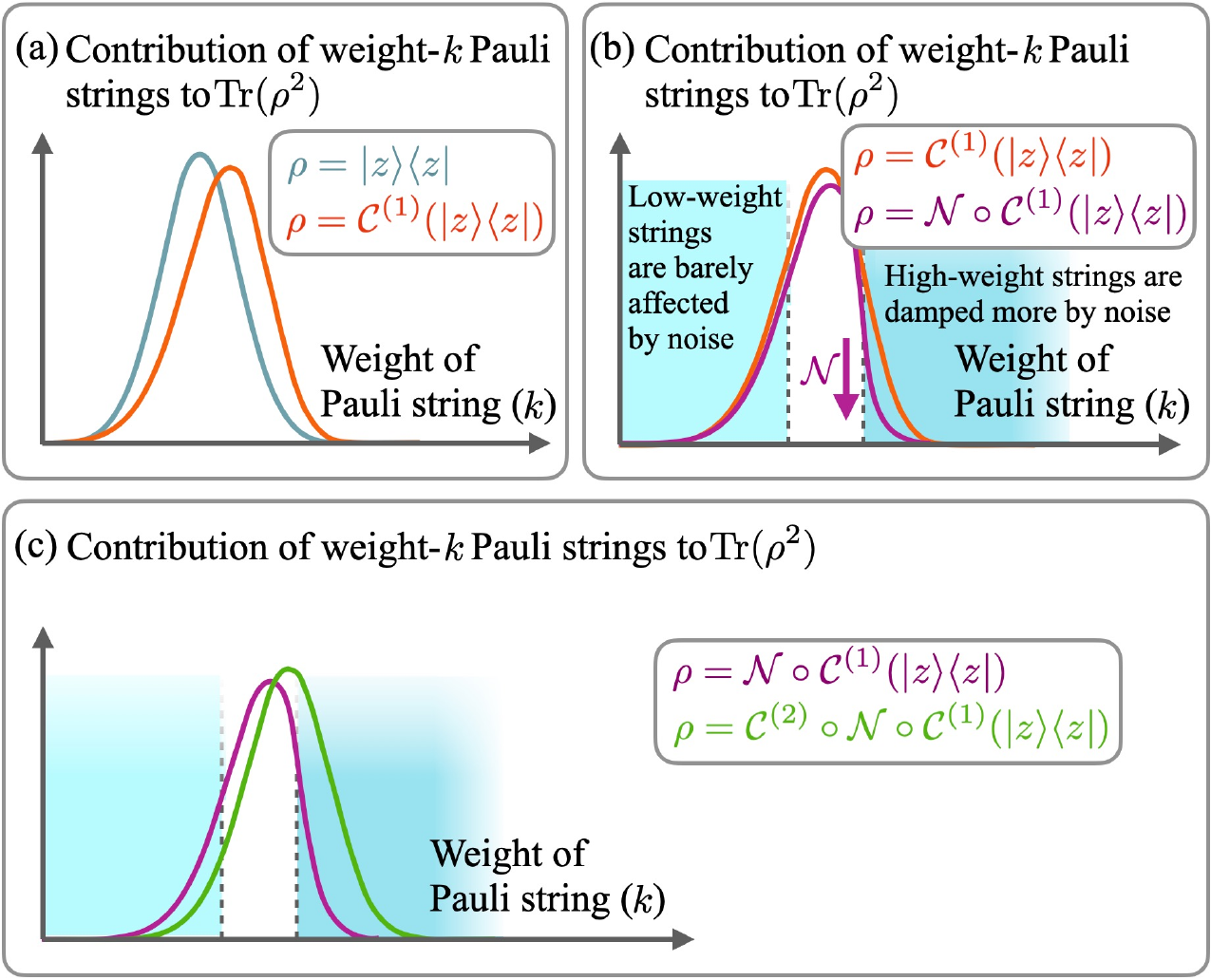}
    \caption{Pauli-weight origin of generic large-system mitigation lower bounds. Local noise contracts high-weight Pauli components more strongly than low-weight components. Reconstructing the original state or observable therefore requires the mitigation estimator to amplify precisely the sectors that have been most strongly suppressed, producing a rapidly growing sampling cost for generic many-body information. Adapted from Ref.~\cite{quek2024tighter}.}
    \label{fig:pec_pauli_weight_limit}
\end{figure*}

A related information-theoretic perspective is shown in
Fig.~\ref{fig:pec_pauli_weight_limit}. Under local noise, low-weight Pauli
components are typically less strongly attenuated than high-weight components.
Recovering strongly damped high-weight information therefore requires larger
statistical amplification. This helps explain why PEC can remain practical for
certain local observables while becoming much more costly for generic
many-body information
\cite{quek2024tighter}.

\subsubsection{Fundamental limits and model violation}

The rapid growth of PEC overhead is not specific to the bit-flip example.
Resource-theoretic and information-theoretic analyses establish general lower
bounds on the sampling cost of unbiased error mitigation
\cite{takagi2021optimal,takagi2022fundamental,
tsubouchi2023universalcost,takagi2023samplinglowerbounds,quek2024tighter}.
Takagi \textit{et al.} showed that, under local noise, maintaining fixed
estimation accuracy can require a cost that grows exponentially with circuit
depth
\cite{takagi2022fundamental}.
Related estimation-theoretic and information-theoretic approaches have derived
broader lower bounds that remain applicable even when general classical
post-processing is allowed
\cite{tsubouchi2023universalcost,takagi2023samplinglowerbounds,
quek2024tighter}.

A separate limitation is \emph{model mismatch}. PEC is unbiased only when the
noise model used to construct the inverse accurately describes the operations
implemented during the target experiment. If the actual channel is
$\mathcal{E}$ but PEC is constructed from an approximate model
$\widetilde{\mathcal{E}}$, then
\begin{equation}
    \widetilde{\mathcal{E}}^{-1}
    \circ
    \mathcal{E}
    \neq
    \mathcal{I},
\end{equation}
and a residual systematic error remains. Calibration drift, context-dependent
noise, non-Markovian effects, and imperfections in the cancellation operations
can all produce such deviations
\cite{endo2018practical,govia2025modelviolation,jin2025noisypec}.

Govia \textit{et al.} quantified how inaccuracies in the learned circuit-noise
model propagate into mitigation bias
\cite{govia2025modelviolation}, while Jin \textit{et al.} generalized PEC to
settings in which the cancellation operations themselves are noisy or
restricted
\cite{jin2025noisypec}.
Thus, PEC scalability depends on two resources simultaneously: the number of
samples required by the inverse representation and the accuracy with which the
relevant effective noise can be characterized.

For continuous open-system evolution, an additional issue can arise from the
time discretization used to construct stepwise inverse channels. Ma and Kim
showed that such discretization can introduce residual approximation errors
even in the infinite-sampling limit
\cite{ma2024limitationsopenpec}. For these problems, the segmentation used for
PEC must therefore remain physically consistent with the underlying open
dynamics rather than being chosen only for computational convenience.

PEC is therefore most useful when the effective inverse channel can be
represented with moderate quasiprobability negativity and when the corresponding
noise model remains stable during execution. Favorable regimes include short
circuits, local observables with small backward light cones, sparse or strongly
structured noise, biased-noise architectures, and workflows in which PEC is
applied only to the most important subset of noisy operations
\cite{cai2023quantum,tran2023locality,vandenberg2023sparsepec}.

These observations do not diminish the conceptual importance of PEC. Rather,
they clarify its practical role. PEC provides a systematic route to unbiased
expectation-value estimation under an accurate noise model, but its usefulness
depends sensitively on circuit structure, observable locality, noise
characterization, and available sampling resources. In realistic workflows,
PEC is therefore often most effective when combined with error suppression or
other mitigation layers that first reduce or simplify the residual noise.

\section{Circuit randomization, twirling, and coherent-error suppression}\label{twirling}

Many experimentally relevant hardware errors are not purely stochastic. Imperfect control, residual couplings, spectator effects, low-frequency fluctuations, and calibration drift can introduce coherent error components that accumulate systematically across a circuit
\cite{cai2023quantum,wallman2016noise,suter2016protecting,wallman2015estimating,sanders2016bounding,kueng2016comparing}.
As a result, two error channels with similar average gate infidelity can lead to markedly different algorithmic or logical performance, depending on how much coherent structure remains in the noise
\cite{wallman2016noise,wallman2015estimating,sanders2016bounding,beale2018decoheres,huang2019performance,jain2023improved}.
For error reduction, it is therefore important to control not only the overall noise strength, but also the form in which that noise accumulates.

The methods discussed in this section address this problem through two complementary mechanisms. Randomized compiling (RC), Pauli-frame randomization (PFR), and related twirling protocols suppress the systematic buildup of coherent errors by averaging over logically equivalent circuit realizations, thereby tailoring the effective noise toward a more stochastic and often Pauli-like form. Dynamical decoupling (DD), by contrast, acts directly on unwanted coherent evolution and seeks to cancel it through refocusing sequences, particularly during idle periods. These methods are therefore most naturally viewed as front-end error-suppression or noise-engineering techniques: they modify the noise before observable-level or estimator-level mitigation is applied, rather than reconstructing an ideal result from noisy measurement data after execution
\cite{cai2023quantum,kimScalableErrorMitigation2023a}.

\begin{itemize}

\item \textbf{Randomized compiling.}
Randomized compiling replaces a target circuit with an ensemble of logically equivalent randomized realizations, typically by inserting random single-qubit gates together with compensating corrections
\cite{wallman2016noise,hashim2021randomized}.
The randomization prevents coherent control errors from accumulating in the same direction from layer to layer. After averaging over the ensemble, the resulting effective noise is tailored toward a stochastic channel, often with a Pauli-like structure. RC therefore does not remove the underlying stochastic error probability, but reduces the coherent accumulation that can otherwise produce disproportionately large circuit-level errors.

\item \textbf{Pauli-frame randomization and twirling.}
Pauli-frame randomization implements the same general noise-tailoring principle using random Pauli operations around selected gates, with the corresponding corrections either applied physically or tracked in software
\cite{ware2021pauliframe,perrin2024crosstalk}.
Averaging over the random Pauli frames suppresses coherent off-diagonal components of the error channel in the Pauli basis and yields a more stochastic effective description. More generally, Pauli and Clifford twirling can be understood as channel-symmetrization procedures that remove selected coherent structure while preserving the desired ideal operation. Their main effect is therefore to simplify the \emph{form} of the residual noise rather than to eliminate its total strength.

\item \textbf{Dynamical decoupling and echo sequences.}
Dynamical decoupling addresses coherent errors through a different physical mechanism. Refocusing pulses are inserted during idle intervals so that unwanted Hamiltonian evolution accumulated during one part of the sequence is approximately cancelled by evolution during another
\cite{viola1998dynamical,viola1999dynamical,suter2016protecting}.
DD is particularly effective against quasi-static dephasing, coherent frequency offsets, and slowly varying unwanted couplings. In contrast, rapidly fluctuating or strongly Markovian noise is generally not removed by the same refocusing mechanism.

\end{itemize}

This distinction between coherent and stochastic noise is also important for subsequent QEM. A residual channel that is approximately stochastic and Pauli-like is generally easier to characterize and incorporate into mitigation procedures
\cite{cai2023quantum,kuritaSynergeticQuantumError2023,ville2022leveraging,kimScalableErrorMitigation2023a}.
For PEC, Pauli-like channels admit simpler characterization and quasi-probability decompositions; for ZNE, suppressing strongly coherent circuit-dependent errors can produce a more regular dependence on the noise-scaling parameter; and for learning-based mitigation, reducing coherent structure can lessen the mismatch between calibration or training circuits and the target computation. RC, PFR, twirling, and DD therefore serve as complementary front-end tools that reshape or suppress coherent errors and leave a residual noise channel that is more amenable to downstream mitigation.

This section develops these ideas in more detail. We first discuss RC as a circuit-level framework for tailoring coherent gate errors into an effective stochastic description. We then formulate Pauli and Clifford twirling more generally as channel-symmetrization procedures, including extensions beyond standard qubit-Clifford settings. Finally, we discuss DD and echo-based suppression, with particular attention to idle-time errors, crosstalk, digital circuit structure, and hardware scheduling constraints
\cite{niu2022analyzing,niu2022effects,ezzell2023dynamical,coote2025graphdd,rahman2024learning}.

\subsection{Randomized compiling and Pauli-frame randomization}

A principal motivation for randomization-based error suppression is that coherent errors can be substantially more damaging than stochastic errors with comparable average strength
\cite{wallman2016noise,wallman2015estimating,sanders2016bounding,kueng2016comparing,beale2018decoheres,huang2019performance,jain2023improved}.
For example, a repeated coherent over-rotation generated by a Pauli operator $P$ gives
\begin{equation}
    U_{\epsilon}
    =
    e^{-i\epsilon P/2},
    \qquad
    U_{\epsilon}^{L}
    =
    e^{-iL\epsilon P/2},
    \label{eq:coherent_accumulation_rc}
\end{equation}
so that a small systematic error $\epsilon$ can accumulate coherently over $L$ layers. Stochastic Pauli errors, by contrast, accumulate probabilistically and are generally less adversarial for deep circuits. This distinction also persists at the logical level, where coherent errors can degrade error-correction performance until they are randomized or otherwise converted into a more stochastic form
\cite{beale2018decoheres,huang2019performance,jain2023improved}.

RC addresses this problem by replacing a target circuit with an ensemble of logically equivalent randomized circuits
\cite{wallman2016noise,hashim2021randomized}.
The circuit is typically decomposed into cycles containing easily implemented single-qubit gates and more noise-sensitive operations such as entangling gates. Random ``twirling'' gates are inserted into each cycle, together with compensating gates chosen so that the ideal unitary implemented by the circuit is unchanged. Different randomized instances therefore implement the same ideal computation but experience different realizations of the coherent hardware error. Averaging their measurement outcomes prevents these coherent contributions from accumulating in a fixed direction across the circuit.

The underlying mechanism can be understood through Pauli twirling. For a noise channel $\mathcal{E}$, averaging over the $n$-qubit Pauli group $\mathcal{P}_n$ gives
\begin{equation}
    \mathcal{E}_{\mathrm{PT}}
    =
    \frac{1}{|\mathcal{P}_n|}
    \sum_{P\in\mathcal{P}_n}
    \mathcal{P}^{\dagger}
    \circ
    \mathcal{E}
    \circ
    \mathcal{P},
    \qquad
    \mathcal{P}(\rho)=P\rho P^{\dagger},
    \label{eq:pauli_twirling_channel}
\end{equation}
which removes the off-diagonal components of the channel in the Pauli-transfer representation and produces a Pauli channel of the form
\begin{equation}
    \mathcal{E}_{\mathrm{PT}}(\rho)
    =
    \sum_{P\in\mathcal{P}_n}
    p_P\,P\rho P^{\dagger}.
    \label{eq:pauli_channel_rc}
\end{equation}
Thus, twirling does not in general reduce the average physical error probability. Instead, it removes coherent structure and replaces the original error channel by a stochastic Pauli mixture with the same ideal circuit action. This distinction is central to RC: its primary benefit is \emph{noise tailoring}, rather than direct cancellation of the underlying physical error
\cite{wallman2016noise}.

Pauli-frame randomization (PFR) provides a particularly direct implementation of this principle. Random Pauli operators are applied around selected circuit operations, while compensating Paulis are either implemented physically or propagated through the circuit and tracked classically
\cite{ware2021pauliframe,perrin2024crosstalk}.
For Clifford operations, this propagation is especially convenient because conjugation maps Pauli operators to Pauli operators. The additional randomization can therefore often be incorporated with little circuit-depth overhead. Averaging over different Pauli frames suppresses coherent, repeatable components of control error and yields a more Pauli-like effective channel.

\subsubsection{Implementation on quantum hardware}

The practical implementation of RC requires the randomization to introduce little additional circuit error or depth. Random single-qubit twirling gates are inserted around computational cycles, while compensating operations preserve the ideal circuit action
\cite{wallman2016noise,hashim2021randomized}. For Pauli-based schemes, these operations can often be absorbed into existing single-qubit gates or tracked in software, so the two-qubit gate count and circuit depth are only weakly affected
\cite{ware2021pauliframe}.

The main overhead instead comes from averaging over multiple randomized circuit realizations. Each realization implements the same ideal computation but experiences a different representation of the coherent hardware error. Ensemble averaging suppresses circuit-dependent coherent contributions, trading additional sampling for a more stochastic effective noise channel. This behavior has been demonstrated experimentally on superconducting processors. Hashim \textit{et al.} showed that RC tailored coherent errors into an effective stochastic model and improved the predictability of algorithmic performance from independently measured error rates
\cite{hashim2021randomized}. Ware \textit{et al.} further demonstrated that Pauli-frame randomization suppresses signatures of non-Markovian behavior and yields noise that is better described by an effective Pauli model, without a substantial fidelity penalty
\cite{ware2021pauliframe}. RC has also been extended to structured crosstalk. Perrin \textit{et al.} introduced additional twirling on neighboring spectator qubits, converting coherent crosstalk into a more stochastic effective channel and improving subsequent noise estimation
\cite{perrin2024crosstalk}. Thus, RC can often be implemented with modest circuit overhead while reducing coherent and correlated error structures that would otherwise complicate later mitigation.

\begin{figure*}[t]
    \centering
    \includegraphics[width=0.95\textwidth]{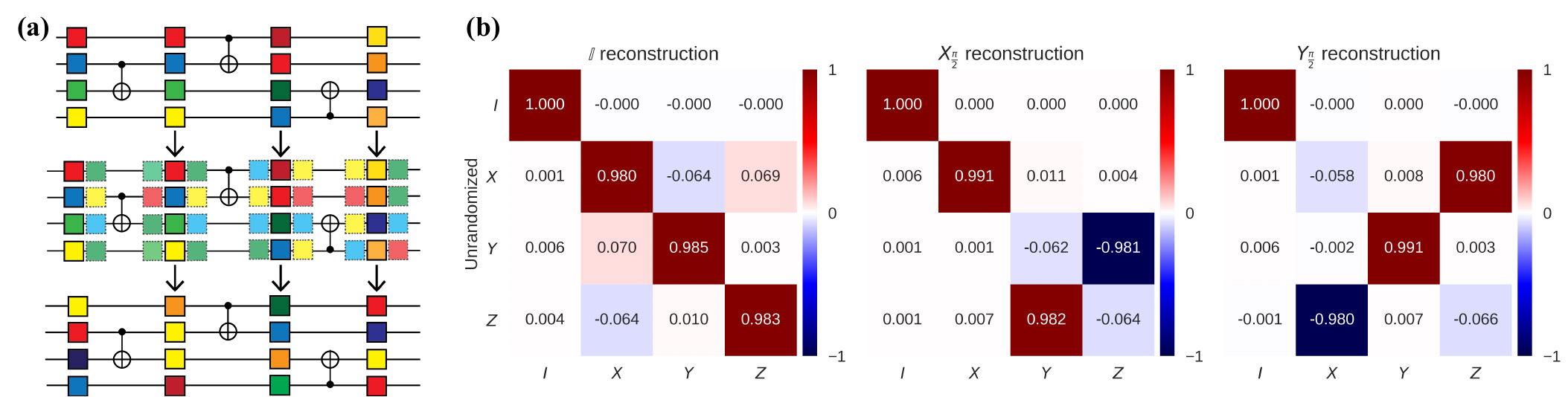}
    \caption{How randomized compiling tailors coherent noise. (a) A bare circuit cycle is converted into an ensemble of dressed cycles by inserting random local frames and compiling the corresponding corrections into adjacent layers. The ideal circuit action is unchanged, while coherent errors are averaged into a stochastic effective channel. (b) Reconstructed process representations before and after Pauli-frame randomization. Suppression of off-diagonal components shows the conversion of coherent and non-Markovian contributions into a more Pauli-like channel. Panels (a) and (b) are adapted from Refs.~\cite{wallman2016noise} and \cite{ware2021pauliframe}, respectively.}
    \label{fig:rc_mechanism}
\end{figure*}

A further practical advantage is that RC can be integrated naturally into the
compilation stack
\cite{wallman2016noise,hashim2021randomized,ware2021pauliframe,
perrin2024crosstalk,cai2023quantum}.
Because the randomization preserves the ideal circuit action, it can often be
introduced after logical synthesis and qubit mapping, with the resulting frame
updates absorbed into native single-qubit operations or accounted for during
scheduling. Its effectiveness nevertheless requires accurate sampling of the
intended random ensemble and sufficiently stable hardware conditions across the
randomized instances. Device drift, or randomizations that substantially alter
pulse timing or crosstalk conditions, can reduce the effectiveness of the
ensemble-averaged noise tailoring. The gain from RC depends strongly on the structure of the underlying noise
\cite{wallman2015estimating,sanders2016bounding,wallman2016noise,
hashim2021randomized}. Its largest benefits arise when coherent calibration
errors, spectator effects, or coherent crosstalk produce systematic
circuit-dependent accumulation. If the residual noise is already predominantly
stochastic, RC can still homogenize the effective channel and improve the
predictability of circuit performance, but the improvement in measured
observables is generally smaller. RC is therefore best viewed as a
noise-tailoring strategy for suppressing the harmful effects of coherent errors,
rather than as a generic circuit-optimization technique.

\begin{figure*}[t]
    \centering
    \includegraphics[width=0.95\textwidth]{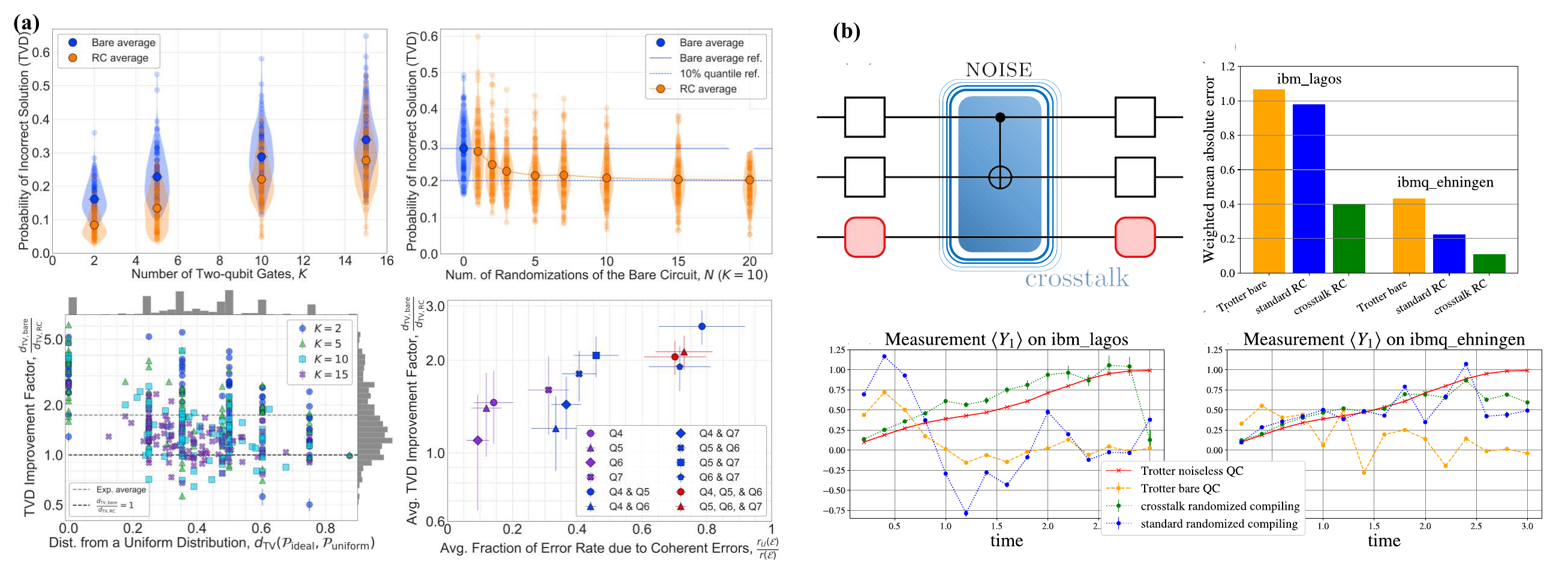}
    \caption{Hardware consequences of randomized compiling. (a) Superconducting-processor benchmarks showing reduced incorrect-solution probability, smaller circuit-to-circuit variability, and improved total-variation distance after averaging randomized circuit instances; the gain is largest when coherent errors constitute a significant fraction of the total error. (b) Crosstalk-aware RC, in which spectator configurations are randomized around entangling gates. The benchmark bars and Trotter-dynamics traces show lower observable error than bare or standard randomized compiling. Panels (a) and (b) are adapted from Refs.~\cite{hashim2021randomized} and \cite{perrin2024crosstalk}, respectively.}
    \label{fig:rc_hardware_results}
\end{figure*}

The mechanism of randomization-based noise tailoring is separated from the hardware benchmarks in Figs.~\ref{fig:rc_mechanism} and \ref{fig:rc_hardware_results}. Fig.~\ref{fig:rc_mechanism}(a) shows the dressed-circuit construction of RC. Random local frame changes are inserted around the hard computational gates, and compensating corrections are compiled into neighbouring cycles so that every sampled circuit implements the same ideal unitary. Averaging over these logically equivalent instances removes the phase coherence associated with a fixed control error and replaces it with an effective stochastic channel \cite{wallman2016noise}. Fig.~\ref{fig:rc_mechanism}(b) shows the same effect at the reconstructed-channel level. Before Pauli-frame randomization, coherent error appears as appreciable off-diagonal components of the process representation; after randomization, these components are strongly reduced, and the channel is closer to a Pauli-diagonal, effectively Markovian model \cite{ware2021pauliframe}.

The resulting hardware-level effects are summarized in Fig.~\ref{fig:rc_hardware_results}. Panel (a) compares ensembles of bare and randomized circuits on a superconducting processor. RC lowers the probability of an incorrect solution, reduces the spread among logically equivalent circuit instances, and yields larger total-variation-distance improvement when a substantial fraction of the physical error is coherent; averaging over more randomizations causes the estimator to converge toward the tailored-channel result \cite{hashim2021randomized}. Panel (b) extends the same logic to spectator-induced crosstalk. Randomizing the spectator configuration around an entangling operation converts a coherent conditional interaction into a stochastic contribution. The reduced weighted mean absolute error and the improved Trotter-dynamics traces show that crosstalk-aware RC outperforms both the bare execution and standard RC when spectator coupling is a leading error source \cite{perrin2024crosstalk}.

\subsection{Twirling techniques: Pauli and Clifford twirling}

Twirling is a channel-symmetrization procedure in which a noise channel is
averaged under conjugation by a chosen set of unitary operations
\cite{cai2023quantum,cai2019constructing}. Let $\mathcal{E}$ denote a noise
channel and $\mathcal{G}$ a finite twirling group. For each
$G\in\mathcal{G}$, define
$\mathcal{U}_{G}(\rho)=G\rho G^{\dagger}$. The corresponding twirled channel is
\begin{equation}
    \mathcal{E}_{\mathcal{G}}
    =
    \frac{1}{|\mathcal{G}|}
    \sum_{G\in\mathcal{G}}
    \mathcal{U}_{G}^{\dagger}
    \circ
    \mathcal{E}
    \circ
    \mathcal{U}_{G},
    \label{eq:twirled_channel}
\end{equation}
where $\mathcal{U}_{G}^{\dagger}(\rho)=G^{\dagger}\rho G$. The purpose of
twirling is generally not to reduce the average error strength, but to simplify
the structure of the ensemble-averaged noise channel by projecting out
components that are not invariant under the chosen symmetry. This can reduce
the number of parameters required to characterize the effective noise and make
it easier to model, simulate, or mitigate.

Two important examples are Pauli and Clifford twirling. For the full Pauli
group, Pauli twirling removes off-diagonal components of the channel in the
Pauli basis and maps an arbitrary noise channel to a stochastic Pauli channel
\cite{cai2019constructing,wallman2016noise,ware2021pauliframe}. For example,
consider a coherent single-qubit $Z$ over-rotation, $U_{\epsilon}
    =
    e^{-i\epsilon Z/2},
    \mathcal{E}_{\epsilon}(\rho)
    =
    U_{\epsilon}\rho U_{\epsilon}^{\dagger}.$
After Pauli twirling, the ensemble-averaged channel becomes
\begin{equation}
    \mathcal{T}_{\mathcal{P}}(\mathcal{E}_{\epsilon})(\rho)
    =
    \cos^2\!\left(\frac{\epsilon}{2}\right)\rho
    +
    \sin^2\!\left(\frac{\epsilon}{2}\right)Z\rho Z,
    \label{eq:pauli_twirl_example}
\end{equation}
so that the coherent over-rotation is replaced by a stochastic $Z$ error with
probability $p_Z=\sin^2(\epsilon/2)$. Importantly, this stochastic description
emerges only after averaging over the Pauli-randomized circuit ensemble; each
individual circuit still experiences a particular realization of the underlying
hardware noise. Pauli twirling is experimentally attractive because the
randomizing operations are local and can often be absorbed into neighboring
single-qubit gates or efficiently propagated through a Pauli frame.

Clifford twirling imposes a stronger symmetry. Because the Clifford group forms
a unitary $2$-design, averaging an arbitrary channel over the full Clifford
group produces a depolarizing channel
\cite{dankert2009exact}. For example, an anisotropic Pauli channel
\begin{equation}
    \mathcal{E}_{P}(\rho)
    =
    p_I\rho
    +
    p_X X\rho X
    +
    p_Y Y\rho Y
    +
    p_Z Z\rho Z
\end{equation}
is transformed by a full single-qubit Clifford twirl into
\begin{equation}
    \mathcal{T}_{\mathcal{C}}(\mathcal{E}_{P})(\rho)
    =
    p_I\rho
    +
    \frac{1-p_I}{3}
    \left(
        X\rho X+Y\rho Y+Z\rho Z
    \right),
    \label{eq:clifford_twirl_example}
\end{equation}
which treats the three nonidentity Pauli errors symmetrically. Thus, Pauli
twirling removes coherent coupling between different Pauli components, whereas
Clifford twirling further homogenizes the remaining Pauli error directions.

This stronger standardization can be useful for noise characterization and
benchmarking, but implementing a full multi-qubit Clifford twirl can require
substantially more circuit resources than Pauli twirling. In practical settings,
one therefore often restricts the randomization to smaller gate sets or to the
circuit elements whose noise is of interest. Recent work has further introduced
symmetric Clifford twirling and its $k$-sparse variant, which restrict the
allowed Clifford operations according to the symmetries of the target
computation and can tailor selected logical noise toward an approximately
global white-noise form
\cite{tsubouchi2025symmetricclifford}.

For error mitigation, the choice of twirling group therefore reflects a
trade-off between the degree of noise simplification and the implementation
cost. Pauli twirling is often sufficient when the objective is to suppress
coherent accumulation and obtain a Pauli-like noise model that can be
characterized or inverted efficiently. Full Clifford twirling produces the
stronger depolarizing symmetry, but this level of randomization is often
unnecessary for a specific circuit or observable and can introduce greater
implementation overhead. Consequently, hardware implementations commonly use
Pauli twirling or randomized compiling around selected gates rather than
twirling the entire computation over the full Clifford group
\cite{wallman2016noise,hashim2021randomized,ware2021pauliframe,cai2023quantum}.

Recent work has extended twirling beyond the standard Pauli and Clifford settings. Pauli conjugation selects suitable Pauli transformations to exploit, rather than completely remove, coherent noise structure and improve logical performance
\cite{cai2020pauliconjugation}. Pseudo twirling extends coherent-error tailoring to native non-Clifford gates, for which standard Pauli twirling or randomized compiling is not directly applicable
\cite{santos2024pseudotwirling}. Symmetric Clifford twirling restricts the randomization to Clifford operations compatible with selected symmetries, enabling noise tailoring for non-Clifford operations
\cite{tsubouchi2025symmetricclifford}. Similar ideas have also been generalized beyond qubits using Weyl--Heisenberg randomization for qudit and qutrit systems
\cite{goss2024qutrit}. These developments show that twirling is better viewed as a broader family of symmetry-based noise-tailoring methods rather than a single fixed protocol.

\subsubsection{Channel twirling and noise simplification}
At the formal level, twirling a noise channel $\mathcal{N}$ over a finite gate
group $\mathbb{G}$ means averaging over conjugation by the group elements
\cite{cai2023quantum,cai2019constructing}. A more representative example is a two-qubit coherent error containing both
local phase offsets and a residual correlated coupling. Suppose that, after an
ideal entangling gate, the residual unitary error is
\begin{equation}
U_{\rm err}
=
\exp\left[
-\frac{i}{2}
\left(
\epsilon\, Z\otimes I
+
\eta\, I\otimes Z
+
\xi\, Z\otimes Z
\right)
\right].
\label{eq:twirl_correlated_unitary}
\end{equation}
Because the three generators commute, this unitary can be expanded in the
two-qubit Pauli basis as
\begin{equation}
U_{\rm err}
=
a_I I
+
a_{ZI} ZI
+
a_{IZ} IZ
+
a_{ZZ} ZZ ,
\end{equation}
where, defining
$c_\mu=\cos(\mu/2)$ and $s_\mu=\sin(\mu/2)$,
\begin{align}
a_I
&=
c_\epsilon c_\eta c_\xi
+
i s_\epsilon s_\eta s_\xi,
\\
a_{ZI}
&=
-i s_\epsilon c_\eta c_\xi
-
c_\epsilon s_\eta s_\xi,
\\
a_{IZ}
&=
-i c_\epsilon s_\eta c_\xi
-
s_\epsilon c_\eta s_\xi,
\\
a_{ZZ}
&=
-s_\epsilon s_\eta c_\xi
-
i c_\epsilon c_\eta s_\xi .
\end{align}
The corresponding coherent channel,
$\mathcal{N}_{\rm err}(\rho)=U_{\rm err}\rho U_{\rm err}^{\dagger}$,
contains not only the diagonal terms
$|a_P|^2P\rho P$, but also interference terms of the form
$a_Pa_Q^\ast P\rho Q$ with $P\neq Q$.

Twirling over the full two-qubit Pauli group removes these off-diagonal
interference terms and gives
\begin{equation}
\mathcal{T}_{\mathcal{P}_2}
(\mathcal{N}_{\rm err})(\rho)
=
p_I\rho
+
p_{ZI}ZI\rho ZI
+
p_{IZ}IZ\rho IZ
+
p_{ZZ}ZZ\rho ZZ ,
\label{eq:twirl_correlated_pauli}
\end{equation}
with
\begin{align}
p_I
&=
c_\epsilon^2 c_\eta^2 c_\xi^2
+
s_\epsilon^2 s_\eta^2 s_\xi^2,
\\
p_{ZI}
&=
s_\epsilon^2 c_\eta^2 c_\xi^2
+
c_\epsilon^2 s_\eta^2 s_\xi^2,
\\
p_{IZ}
&=
c_\epsilon^2 s_\eta^2 c_\xi^2
+
s_\epsilon^2 c_\eta^2 s_\xi^2,
\\
p_{ZZ}
&=
s_\epsilon^2 s_\eta^2 c_\xi^2
+
c_\epsilon^2 c_\eta^2 s_\xi^2 .
\end{align}
For weak coherent errors,
$|\epsilon|,|\eta|,|\xi|\ll1$, these probabilities reduce approximately to
\begin{equation}
p_{ZI}\simeq\frac{\epsilon^2}{4},
\qquad
p_{IZ}\simeq\frac{\eta^2}{4},
\qquad
p_{ZZ}\simeq\frac{\xi^2}{4}.
\end{equation}
Thus, the original error contains coherent interference between local and
correlated error components, whereas Pauli twirling converts the
ensemble-averaged channel into a stochastic mixture of $ZI$, $IZ$, and $ZZ$
faults. The twirl does not eliminate the underlying error strength; rather, it
removes its coherent off-diagonal structure and produces a noise model that is
typically easier to characterize and incorporate into subsequent mitigation
procedures.

The preceding example illustrates the general action of Pauli twirling: coherent interference between different Pauli components is removed, while the corresponding diagonal Pauli-error weights are retained in the ensemble-averaged channel. More generally, when $\mathbb{G}$ is the full Pauli group, the twirl eliminates the off-diagonal components of the channel in the Pauli basis and produces a stochastic Pauli channel. This is particularly useful when coherent errors would otherwise accumulate systematically with circuit depth
\cite{wallman2016noise,sanders2016bounding}. The resulting Pauli channel is also typically easier to characterize, model, and incorporate into subsequent mitigation or inversion procedures
\cite{cai2023quantum}. When $\mathbb{G}$ is the Clifford group, the symmetrization is stronger. Because the Clifford group forms a unitary $2$-design and acts transitively on the nonidentity Pauli operators, a full Clifford twirl further averages the remaining Pauli-error directions and maps the channel to a depolarizing form
\cite{dankert2009exact}. Thus, Pauli twirling mainly removes coherent structure while preserving the anisotropy of the stochastic Pauli errors, whereas Clifford twirling additionally homogenizes the nonidentity error components. The stronger symmetry can simplify benchmarking and noise characterization, but it is not always necessary for mitigation and generally carries a larger implementation overhead than Pauli twirling.

Importantly, twirling changes the \emph{structure} of the effective noise rather
than simply reducing its average strength. In particular, Pauli twirling
preserves the average fidelity of the underlying error channel while converting
coherent components into stochastic Pauli errors. Its advantage therefore
appears primarily in how errors accumulate across a circuit and in the
predictability of the resulting effective noise, rather than as an improvement
in average gate fidelity
\cite{sanders2016bounding,wallman2016noise}. This simpler effective channel can
in turn provide a more favorable starting point for downstream error mitigation.

Recent work also shows that useful noise tailoring can lie between leaving the
noise unchanged and fully twirling it. Pauli conjugation can improve logical
fidelity over full twirling for certain coherent-noise models by selecting Pauli
conjugations that retain beneficial parts of the coherent structure
\cite{cai2020pauliconjugation}. Symmetric Clifford twirling provides a related
extension for non-Clifford operations by restricting the randomization to
Clifford operators that commute with a relevant Pauli subgroup, thereby
scrambling selected noise components while preserving the symmetry of the
target operation
\cite{tsubouchi2025symmetricclifford}. These developments illustrate that the
appropriate twirling strategy depends on which noise structure should be
removed, which symmetries must be preserved, and which randomizing operations
can be implemented efficiently on the target hardware.

\subsubsection{Implementation on quantum hardware}
The hardware implementation of Pauli twirling is particularly convenient when
the circuit is dominated by Clifford operations or by native entangling gates
surrounded by low-overhead local rotations
\cite{wallman2016noise,cai2023quantum}. To see this explicitly, consider an
ideal gate $U$ and insert a Pauli operator $P$ before it. The corresponding
post-gate correction $P'$ is chosen such that
\begin{equation}
    P'^{\dagger} U P = U,
\end{equation}
which requires $P' = U P U^{\dagger}$.
If $U$ is a Clifford gate, then by definition $ U P U^{\dagger}\in\mathcal{P}_n
    \qquad
    \forall\,P\in\mathcal{P}_n$,
so both the pre- and post-gate corrections remain Pauli operators. They can
therefore often be merged into neighboring single-qubit layers or propagated
through the circuit as software Pauli-frame updates.

For example, for a controlled-$Z$ gate,
\begin{equation}
\begin{aligned}
    \mathrm{CZ}(X\otimes I)\mathrm{CZ}
    &=X\otimes Z,\\
    \mathrm{CZ}(I\otimes X)\mathrm{CZ}
    &=Z\otimes X,\\
    \mathrm{CZ}(Z\otimes I)\mathrm{CZ}
    &=Z\otimes I,\\
    \mathrm{CZ}(I\otimes Z)\mathrm{CZ}
    &=I\otimes Z.
\end{aligned}
\label{eq:cz_pauli_propagation}
\end{equation}
Thus, Pauli randomizations placed around the entangling gate remain entirely
within the two-qubit Pauli group. This makes local twirling of native
entangling gates especially convenient on present-day hardware, where
two-qubit operations are often an important source of coherent control error
and crosstalk.

The situation is more subtle for non-Clifford operations. In general,
\begin{equation}
    U_{\rm NC} P U_{\rm NC}^{\dagger}
    \notin \mathcal{P}_n ,
    \label{eq:nonclifford_pauli_propagation}
\end{equation}
so propagating a Pauli correction through a non-Clifford gate can generate a
more complicated operation. A simple example is the $T$ gate, $ T=e^{-i\pi Z/8}$
for which $T X T^{\dagger}
    =
    \frac{X+Y}{\sqrt{2}}
    =
    e^{-i\pi/4}SX $. 
The propagated correction is therefore no longer a Pauli operator, although in
this particular case it remains a Clifford operation. For more general
non-Clifford gates, the required compensation can itself be non-Clifford and
may introduce additional circuit depth or control error.

For this reason, the most extensive possible twirling is not necessarily the
most practical. In experiments and software stacks, randomization is often
applied selectively: native entangling gates are twirled while surrounding
single-qubit corrections are absorbed into the compiled layers, Clifford-
dominated portions of the circuit are randomized, or the allowed
randomizations are restricted to a provider-native frame in which the
compensating operations remain low overhead. This selectivity reflects the
hardware-aware nature of twirling: the useful randomization group is determined
not only by its formal noise-symmetrization properties, but also by how
efficiently the corresponding frame corrections can be implemented on the
target processor.

Recent developments also make this hardware dependence even clearer. Pseudo twirling provides a route to coherent-error mitigation for compact native non-Clifford gates, where one would like to preserve the shorter native implementation rather than replace it by a longer Clifford decomposition \cite{santos2024pseudotwirling}. Symmetric Clifford twirling provides a complementary route when a relevant Pauli symmetry subgroup is known: one twirls using only Clifford operators compatible with the symmetry, allowing certain non-Clifford layers to be noise-tailored without unrestricted Clifford randomization \cite{tsubouchi2025symmetricclifford}. In higher-dimensional devices, generalized Weyl--Heisenberg frames play the role of Pauli frames. The superconducting-qutrit experiment of Goss \textit{et al.} demonstrates this principle by tailoring qudit Markovian noise to stochastic Weyl--Heisenberg channels and combining noise tailoring with mitigation on a transmon qutrit processor \cite{goss2024qutrit}.
A related practical idea is measurement-side symmetrization. Randomized pre-measurement bit flips together with classical relabeling do not twirl gate noise, but they apply the same general philosophy to detector bias: an asymmetric readout channel is averaged into a simpler effective model \cite{cai2023quantum}. Since this manuscript already contains a dedicated section on readout mitigation, it is enough here to note that this is another instance of symmetry-based noise simplification. The broader lesson is that twirling is valuable when a complicated but structured error can be replaced by a simpler symmetric one without large physical overhead.

\begin{figure*}[t]
    \centering
    \includegraphics[width=\linewidth]{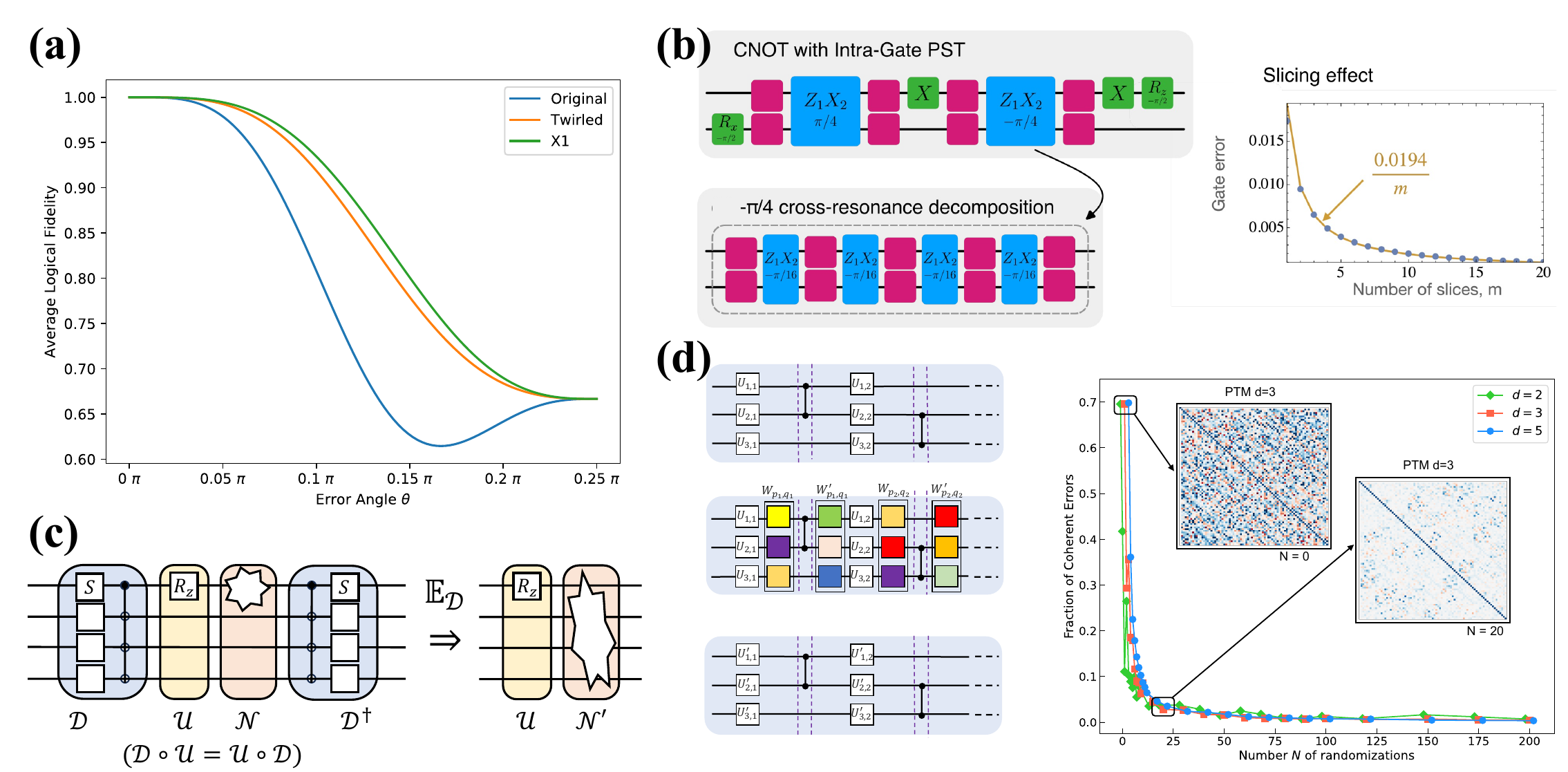}
    \caption{Beyond standard Pauli twirling: recent generalizations of coherent-noise tailoring. (a)
    Pauli conjugation can outperform full twirling for certain coherent-noise models by exploiting,
    rather than fully erasing, part of the channel coherence. Adapted from Fig.~2 of
    Ref.~\cite{cai2020pauliconjugation}. (b) Pseudo twirling for native non-Clifford gates,
    illustrating experimental coherent-error mitigation without reverting to longer Clifford
    decompositions. Adapted from Fig.~3 of Ref.~\cite{santos2024pseudotwirling}. (c) Symmetric
    Clifford twirling, where only symmetry-compatible Clifford operations are used to scramble noise
    affecting non-Clifford layers. Adapted from Fig.~2 of
    Ref.~\cite{tsubouchi2025symmetricclifford}. (d) RC for qudit circuits on a
    superconducting qutrit processor, highlighting that the same basic noise-tailoring idea extends
    beyond qubits through generalized Weyl frames. Adapted from Fig.~2 of
    Ref.~\cite{goss2024qutrit}.}
    \label{figS7_2}
\end{figure*}

Fig.~\ref{figS7_2} shows that modern twirling has evolved beyond the textbook qubit-Pauli setting. Panel (a) illustrates Pauli conjugation as a partial alternative to full twirling: for certain coherent-noise models, selectively conjugating the error by low-weight Pauli operators can preserve useful structure and yield higher logical fidelity than either the original noisy channel or a fully twirled channel \cite{cai2020pauliconjugation}. Panel (b) shows pseudo-twirling for compact native non-Clifford gates. By inserting randomized frame changes within a sliced native gate implementation, coherent gate errors can be suppressed without replacing the native operation by a much longer Clifford decomposition, thereby preserving the depth advantage of the hardware-native gate \cite{santos2024pseudotwirling}. Panel (c) illustrates symmetric Clifford twirling, where only Clifford operations compatible with a chosen Pauli symmetry are used. This allows noise affecting non-Clifford layers to be scrambled while preserving the symmetry structure of the target circuit \cite{tsubouchi2025symmetricclifford}. Panel (d) shows the corresponding generalization beyond qubits: randomized compiling for qudit circuits replaces Pauli frames by generalized Weyl frames, and the superconducting-qutrit implementation demonstrates that repeated randomization can reduce coherent components in the reconstructed process matrix \cite{goss2024qutrit}. Taken together, these examples show that Pauli twirling is best viewed as the simplest member of a broader family of symmetry-based coherent-noise tailoring methods, whose practical form depends on the native gates, available frame updates, protected symmetries, and local Hilbert-space dimension.

\subsection{Dynamical decoupling and echo-based coherent-error suppression}

Unlike circuit randomization and twirling, which mainly reshape the effective error channel, dynamical decoupling (DD) and echo-based methods aim to suppress unwanted coherent evolution directly within each circuit run
\cite{viola1998dynamical,viola1999dynamical,suter2016protecting}. Their natural targets are slowly varying errors such as quasi-static dephasing, low-frequency detuning fluctuations, residual always-on couplings, and spectator-induced coherent interactions that accumulate during idle intervals. This makes DD particularly relevant to digital quantum hardware, where compiled circuits often contain substantial idle windows due to routing, synchronization, or nonuniform gate durations
\cite{pokharel2018demonstration,niu2022analyzing,niu2022effects,ezzell2023dynamical}.

The basic mechanism is most transparent in the toggling frame. Let the unwanted evolution be generated by a noise Hamiltonian $H_{\mathrm{err}}(t)$, and let $P(t)$ denote the control sequence generated by the DD pulses. In the toggling frame, the error Hamiltonian becomes
\begin{equation}
    \widetilde{H}_{\mathrm{err}}(t)
    =
    P^{\dagger}(t)
    H_{\mathrm{err}}(t)
    P(t).
    \label{eq:dd_toggling}
\end{equation}
Over a DD cycle of duration $T$, the corresponding error evolution is
\begin{equation}
    U_{\mathrm{err}}(T)
    =
    \mathcal{T}
    \exp\!\left[
        -i\int_{0}^{T}
        \widetilde{H}_{\mathrm{err}}(t)\,dt
    \right].
    \label{eq:dd_error_evolution}
\end{equation}
To leading order in average-Hamiltonian theory, DD suppresses the coherent error when
\begin{equation}
    \overline{H}_{\mathrm{err}}^{(0)}
    =
    \frac{1}{T}
    \int_{0}^{T}
    \widetilde{H}_{\mathrm{err}}(t)\,dt
    \approx 0.
    \label{eq:dd_average_hamiltonian}
\end{equation}
Thus, DD does not merely randomize the error; it engineers the control sequence so that unwanted Hamiltonian contributions cancel over the cycle
\cite{viola1999dynamical,suter2016protecting}.

The simplest example is a spin echo for quasi-static dephasing,
\begin{equation}
    H_{\mathrm{err}}
    =
    \frac{\delta}{2}Z.
\end{equation}
An intermediate $X$ pulse reverses the sign of the dephasing Hamiltonian because
$XZX=-Z$, giving
\begin{equation}
    e^{-i(\delta Z/2)\tau}
    X
    e^{-i(\delta Z/2)\tau}
    X
    =
    I,
    \label{eq:spin_echo}
\end{equation}
for ideal instantaneous pulses and constant $\delta$. More general DD sequences extend this refocusing principle to multiple axes and more complicated slowly varying couplings.

From a conceptual standpoint, DD therefore belongs to a different branch of coherent-error suppression than RC. Randomization accepts that the error is present and seeks to standardize its effective form, whereas DD attempts to cancel the unwanted Hamiltonian itself. Twirling is especially effective when coherent errors are associated with gate operations and can be efficiently randomized; DD is most effective when the dominant error accumulates during idling or arises from slowly varying control offsets that can be refocused. In superconducting architectures, both mechanisms can be relevant because coherent errors may arise from idle dephasing as well as residual or tunable-coupler-mediated interactions
\cite{qiu2021suppressing,tripathi2022suppression}.

This distinction also explains why the two approaches can be used simultaneously. A processor may apply DD to suppress idle-time dephasing while using Pauli twirling or RC to tailor coherent gate errors. The appropriate strategy therefore depends on the dominant coherent error mechanism in the execution regime of interest rather than on one protocol being universally superior
\cite{cai2023quantum,suter2016protecting,qiu2021suppressing}.

\subsubsection{Sequences and insertion strategies}

Building on the refocusing picture introduced above, practical DD differs mainly in the choice of pulse sequence and in how that sequence is embedded into the compiled circuit. For a sequence of $m$ control pulses $P_j$ separated by free-evolution intervals $\tau_j$, the total evolution can be written schematically as
\begin{equation}
    U_{\rm DD}(T)
    =
    P_m e^{-iH\tau_m}
    \cdots
    P_2 e^{-iH\tau_2}
    P_1 e^{-iH\tau_1},
    \qquad
    T=\sum_j \tau_j ,
    \label{eq:dd_sequence}
\end{equation}
where $H$ contains both the desired and unwanted Hamiltonian terms. In the toggling frame defined by the applied pulses, the leading effective error Hamiltonian is
\begin{equation}
    \overline{H}_{\rm err}^{(0)}
    =
    \frac{1}{T}
    \sum_j
    \tau_j
    Q_j^\dagger H_{\rm err} Q_j,
    \qquad
    Q_j=P_jP_{j-1}\cdots P_1 .
    \label{eq:dd_discrete_average}
\end{equation}
A DD sequence is effective when the pulse pattern makes the unwanted contributions in Eq.~\eqref{eq:dd_discrete_average} cancel or become strongly reduced.

Different sequences realize this cancellation in different ways. Common choices include CPMG, XY4, XY8, concatenated DD, universally robust sequences, quadratic DD, and Uhrig DD
\cite{suter2016protecting,uhrig2007keeping,ezzell2023dynamical}.
For dephasing noise, their action can also be understood through a modulation function $y(t)=\pm1$ that changes sign at each refocusing pulse. The accumulated phase becomes
\begin{equation}
    \phi(T)
    =
    \int_0^T y(t)\,\delta(t)\,dt ,
    \label{eq:dd_phase_filter}
\end{equation}
so that the pulse timing determines which frequency components of the noise are averaged away. In this language, DD acts as a spectral filter: different sequences produce different filter functions and are therefore optimized for different noise spectra
\cite{suter2016protecting,ezzell2023dynamical}. For example, the $N$ pulses of Uhrig DD are placed at
\begin{equation}
    t_j
    =
    T\sin^2\!\left(
    \frac{j\pi}{2N+2}
    \right),
    \qquad
    j=1,\ldots,N,
    \label{eq:udd_timing}
\end{equation}
rather than at uniformly spaced times, suppressing low-frequency dephasing to high order under the assumptions of the protocol
\cite{uhrig2007keeping}.

For quantum hardware, however, the formally highest-order sequence is not necessarily the most effective. The optimal choice depends on the dominant noise spectrum, pulse fidelity, and available idle duration. On superconducting processors, relatively simple DD sequences can already reduce idle-time dephasing
\cite{pokharel2018demonstration}, while more elaborate sequences can provide additional improvement when the idle window is sufficiently long and the additional control pulses are well calibrated
\cite{niu2022analyzing,ezzell2023dynamical}. Related refocusing strategies have also been used to suppress coherent errors associated with tunable-coupler interactions
\cite{qiu2021suppressing}.

Echoed native gates apply the same principle during active operations rather than idle periods. In this case, the control sequence is chosen so that an unwanted interaction $H_{\rm err}$ changes sign under an echo operation $P$, $P^\dagger H_{\rm err}P=-H_{\rm err},$
while the desired interaction is preserved. The unwanted evolution then cancels to leading order across the echoed gate sequence
\cite{tripathi2022suppression}. Idle-window DD and echoed gate constructions can therefore be understood within the same toggling-frame framework.

Finally, the insertion strategy is as important as the pulse sequence itself. In compiled digital circuits, DD is normally inserted only into idle windows that are long enough to accommodate the required pulses after routing and scheduling. Its effectiveness depends on the available window length, pulse spacing, neighboring operations, and errors introduced by the additional pulses
\cite{niu2022analyzing,niu2022effects,ezzell2023dynamical}. Effective hardware-level DD therefore requires joint optimization of the sequence, pulse timing, and placement within the compiled schedule.

\subsubsection{Compatibility with digital algorithms and compilation constraints}

The effectiveness of dynamical decoupling (DD) depends not only on the pulse sequence itself, but also on whether that sequence is compatible with the physical noise, the compiled schedule, and the structure of the target algorithm. In practice, DD should therefore be treated as a hardware-aware scheduling problem rather than as an isolated control protocol
\cite{niu2022analyzing,niu2022effects,ezzell2023dynamical,coote2025graphdd}.

\begin{itemize}

\item \textbf{Physical applicability.}
DD is useful only when the dominant error can be refocused and when the circuit contains sufficiently long idle intervals to accommodate the required pulses. It is most effective against slowly varying coherent errors such as quasi-static dephasing, low-frequency detuning, and residual couplings. By contrast, irreversible processes such as amplitude damping and leakage cannot be cancelled by basis flips in the same way
\cite{suter2016protecting,pokharel2018demonstration}.
Thus, the first question is whether the relevant noise mechanism and idle timescale are compatible with DD at all.

\item \textbf{Hardware and compilation constraints.}
Even when DD is physically appropriate, its implementation is restricted by the compiled backend schedule. The inserted pulses must satisfy hardware timing rules, pulse-duration constraints, channel-sharing requirements, forbidden overlaps, and crosstalk-management conditions
\cite{tripathi2022suppression,niu2022analyzing}.
Because these constraints are device-dependent, the same nominal DD sequence can perform differently on different processors or even under different compilations of the same circuit.

\item \textbf{Sequence placement and optimization.}
Within the available schedule, one must decide which idle qubits should be protected, where the refocusing pulses should be placed, and how the pulse spacing should balance error cancellation against imperfections introduced by the additional control operations
\cite{niu2022effects,ezzell2023dynamical}.
Context-aware methods such as GraphDD address this problem by tailoring the DD placement to the circuit structure and device connectivity
\cite{coote2025graphdd}. Learning-based approaches provide a complementary strategy by optimizing pulse choices or sequence parameters directly from hardware data
\cite{rahman2024learning}.
The optimal DD protocol is therefore determined jointly by the pulse sequence and its placement within the compiled circuit.

\item \textbf{Algorithm dependence.}
The usefulness of DD also depends on the idle structure generated by the target algorithm. Shallow or densely scheduled circuits may contain only short and fragmented idle intervals, limiting the benefit of additional refocusing pulses. By contrast, digital simulation, phase estimation, and routed many-qubit circuits can contain longer spectator windows caused by asynchronous entangling operations, connectivity constraints, or synchronization between different parts of the processor
\cite{pokharel2018demonstration,tripathi2022suppression}.
Such circuits can provide substantially more opportunity for DD-based error suppression.

\end{itemize}

\begin{figure*}[t]
    \centering
    \includegraphics[width=0.6\textwidth]{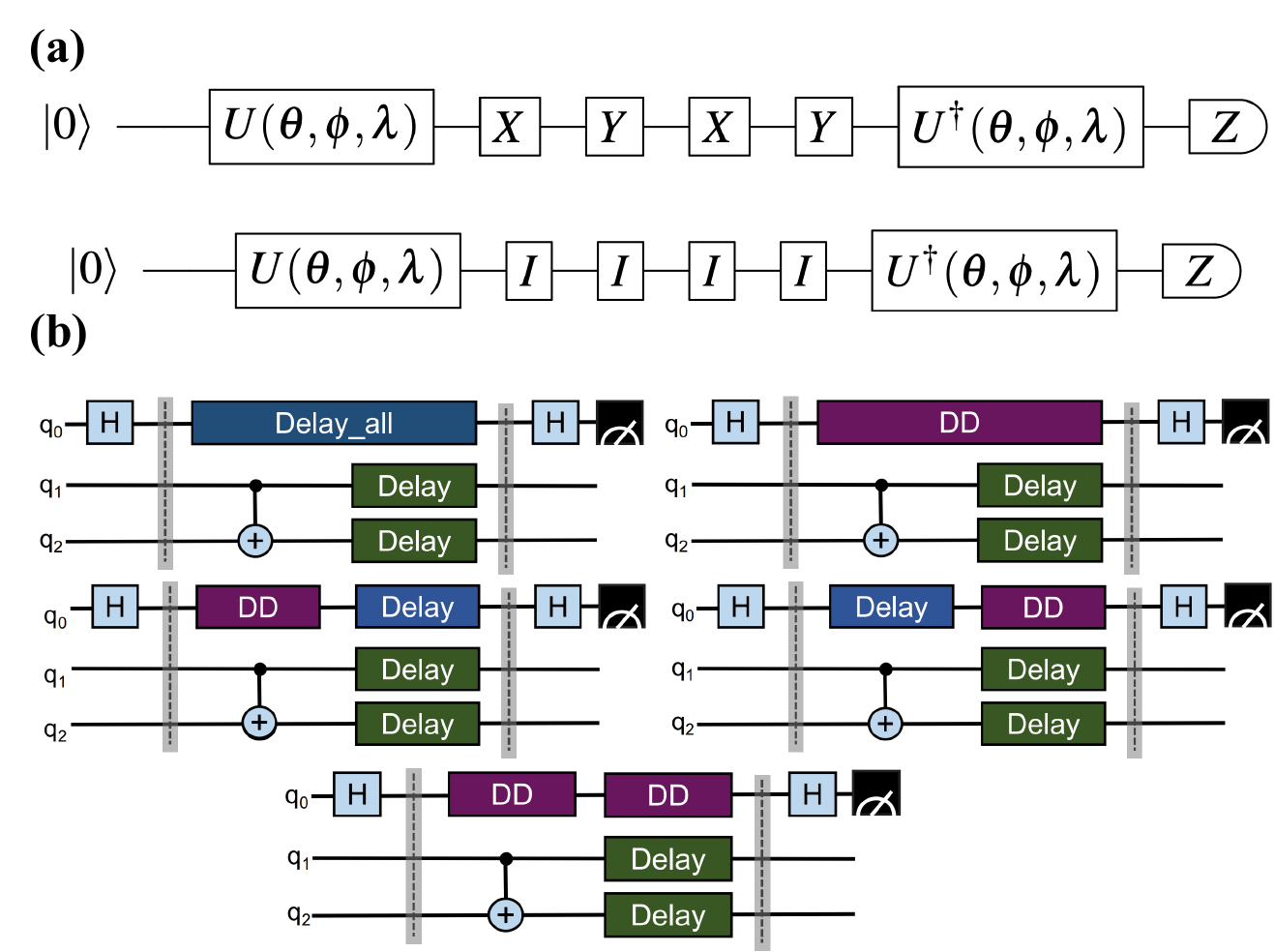}
    \caption{DD mechanism and insertion choices in digital circuits. (a) Free evolution is compared with a refocusing sequence inserted between a circuit and its inverse. Alternating pulses reverse slow phase accumulation and improve recovery of the initial state relative to an equal-duration identity sequence. (b) Representative compiler-level insertion patterns. A DD sequence may fill, precede, follow, or be split across idle windows, and the resulting performance depends on the surrounding entangling operations and delay structure. Panels (a) and (b) are adapted from Refs.~\cite{pokharel2018demonstration} and \cite{niu2022analyzing}, respectively.}
    \label{fig:dd_sequences_and_insertion}
\end{figure*}

The importance of sequence placement is illustrated in Fig.~\ref{fig:dd_sequences_and_insertion}. Panel (a) shows the basic refocusing mechanism, comparing an unprotected idle interval with one containing alternating control pulses. The pulses reverse the sign of slowly accumulated phase errors so that contributions from different parts of the idle interval approximately cancel
\cite{pokharel2018demonstration}. Panel (b) emphasizes the corresponding compilation problem: the same DD sequence can occupy an entire idle window, be shifted relative to neighboring operations, or be divided across multiple idle segments. These choices change the pulse-error and crosstalk environment experienced by the qubit and can therefore lead to different experimental performance
\cite{niu2022analyzing}. The practical performance of DD is therefore determined by four coupled factors: whether the dominant noise is refocusable, whether the hardware schedule can accommodate the sequence, how the pulses are placed within the available idle windows, and whether the target algorithm provides sufficient idle structure. Formal cancellation order alone is therefore not a reliable criterion for selecting DD on real quantum hardware.

\begin{figure*}[t]
    \centering
    \includegraphics[width=0.90\textwidth]{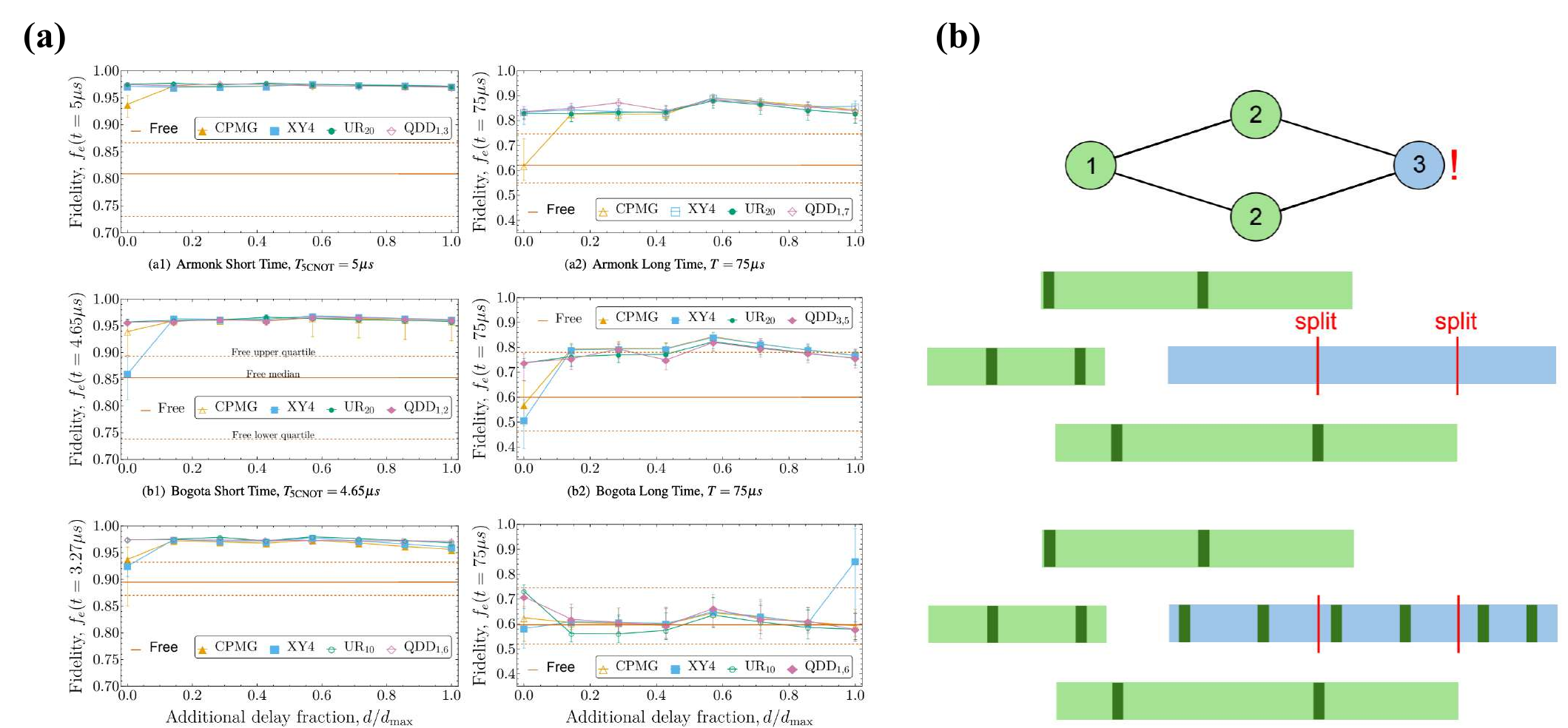}
    \caption{Hardware dependence and compiler-aware DD. (a) Experimental fidelity versus additional idle fraction for several DD families and device settings. The relative ordering changes across backends and delay regimes, demonstrating that there is no universally optimal sequence. (b) Graph-aware DD placement: the compiler locates and, when necessary, splits idle windows generated by the routed circuit before inserting refocusing pulses. This context-aware procedure avoids pulse placement that conflicts with neighbouring operations or schedule boundaries. Panels (a) and (b) are adapted from Refs.~\cite{ezzell2023dynamical} and \cite{coote2025graphdd}, respectively.}
    \label{fig:dd_hardware_and_compiler}
\end{figure*}

Hardware comparisons and context-aware placement are separated in Fig.~\ref{fig:dd_hardware_and_compiler}. Panel (a) compares CPMG, XY4, Uhrig-type, and quadratic DD families over different idle fractions and superconducting backends. No sequence is uniformly optimal: the fidelity gain changes with the available delay, the device time scales, and the accumulated pulse error, and an overlong or poorly matched sequence can remove the benefit \cite{ezzell2023dynamical}. Panel (b) illustrates graph- and compiler-aware insertion. The routed dependency graph identifies idle intervals created by neighbouring operations; long windows are split at schedule boundaries, and DD pulses are inserted only where they fit the connectivity and timing constraints. The method therefore acts on the actual compiled schedule rather than applying one global sequence blindly \cite{coote2025graphdd}. Together, the two panels show that the effect of DD is determined jointly by the noise spectrum, pulse calibration, idle duration, routing graph, and target circuit.

\subsection{Limitations and scalability}

The benefits of circuit randomization, twirling, and DD are ultimately limited by different resources. Randomization suppresses the coherent accumulation of errors but leaves a residual stochastic channel, whereas DD suppresses selected coherent evolution only when the gain from refocusing exceeds the additional control cost. These distinctions determine how the two approaches scale with circuit depth and hardware size.

\begin{itemize}

\item \textbf{Residual stochastic errors after randomization.}
Circuit randomization and twirling can substantially reduce coherent bias, but they do not eliminate the underlying stochastic error strength 
\cite{wallman2016noise,cai2023quantum}. Consider a repeated coherent $Z$ over-rotation, $U_{\epsilon}=e^{-i\epsilon Z/2}.$
Preparing $\ket{+}$ and measuring $X$ after $L$ identical cycles gives
\begin{equation}
    \langle X\rangle_{\rm bare}
    =
    \cos(L\epsilon),
    \qquad
    1-\langle X\rangle_{\rm bare}
    \simeq
    \frac{L^{2}\epsilon^{2}}{2},
    \qquad
    L|\epsilon|\ll1.
    \label{eq:coherent_accumulation_bare}
\end{equation}
The error therefore accumulates coherently as $O(L^{2}\epsilon^{2})$.

If random Pauli frames independently reverse the sign of the over-rotation, the ensemble-averaged channel becomes
\begin{align}
    \mathcal{T}_{\epsilon}(\rho)
    =
    \frac{1}{2}
    \left(
        U_{\epsilon}\rho U_{\epsilon}^{\dagger}
        +
        U_{-\epsilon}\rho U_{-\epsilon}^{\dagger}
    \right)
    \nonumber=
    \cos^{2}\!\left(\frac{\epsilon}{2}\right)\rho
    +
    \sin^{2}\!\left(\frac{\epsilon}{2}\right)Z\rho Z,
    \label{eq:sign_twirled_channel}
\end{align}
which is a stochastic $Z$ channel with
$q=\sin^{2}(\epsilon/2)$. The same observable then follows
\begin{equation}
    \langle X\rangle_{\rm rand}
    =
    (1-2q)^L
    =
    (\cos\epsilon)^L
    \simeq
    e^{-L\epsilon^{2}/2},
    \label{eq:randomized_accumulation}
\end{equation}
and hence $1-\langle X\rangle_{\rm rand}
    \simeq
    \frac{L\epsilon^{2}}{2}$. 
Randomization therefore changes the leading accumulation from
$O(L^{2}\epsilon^{2})$ to $O(L\epsilon^{2})$ in this example, without removing the single-cycle error strength
\cite{wallman2016noise,hashim2021randomized}. At sufficiently large depth, however, the randomized signal still decays. Twirling therefore converts coherent accumulation into stochastic accumulation rather than eliminating noise altogether.

\item \textbf{Sampling and implementation overhead.}
Randomization introduces an additional ensemble cost. Even when Pauli-frame updates can be absorbed into neighboring gates or tracked virtually, randomized compiling requires several logically equivalent circuit instances to be executed and averaged
\cite{hashim2021randomized,wallman2016noise}. The hardware should also remain sufficiently stable during acquisition; otherwise, calibration drift across randomized instances can spoil the intended ensemble averaging.

The physical overhead depends strongly on the twirling group. Pauli twirling is comparatively inexpensive because the corrections can often be propagated through a Pauli frame, whereas full Clifford twirling or noise tailoring for non-Clifford gates can require additional compilation and control resources
\cite{cai2023quantum,dankert2009exact,santos2024pseudotwirling,
tsubouchi2025symmetricclifford}. Similar considerations apply to qudit systems, where generalized Weyl--Heisenberg frames remain formally simple but require more demanding calibration and control
\cite{goss2024qutrit}. Scalability therefore depends on both the number of randomized circuit instances and the physical cost of implementing the corresponding frame transformations.

\item \textbf{Break-even condition for dynamical decoupling.}
Dynamical decoupling has a different limitation: the error removed by the refocusing sequence must exceed the errors introduced by the additional pulses. Let $\chi_0(T)$ denote the dephasing exponent accumulated during an unprotected idle interval of duration $T$,
\begin{equation}
    W_0(T)
    =
    \exp\!\left[
        -\chi_0(T)
        -\frac{T}{2T_1}
    \right].
    \label{eq:dd_unprotected_coherence}
\end{equation}
Suppose that a DD sequence reduces this to $\chi_{\rm DD}(T)$, uses $m$ imperfect refocusing pulses with coherence factor $\eta_\pi$ per pulse, and increases the total duration by $\Delta T$. A simple phenomenological model is
\begin{equation}
    W_{\rm DD}(T)
    =
    \exp\!\left[
        -\chi_{\rm DD}(T)
        -\frac{T+\Delta T}{2T_1}
    \right]
    \eta_\pi^m.
    \label{eq:dd_protected_coherence}
\end{equation}
The sequence is beneficial only if $W_{\rm DD}>W_0$, giving $ \chi_0(T)-\chi_{\rm DD}(T)
    >
    -m\ln\eta_\pi
    +
    \frac{\Delta T}{2T_1}$.
For $\eta_\pi=1-r_\pi$ with $r_\pi\ll1$, $\chi_0(T)-\chi_{\rm DD}(T)
    \gtrsim
    mr_\pi
    +
    \frac{\Delta T}{2T_1}$
The left-hand side is the dephasing removed by DD, whereas the right-hand side represents the additional cost of pulse errors and increased relaxation exposure.

For quasi-static Gaussian detuning with variance $\sigma^2$, $\chi_0(T)
    =
    \frac{\sigma^2T^2}{2}$, 
and an ideal echo cancels this static contribution. The approximate break-even condition becomes
\begin{equation}
    \frac{\sigma^2T^2}{2}
    \gtrsim
    mr_\pi
    +
    \frac{\Delta T}{2T_1}.
    \label{eq:dd_quasistatic_threshold}
\end{equation}
This explains why DD may fail for short idle windows, why too many pulses can become counterproductive, and why irreversible processes such as $T_1$ relaxation and leakage cannot be refocused
\cite{suter2016protecting,ezzell2023dynamical,niu2022analyzing}.

\item \textbf{Scalability on compiled hardware.}
At larger scale, the effectiveness of suppression increasingly depends on the compiled hardware schedule. Highly parallel circuits may leave little idle time for DD, while additional refocusing pulses can introduce crosstalk comparable to the coherent error they are intended to remove
\cite{tripathi2022suppression,coote2025graphdd}. Context-aware and learning-based methods can improve pulse placement and sequence selection, but they do not remove the underlying break-even condition
\cite{rahman2024learning,coote2025graphdd}.

Randomization and DD therefore have different scalability requirements. Randomization remains attractive when frame updates have low physical overhead, and the ensemble size remains affordable. DD is useful when sufficiently long idle periods are available. Neither method is universally advantageous; their useful regimes are determined by the balance between the coherent structure suppressed and the additional stochastic, control, timing, and sampling costs introduced by the suppression layer
\cite{cai2023quantum,suter2016protecting,jain2023improved}.

\end{itemize}

\subsection{Connections to other mitigation workflows}

Circuit randomization, twirling, and DD are most useful when incorporated as front-end suppression layers within broader mitigation workflows, rather than treated as alternatives to ZNE, PEC, symmetry verification, or learning-based mitigation
\cite{cai2023quantum,ville2022leveraging,kimScalableErrorMitigation2023a}.
Their role is to reduce or simplify coherent error structure before the downstream correction is performed. RC and twirling primarily tailor circuit-dependent coherent errors toward a more stochastic effective channel, whereas DD suppresses slowly varying coherent evolution during idle periods. Both can therefore improve the noise conditions under which subsequent mitigation operates.

\begin{itemize}

\item \textbf{Zero-noise extrapolation.}
ZNE relies on a sufficiently regular dependence of the measured observable on the applied noise-scaling procedure. Strong coherent and circuit-dependent errors can complicate this dependence, whereas prior randomization or coherent-error suppression can make the effective noise behavior more reproducible across scaled circuits
\cite{cai2023quantum,kimScalableErrorMitigation2023a}.

\item \textbf{Probabilistic error cancellation.}
PEC requires characterization and inversion of the effective noise channel. Pauli twirling and RC can reduce a generic coherent channel to a simpler Pauli-like description, for which the noise parameters and quasi-probability decomposition are generally easier to construct
\cite{cai2023quantum}.

\item \textbf{Symmetry-based mitigation.}
Coherent rotations can bias observables without necessarily producing a clear violation of the measured symmetry constraint. Suppressing or randomizing these coherent components can leave a larger fraction of the residual error in stochastic forms that are more naturally detected by symmetry checks and post-selection
\cite{cai2023quantum}.

\item \textbf{Learning-based mitigation.}
Learning-based methods rely on calibration or training circuits whose noise should resemble that of the target computation. Randomization can reduce circuit-dependent coherent variations and thereby make the effective noise more homogeneous across the training and target circuit ensembles
\cite{ville2022leveraging,cai2023quantum}. DD can play a complementary role by suppressing idle-time coherent evolution before the learned correction is applied.

\end{itemize}

A practical requirement is that the suppression layer be included consistently during calibration and execution. If RC, twirling, or DD changes the effective noise experienced by the target circuit, then characterization or training data used by the downstream mitigation method should be collected under the same conditions. Otherwise, the correction may be constructed for a noise model that differs from the one present during the final computation
\cite{cai2023quantum}.

Circuit randomization, twirling, and DD therefore connect to downstream QEM through noise preparation rather than direct estimator correction. By suppressing or simplifying coherent error structure before mitigation, they can make the residual channel easier to characterize and more compatible with ZNE, PEC, symmetry-based, and learning-based procedures
\cite{cai2023quantum,wallman2016noise,suter2016protecting,kimScalableErrorMitigation2023a}.
In practice, these suppression layers are increasingly integrated into software and runtime workflows, where randomized compiling, Pauli twirling, and dynamical-decoupling sequences can be inserted automatically during circuit compilation or execution; representative software implementations are discussed in Sec.~\ref{software}.

\section{Error-detection and post-selection methods}\label{verification}

Error-detection and post-selection methods mitigate noise by identifying outcomes
that are inconsistent with properties of the ideal computation and removing or
reweighting them in classical post-processing. Unlike methods such as ZNE or
PEC, they do not attempt to reconstruct the zero-noise result from an explicit
noise-scaling procedure or inverse noise model. Instead, they exploit
information that determines whether a measured state or trajectory is
compatible with the intended computation. This information may originate from
physical symmetries, conservation laws, local constraints, or additional
circuit-dependent checks
\cite{bonetmonroig2018lowcost,cai2021symmetry,cai2023quantum}.

The most common example is \emph{symmetry verification} (SV). Many quantum
algorithms ideally remain within a known symmetry sector of the Hilbert space.
Measurements that violate this sector can therefore be identified as erroneous
and discarded or appropriately reweighted. In quantum chemistry and fermionic
simulation, relevant quantities include particle number, fermionic parity,
spin projection, and total spin
\cite{gard2020efficient,barron2021preserving,cai2023quantum}. In
combinatorial optimization, similar verification can exploit symmetries of the
cost function, including global bit-flip symmetry or graph automorphisms in
QAOA-type circuits
\cite{shaydulin2021problem,kakkar2022qaoa}. In lattice gauge theories, local
gauge constraints provide an especially direct form of error detection:
outcomes violating Gauss-law-type conditions lie outside the physical Hilbert
space and can therefore be rejected
\cite{stryker2019gauss,raychowdhury2020lsh,ballini2025nonabelian}.

These methods are attractive for near-term hardware because the validity
condition is often known much more reliably than the microscopic noise model.
Their effectiveness, however, is intrinsically selective. Only errors that
violate the chosen symmetry, constraint, or check can be detected, while errors
that remain within the accepted sector are left unchanged. In addition,
post-selection reduces the acceptance probability and therefore increases the
number of circuit executions required for a fixed statistical precision.
Error-detection and post-selection methods are therefore most effective when a
large fraction of relevant errors can be identified using low-overhead checks
\cite{bonetmonroig2018lowcost,cai2021symmetry,cai2023quantum}.

\subsection{Motivation}

The motivation for symmetry- and constraint-based mitigation is that many quantum algorithms come with useful structure known before the experiment is run. In electronic-structure calculations, for example, the target state is often restricted by electron number, fermionic parity, spin projection, total spin, or other symmetry labels inherited from the molecular Hamiltonian and the chosen ansatz \cite{gard2020efficient,barron2021preserving,cai2023quantum}. In spin and lattice-fermion models, conserved quantities such as total
magnetization, particle filling, excitation number, or fermion parity can often
be known in advance and used to identify the physically relevant symmetry sector
\cite{bonetmonroig2018lowcost,mcardle2019errormitigated,cai2021symmetry}.
Similarly, in combinatorial optimization, the quantum approximate optimization
algorithm (QAOA) encodes the structure and constraints of a classical cost
function into a parametrized quantum circuit
\cite{farhi2014quantum,hadfield2019quantum,shen2026benchmarking}.
For example, QAOA has been applied to constrained portfolio-optimization
problems on noisy digital quantum processors
\cite{shen2026benchmarking}.
Depending on the problem, the cost Hamiltonian and mixer can also preserve exact
symmetries, including global bit-flip symmetry, permutation symmetry, or graph
automorphisms, which can be exploited to restrict the accessible state space or
identify symmetry-violating errors
\cite{shaydulin2021problem,kakkar2022qaoa}. In lattice gauge theories, the physical Hilbert space is defined by local gauge constraints, so violations of Gauss-law-type conditions correspond to unphysical states rather than merely small numerical errors \cite{stryker2019gauss,ballini2025nonabelian,carena2024gaugeredundant}. In all these settings, the relevant sector is not learned from noisy data; it is fixed by the problem itself.
This prior knowledge gives SV its appeal. Instead of reconstructing a full noise model or extrapolating to a zero-noise limit, one checks whether the noisy output remains in the allowed sector and discards or downweights the components that violate it \cite{bonetmonroig2018lowcost,cai2021symmetry}. When the required symmetry measurements overlap with observables already measured in the experiment, the extra quantum cost can be small, and the correction can be implemented largely in post-processing \cite{bonetmonroig2018lowcost,sagastizabal2019experimental}. This low overhead is one reason why SV became an early and practically useful structure-aware mitigation method on NISQ hardware.
A second advantage is interpretability. General mitigation protocols often infer a corrected value through a fitted response model or an extrapolation ansatz. In SV, the rejected contribution usually has a direct meaning: population outside the target particle-number sector, violation of a conserved parity, violation of a logical or problem-specific constraint, or leakage from the gauge-invariant subspace \cite{botelho2022midcircuit,stryker2019gauss,ballini2025nonabelian}. This makes the method scientifically transparent: if mitigation changes the result, the change can often be traced to a specific symmetry or constraint violated by the noisy device.
The strongest use case arises when the dominant hardware errors actually move the state into forbidden sectors. In quantum chemistry, relaxation and excitation errors can populate states with the wrong particle number or parity, which SV can detect \cite{sagastizabal2019experimental}. In optimization circuits, noise can break symmetries of the cost function and mix equivalent or forbidden sectors \cite{shaydulin2021problem,kakkar2022qaoa}. In lattice gauge simulations, control and digitization errors can couple the physical gauge-invariant subspace to unphysical states \cite{stryker2019gauss,ballini2025nonabelian}. The limitation is equally important: SV removes only the component of the error that violates the chosen symmetry or constraint. Symmetry-preserving errors remain invisible and must be addressed by other mitigation methods or by improved hardware control.

\subsection{Symmetry verification and symmetry post-selection}
\subsubsection{Principle}
SV begins from the observation that the ideal state is known to lie in a target symmetry sector. For a unitary symmetry operator $S$ with target eigenvalue $s=\pm 1$, the projector onto that sector is
\begin{equation}
    \Pi_s
    =
    \frac{I+sS}{2},
    \label{eq:sv_single_projector}
\end{equation}
where $I$ is the identity operator. Given a noisy state $\rho$ and an observable $O$, the symmetry-verified estimator can be written as
\begin{equation}
    \langle O\rangle_{\mathrm{SV}}
    =
    \frac{\Tr(\Pi_s O \Pi_s \rho)}
    {\Tr(\Pi_s \rho)} .
    \label{eq:sv_estimator}
\end{equation}
The denominator is the probability that the noisy state passes the symmetry check. When $O$ commutes with $S$, this expression is simply the expectation value conditioned on the accepted symmetry sector. When $O$ does not commute with $S$, the result should instead be interpreted as the expectation value of the projected observable, rather than as a simple correction factor applied to the original measurement.
For several commuting symmetries $\{S_\alpha\}$ with target eigenvalues $\{s_\alpha\}$, the target-sector projector is
\begin{equation}
    \Pi_{\mathrm{tar}}
    =
    \prod_\alpha
    \frac{I+s_\alpha S_\alpha}{2}.
    \label{eq:sv_multiple_projector}
\end{equation}
These equations summarize the central idea of SV: noisy contributions outside the known physical or logical sector are removed, while the accepted component is renormalized \cite{bonetmonroig2018lowcost,sagastizabal2019experimental,cai2021symmetry}.

In practice, one should distinguish direct post-selection from post-processed SV. In direct post-selection, the symmetry is measured explicitly, for example by an ancilla-assisted parity check, and only shots with the correct symmetry outcome are retained. This is natural when the symmetry check is cheap or when violations should be detected during the circuit. In post-processed SV, the same projection is reconstructed classically from measured Pauli data, without necessarily performing an explicit accept-reject measurement on hardware. This is especially useful when the symmetry operators overlap with Pauli strings already measured for the target observable \cite{bonetmonroig2018lowcost,sagastizabal2019experimental,cai2021symmetry}.
A useful generalization is symmetry expansion. In ordinary SV, one projects uniformly onto the target symmetry sector. Symmetry expansion instead allows weighted combinations of symmetry operators or group elements, producing a family of estimators that trade bias reduction against sampling cost \cite{cai2021symmetry}. This viewpoint is useful in a review because it places SV, symmetry post-selection, and related symmetry-based estimators within a single framework rather than treating them as separate tricks.
SV is also closely related to quantum subspace methods. Bonet-Monroig \textit{et al.} showed that a zero-cost post-processing version of SV can be interpreted as a variant of quantum subspace expansion \cite{bonetmonroig2018lowcost,mcclean2017hybrid}. Related parity-check and stabilizer-like methods, such as error-mitigated digital simulation and Pauli-check sandwiching, extend the same logic by inserting or measuring checks that diagnose whether the state remains in a protected sector \cite{mcardle2019errormitigated,gonzales2023paulicheck}. These methods go beyond inherent problem symmetries, but they clarify the broader conceptual neighborhood of SV.
The main statistical cost is the loss of samples. If
$p_{\mathrm{pass}}=\Tr(\Pi_{\mathrm{tar}}\rho)$
is the probability of passing the verification test, then direct post-selection requires roughly a factor $p_{\mathrm{pass}}^{-1}$ more shots to obtain the same number of accepted samples \cite{cai2021symmetry,cai2023quantum}. For ratio estimators such as Eq.~\eqref{eq:sv_estimator}, the precise variance also depends on the observable fluctuations within the accepted sector and on correlations between the observable value and the pass/fail event. SV is therefore most useful when symmetry-breaking errors are common enough to remove, but not so frequent that the acceptance probability becomes too small. This bias--variance trade-off is what ultimately determines whether symmetry verification is practical for a given workload.

\begin{figure*}[t]
    \centering
    \includegraphics[width=0.7\textwidth]{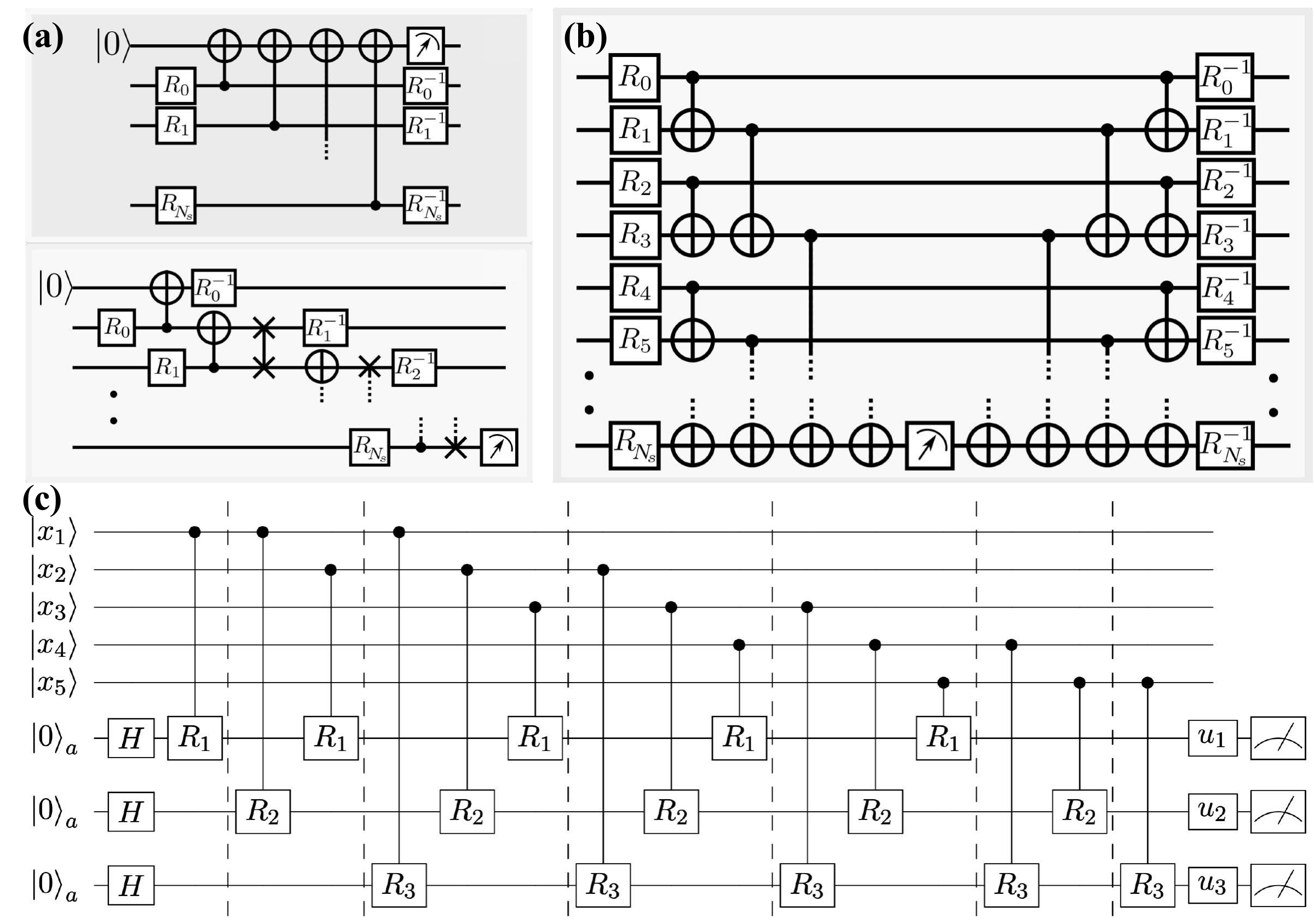}
    \caption{Circuit implementations of symmetry and constraint verification. (a) Ancilla-assisted full SV, where a controlled parity or symmetry measurement distinguishes the target sector and invalid outcomes are rejected. (b) In-line and locality-aware verification circuits, showing how the same projector can be decomposed and embedded into the native circuit structure. (c) Mid-circuit validation of a constrained encoding, where repeated ancillary checks detect invalid configurations during rather than only after the computation. Panels (a) and (b) are adapted from Ref.~\cite{bonetmonroig2018lowcost}, and panel (c) is adapted from Ref.~\cite{botelho2022midcircuit}.}
    \label{fig:sv_circuits}
\end{figure*}

\begin{figure*}[t]
    \centering
    \includegraphics[width=0.66\textwidth]{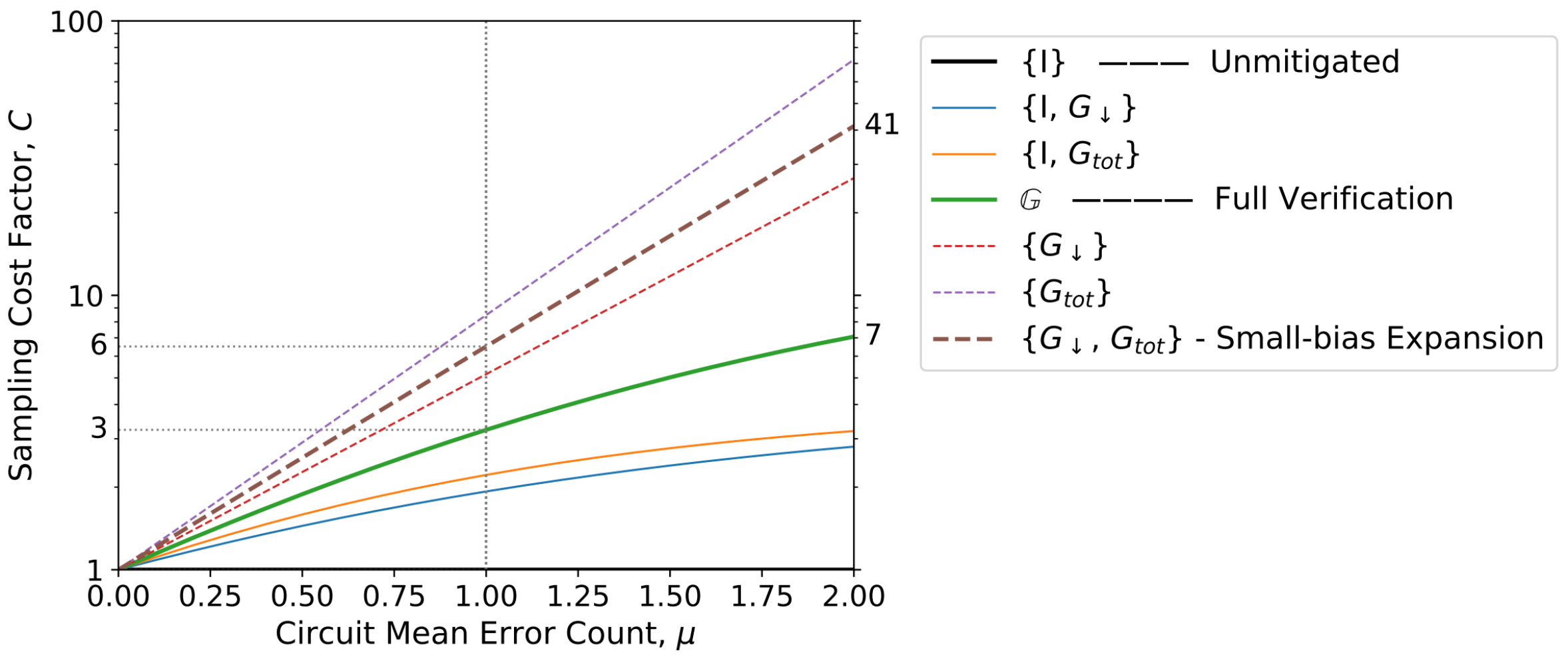}
    \caption{Sampling-cost trade-off in symmetry expansion. The curves compare the unmitigated estimator, partial symmetry expansions, full verification, and a small-bias expansion as the circuit mean error count increases. Stronger projection suppresses more symmetry-violating bias but increases the sampling cost, whereas selected weighted expansions interpolate between these limits. Adapted from Ref.~\cite{cai2021symmetry}.}
    \label{fig:symmetry_expansion_cost}
\end{figure*}

Circuit-level realizations of symmetry verification are summarized in Fig.~\ref{fig:sv_circuits}. Panel (a) shows a direct ancilla-assisted check: controlled operations map the symmetry eigenvalue onto an ancilla, the ancilla is measured, and shots with the wrong eigenvalue are rejected. This implements the projector explicitly but requires an ancilla and controlled interactions \cite{bonetmonroig2018lowcost}. Panel (b) shows locality-aware and in-line constructions in which rotations, controlled checks, and inverse rotations are embedded into the circuit. Such decompositions reduce the need for one fully nonlocal parity check, although their depth and routing overhead remain hardware dependent \cite{bonetmonroig2018lowcost}. Panel (c) shows mid-circuit validation for a constrained encoding. Ancillas interrogate the data register at several points in the evolution, so invalid configurations can be identified before subsequent gates propagate them; the method exchanges additional checks and dynamic-circuit operations for earlier detection of leakage \cite{botelho2022midcircuit}.
The complementary estimator-level trade-off is isolated in Fig.~\ref{fig:symmetry_expansion_cost}. Full verification applies the sharpest projector and can give the strongest bias suppression, but the normalization by a small pass probability increases the sampling cost. Symmetry expansion replaces the sharp projector by selected weighted symmetry operators. The plotted cost factors show how partial and small-bias choices interpolate between the unmitigated estimator and full verification: weaker projection retains more residual bias but can require substantially fewer samples \cite{cai2021symmetry}. This separation makes clear that circuit implementation cost and statistical reconstruction cost are distinct design choices.

The most common symmetries used in NISQ mitigation are global conserved quantities. In quantum chemistry and fermionic simulation, the relevant sector is often fixed by particle number, fermionic parity, spin projection, total spin, or time-reversal symmetry \cite{gard2020efficient,barron2021preserving,xu2026fractional}. These symmetries are useful because they both restrict the physical search space and provide simple checks for noise-induced leakage. For example, relaxation, excitation, or mapping errors can move population out of the target particle-number or parity sector, making chemistry a natural setting for symmetry-based mitigation \cite{sagastizabal2019experimental,barron2021preserving}.

A closely related situation appears in spin and lattice-fermion models, where the ideal dynamics may preserve magnetization, excitation number, filling, or global parity. In many cases, the useful structure is not a complicated group-theoretic symmetry, but a simple constraint on the allowed bit strings. This viewpoint is especially important for constrained encodings such as one-hot, binary, gray-code, and domain-wall encodings, where valid computational states occupy only a structured subset of the full Hilbert space \cite{botelho2022midcircuit}. In such settings, post-selection or mid-circuit checks can be interpreted either as symmetry verification or as validation of the chosen encoding.
Optimization circuits provide another important class. In QAOA and related variational optimization algorithms, the relevant symmetries often come from the classical objective function rather than from a physical Hamiltonian. Examples include the global $\mathbb{Z}_2$ bit-flip symmetry of MaxCut-type objectives and graph automorphisms of the problem instance, both of which can induce exact symmetries of the ideal QAOA evolution \cite{shaydulin2021problem,kakkar2022qaoa}. This shows that symmetry-based mitigation is not limited to quantum simulation; it can apply whenever the noiseless algorithm is known to remain inside a symmetry-restricted or constraint-restricted subspace.

Symmetry-preserving circuit design and symmetry-based mitigation should be viewed as complementary. In chemistry and variational simulation, one often first chooses an ansatz that respects the desired symmetry sector, and then uses symmetry verification to filter residual symmetry-breaking noise \cite{gard2020efficient,barron2021preserving}. The ansatz reduces unphysical motion in the ideal circuit, while mitigation removes the sector-violating components introduced by imperfect hardware.
Lattice gauge theories represent the strongest version of this idea. There, the physical Hilbert space is defined by local Gauss-law constraints, so a gauge-violating state is not merely inaccurate but unphysical \cite{stryker2019gauss,raychowdhury2020lsh,ballini2025nonabelian}. Gauge simulations are therefore a natural target for symmetry- and constraint-based mitigation. They are also demanding, because the number of local checks grows with system size and, in non-Abelian theories, the constraint structure can be more complex than a set of independent commuting global symmetries.

\subsubsection{Implementation  strategies on quantum hardware}
On quantum hardware, symmetry-based mitigation usually appears in three recurring forms. 
\begin{itemize}
    \item The first is end-of-circuit verification. One prepares the state using a symmetry-preserving or symmetry-aware circuit, measures the relevant parity or symmetry at the end, and keeps only the shots consistent with the target sector \cite{bonetmonroig2018lowcost,sagastizabal2019experimental,shaydulin2021problem}. This approach is natural in VQE-like experiments, where the same state is prepared many times to estimate Pauli observables. If the symmetry check can be measured together with, or reconstructed from, the required Pauli data, the additional quantum overhead can be small.
    \item The second form is post-processed verification. Instead of adding an explicit ancilla check, one reconstructs the symmetry-verified estimator from the measured Pauli data. This was a key advantage of early chemistry applications: when the symmetry operator is aligned with the measured observable set, symmetry verification can be implemented largely in classical post-processing \cite{bonetmonroig2018lowcost,sagastizabal2019experimental}. A representative example is the superconducting-transmon experiment of Sagastizabal \textit{et al.}, where the odd-parity symmetry of a two-qubit mapped $\mathrm{H}_2$ Hamiltonian was used to shift the measured VQE energy curve toward the exact ground-state result and improve the reconstructed state estimate \cite{sagastizabal2019experimental}. The broader lesson is that SV is most efficient when the relevant symmetry is naturally compatible with the encoding, ansatz, and measurement basis.
    \item The third form is repeated or mid-circuit post-selection. This becomes useful when an early symmetry-violating event would otherwise propagate through the rest of the circuit, making final-shot filtering less effective. In QAOA and constrained encodings, one can insert checks during the computation or between layers, so that invalid states are removed before they affect later evolution \cite{shaydulin2021problem,botelho2022midcircuit}. This pattern is especially relevant for fixed-Hamming-weight, one-hot, domain-wall, or related encodings, provided that the hardware supports mid-circuit measurement and reset with sufficiently low overhead.
\end{itemize}
Optimization experiments illustrate how these patterns appear on current devices. Shaydulin and Galda showed on IBM Quantum hardware that symmetry projection, applied either at the end of the circuit or throughout the evolution, can improve QAOA state fidelity and objective estimates for MaxCut by exploiting global bit-flip symmetry \cite{shaydulin2021problem}. Kakkar \textit{et al.} further analyzed SV for QAOA and demonstrated it on IonQ trapped-ion hardware, reporting up to a 19.2\% improvement in the QAOA objective for small graph instances \cite{kakkar2022qaoa}. These studies show that the benefit of SV depends not only on the existence of a symmetry, but also on circuit depth, noise strength, hardware connectivity, and the cost of the verification circuit.

\begin{figure*}[t]
    \centering
    \includegraphics[width=0.76\textwidth]{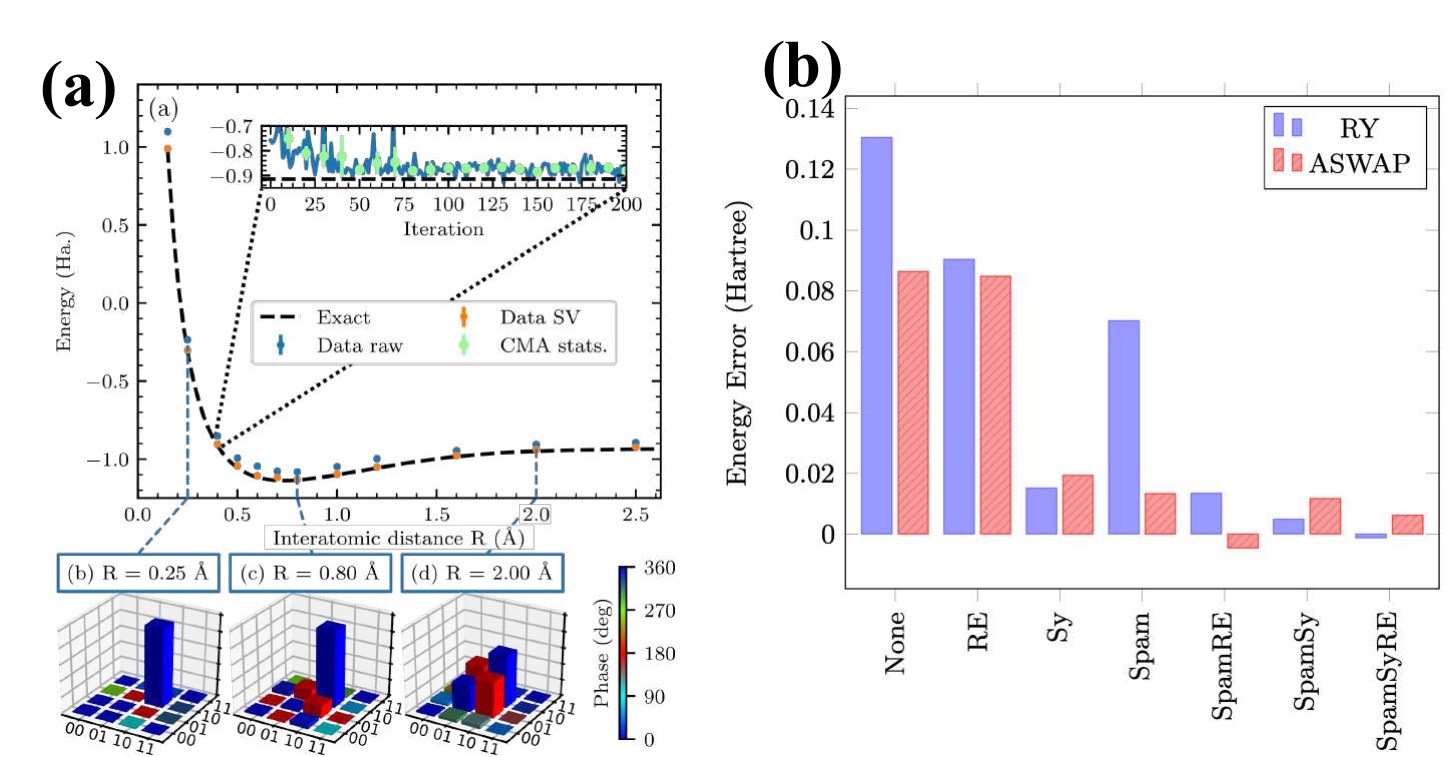}
    \caption{Symmetry-based mitigation in quantum chemistry. (a) Molecular VQE for $\mathrm{H}_2$: parity verification removes sector-violating outcomes and shifts the measured energy toward the exact curve; the state-population plots show the configurations affected at representative bond lengths. (b) Comparison of partially symmetry-preserving chemistry ansätze, illustrating that circuit-level symmetry preservation reduces the raw error and that symmetry-based post-processing can further improve the estimate. Panels (a) and (b) are adapted from Refs.~\cite{sagastizabal2019experimental} and \cite{barron2021preserving}, respectively.}
    \label{fig:sv_chemistry}
\end{figure*}

Platform details therefore matter. On trapped-ion devices with high connectivity, nonlocal parity checks and global symmetry measurements can be comparatively natural \cite{kakkar2022qaoa}. On sparse-connectivity superconducting processors, the same nonlocal checks may require additional routing and entangling gates, so one must choose symmetries whose verification cost is small compared with the algorithmic circuit itself \cite{shaydulin2021problem}. Thus, in practice, SV is not defined only by the abstract projector; it is defined by how that projector is measured, reconstructed, or embedded into the hardware schedule.
Chemistry applications are shown separately in Fig.~\ref{fig:sv_chemistry}. Panel (a) presents a molecular VQE experiment for $\mathrm{H}_2$. Parity-based post-selection removes outcomes outside the target sector, moving the measured energy curve toward the exact ground-state energy; the accompanying state-population plots show which computational-basis components are removed by the check \cite{sagastizabal2019experimental}. Panel (b) compares ansätze with different degrees of symmetry preservation. Enforcing the appropriate sector in the circuit design reduces the raw energy error, while residual symmetry-based mitigation provides an additional correction. The result demonstrates that symmetry-preserving ansatz construction and symmetry verification are complementary rather than interchangeable \cite{barron2021preserving}.

\begin{figure*}[t]
    \centering
    \includegraphics[width=0.85\textwidth]{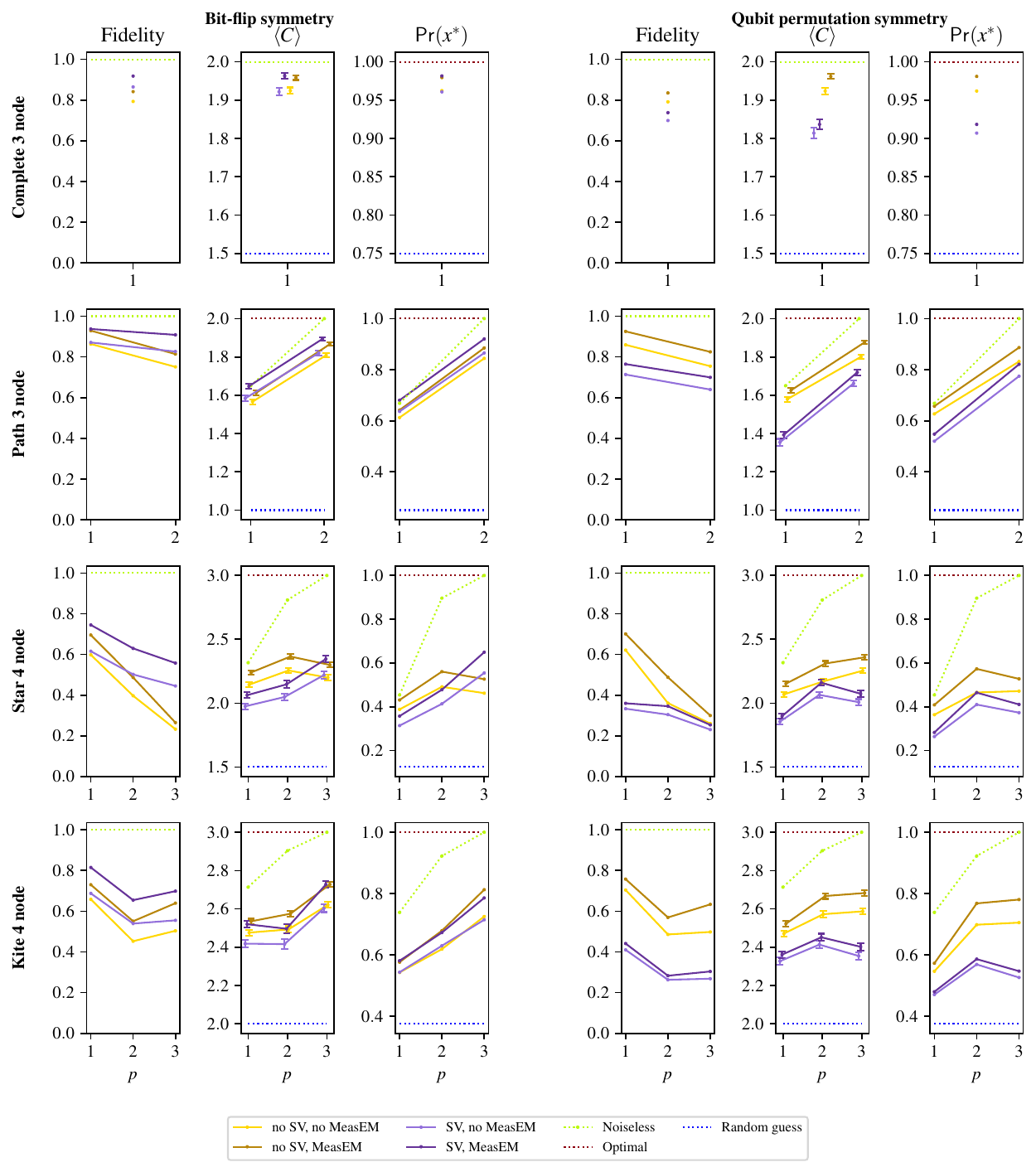}
    \caption{Symmetry verification for QAOA on a superconducting processor. Bit-flip and qubit-permutation symmetries are tested for several graph families and depths. The benchmark compares fidelity, objective-value estimation, and optimal-solution probability for unmitigated data, measurement-error mitigation, SV, and their combination. The improvement depends on the graph and depth because only symmetry-violating components are removed. Adapted from Ref.~\cite{shaydulin2021problem}.}
    \label{fig:sv_qaoa_superconducting}
\end{figure*}

Superconducting-hardware QAOA results are collected in Fig.~\ref{fig:sv_qaoa_superconducting}. Bit-flip and qubit-permutation symmetries are applied across several graph families and circuit depths. The columns compare state fidelity, the estimated objective value, and the probability of sampling an optimal solution, with and without SV and measurement-error mitigation. SV raises these metrics when noise transfers population into symmetry-incompatible sectors, while the combined treatment shows that readout correction and symmetry filtering address different error components \cite{shaydulin2021problem}.

\begin{figure*}[t]
    \centering
    \includegraphics[width=0.85\textwidth]{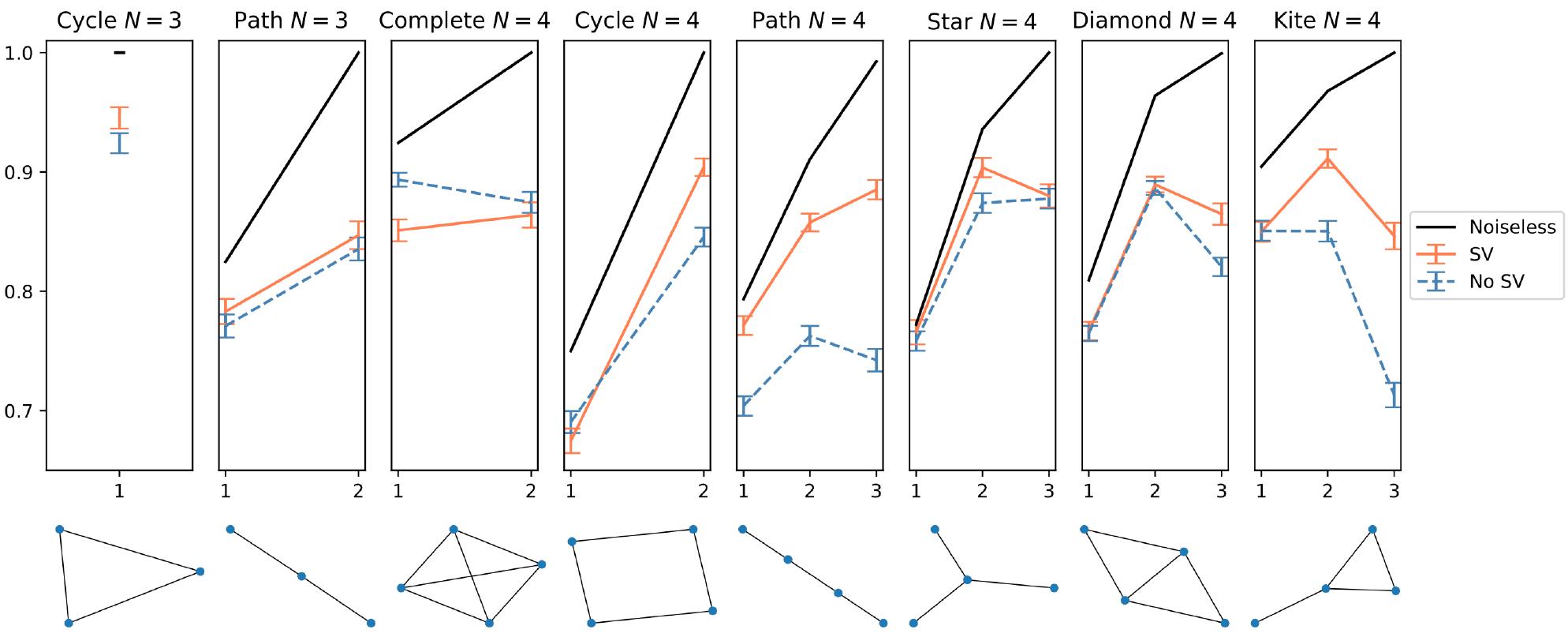}
    \caption{Symmetry verification for QAOA on a trapped-ion processor. Approximation ratios are compared with noiseless and unmitigated results for several small graph instances and QAOA depths. SV improves the result for most instances, while the nonuniform gain across graph families and depths reflects the fraction of the hardware error that violates the verified symmetry. Adapted from Ref.~\cite{kakkar2022qaoa}.}
    \label{fig:sv_qaoa_trapped_ion}
\end{figure*}

A trapped-ion counterpart is shown in Fig.~\ref{fig:sv_qaoa_trapped_ion}. For cycle, path, complete, star, diamond, and kite graph instances, the approximation ratio is plotted as the QAOA depth increases. The SV curves generally lie above the unmitigated data and closer to the noiseless result, but the gain is graph- and depth-dependent. This variation is consistent with the selective action of the method: SV helps only to the extent that the dominant faults populate sectors excluded by the problem symmetry \cite{kakkar2022qaoa}.

Together, Figs.~\ref{fig:sv_chemistry}--\ref{fig:sv_qaoa_trapped_ion} show that symmetry-based mitigation is not tied to a single application area or hardware platform. It is useful whenever the ideal computation is known to occupy a symmetry- or constraint-restricted sector that realistic noise can populate incorrectly, but its benefit remains controlled by the amount of symmetry-breaking error present in the workload.

\subsection{Gauge constraints and projector-based filtering}

\subsubsection{Physical-subspace constraints and Gauss-law projectors}
Gauge constraints provide a particularly strong form of structure for error mitigation. For an ordinary global symmetry, the target state belongs to one selected sector of the Hilbert space. In a lattice gauge theory, by contrast, the physical Hilbert space is defined by satisfying local Gauss-law constraints at every site or vertex. Gauge-redundant formulations keep the local gauge degrees of freedom because this often preserves locality of the Hamiltonian and of the digitized dynamics. The cost is that the computational Hilbert space is larger than the physical one, and noisy evolution can populate states that violate Gauss's law \cite{stryker2019gauss,carena2024gaugeredundant}. In this setting, symmetry verification is not only a way to sharpen an estimate; it removes components that are outside the target theory.

For commuting Abelian constraints, the physical-sector projector has a simple product form. If $G_x$ is a local Gauss-law operator with target eigenvalue $+1$, then
\begin{equation}
    \Pi_{\mathrm{phys}}
    =
    \prod_x
    \frac{I+G_x}{2}.
    \label{eq:gauge_projector_abelian}
\end{equation}
More generally, when the constraint is expressed through local gauge transformations $\Theta_{x,g}$ with group elements $g\in\mathcal{G}_x$, the local projector can be written as a group average,
\begin{equation}
    \Pi_{\mathrm{phys}}
    =
    \prod_x
    \frac{1}{|\mathcal{G}_x|}
    \sum_{g\in\mathcal{G}_x}
    \Theta_{x,g}.
    \label{eq:gauge_projector_group}
\end{equation}
These expressions show that gauge filtering is mathematically close to ordinary symmetry verification, but applied to an extensive set of local constraints rather than to one or a few global symmetries \cite{stryker2019gauss,carena2024gaugeredundant}.

For Abelian gauge theories, this logic can be implemented relatively directly. Stryker constructed Gauss-law oracles for digital quantum computers that distinguish physical from unphysical wave functions in Abelian lattice gauge theories, providing an explicit accept-reject mechanism for detecting gauge-violating errors \cite{stryker2019gauss}. Related ideas appear in loop-string-hadron digitizations, where the representation makes the physicality constraints more natural to encode and check \cite{raychowdhury2020lsh}. These works show that projector-based filtering is not only a formal idea, but can be translated into concrete quantum circuits.
Projector-based filtering is one part of a broader toolbox for protecting gauge symmetry. Other approaches add energy penalties or gauge-protection terms to suppress leakage dynamically, or use control-based constructions related to dynamical decoupling to reduce gauge-violating evolution during the simulation \cite{halimeh2021singlebody,kasper2023dd}. The focus here is narrower: verification and filtering, where measurements or projected estimators are used to identify and remove unphysical components after or during the evolution.

The non-Abelian case is more subtle. Local gauge transformations at a vertex generally do not reduce to a simple set of commuting parity checks, so the Abelian projector picture cannot be applied without modification \cite{ballini2025nonabelian}. Ballini \textit{et al.} addressed this by introducing two symmetry-verification schemes for noisy non-Abelian lattice gauge simulations on qudit hardware: dynamical post-selection (DPS), based on repeated mid-circuit validation of local gauge charges, and post-processed symmetry verification (PSV), based on reconstructing gauge-invariant observables through correlations with gauge transformations \cite{ballini2025nonabelian}. This shows that symmetry-based mitigation can still be useful even when the local constraint structure is noncommuting and cannot be checked by a single set of simultaneous projectors.
The broader lesson is that gauge theories turn symmetry-based mitigation into physical-subspace filtering. In ordinary global-SV, the selected sector is one useful block of a larger Hilbert space. In lattice gauge theories, the physical sector is often the only sector with direct physical meaning. Gauge-violating states are therefore not just quantitatively inaccurate; they are outside the intended theory \cite{stryker2019gauss,ballini2025nonabelian,carena2024gaugeredundant}. This makes constraint-based mitigation especially compelling, while also making its overhead and scalability central practical concerns.

\subsubsection{Experimental overhead}
The main overhead in symmetry- and constraint-based mitigation is statistical. In direct post-selection, only shots that pass the required symmetry or constraint checks are retained. The usable sample fraction can therefore decrease with circuit depth, evolution time, or the number of enforced checks \cite{cai2023quantum,cai2021symmetry,ballini2025nonabelian}. This loss is already important for global symmetries when the pass probability becomes small, and it is more severe for local-constraint systems such as lattice gauge theories, where many Gauss-law conditions must be satisfied simultaneously.

A second overhead is circuit-level implementation. Ancilla-assisted parity checks and Gauss-law oracles require controlled operations, additional work registers, and in some cases mid-circuit measurement or reset \cite{bonetmonroig2018lowcost,stryker2019gauss}. These costs are strongly hardware dependent. On highly connected trapped-ion devices, long-range parity checks can be comparatively natural, whereas on sparse superconducting processors the same logical check may require substantial routing and additional entangling gates. For dynamic-circuit implementations, the latency and fidelity of mid-circuit measurement become further practical bottlenecks \cite{botelho2022midcircuit,ballini2025nonabelian}.
Post-processed schemes avoid some of this quantum overhead, but they are not free. Classical projection or symmetry expansion can reduce the need for ancillas and in-circuit checks, yet they may require additional measurement settings and more elaborate classical aggregation of the data \cite{bonetmonroig2018lowcost,cai2021symmetry}. For commuting global symmetries, this cost can be mild, especially when the relevant Pauli strings are already measured for the target observable. For non-Abelian post-processed symmetry verification, however, one must estimate correlations with multiple gauge transformations, making the measurement and classical-processing burden more substantial \cite{ballini2025nonabelian}.
Gauge simulations introduce an additional architectural trade-off between gauge-fixed and gauge-redundant encodings. Gauge fixing can reduce the number of degrees of freedom, but it may also remove the simple local projector structure that makes Gauss-law checking straightforward. Gauge-redundant encodings use a larger Hilbert space, but expose the gauge constraints as checkable conditions and can provide a code-like redundancy useful for mitigation or partial correction \cite{carena2024gaugeredundant}. Carena \textit{et al.} made this trade-off explicit through a threshold-style analysis: redundancy costs additional qudits, but below suitable error thresholds the redundant encoding with Gauss-law correction can outperform a gauge-fixed digitization \cite{carena2024gaugeredundant}. This connection is useful for a review because it links NISQ constraint filtering to the broader logic of fault-tolerant protection.

\begin{figure*}[t]
    \centering
    \includegraphics[width=1.\textwidth]{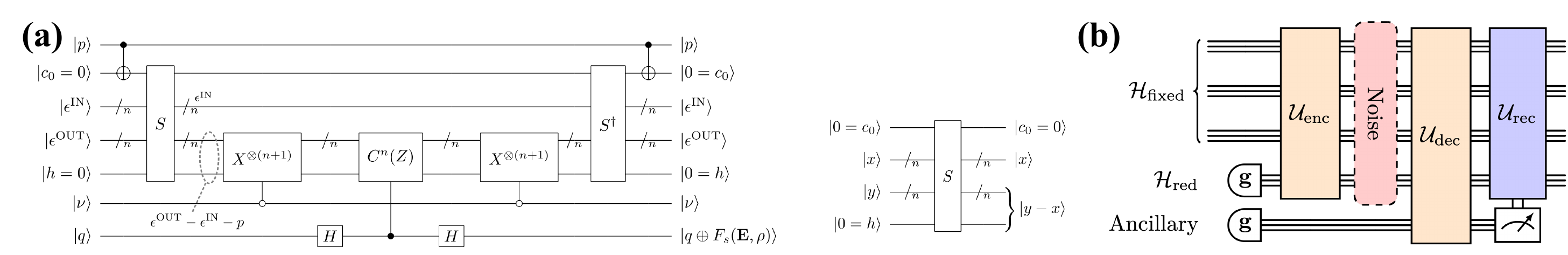}
    \caption{Abelian gauge constraints as explicit checks and code-space redundancy. (a) A Gauss-law oracle coherently distinguishes physical and unphysical configurations, enabling ancilla-based post-selection on the local constraint. (b) Gauge-redundant digitization viewed as a code-space construction: additional degrees of freedom raise the hardware cost but convert gauge violations into detectable syndrome information that can be used for mitigation or protection. Panels (a) and (b) are adapted from Refs.~\cite{stryker2019gauss} and \cite{carena2024gaugeredundant}, respectively. }
    \label{fig:gauge_abelian}
\end{figure*}

The Abelian circuit and encoding viewpoints are separated in Fig.~\ref{fig:gauge_abelian}. Panel (a) shows a Gauss-law oracle that coherently evaluates a local constraint and records whether the state belongs to the physical sector. Measuring the oracle ancilla provides an explicit accept--reject test for gauge-violating errors \cite{stryker2019gauss}. Panel (b) shows gauge-redundant digitization as a code-space construction. The redundant degrees of freedom increase the Hilbert-space and hardware cost, but they also expose local Gauss-law violations as detectable syndromes; below a suitable error regime, this redundancy can therefore support mitigation or partial correction rather than acting only as overhead \cite{carena2024gaugeredundant}.

\begin{figure*}[t]
    \centering
    \includegraphics[width=1.\textwidth]{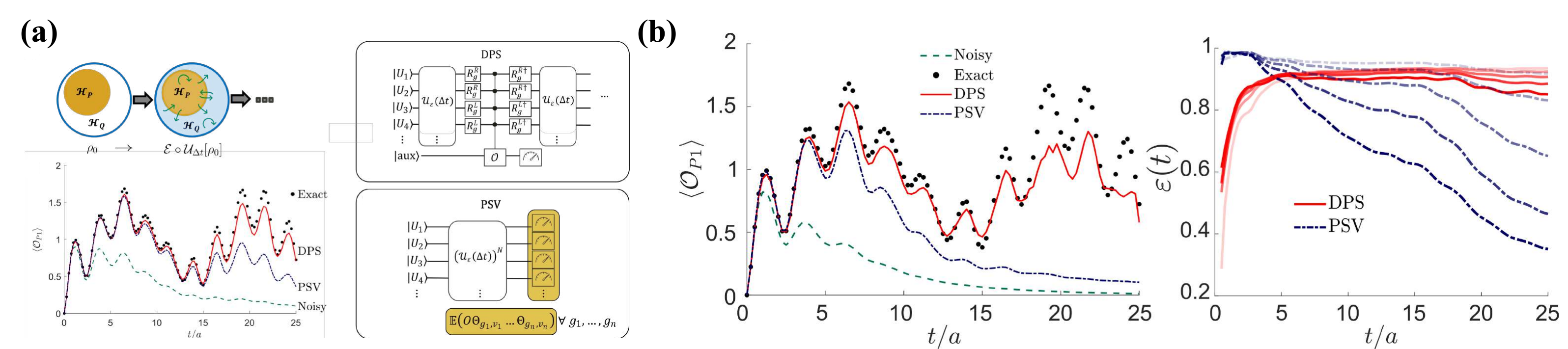}
    \caption{Constraint mitigation in non-Abelian lattice gauge simulations. (a) DPS repeatedly checks local gauge charges during time evolution, whereas PSV reconstructs gauge-invariant observables from post-processed correlations with gauge transformations. The representative trace shows the recovery relative to exact and noisy dynamics. (b) Time-dependent comparison of the recovered observable and gauge-symmetry metric. DPS gives stronger protection in the displayed regime, while PSV avoids repeated in-circuit validation but incurs greater measurement and reconstruction overhead. Panels (a) and (b) are adapted from Ref.~\cite{ballini2025nonabelian}.}
    \label{fig:gauge_nonabelian}
\end{figure*}

Non-Abelian strategies and their measured effect are summarized in Fig.~\ref{fig:gauge_nonabelian}. Panel (a) contrasts dynamical post-selection (DPS), which repeatedly validates local gauge charges during the evolution, with post-processed symmetry verification (PSV), which reconstructs gauge-invariant observables from measurements of gauge transformations after the circuit. The accompanying time trace shows that both methods improve over the noisy evolution, with DPS following the exact dynamics more closely in the displayed regime \cite{ballini2025nonabelian}. Panel (b) compares the recovered observable and gauge-symmetry metric over time. DPS provides stronger protection because violations are detected before they propagate, whereas PSV avoids repeated in-circuit checks but loses accuracy at later times and requires a larger measurement and reconstruction burden. The figure therefore makes the implementation trade-off explicit: stronger in-circuit protection costs ancillas, checks, and dynamic operations, while post-processing reduces circuit overhead at the expense of statistical and classical complexity.

\subsection{Ancilla-assisted error detection}

Ancilla-assisted error detection adds one or more auxiliary qubits that act as
\emph{error flags}. Rather than measuring the data qubits directly, the ancilla
is coherently coupled to the computation before and after a target circuit.
The two couplings are designed to cancel for an ideal execution. If an error
changes the relation between them, the ancilla can flip to a different
measurement outcome, allowing the corresponding shot to be discarded. This
provides error detection without encoding the computational state into a full
quantum error-correcting code
\cite{debroy2020extended,gonzales2023paulicheck,vandenberg2023coherent}.

To see the mechanism explicitly, consider a target circuit $U$ and two check
operators $\widetilde C_1$ and $\widetilde C_2$ chosen such that
\begin{equation}
    \widetilde C_2 U \widetilde C_1 = U .
    \label{eq:pcs_identity}
\end{equation}
An ancilla is initialized in $\ket{0}$ and transformed to $\ket{+}$ by a
Hadamard gate. The ancilla then controls $\widetilde C_1$ before $U$ and
$\widetilde C_2$ after $U$. Immediately before the final Hadamard, the joint
state in an ideal execution is
\begin{equation}
    \frac{1}{\sqrt{2}}
    \left[
        U\ket{\psi}\ket{0}
        +
        \widetilde C_2 U\widetilde C_1
        \ket{\psi}\ket{1}
    \right]
    =
    U\ket{\psi}\ket{+}.
    \label{eq:pcs_ideal_interference}
\end{equation}
The final Hadamard therefore returns the ancilla to $\ket{0}$. Thus, in the
absence of a detectable error, the check outcome is deterministically zero.

The role of the check becomes clear when an error $E$ occurs. For simplicity,
suppose the effective error is placed after $U$. The two ancilla branches then
contain $EU,
    \widetilde C_2 E U\widetilde C_1$.
If $E$ anticommutes with the second check,
$\{\widetilde C_2,E\}=0$, Eq.~\eqref{eq:pcs_identity} gives
\begin{equation}
    \widetilde C_2 E U\widetilde C_1
    =
    -E\widetilde C_2U\widetilde C_1
    =
    -EU.
\end{equation}
The relative sign changes the ancilla from $\ket{+}$ to $\ket{-}$, so that the
final Hadamard produces the measurement outcome $1$. The shot is therefore
rejected:
\begin{equation}
    0\rightarrow\text{accept},
    \qquad
    1\rightarrow\text{reject}.
\end{equation}
By contrast, an error that commutes with the check does not change the ancilla
outcome and remains undetected. Errors occurring at other positions in the
circuit can be propagated to the check boundary, where the same commutation
argument applies. Ancilla-assisted checking therefore removes selected
components of the error channel rather than correcting every possible error
\cite{gonzales2023paulicheck}.

A particularly useful implementation is obtained for Clifford circuits. Since
a Clifford unitary maps Pauli operators to Pauli operators, one may choose a
Pauli operator $L$ before the circuit and construct the corresponding
right-side check as $R=ULU^\dagger$,
which immediately gives $RUL=U$. Both $L$ and $R$ are Pauli strings and can
therefore be implemented as ancilla-controlled Pauli operations. This is the
basis of \emph{coherent Pauli checks} (CPCs)
\cite{debroy2020extended,vandenberg2023coherent}. If $L$ is sampled randomly
from the nontrivial $n$-qubit Pauli group, a given nonidentity Pauli error
anticommutes with a random check with probability approximately $1/2$. Adding
several independent checks decreases the probability that an error commutes
with every check, although only shots for which all check ancillas return zero
are retained.

For circuits followed immediately by computational-basis measurement,
Ref.~\cite{vandenberg2023coherent} introduced a particularly hardware-efficient
\emph{one-sided} version. In this case the right check can be chosen from
$Z$-type Paulis,
\begin{equation}
    R\in\{I,Z\}^{\otimes n},
    \qquad
    L=U^\dagger R U.
\end{equation}
Because the data qubits are eventually measured in the $Z$ basis, the parity
associated with $R$ can be calculated directly from the measured bit string.
The controlled right-side Pauli gates therefore do not need to be physically
executed: only the left check is implemented on the quantum processor, while
the corresponding right check is evaluated classically after measurement.
This substantially reduces the additional two-qubit-gate count and avoids the
long idle period that an ancilla would otherwise experience while waiting
between the two sides of the check. This one-sided construction is especially
suited to single-shot sampling tasks and readout-error detection
\cite{vandenberg2023coherent}.

\emph{Pauli Check Sandwiching} (PCS) generalizes this principle by introducing
multiple pairs of circuit-dependent checks,
\begin{equation}
    \widetilde C_{2,k}U\widetilde C_{1,k}=U,
    \qquad k=1,\ldots,m,
\end{equation}
with one ancilla for each pair
\cite{gonzales2023paulicheck}. The target circuit is therefore ``sandwiched'' between
the checks. For Clifford circuits, suitable Pauli checks can be obtained simply
by propagating Pauli operators through $U$. For circuits containing
non-Clifford gates, finding check pairs that remain Pauli operators is more
restrictive. Gonzales \emph{et al.} proposed a search procedure that pushes
candidate Pauli checks through circuits composed of Clifford and diagonal
gates and preferentially selects low-weight checks so that the additional
gate noise is kept small. The original PCS study demonstrated this procedure
mainly through numerical simulations rather than experiments on quantum
hardware
\cite{gonzales2023paulicheck}.

The experimentally demonstrated CPC protocol is therefore particularly
important for assessing the practical status of ancilla-assisted mitigation.
Van den Berg \emph{et al.} implemented CPCs on IBM superconducting processors
for Clifford-dominated circuits with up to ten data qubits and more than one
hundred logical CNOT gates
\cite{vandenberg2023coherent}. The experiments showed that post-selection on
the check syndromes can reduce the logical error probability. At the same
time, they exposed the central hardware trade-off: every additional check
requires controlled Pauli operations, ancilla preparation and measurement,
and potentially additional idle time. These operations are themselves noisy.
Consequently, increasing the number of checks eventually provides diminishing
returns and can even worsen the result when errors introduced by the checking
circuit exceed the errors that it detects.

Ancilla-assisted error detection should therefore presently be viewed as a
promising but comparatively less mature QEM strategy. Its advantage is that it
can construct verification conditions from the circuit itself and does not
require a physical symmetry of the target state. Its limitation is that the
error detector requires additional quantum operations precisely on hardware
where such operations are imperfect. The method is most favorable when the
payload circuit is sufficiently noisy or deep, the checks can be implemented
with low-weight Pauli operators, and the additional ancilla and entangling-gate
errors remain substantially smaller than the errors removed by post-selection.

\subsection{Limitations and scalability}
Symmetry- and constraint-based mitigation is powerful because it uses prior knowledge of the target problem, but the same selectivity limits what it can correct. A two-qubit example makes the distinction between detectable and invisible errors explicit. Let the target state be
\begin{equation}
\lvert\Phi^{+}\rangle
=
\frac{\lvert00\rangle+\lvert11\rangle}{\sqrt{2}},
\end{equation}
with parity symmetry $S=Z_1Z_2$, target projector $\Pi_{+}=(I+S)/2$, and observable $O=X_1X_2$. The ideal expectation is $\langle O\rangle_{\rm id}=1$. Consider the noisy mixture
\begin{align}
\rho={}&(1-q-r)\lvert\Phi^{+}\rangle\!\langle\Phi^{+}\rvert
+q\lvert\Psi^{-}\rangle\!\langle\Psi^{-}\rvert +r\lvert\Phi^{-}\rangle\!\langle\Phi^{-}\rvert,
\label{eq:sv_bell_noise_model}
\end{align}
where $q,r\geq0$, $q+r\leq1$, $\lvert\Psi^{-}\rangle=(\lvert01\rangle-\lvert10\rangle)/\sqrt{2}$, and $\lvert\Phi^{-}\rangle=(\lvert00\rangle-\lvert11\rangle)/\sqrt{2}$. Up to an irrelevant phase, a $Y_1$ error maps $\lvert\Phi^{+}\rangle$ to the odd-parity state $\lvert\Psi^{-}\rangle$, so the $q$ component is detected by $S$. By contrast, a $Z_1$ error maps $\lvert\Phi^{+}\rangle$ to $\lvert\Phi^{-}\rangle$, which remains in the $S=+1$ sector and is invisible to the parity check. Since both error states have eigenvalue $-1$ under $X_1X_2$,
\begin{align}
\langle O\rangle_{\rm raw}
&=1-2(q+r), \\
p_{\rm pass}
&=\operatorname{Tr}(\Pi_{+}\rho)=1-q, \\
\langle O\rangle_{\rm SV}
&=\frac{\operatorname{Tr}(\Pi_{+}O\rho)}{\operatorname{Tr}(\Pi_{+}\rho)}
=\frac{1-q-2r}{1-q}
=1-\frac{2r}{1-q}.
\label{eq:sv_bell_result}
\end{align}
The symmetry-violating contribution $q$ is removed exactly in this model, whereas the symmetry-preserving contribution $r$ survives and is renormalized within the accepted sector. For $q=0.10$ and $r=0.02$, the raw value is $0.76$, and the verified value is $0.956$, a substantial improvement but not a recovery of the ideal value. This example gives a concrete version of the general limitation: particle-number-, parity-, stabilizer-, or gauge-preserving faults can bias an observable while passing every enforced check \cite{bonetmonroig2018lowcost,cai2021symmetry,cai2023quantum}.

The second limitation is the statistical cost of post-selection. If $p_{\rm pass}$ denotes the probability that a shot satisfies all imposed checks, only $N_{\rm eff}\simeq p_{\rm pass}N_{\rm raw}$ of the acquired shots contribute to the verified estimator. For a bounded observable,
\begin{equation}
\operatorname{Var}\!\left(\widehat{O}_{\rm SV}\right)
\sim
\frac{\operatorname{Var}(O\mid {\rm pass})}
     {N_{\rm raw}p_{\rm pass}},
\label{eq:symmetry_variance}
\end{equation}
so maintaining the same statistical uncertainty as an unfiltered estimate requires an acquisition overhead of order $1/p_{\rm pass}$. A simple independent-check model shows how this can become a scalability problem. If each of $K$ required checks is passed with probability $1-\lambda$, independently of the others, then
\begin{equation}
p_{\rm pass}=(1-\lambda)^K\simeq e^{-\lambda K},
\quad
\frac{N_{\rm raw}^{\rm SV}}{N_{\rm raw}^{(0)}}
\simeq
(1-\lambda)^{-K}
\simeq e^{\lambda K}.
\label{eq:sv_many_check_cost}
\end{equation}
For $\lambda=1\%$ and $K=100$, $p_{\rm pass}=0.366$ and the acquisition overhead is $2.73$. For $\lambda=5\%$ and the same number of checks, $p_{\rm pass}=5.92\times10^{-3}$ and the overhead is approximately $169$. If $K_{\rm loc}$ local constraints are checked at each of $L$ time steps, the inspected spacetime volume is approximately $K=K_{\rm loc}L$, giving $p_{\rm pass}\simeq e^{-\lambda K_{\rm loc}L}$ in this toy model. Repeated mid-circuit checks can therefore detect violations before they propagate, but the accepted data set can shrink rapidly with depth. Correlated faults, redundant checks, and locality can change the numerical scaling, so Eq.~\eqref{eq:sv_many_check_cost} is an illustration rather than a universal law; it isolates the basic post-selection pressure relevant to large constraint sets \cite{botelho2022midcircuit,cai2021symmetry,ballini2025nonabelian}.

A third limitation is that the enforced structure must be physically justified. When the symmetry or constraint is exact, projection onto the target sector is well defined. If it is only approximate, weakly broken, or not exactly preserved by the ansatz or effective model, aggressive projection can introduce systematic bias by forcing the state into an incorrect sector
\cite{barron2021preserving,cai2023quantum}.
This distinction is important in practical applications: symmetry-based mitigation is most reliable for exact conserved quantities and exact physical constraints, whereas emergent or approximate symmetries require a careful comparison between the reduction of noise-induced leakage and the bias introduced by projection.

Hardware overhead provides a further scalability constraint. Nonlocal parity checks may be comparatively natural on highly connected trapped-ion processors, but they can require substantial routing and additional entangling gates on sparsely connected superconducting architectures. The check circuit can then introduce errors comparable to, or larger than, those it is intended to detect. Mid-circuit post-selection additionally requires dynamic-circuit support, low-latency measurement and feedforward, reliable qubit reset, and sufficiently high mid-circuit readout fidelity
\cite{botelho2022midcircuit,kakkar2022qaoa,ballini2025nonabelian}.
The challenge is greater for non-Abelian gauge theories, where the local constraint structure generally cannot be reduced to a simple collection of commuting parity checks and may instead require more involved procedures, such as dynamical post-selection or post-processed symmetry verification
\cite{ballini2025nonabelian}.

Symmetry- and constraint-based mitigation is therefore best viewed as a selective first layer within a broader mitigation workflow rather than as a complete solution. It can efficiently remove clearly forbidden components of the noisy output when the target sector is known, the checks are inexpensive, and $p_{\rm pass}$ remains sufficiently large. The residual symmetry-preserving errors must still be addressed through complementary approaches, such as readout mitigation, ZNE, PEC, circuit-level noise suppression, or improved hardware control
\cite{cai2023quantum,cai2021symmetry}.
Its practical scalability is consequently determined not only by the number of qubits, but by the combined balance among pass probability, sampling overhead, check-circuit cost, hardware connectivity, constraint locality, and the extent to which the chosen symmetry captures the dominant noise channels.

\section{Software ecosystems for quantum error mitigation}\label{software}

The practical use of quantum error mitigation depends not only on the underlying protocol, but also on how that protocol is exposed to users through software for cloud quantum hardware
\cite{laroseMitiqSoftwarePackage2022a,cirstoiu2023qermit,javadi2024quantum,beisel2022configurable}.
Most mitigation methods require additional circuit generation, calibration data, and repeated backend execution
\cite{temmeErrorMitigationShortDepth2017a,endo2018practical,giurgica-tironDigitalZeroNoise2020,nation2021scalable,vandenberg2023sparsepec,qiskitdocs_techniques_2026}.
As a result, the software layer determines how easily a method can be deployed, reproduced, combined with other mitigation steps, and compared across devices
\cite{laroseMitiqSoftwarePackage2022a,cirstoiu2023qermit,beisel2022configurable,qiskitFunctions2026}.
This role is especially important on current pre-fault-tolerant and early-logical hardware, where users often access processors through cloud services
\cite{javadi2024quantum,qiskitdocs_mitigation_2026,qiskitdocs_techniques_2026,braket_errmit_2026,braket_ionq_debias_2026,ionq_qiskit_debias_2026}.
In this section, we review the software ecosystem that supports practical error mitigation.
We first discuss cross-platform toolkits, which aim to provide mitigation workflows that can be used across different SDKs and hardware backends
\cite{laroseMitiqSoftwarePackage2022a,cirstoiu2023qermit,mitiqdocs2026,qermitdocs2026,pennylanezne2026,cudaq_readout_2026}.
We then turn to provider-native mitigation in hardware runtimes
\cite{qiskitdocs_mitigation_2026,qiskitdocs_techniques_2026,braket_errmit_2026,braket_ionq_debias_2026,ionq_qiskit_debias_2026,qiskitFunctions2026}.
Finally, we discuss tensor-network error mitigation (TEM), including the provider-facing implementation offered by Algorithmiq as a Qiskit Function on IBM Quantum, which uses classical tensor-network processing to assist, benchmark, and supplement error mitigation
\cite{filippov2023tem,ibm2026tem}.

\subsection{Cross-platform toolkits}

Cross-platform error-mitigation toolkits are designed to make the same mitigation workflow usable with different quantum-programming frameworks and hardware providers. They do not replace software such as Qiskit, Cirq, or PennyLane, nor do they replace the runtime system of a hardware provider. Instead, they provide an additional software layer that prepares the circuits needed for mitigation, sends them for execution, and processes the returned data
\cite{laroseMitiqSoftwarePackage2022a,cai2023quantum}.

A typical workflow starts from a user circuit $C$. Depending on the mitigation method, the toolkit generates one or more modified circuits $\{T[C]\}$, executes them on the chosen backend, and combines the results to produce a mitigated estimate. For example, ZNE generates circuits with different effective noise levels, randomized methods generate logically equivalent circuit variants, and readout mitigation generates calibration circuits.

It is useful to distinguish three parts of this workflow. The \emph{frontend} is the software in which the user writes the circuit, such as Qiskit, Cirq, or PennyLane. The \emph{backend} is the simulator or physical quantum processor on which the circuit is executed. Between them is an execution interface that sends circuits to the backend and returns measurement data. In \texttt{Mitiq}, this interface is called an \emph{executor}
\cite{laroseMitiqSoftwarePackage2022a,mitiqdocs2026}. The returned data may consist of individual measurement outcomes, bit-string counts, or expectation values.

The main advantage of this separation is portability. The mitigation toolkit handles circuit modification and classical post-processing, while the frontend and backend can be changed with relatively little modification to the mitigation procedure. In this way, cross-platform toolkits provide a common software structure for implementing and comparing error-mitigation methods across different quantum devices.

\subsubsection{Dedicated quantum error mitigation libraries}

Dedicated QEM libraries provide mitigation functions that can be inserted between a user circuit and the hardware backend. Two representative examples are \texttt{Mitiq} and \texttt{Qermit}, which follow different software designs
\cite{laroseMitiqSoftwarePackage2022a,cirstoiu2023qermit}.

\begin{itemize}

\item \textbf{\texttt{Mitiq}: executor-based and cross-platform mitigation.}
\texttt{Mitiq} is an open-source Python library designed to work with several quantum software frameworks and hardware backends
\cite{laroseMitiqSoftwarePackage2022a,mitiqdocs2026,mitiqgithub2026}.
Its central abstraction is the \emph{executor}, namely a user-provided function that takes a quantum circuit, runs it on a simulator or quantum processor, and returns the measured result. This allows \texttt{Mitiq} to add mitigation without controlling the complete provider workflow.

For example, in ZNE [see Sec.~\ref{sec:zne}], \texttt{Mitiq} generates several logically equivalent circuits with amplified noise, sends them to the executor, and then extrapolates the returned expectation values toward the zero-noise limit. The same basic pattern is used for other supported techniques, including readout mitigation, Clifford data regression, digital dynamical decoupling, quantum subspace expansion, and Pauli twirling
\cite{mitiqdocs2026}. Circuits from supported frontends can be translated internally into a common representation, which enables the same mitigation procedure to be reused across different software and hardware platforms.

This organization is shown in \cref{fig:cross_platform_toolkits}(a,c): the mitigation library sits between the user circuit and the backend, automatically generating the additional circuits and classical post-processing required by the chosen QEM method.

\item \textbf{\texttt{Qermit}: task-graph-based mitigation.}
\texttt{Qermit} is built on the \texttt{pytket} ecosystem and organizes mitigation as a sequence of modular tasks
\cite{cirstoiu2023qermit,qermitdocs2026}.
Its basic component, a \texttt{MitTask}, represents one step of an experiment, such as compiling a circuit, generating calibration circuits, submitting jobs to a backend, collecting results, or performing regression. These tasks are connected into a \texttt{TaskGraph}, which specifies the order in which the different steps are performed.

This structure makes multi-stage mitigation workflows explicit and easy to combine. For example, a workflow may first compile a circuit, generate several noise-scaled copies for ZNE, execute them on the hardware, and finally extrapolate the results. Other workflows can incorporate Clifford data regression, SPAM correction, frame randomization, or post-selection
\cite{cirstoiu2023qermit,qermitdocs2026}.

\texttt{Qermit} also distinguishes between two common outputs. A \texttt{MitRes} workflow returns a mitigated measurement distribution, whereas a \texttt{MitEx} workflow returns a mitigated expectation value. This allows the mitigation procedure to be matched directly to whether the application needs complete output statistics or only selected observables. The corresponding task-graph structure is illustrated in \cref{fig:cross_platform_toolkits}(b).

\end{itemize}

The main difference is therefore organizational. \texttt{Mitiq} emphasizes portability: the user mainly supplies a circuit and an executor, while the library generates the required mitigation circuits and post-processing. \texttt{Qermit} emphasizes composability: the mitigation workflow is decomposed into explicit tasks that can be inspected and combined within the \texttt{pytket} framework. Both approaches aim to make QEM reusable across experiments, but they expose different levels of control over the mitigation workflow.

\subsubsection{Mitigation within broader software development kits}

Error mitigation is increasingly integrated into general-purpose quantum
software development kits (SDKs), rather than being available only through
dedicated QEM packages. In this setting, mitigation becomes part of the same
software abstraction used to construct circuits, select backends, estimate
observables, and process measurement results.

\begin{figure*}[t]
    \centering
    \includegraphics[width=\linewidth]{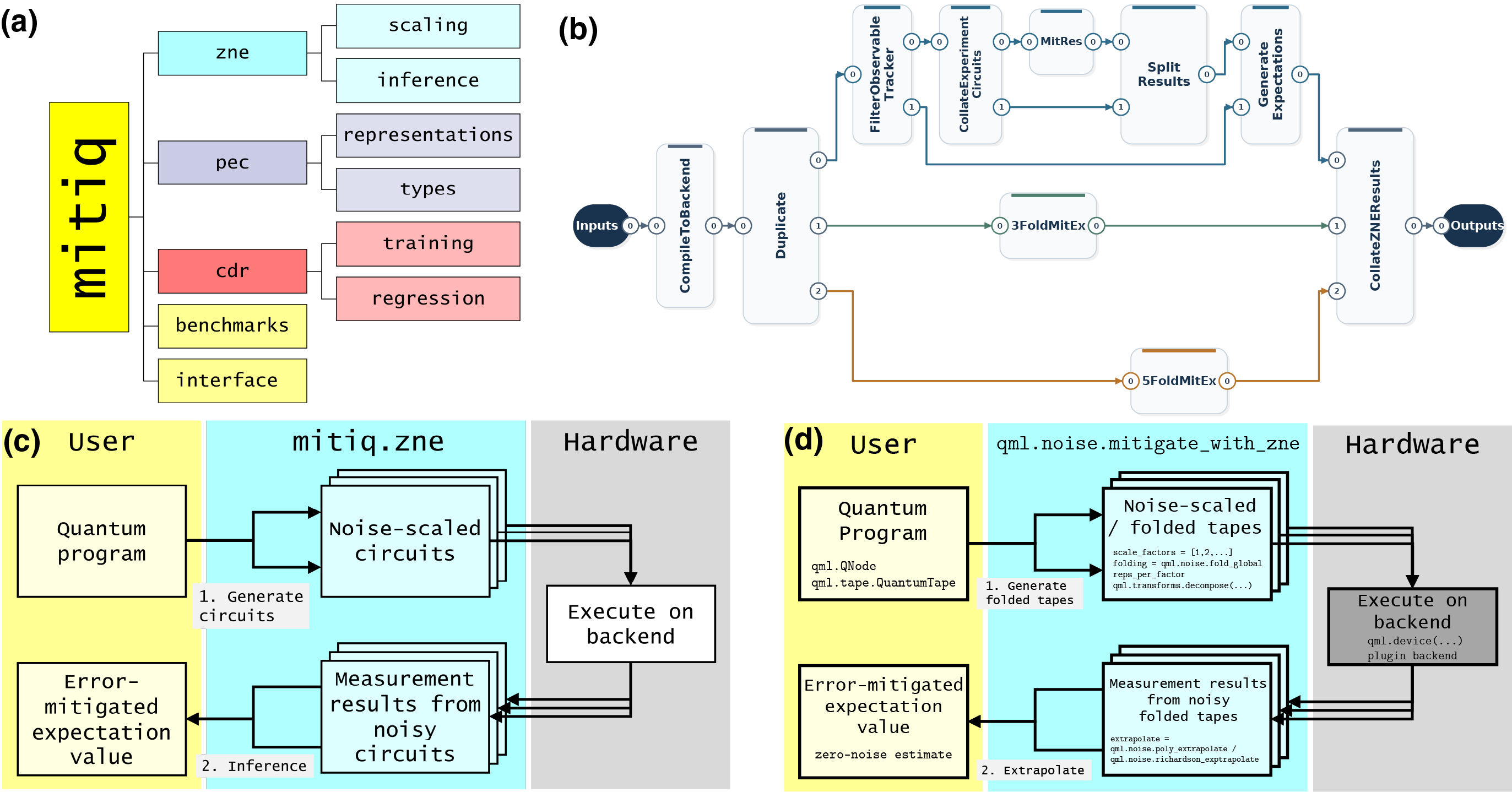}
    \caption{Cross-platform software abstractions for error mitigation.
    (a) Overall design of \texttt{Mitiq} as a dedicated error-mitigation package, with separate method modules, benchmarking tools, and circuit-interface support.
    (b) Graph-based composition of a ZNE workflow in \texttt{Qermit}, where mitigation is assembled from explicit tasks such as compilation, circuit duplication, result collation, and expectation-value generation.
    (c) Dedicated-toolkit ZNE workflow in \texttt{Mitiq}: a user program is wrapped by \texttt{mitiq.zne}, which generates noise-scaled circuits, sends them through a user-provided execution path to hardware or a simulator, and infers an error-mitigated expectation value.
    (d) SDK-native ZNE workflow in PennyLane: the same generate--execute--extrapolate loop is expressed inside \texttt{qml.noise.mitigate\_with\_zne} using \texttt{QNode}/\texttt{QuantumTape} and \texttt{qml.device} abstractions, so mitigation remains within one software stack.
    The common pattern is to separate circuit rewriting, execution, and classical post-processing; the software distinction is whether this pattern is provided by a dedicated external toolkit, a task graph, or a native SDK transform.
    Panels (a) and (c) are adapted from \cite{laroseMitiqSoftwarePackage2022a}; panel (b) is adapted from \cite{cirstoiu2023qermit}; panel (d) is based on \cite{pennylanezne2026,pennylanedocs2026}.}
    \label{fig:cross_platform_toolkits}
\end{figure*}

\begin{itemize}

\item \textbf{PennyLane: transform-based ZNE.}
PennyLane provides ZNE through
\texttt{qml.noise.mitigate\_with\_zne}, implemented as a circuit
\emph{transform}
\cite{pennylanezne2026,pennylanedocs2026}.
A transform takes an input quantum circuit and generates a set of modified
circuits together with a classical post-processing rule. The ZNE transform can
be applied to a \texttt{QNode}, PennyLane's executable quantum-circuit object,
or to its lower-level circuit representation, a \texttt{QuantumTape}. The user
specifies the noise scale factors, a circuit-folding rule, and an extrapolation
function. PennyLane provides native global folding together with polynomial,
Richardson, and exponential extrapolation routines. The resulting folded
circuits are executed on the selected backend, and their results are combined to
estimate the zero-noise observable.

This workflow is illustrated in
\cref{fig:cross_platform_toolkits}(d). In contrast to an external mitigation
wrapper, circuit transformation, backend execution, and extrapolation remain
within the PennyLane programming model. PennyLane can also use compatible
folding and extrapolation routines from \texttt{Mitiq}; however, its
documentation notes that differentiation through the mitigated circuit is not
supported when Mitiq is used for these components
\cite{pennylanezne2026}.

\item \textbf{QURI Parts: mitigation within the estimator stack.}
\texttt{QURI Parts} integrates several QEM methods into a broader
platform-independent framework for circuit construction, sampling, expectation-
value estimation, transpilation, and backend execution
\cite{quriparts_errmit2024}.
Its current mitigation modules include readout-error mitigation, ZNE, and
Clifford data regression (CDR). In particular, ZNE can be incorporated through
a mitigated estimator that generates noise-scaled circuits and extrapolates
their expectation values, while CDR combines a noisy estimator with a
classically tractable reference estimator to learn a correction from
near-Clifford training circuits.

The important software feature is that these methods use the same
\emph{sampler} and \emph{estimator} abstractions as ordinary circuit execution.
A sampler returns measurement statistics from repeated circuit executions,
whereas an estimator converts such information into expectation values of
observables. Mitigation can therefore be inserted into the estimation pipeline
without requiring the application to implement an independent QEM layer.

\item \textbf{CUDA-Q: readout-mitigation workflow.}
NVIDIA's \texttt{CUDA-Q} documentation provides a readout-error-mitigation
workflow based on detector calibration [see Sec.~\ref{sec:readout_models}]
\cite{cudaq_readout_2026}.
Here a confusion matrix $A$ is constructed from computational-basis calibration
experiments, with matrix element $A_{y,x}=P(y|x)$ describing the probability
of reporting outcome $y$ when state $x$ was prepared. Its Moore--Penrose
pseudoinverse $A^{+}$ is then applied to the noisy measurement distribution to
estimate a corrected distribution.

The tutorial considers three levels of detector modeling: a common
single-qubit model, separate local confusion matrices, and the complete
$2^n\times2^n$ confusion matrix. The local approaches are scalable but neglect
some correlated readout errors, whereas the full matrix captures such
correlations at an exponentially increasing calibration cost. The present
CUDA-Q example should therefore be viewed primarily as a documented
SDK-level implementation pattern for readout correction, rather than as a
general built-in QEM service.

\item \textbf{Cross-platform deployment and benchmarking.}
A major advantage of software-level abstractions is that the same mitigation
protocol can be tested across different hardware platforms. Using
\texttt{Mitiq}, Russo \emph{et al.} applied ZNE and probabilistic error
cancellation to benchmark problems executed on IBM, IonQ, and Rigetti
processors
\cite{russo2023platformindependent}. They introduced a resource-normalized
improvement factor and found that mitigation was beneficial on average even
after accounting for its additional computational cost, while the magnitude of
the improvement remained strongly hardware dependent.
\end{itemize}

These examples illustrate a broader trend toward expressing QEM as reusable
software objects, such as circuit transforms, mitigated estimators, samplers,
or calibration-and-post-processing workflows. Such abstractions improve
portability and reproducibility by standardizing how a mitigation protocol is
implemented, while still allowing its effectiveness to depend on the noise,
connectivity, calibration quality, and native operations of the underlying
hardware.

\subsection{Provider-native mitigation in quantum-computing runtimes}
\label{provider}

A complementary direction is to build error suppression and error mitigation directly into the cloud service that runs jobs on a provider's hardware.
Here, a \emph{runtime} refers to the managed execution layer operated by the hardware provider.
Instead of asking the user to assemble every mitigation step manually, the runtime allows the user to submit a circuit together with a small set of options, and then handles much of the execution workflow internally.
This may include translating the circuit into hardware-native instructions, scheduling it on the device, running supported calibration or randomization routines, and classically post-processing the measured data.

In this model, mitigation becomes part of the execution service rather than a separate external workflow.
For example, the provider may automatically run additional calibration experiments, generate randomized or noise-amplified circuit variants, discard shots flagged as faulty, or apply a built-in readout-correction procedure.
The returned result is therefore often not just a raw collection of measurement outcomes.
Instead, the user may receive a processed estimate together with diagnostic information, such as uncertainty bars, sampling overhead, or metadata describing which mitigation and suppression steps were applied.

This provider-native approach is important because the hardware provider has direct access to information that is difficult for an external toolkit to reproduce, including current calibration data, device scheduling constraints, pulse-level options, and backend-specific noise diagnostics.
At the same time, it makes the mitigation workflow more platform-dependent: the available options, their implementation details, and their effectiveness can differ substantially from one provider or device generation to another.
\subsubsection{Mitigation methods for IBM Quantum hardware}

IBM Quantum provides a particularly useful example of \emph{provider-native}
error mitigation through \texttt{Qiskit Runtime}. Unlike a cross-platform QEM
library that only receives circuits and measurement results, the Runtime is
integrated with the IBM backend and can combine circuit compilation, hardware
calibration, noise learning, circuit randomization, and statistical
post-processing within the same execution workflow
\cite{qiskitdocs_mitigation_2026,qiskitdocs_techniques_2026}.
This allows different error mechanisms to be addressed at different stages of
the computation.

\begin{itemize}[itemsep=0.5ex]

\item \textbf{Suppressing errors before measurement.}
The Runtime supports \emph{Pauli twirling} and \emph{dynamical decoupling}
(DD). Gate twirling randomizes coherent errors associated with entangling gates
into a more stochastic effective form, reducing systematic coherent
accumulation. DD instead inserts refocusing pulses into suitable idle periods
to suppress slowly varying phase errors and residual coherent interactions.
These methods act during circuit execution and therefore reduce or simplify the
noise before subsequent mitigation is applied
\cite{wallman2016noise,qiskitdocs_techniques_2026}.

\item \textbf{Mitigating measurement and circuit errors.}
For readout errors, IBM provides Twirled Readout Error eXtinction (TREX), which
uses measurement randomization and calibration data to reduce bias in Pauli
expectation values. For accumulated circuit errors, the Runtime provides
zero-noise extrapolation (ZNE), in which related circuits are executed at
several effective noise levels and extrapolated toward the zero-noise limit
\cite{qiskitdocs_techniques_2026}.
The two techniques therefore address different parts of the experiment: TREX
acts on measurement bias, whereas ZNE targets errors accumulated during circuit
execution.

\item \textbf{Hardware-informed noise learning.}
A distinctive capability of the IBM workflow is that mitigation can use a
noise model learned automatically from the target hardware. In probabilistic
error amplification (PEA), the Runtime characterizes the effective noise of
entangling layers and uses the learned model to amplify that noise before ZNE.
Probabilistic error cancellation (PEC) uses related characterization to
construct signed combinations of noisy circuits that compensate the learned
effective noise
\cite{kimEvidenceUtilityQuantum2023c,vandenberg2023sparsepec,
qiskitdocs_techniques_2026}.
This backend-level noise learning is especially useful for large circuits,
where manually characterizing every relevant error channel would be difficult.

\item \textbf{Simple presets with optional fine control.}
For general users, \texttt{EstimatorV2} exposes \emph{resilience levels} that
bundle commonly useful methods. Level~0 applies no built-in mitigation;
level~1 uses TREX for readout mitigation; and level~2 adds gate twirling and
ZNE. More advanced users can instead configure individual techniques such as
DD, PEA, or PEC directly
\cite{qiskitdocs_mitigation_2026,qiskitdocs_techniques_2026}.
This provides a practical trade-off between ease of use and detailed control
over the mitigation workflow.

\end{itemize}
The main advantage of the IBM approach is therefore not a single mitigation
method, but the integration of several complementary methods within the same
hardware-aware runtime. Coherent errors can first be suppressed or randomized,
measurement bias can be calibrated separately, and residual circuit noise can
then be treated using ZNE, PEA, or PEC. This layered structure makes
provider-native mitigation particularly useful for experiments in which the
dominant error mechanism varies across circuit layers and hardware
calibrations.

Beyond the native Runtime presets, IBM Quantum also exposes partner-developed services through the Qiskit Functions catalog. These hosted functions run on IBM hardware but implement vendor-specific workflows, and therefore should not be identified with an additional IBM \emph{resilience level} or with the native Runtime settings described above \cite{qiskitFunctions2026}. Two representative examples are Qedma's QESEM and Q-CTRL's Fire Opal Performance Management.

\medskip
\noindent\textbf{Qedma QESEM.}\par
\nobreak\noindent Qedma's Quantum Error Suppression and Error Mitigation service is loaded from the catalog as \texttt{qedma/qesem} and combines characterization, compilation, suppression, and estimator-level mitigation in one managed workflow \cite{qiskitFunctions2026,ibm_qedma_qesem_2026,aharonovReliableHighAccuracy2025}. The user supplies a circuit, one or more observables, a target backend, and a target statistical precision. Before committing the full experimental budget, the service can estimate the required QPU time. During an execution job, QESEM performs circuit-specific device characterization, uses the resulting effective-noise information for noise-aware layout and transpilation, and selects physical gates and measurement bases that reduce the error affecting the requested observables. It then executes characterization-informed error-suppression and quasi-probabilistic mitigation circuits and classically combines the data to return expectation values with a statistical error bar.

The inverse-channel component of QESEM is closely connected to the quasiprobability framework in Sec.~\ref{pec}. When the characterized effective channel remains valid during acquisition, the signed estimator is designed to remove the corresponding modelled bias rather than merely fit a heuristic correction curve. This guarantee is conditional, however, on the accuracy and stability of the characterization model and on reaching the requested statistical precision. Its cost is also workload dependent: it grows strongly with the requested precision and with the observable-dependent active volume, namely the entangling operations lying in the backward light cone of the measured observable, rather than with the nominal qubit count alone.

\medskip
\noindent\textbf{Q-CTRL Fire Opal Performance Management.}\par
\begin{figure*}[t]
    \centering
    \includegraphics[width=0.60\textwidth]{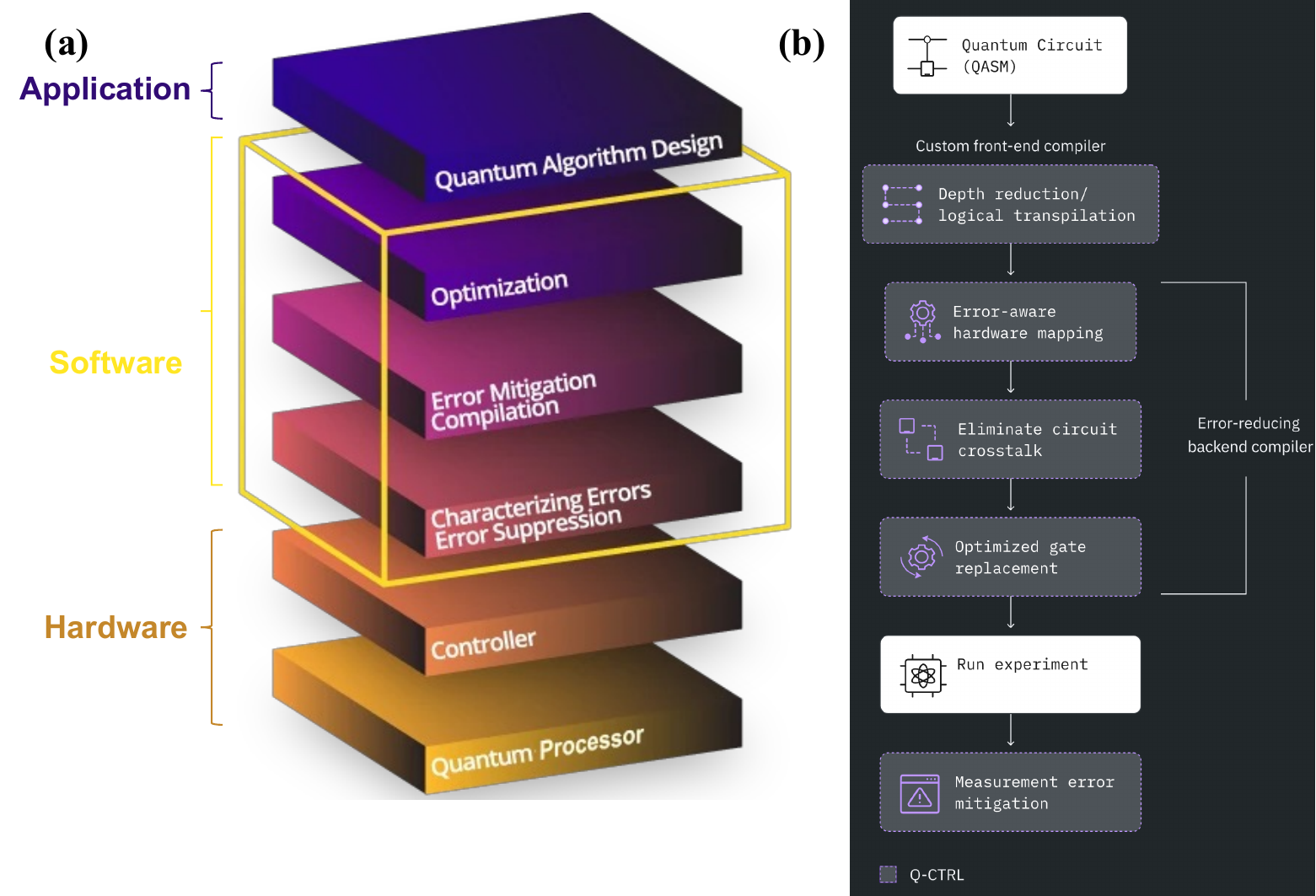}
    \caption{Cross-stack error management and the automated Q-CTRL Fire Opal workflow on supported IBM Quantum hardware. (a) Error reduction spans the application, software, and hardware layers: algorithm design is coupled to circuit optimization, error-mitigation-aware compilation, and error characterization and suppression above the controller and quantum processor. (b) The Fire Opal pipeline accepts a quantum circuit, applies depth reduction and logical transpilation, error-aware hardware mapping, circuit-crosstalk suppression, and optimized gate replacement before execution, and then applies measurement-error mitigation to the returned data. The highlighted stages are managed by Q-CTRL. In contrast, QESEM additionally uses circuit-specific characterization to construct a quasi-probabilistic reconstruction of residual error, as described in the surrounding text. The workflow in panel (b) follows Ref.~\cite{mundada2023fireopal}.}
    \label{fig:ibm_partner_error_management}
\end{figure*}

\nobreak\noindent Q-CTRL's service is loaded as \texttt{q-ctrl/performance-management} and exposes Sampler-like and Estimator-like interfaces for abstract input circuits \cite{qiskitFunctions2026,mundada2023fireopal}. Fire Opal follows an automated deterministic error-suppression strategy. Its front-end compiler performs depth reduction and logical transpilation; the back-end compiler then performs error-aware hardware mapping, crosstalk-suppression through context-aware control and dynamical-decoupling choices, and replacement of selected operations by optimized gate implementations. After hardware execution, a calibrated measurement-error mitigation stage processes the returned data. These steps connect directly to the circuit-optimization methods of Sec.~\ref{ansatz}, the coherent-error-suppression methods of Sec.~\ref{twirling}, and the detector corrections of Sec.~\ref{read}. The resulting cross-stack organization and execution sequence are summarized in Fig.~\ref{fig:ibm_partner_error_management}.

Unlike a quasiprobability protocol, Fire Opal does not use a signed ensemble of modified circuits as its central correction mechanism. Its objective is instead to reduce the physical error entering each sample by changing how the circuit is compiled, mapped, controlled, and measured. This does not eliminate ordinary shot requirements or classical processing time, and its benefit remains backend- and calibration-dependent, but it avoids the characteristic precision-dependent negativity overhead of inverse-channel sampling.

The distinction between the two services is operational. Fire Opal primarily suppresses errors before and during a hardware run so that each execution has a larger usable signal, whereas QESEM combines front-end suppression with characterization-based statistical reconstruction of the residual channel. Accordingly, QESEM is most naturally compared with the inverse-channel methods of Sec.~\ref{pec}, while Fire Opal is a cross-stack realization of the compilation, coherent-error suppression, and readout-correction layers developed in Secs.~\ref{ansatz}, \ref{twirling}, and \ref{read}. For reproducibility, experiments using either service should report the function provider and version, backend, circuit layout or transpilation mode, target precision or shot budget, and the execution and uncertainty metadata returned by the service.

\subsubsection{Mitigation methods for Google quantum hardware}

Google Quantum AI follows a relatively modular approach to error suppression and
mitigation through \texttt{Cirq}, \texttt{cirq-google}, and experiment-specific
workflows in \texttt{ReCirq}. Rather than exposing a single runtime-level
``mitigation mode,'' the software stack provides hardware-aware compilation,
calibration, dynamical decoupling, readout correction, and gate-characterization
tools that can be combined according to the target experiment.

\begin{itemize}[itemsep=0ex]

\item \textbf{Calibration-aware circuit optimization and qubit selection.}
Google hardware exposes calibration information including single-qubit
randomized-benchmarking errors, two-qubit cross-entropy-benchmarking (XEB)
errors, and readout errors. These metrics can be used to avoid poorly performing
qubits and couplers and to choose a physical layout with lower expected error
\cite{google_cirq_calibration_2026}.
The software workflow also emphasizes reducing circuit depth, aligning
single- and two-qubit layers, and propagating virtual $Z$ rotations so that they
do not introduce additional physical pulses. For native two-qubit gates,
characterization and gate refitting can compensate coherent errors, calibration
drift, and unwanted crosstalk by estimating the realized FSim-type interaction
and correcting selected phase parameters using virtual $Z$ rotations
\cite{google_cirq_best_practices_2026}.

\item \textbf{Dynamical decoupling and spin echoes.}
For qubits that remain idle during part of a compiled circuit, Cirq provides
\texttt{cirq.add\_dynamical\_decoupling}, which inserts spin-echo or related
pulse sequences into idle windows
\cite{google_cirq_best_practices_2026}.
The primary purpose is to suppress phase accumulation and other slowly varying
coherent memory errors during idle evolution. Because subsequent circuit
optimization could otherwise remove these cancelling pulses, Google recommends
performing the DD insertion after optimization for the target native gateset or
protecting the inserted operations from later compiler passes.

\item \textbf{Readout-error mitigation.}
Cirq provides explicit tools for measurement-error characterization and
correction. The routine \texttt{cirq.measure\_confusion\_matrix} estimates $A_{y,x}
    =
    \Pr(y|x),$
from prepared computational-basis states, producing a
\texttt{TensoredConfusionMatrices} object
\cite{google_cirq_readout_2026}.
The calibrated detector model can then be used to correct measured
distributions. Cirq also provides
\texttt{readout\_mitigation\_pauli\_uncorrelated}, which directly mitigates
Pauli expectation values under a tensor-product readout-noise model without
constructing an exponentially large global response matrix. This makes the
observable-level approach particularly useful for larger experiments.

\end{itemize}

The \texttt{ReCirq} repository provides a complementary research-oriented layer,
containing workflows used in Google Quantum AI experiments together with their
experiment-specific calibration and error-mitigation procedures
\cite{google_recirq_2024}. Thus, mitigation on Google hardware is best viewed as
a collection of composable, hardware-aware software tools: calibration and
layout selection reduce exposure to poorly performing components, gate
refitting suppresses coherent two-qubit control errors, dynamical decoupling
reduces idle-time dephasing, and detector calibration corrects measurement
errors. At present, this software model is more modular than provider-native
ZNE or PEC services, with the user assembling the appropriate error-reduction
workflow for the experiment.

\subsubsection{Mitigation methods for trapped-ion quantum hardware}
IonQ illustrates a different design philosophy through its runtime-native
\emph{debiasing} workflow
\cite{ionq_debias_2025,ionq_qiskit_debias_2026}.
Rather than estimating an explicit detector matrix or noise channel, the service
uses the compiler to generate several logically equivalent implementations of
the same circuit. These variants can differ in physical-qubit assignment, native
gate decomposition, or pulse-level realization while preserving the ideal
computation. If the $r$th implementation realizes
\begin{equation}
    \widetilde{\mathcal{C}}_r
    =
    \mathcal{N}_r\circ\mathcal{C},
\end{equation}
the ideal channel $\mathcal{C}$ is common to all variants, whereas the effective
noise $\mathcal{N}_r$ changes with the particular hardware realization.
Aggregating over the variants therefore suppresses error components that depend
systematically on mapping or implementation details
\cite{maksymov2023symmetrization}.

The method is primarily aimed at \emph{implementation-dependent bias}. Examples
include coherent control errors associated with a particular native-gate
realization, qubit- or interaction-dependent imperfections, calibration
inhomogeneity, and other systematic errors whose sign or magnitude changes
between equivalent circuit realizations. It does not directly invert a
characterized noise channel, and errors that are essentially identical across
all variants are not removed simply by averaging. Decoherence and stochastic
gate errors can still remain as a residual noise floor.

IonQ provides two aggregation strategies:
\begin{itemize}[itemsep=0ex]

    \item \textbf{Component-wise averaging.}
    If $p_r(x)$ is the measured probability of bit string $x$ for variant $r$,
    the debiased distribution is
    \begin{equation}
        p_{\rm avg}(x)
        =
        \frac{1}{R}
        \sum_{r=1}^{R}p_r(x).
    \end{equation}
    This is the more general option and is appropriate for broad probability
    distributions, such as those occurring in VQE, sampling, and many
    simulation tasks. Variant-dependent coherent and calibration biases can
    partially cancel under the average, while stochastic fluctuations are
    reduced through repeated sampling.

    \item \textbf{Sharpening.}
    For algorithms whose ideal output is concentrated on one or a few
    approximately equiprobable bit strings, IonQ additionally provides a
    plurality-voting procedure across the symmetrized variants. Outcomes that
    recur consistently across different realizations are enhanced, whereas
    variant-specific erroneous outcomes are suppressed. This can be effective,
    for example, for algorithms with a single dominant computational-basis
    answer or a small set of well-defined peaks. It is not generally appropriate
    for broad or strongly nonuniform distributions, because the nonlinear
    voting procedure can distort the physical output probabilities.

\end{itemize}

Thus, averaging and sharpening address somewhat different use cases. Averaging
is a symmetry-based suppression of implementation-dependent bias while
preserving a general output distribution, whereas sharpening additionally uses
prior structure in a peaked ideal distribution to reject inconsistent outcomes.
Neither procedure should be interpreted as complete correction of relaxation,
dephasing, leakage, or readout error; rather, they reduce the part of the
observed error that varies across logically equivalent hardware realizations. The practical advantage is that the randomization and aggregation are handled
on the service side and require no additional logical qubits or deliberate
increase in circuit depth, although the requested shots must be distributed
among multiple circuit variants. The same basic symmetrization principle is
analyzed more generally by Maksymov \emph{et al.}
\cite{maksymov2023symmetrization}.
IonQ's native cloud interface, Qiskit integration, and Amazon Braket expose this
workflow through somewhat different user-facing controls, including different
shot requirements and result formats
\cite{ionq_qiskit_debias_2026,braket_ionq_debias_2026,
braket_sdk_errmit_2026}.

Quantinuum provides another pattern in which error suppression and mitigation
are closely integrated with device-specific control, compilation, and
measurement capabilities. On System Model H2, for example,
\emph{dynamical decoupling} (DD) is available as a server-side error-suppression
option
\cite{quantinuum_dd_2026}.
The hardware compiler identifies idle intervals longer than a specified
threshold and opportunistically inserts DD pulses. The main target is
\emph{coherent memory error}, including phase accumulation caused by slowly
varying qubit-frequency offsets, magnetic-field fluctuations, reference-clock
errors, and finite calibration precision. DD therefore suppresses coherent
errors accumulated during idle evolution, rather than correcting irreversible
relaxation or leakage.

Quantinuum also provides several distinct mechanisms for handling
\emph{leakage}, in which population leaves the computational qubit subspace.
For H2, the Nexus/\texttt{pytket-quantinuum} workflow supports an
ancilla-assisted leakage-detection gadget through
\texttt{leakage\_detection=True}
\cite{quantinuum_nexus_backend_2026}.
The compiler adds auxiliary qubits and measurements that flag leakage events,
and the returned dataset can subsequently be processed with
\texttt{prune\_shots\_detected\_as\_leaky} to discard the corresponding shots.
This is therefore a detection-and-post-selection strategy: leakage is not
inverted or corrected after it occurs, but contaminated experimental samples
are identified and removed. H2 additionally supports a server-side
\emph{leakage-repump} option, which uses optical pumping during execution to
suppress accumulated leakage population.

Helios provides a more direct hardware-level capability. Its native heralded
leakage measurement shelves the ${}^{137}\mathrm{Ba}^{+}$ qubit states into
metastable manifolds and returns three possible outcomes,
\begin{equation}
    m\in\{0,1,L\},
\end{equation}
where $L$ explicitly heralds population outside the computational subspace
\cite{quantinuum_helios_2026,quantinuum_leakage_2026}.
In this sense, an otherwise hidden leakage error is converted into a detectable
erasure-type event that can be handled by post-selection, conditional logic, or
error-correction protocols.

Helios also supports \emph{protected measurement} to suppress
measurement-induced crosstalk. In the standard measurement procedure,
fluorescence from measured ions can disturb other ions in the same operation
region. Protected measurement instead shelves the relevant hyperfine-state
population into metastable $D_{5/2}$ states before readout, reducing the effect
of this measurement crosstalk
\cite{quantinuum_helios_2026}.
Protected measurement and heralded leakage measurement can also be combined
when an entire batch of ions is measured.

These provider-native capabilities therefore address distinct components of
the error budget:
\begin{itemize}[itemsep=0ex]
    \item \textbf{Dynamical decoupling:} suppresses slowly varying coherent
    memory errors during idle periods.
    
    \item \textbf{Leakage detection and repumping:} identifies leakage for
    post-selection or actively reduces leakage population during execution.
    
    \item \textbf{Protected measurement:} suppresses measurement-induced
    crosstalk.
    
    \item \textbf{Heralded leakage measurement:} converts otherwise hidden
    leakage into a directly detectable erasure-type outcome.
\end{itemize}

These examples illustrate why provider-native error reduction can go beyond
generic software-level QEM. The runtime has access to information and control
primitives---including pulse timing, ion transport, internal atomic levels,
measurement sequencing, and compiler-generated idle windows---that are not
generally available to an external mitigation library.

\begin{figure*}[t]
    \centering
    \includegraphics[width=0.95\textwidth]{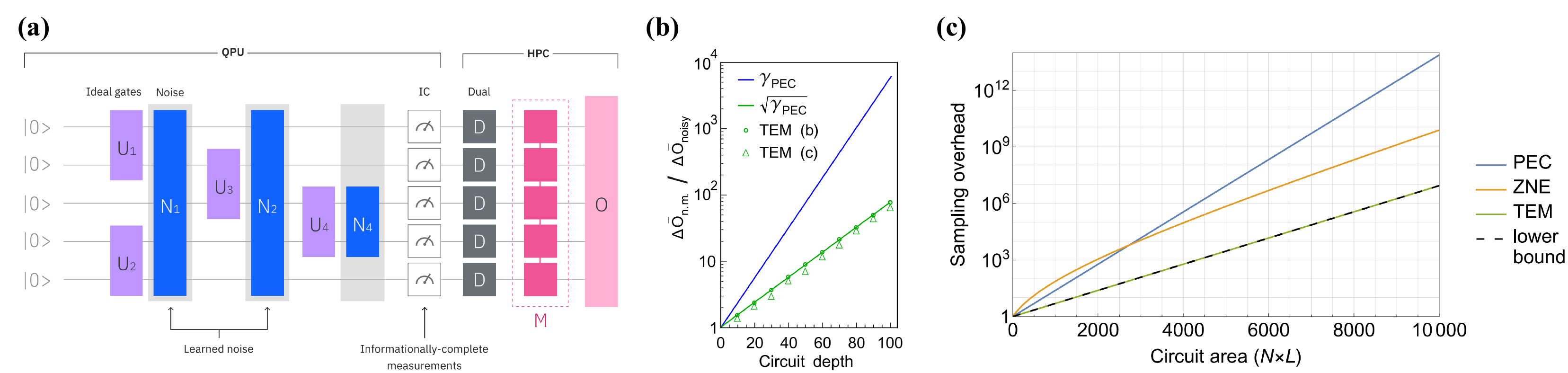}
    \caption{
    Tensor-network error mitigation (TEM).
    (a) Hybrid QPU--HPC workflow for TEM, where the quantum processor prepares and measures the noisy
    state, while the inverse-noise correction is implemented classically through tensor-network
    post-processing of informationally complete measurement data.
    (b) Comparison of quantum-side overhead as a function of circuit depth, showing that TEM is
    substantially less measurement-intensive than PEC.
    (c) Comparison of sampling overhead versus circuit area, indicating that TEM can outperform PEC and
    ZNE and approach the fundamental lower bound under realistic sparse-noise assumptions.
    Panel (a) is adapted from Ref.~\cite{ibm2026tem}; panels (b) and (c) are adapted from
    Refs.~\cite{filippov2023tem,filippov2024scalability}.
    }
    \label{figS10_10_1}
\end{figure*}

\subsection{Tensor-network error mitigation}\label{tensor}
The provider-facing implementation discussed in this subsection is \textit{Algorithmiq's Tensor-network Error Mitigation Qiskit Function}, loaded from the IBM Quantum catalog as \texttt{algorithmiq/tem} \cite{qiskitFunctions2026,ibm2026tem}. We use ``TEM'' for the underlying mitigation method and \texttt{algorithmiq/tem} for this current hosted implementation. The underlying method provides a software-centered approach to quantum error mitigation in which the main correction step is shifted from additional quantum-circuit execution to classical post-processing
\cite{guo2022mpo,filippov2023tem,filippov2024scalability}.
This distinguishes TEM from circuit-level mitigation methods such as ZNE and PEC.
In ZNE, one estimates the noiseless value by executing noise-scaled variants of the target circuit and extrapolating the measured observable to the zero-noise limit.
In PEC, one represents the inverse of an effective noise channel through a quasiprobability decomposition and estimates the ideal observable by sampling from a signed ensemble of modified noisy operations
\cite{temmeErrorMitigationShortDepth2017a,endo2018practical,kimScalableErrorMitigation2023a}.
TEM follows a different route: the target circuit is executed together with informationally complete measurements, and an approximate inverse-noise map is then applied classically using a tensor-network representation of the noisy measurement data and detector or channel model
\cite{guo2022mpo,filippov2023tem,filippov2024scalability}.
The key advantage of this strategy is that the dominant mitigation overhead is moved away from running many different mitigated circuit variants on the quantum processor.
Instead, the cost is concentrated in three resources: the acquisition of many-shot informationally complete data, the characterization or learning of a suitable noise model, and the classical contraction of the resulting tensor network.

Thus, TEM should not be viewed as a resource-free replacement for ZNE or PEC.
Rather, it exchanges quantum execution overhead for classical post-processing overhead and for assumptions about the structure of the noise and the compressibility of the relevant probability or quasi-probability tensors.
This makes the method especially attractive when the effective noise admits a low-bond-dimension tensor-network representation, for example, under sufficiently local or weakly correlated noise, but it also makes the scalability sensitive to correlation growth, noise-model mismatch, and the conditioning of the inverse map
\cite{guo2022mpo,filippov2023tem,filippov2024scalability}.
The basic idea can be expressed in the Heisenberg picture. Let $\rho_{\mathrm{noisy}}$ denote the noisy output state of the executed circuit, and let $O$ be the observable of interest. TEM estimates the ideal expectation value by applying an approximate inverse-noise map $\mathcal{M}$ virtually in post-processing,
\begin{equation}
\langle O\rangle_{\mathrm{TEM}}
=
\operatorname{Tr}\!\left[
\mathcal{M}(\rho_{\mathrm{noisy}})\, O
\right]
=
\operatorname{Tr}\!\left[
\rho_{\mathrm{noisy}}\, \mathcal{M}^{\dagger}(O)
\right],
\label{eq:tem_basic}
\end{equation}
where $\mathcal{M}^{\dagger}$ is the dual mitigation map acting on observables \cite{filippov2023tem}. The tensor-network representation is used to make this dual propagation and contraction feasible. Operationally, the QPU supplies informationally complete measurement data for the noisy circuit output, while the classical backend contracts the tensor-network representation of $\mathcal{M}^{\dagger}(O)$ to obtain corrected estimates of the desired observables.
This formulation makes clear that TEM is not a black-box extrapolation protocol. It relies on a structured model of the noise. In current implementations, the effective noise is commonly represented using sparse Pauli--Lindblad models, which provide a compact description of dominant local errors and selected correlated errors \cite{vandenberg2023sparsepec,filippov2023tem}. Such sparse models are essential because an arbitrary global noise channel would be impossible to learn, store, and invert for large circuits. TEM should therefore be viewed as part of a broader noise-learning and tensor-network post-processing framework, rather than as a purely formal tensor-network identity.
The practical workflow is as follows. 

First, the target circuit is executed on the quantum device with informationally complete measurements. Second, a sparse effective noise model is learned or supplied. Third, an approximate inverse-noise map is encoded as a tensor network. Finally, the desired observables are estimated by contracting the corresponding dual tensor-network object with the measurement data. This workflow is summarized in Fig.~\ref{figS10_10_1} (a): the QPU provides noisy informationally complete data, while the classical side performs noise inversion, tensor-network contraction, and observable reconstruction. An important practical benefit is that the same measurement data can be reused for many observables, which is especially useful for Hamiltonian estimation, many-body correlators, and quantum-simulation workloads where many expectation values are required \cite{filippov2023tem,ibm2026tem}.
The main resource tradeoff in TEM is between quantum sampling overhead and classical contraction cost. Compared with PEC, TEM avoids sampling over signed circuit ensembles. In the original TEM proposal, the quantum-side measurement overhead for representative Pauli observables was shown to be quadratically smaller than the corresponding PEC sampling overhead, scaling effectively as $\sqrt{\gamma_{\mathrm{PEC}}}$ rather than $\gamma_{\mathrm{PEC}}$, where $\gamma_{\mathrm{PEC}}$ denotes the PEC quasiprobability overhead \cite{filippov2023tem}. This behavior is illustrated in Fig.~\ref{figS10_10_1}(b), where TEM requires fewer quantum samples than PEC as the circuit depth increases. A later scalability analysis further compared TEM, PEC, and ZNE under sparse Pauli--Lindblad noise models and found that TEM can have the lowest quantum-side sampling overhead among these methods in favorable regimes, approaching the corresponding lower bound \cite{filippov2024scalability}. This comparison, shown in Fig.~\ref{figS10_10_1} (c), should be interpreted specifically as a statement about quantum sampling cost, not total wall-clock runtime.
The price for this reduced quantum overhead is a potentially demanding classical post-processing step. Tensor-network contraction can become expensive when the effective inverse-noise map generates large operator entanglement or requires a large bond dimension. In practice, TEM must truncate the tensor-network representation to keep the classical computation feasible \cite{filippov2023tem}. This truncation introduces a new accuracy--cost tradeoff: small bond dimension reduces classical cost but can bias the mitigated estimator, while large bond dimension improves accuracy but may become computationally prohibitive. Therefore, TEM does not eliminate mitigation overhead; rather, it transfers much of the overhead from the QPU to the classical backend.

TEM is closely related to earlier tensor-network approaches to error mitigation. In particular, Guo and Yang introduced a matrix-product-operator formulation in which the effect of noise is represented compactly and used to correct measured observables \cite{guo2022mpo}. TEM develops this general idea into a workflow based on informationally complete measurements, sparse noise modeling, Heisenberg-picture inverse-noise propagation, and tensor-network contraction \cite{filippov2023tem,filippov2024scalability}. This connection is useful because it places TEM within a broader family of tensor-network-based mitigation methods rather than treating it as an isolated software feature.
From the software-ecosystem perspective, TEM is particularly notable because it has been exposed as a provider-facing runtime service. IBM provides Algorithmiq's TEM through a Qiskit Function interface, where the user submits circuits and observables and the service performs the informationally complete measurement workflow, tensor-network post-processing, and result aggregation \cite{ibm2026tem}. This represents an important practical step toward packaging advanced mitigation methods as cloud-accessible services. At the same time, the present provider-native implementation has explicit restrictions, including limited support for parametrized circuits, circuit loops, and nonunitary operations such as reset, measurement, and general control flow \cite{ibm2026tem}. These restrictions are important because they show that TEM is not yet a universal black-box mitigation layer for arbitrary circuits.

Overall, TEM illustrates an important direction in the NISQ software stack: error mitigation is increasingly becoming a hybrid runtime workflow that combines quantum data acquisition, structured noise learning, classical tensor-network contraction, and observable reconstruction
\cite{guo2022mpo,filippov2023tem,filippov2024scalability}.
Its main strengths are that it can reduce the need to execute many distinct mitigated circuit variants on the quantum processor, reuse informationally complete measurement data across multiple observables, and integrate naturally with runtime-style workflows in which quantum execution and classical post-processing are coordinated at the software level.
Its main limitations are the need for an accurate and sufficiently stable noise model, the classical cost of tensor-network contraction, possible truncation-induced bias, sensitivity to correlation growth in the effective noise, and present restrictions in provider-native runtime implementations.
TEM is therefore most promising when the circuit noise, measurement data, and target observables admit compact tensor-network representations, and when classical post-processing resources can be traded for a reduced demand on quantum hardware time
\cite{guo2022mpo,filippov2023tem,filippov2024scalability}.

\subsection{Learning-based and data-driven  {error} mitigation}

Learning-based and data-driven mitigation treats QEM as a supervised calibration problem tailored to a family of circuits or observables. Rather than reconstructing the complete device noise channel, one learns a map between noisy hardware outputs and trusted reference values obtained from classically tractable circuits or processes with known ideal behavior
\cite{czarnik2021error,loweUnifiedApproachDataDriven2021,
liaoMachineLearningPractical2024,liaoNoiseAgnosticQuantum2025}. Such workflows can also be supported by software frameworks: current versions of
\texttt{Mitiq} provide CDR- and vnCDR-based routines, while \texttt{Qermit}
supports modular regression-based mitigation within the \texttt{pytket} ecosystem
\cite{mitiqdocs2026,cirstoiu2023qermit}.

For training circuits $C_i$, the data may be written as
\begin{equation}
\mathcal{D}
=
\left\{
(\boldsymbol{x}_i,y_i)
\right\}_{i=1}^{N_{\rm tr}},
\end{equation}
where $\boldsymbol{x}_i$ contains noisy measurements and optional circuit or device descriptors, while $y_i$ is the corresponding trusted reference value. A model $f_\theta$ is trained according to
\begin{equation}
\theta^\star
=
\operatorname*{arg\,min}_{\theta}
\sum_i
\ell\!\left[
f_{\theta}(\boldsymbol{x}_i),y_i
\right],
\end{equation}
and subsequently applied to the target circuit. Depending on the application, the learned output may be an expectation value, a set of observables, or a probability distribution.

\begin{itemize}

\item \textbf{Clifford data regression (CDR).}
An early data-driven approach used neural networks to learn corrections to noisy output distributions from paired ideal and hardware-generated data
\cite{kimQuantumErrorMitigation2020}.
A more structured approach is Clifford data regression (CDR)
\cite{czarnik2021error}. CDR constructs training circuits that remain close to the target circuit but are made classically tractable by replacing selected non-Clifford gates with Clifford gates.

For a target observable $O$, each training circuit provides
\begin{equation}
x_i
=
\langle O\rangle_i^{\rm noisy},
\qquad
y_i
=
\langle O\rangle_i^{\rm exact}.
\end{equation}
Canonical CDR assumes an approximately affine relation,
\begin{equation}
y_i\simeq a x_i+b,
\end{equation}
where the coefficients are obtained from
\begin{equation}
(a^\star,b^\star)
=
\operatorname*{arg\,min}_{a,b}
\sum_i
\left[
y_i-(a x_i+b)
\right]^2.
\end{equation}
For the noisy target value $x_\star$, the mitigated estimate is
\begin{equation}
\widehat{O}_{\rm CDR}
=
a^\star x_\star+b^\star.
\end{equation}
The main challenge is to construct training circuits that are both classically tractable and representative of the target computation. Targeted sampling, including Markov-chain Monte Carlo strategies, can improve this balance
\cite{czarnik2021error}.

\item \textbf{Variable-noise Clifford data regression (vnCDR).}
Variable-noise CDR extends CDR by executing each training circuit at several controlled noise levels
\cite{loweUnifiedApproachDataDriven2021}. The corresponding input becomes
\begin{equation}
\boldsymbol{x}_i
=
\left[
x_i(\lambda_0),
x_i(\lambda_1),
\ldots,
x_i(\lambda_K)
\right],
\end{equation}
and the ideal value is modeled as
\begin{equation}
y_i
\simeq
b+\boldsymbol{a}\cdot\boldsymbol{x}_i.
\end{equation}
The target circuit is measured at the same noise levels and corrected through
\begin{equation}
\widehat{O}_{\rm vnCDR}
=
b^\star+
\boldsymbol{a}^\star\cdot\boldsymbol{x}_\star.
\end{equation}
vnCDR therefore combines classically simulable training circuits with the noise-scaling information used in ZNE. Unlike conventional ZNE, however, the combination of measurements at different noise levels is learned from training data rather than fixed by a prescribed extrapolation model.
\begin{figure*}
    \centering
    \includegraphics[width=0.96\textwidth]{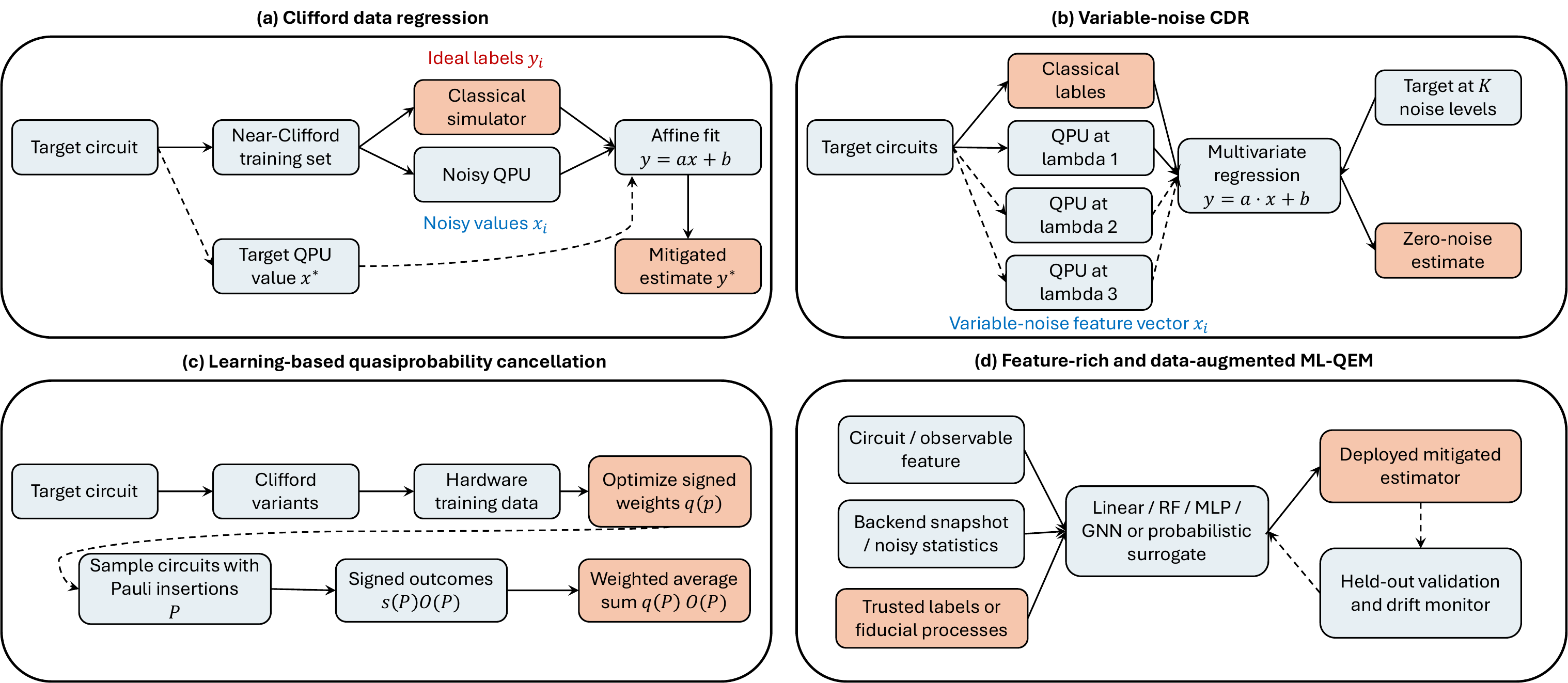}
    \caption{Learning-based and data-driven QEM workflows. (a) CDR constructs near-Clifford training circuits, pairs classically computed ideal labels with noisy hardware values, and fits a regression map for the target observable. (b) vnCDR uses the same reference circuits but supplies noisy values at several effective noise levels to a multivariate regression. (c) Learning-based probabilistic error cancellation optimizes signed quasiprobability weights for sampling circuits with inserted Pauli operations rather than reconstructing a complete process model. (d) Modern ML-QEM and data-augmentation pipelines combine noisy statistics with circuit, observable, and backend features, or with fiducial processes of known ideal action, and deploy the trained model together with validation and drift monitoring. The panel concepts are original schematic redraws based on Fig.~1 of Refs.~\cite{czarnik2021error,loweUnifiedApproachDataDriven2021,liaoMachineLearningPractical2024,liaoNoiseAgnosticQuantum2025}, Fig.~2 of Ref.~\cite{PhysRevResearch.6.013223}, and Fig.~1(b) of Ref.~\cite{strikisLearningBasedQuantum2021}.}
    \label{fig:learning_data_driven_qem}
\end{figure*}

\item \textbf{Learning-based probabilistic error cancellation.}
Learning can also be incorporated into probabilistic error cancellation (PEC)
\cite{strikisLearningBasedQuantum2021}. Instead of reconstructing the complete device noise channel, Pauli operations are inserted at selected circuit locations to generate a family of noisy circuit variants. Classically simulable Clifford circuits are then used to learn the signed quasiprobability coefficients. The resulting estimator takes the form
\begin{equation}
\widehat{O}_{\rm mit}
=
\sum_{\alpha}
q_\alpha
\widehat{O}_{\alpha},
\end{equation}
where $\widehat{O}_{\alpha}$ is measured from a Pauli-modified circuit and $q_\alpha$ is learned from the training data. The method therefore retains the inverse-noise principle of PEC while replacing full process characterization with a task-specific learning problem.

\item \textbf{Nonlinear and scalable machine-learning models.}
More expressive machine-learning approaches generalize the same framework beyond affine regression
\cite{bennewitzNeuralErrorMitigation2022,liaoMachineLearningPractical2024}.
Inputs can include noisy observables together with circuit depth, gate counts, observable support, qubit locations, calibration information, and backend descriptors. Neural networks, random forests, graph neural networks, Gaussian processes, and related models can therefore capture nonlinear and circuit-dependent error patterns.

Neural error mitigation has been applied to energies, fidelities, order parameters, and entanglement diagnostics
\cite{bennewitzNeuralErrorMitigation2022}, while deep-learning methods have also been developed for readout-distribution correction
\cite{kim2022deeplearning}. Gaussian-process regression combined with active learning has been explored for variational algorithms
\cite{jiangErrorMitigationVariational2024}. Data-augmented error mitigation (DAEM) further reduces the need for ideal target-circuit labels by training on noisy fiducial processes whose ideal actions are known
\cite{liaoNoiseAgnosticQuantum2025}.

\end{itemize}

The different learning paradigms are summarized in Fig.~\ref{fig:learning_data_driven_qem}. Panel~(a) shows CDR using near-Clifford training circuits with paired noisy and ideal expectation values. Panel~(b) extends this to vnCDR by measuring each circuit at several noise strengths. Panel~(c) illustrates learning-based PEC, where the learned quantities are quasiprobability weights for Pauli-modified circuit variants. Panel~(d) summarizes more general ML-QEM workflows that combine circuit, observable, backend, and measurement features with trusted or fiducial-process training data
\cite{czarnik2021error,loweUnifiedApproachDataDriven2021,
strikisLearningBasedQuantum2021,liaoMachineLearningPractical2024,
liaoNoiseAgnosticQuantum2025}.

The central limitation of learning-based QEM is not model expressiveness but transferability. A correction trained on one circuit family, observable, qubit layout, compiler configuration, device, or calibration window may become biased when applied outside that training distribution. Clifford and near-Clifford training data can also become less representative as the target circuit contains more non-Clifford structure, while fiducial-process approaches still assume that the relevant noise is sufficiently similar between training and target circuits. Training circuits and shots also introduce additional hardware cost, which must be amortized over enough target evaluations to justify the learning stage. Flexible models may furthermore violate positivity, normalization, symmetries, or other physical constraints unless these are imposed explicitly. Held-out validation, time-separated drift tests, physical constraints, and comparison with unmitigated or conventional-QEM baselines are therefore essential
\cite{takagi2022fundamental,takagi2023samplinglowerbounds,
quek2024tighter,wang2024trainability}.

A promising direction is consequently to combine learned residual corrections with more interpretable mitigation layers. Readout mitigation, randomized compiling, ZNE, PEC, or TEM can first remove a structured component of the noise, after which a learned model corrects the remaining mismatch. Active learning, graph-based models, physics-informed architectures, and continual adaptation may further reduce training cost and improve robustness to hardware drift. In this setting, the learned model should be associated with its training data, circuit family, compiler configuration, qubit layout, calibration window, and uncertainty diagnostics to ensure reproducible deployment.

\section{Error mitigation from NISQ to early fault-tolerant quantum computing}

\subsection{From NISQ devices to early fault-tolerant quantum computing}

The error-mitigation techniques reviewed above are applied primarily to imperfect physical qubits without the protection of scalable quantum error correction
\cite{preskill2018quantum,cai2023quantum}.
In this regime, hardware errors are not continuously detected and corrected throughout the computation.
Instead, error-mitigation methods suppress, reshape, extrapolate, or classically post-process the effects of noise so that useful expectation values can still be estimated from imperfect experimental data
\cite{cai2023quantum}.
These methods are particularly well suited to tasks whose outputs consist of a limited set of physical observables, rather than an exact computational result or a complete reconstruction of the quantum state or process.
For such tasks, controlled approximate estimates can remain scientifically meaningful when their residual bias or statistical uncertainty can be properly quantified, as in zero-noise extrapolation [see \cref{sec:zne} above].
Representative quantities include energies, correlation functions, and order parameters.

Fault-tolerant quantum computing (FTQC) is based on a fundamentally different principle. Quantum information is encoded redundantly into logical qubits formed from multiple physical qubits, and physical errors are repeatedly diagnosed through syndrome measurements during the computation. A classical decoder interprets these syndromes and determines the corrections or frame updates required to preserve the encoded state. When the physical error rate lies below the relevant fault-tolerance threshold, increasing the code distance suppresses the logical error rate, ideally exponentially for suitable codes and noise models. The accessible circuit depth is then determined primarily by the logical error rate and available fault-tolerant resources, rather than directly by the coherence time or error rate of a single physical qubit. NISQ error mitigation therefore aims to extract more reliable information from an imperfect physical computation, whereas FTQC aims to make the encoded computation itself reliable.
Recent experiments have demonstrated repeated error-correction cycles, fault-tolerant logical operations, programmable logical circuits, and below-threshold logical-error suppression in small or specialized systems
\cite{ryananderson2021realtime,krinner2022repeated,postler2022faulttolerant,paetznick2024demonstration,bluvstein2024logical,google2025quantum}.

The intermediate regime between unencoded NISQ computation and large-scale FTQC is commonly described as \emph{early fault-tolerant quantum computing}. In this regime, quantum error correction is actively used: information is encoded, stabilizer or syndrome measurements are repeated, and decoders are employed to detect and manage physical faults. Nevertheless, the available code distances remain modest, only a small number of logical qubits may be accessible, and the overhead of logical state preparation, gates, measurements, magic-state resources, decoding, and real-time feedforward remains substantial. The resulting logical computation is therefore protected, but not yet sufficiently accurate or scalable to support arbitrarily long universal algorithms. Early FTQC should thus be understood not merely as improved NISQ hardware, but as a regime in which genuine quantum error correction is present while logical resources and logical fidelity remain limited.

The boundary between these two regimes is not sharp.
Current and near-future processors increasingly support capabilities needed for fault-tolerant operation, including mid-circuit measurement, qubit reset, real-time classical feedforward, repeated stabilizer measurements, and small error-detecting or error-correcting codes
\cite{ryananderson2021realtime,krinner2022repeated,postler2022faulttolerant,paetznick2024demonstration}.
Nevertheless, these capabilities do not yet provide arbitrarily long and universally protected logical computations.
The resulting intermediate regime is often described as \emph{early fault-tolerant quantum computing}.

For surface codes, the benefit of QEC becomes quantitative once the effective physical error rate $p$ lies below the threshold $p_{\rm th}$, where the logical error per cycle approximately follows
\begin{equation}
    \epsilon_d \propto
    \left(\frac{p}{p_{\rm th}}\right)^{(d+1)/2},
\end{equation}
with $d$ the code distance
\cite{google2025quantum}.
Thus, even a moderate reduction of the effective physical error rate through improved hardware, calibration, or error suppression can be amplified by QEC as the code distance increases. In the first below-threshold Willow experiment, increasing the surface-code distance by two reduced the logical error by a factor $\Lambda=2.14\pm0.02$
\cite{google2025quantum}.
More recently, further improvements in control and decoding reduced the logical error of a distance-$7$ surface-code memory to
$\epsilon_7=7.72(9)\times10^{-4}$ per correction cycle, with reinforcement-learning-based control providing an additional ${\sim}20\%$ suppression beyond conventional calibration
\cite{sivak2026reinforcement}.
These results demonstrate continuing progress in below-threshold QEC, but logical error rates at currently accessible code distances remain much larger than those ultimately required for long fault-tolerant computations. In this intermediate regime, logical resources remain limited, while logical operations, measurements, decoding, and feedforward can introduce additional errors. Residual logical errors may therefore remain sufficiently large that error mitigation can provide a further improvement
\cite{suzuki2022universal,smith2024logical,zhangDemonstratingQuantumError2026}.

\begin{figure}
    \centering
    \includegraphics[width=0.5\linewidth]{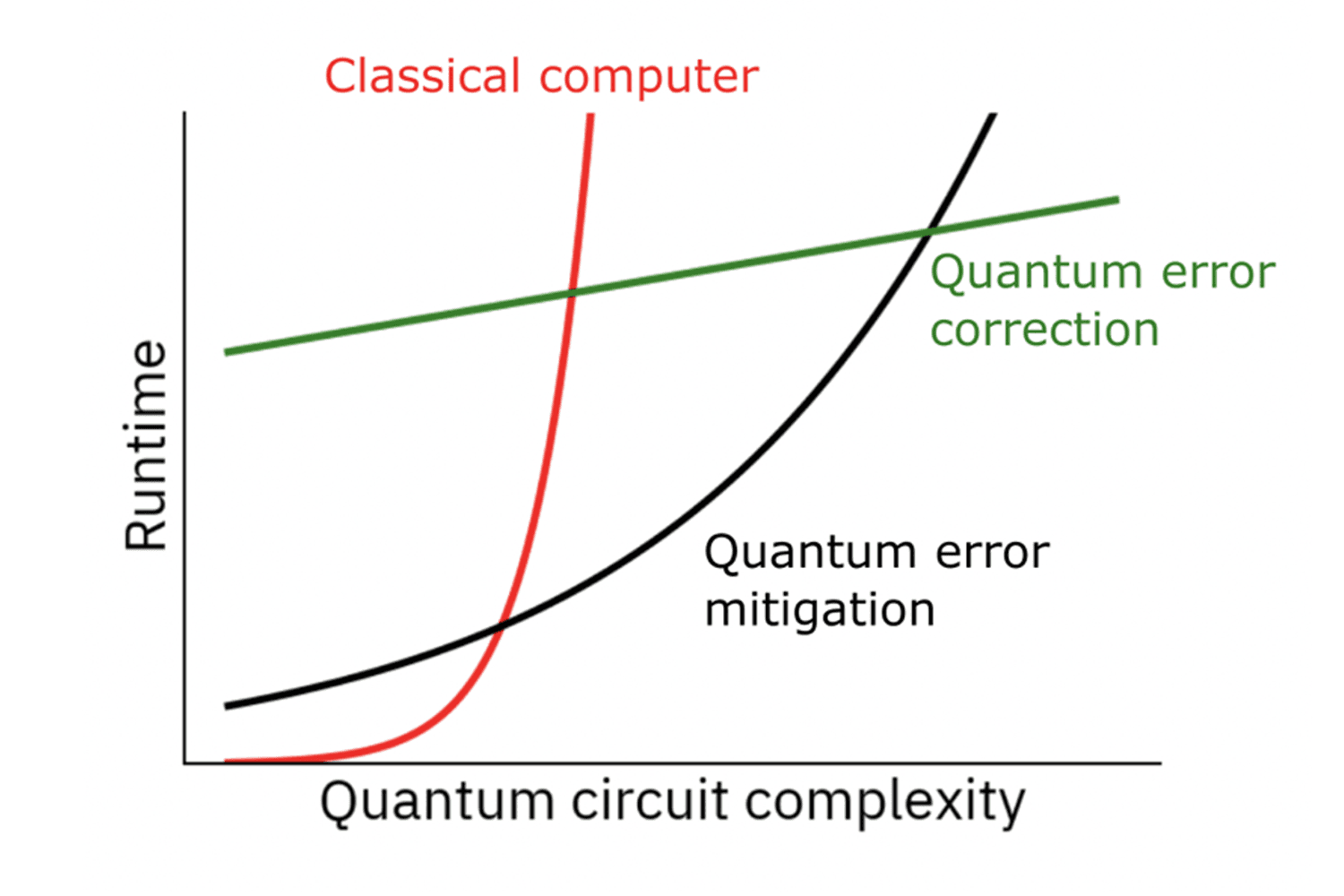}
    \caption{{Schematic runtime regimes for classical simulation, quantum error
    mitigation, and quantum error correction.}
    Classical simulation is efficient for relatively small circuits but becomes
    rapidly more expensive with increasing circuit complexity. QEM can occupy an
    intermediate regime in which noisy quantum computations are already difficult
    to simulate classically but full fault-tolerant QEC remains too costly. QEC
    requires substantial encoding and control overhead, but offers more favorable
    scaling for sufficiently large computations. Adapted from Ref.~\cite{ibm_qem_path_2022}.}
    \label{fig:qem_qec}
\end{figure}

This intermediate regime motivates a hybrid view of quantum error mitigation and
quantum error correction. As illustrated schematically in
Fig.~\ref{fig:qem_qec}, classical simulation is efficient for sufficiently
small circuits but becomes rapidly more costly as circuit complexity grows.
QEM can extend the useful range of noisy quantum computation without the full
hardware overhead of fault-tolerant encoding, although its sampling cost also
increases with circuit size and noise strength. QEC has a much larger initial
resource overhead, but below threshold it provides the favorable scaling needed
for sufficiently long computations
\cite{ibm_qem_path_2022,piveteau2021error}. Importantly, Fig.~\ref{fig:qem_qec} should not be interpreted as implying that
QEM is simply replaced by QEC once logical qubits become available. The two
approaches act at different levels of the computational stack and can be
combined. QEC uses redundant encoding and syndrome measurements to suppress
physical errors and protect logical information, while QEM can be applied to
the residual physical or logical noise that remains after finite-distance error
correction.  A representative example is the hybrid strategy of
Ref.~\cite{piveteau2021error}, in which encoded Clifford operations are protected
by QEC while residual errors associated with costly logical non-Clifford
operations are treated using quasiprobability-based mitigation.

\subsection{Logical-level analogues of NISQ error mitigation}

Many error-mitigation protocols were originally formulated for circuits acting directly on physical qubits. Nevertheless, their underlying logic is not restricted to the unencoded NISQ setting. Once small logical qubits become available, the same ideas can often be lifted to the logical level, where the relevant objects are not only physical gates, physical measurement outcomes, or physical noise channels, but also logical operations, encoded observables, syndrome records, and residual decoder-dependent noise \cite{suzuki2022universal,smith2024logical,zhangDemonstratingQuantumError2026}.

One example is ZNE. In physical NISQ circuits, ZNE typically constructs a noise-scaling axis through pulse stretching, gate folding, or controlled error insertion
\cite{giurgica-tironDigitalZeroNoise2020,heZeronoiseExtrapolationQuantumgate2020}.
In an encoded computation, the same idea can be extended to the logical level. One possibility is to vary the strength of error correction itself, for example through the code distance, and use the resulting logical error rate $p_L$ as the extrapolation coordinate
\cite{wahlZeroNoiseExtrapolation2023}. Schematically, a logical observable may then be written as
\begin{equation}
E_L(p_L)
=
E_L(0)
+
a_1 p_L
+
a_2 p_L^2
+
\cdots ,
\end{equation}
where $E_L(0)$ denotes the ideal logical value. Extrapolating toward $p_L\rightarrow0$ is therefore analogous to conventional ZNE, but the noise axis is determined by the strength of logical protection rather than directly by the physical noise rate.

\begin{figure}
    \centering
    \includegraphics[width=0.8\linewidth]{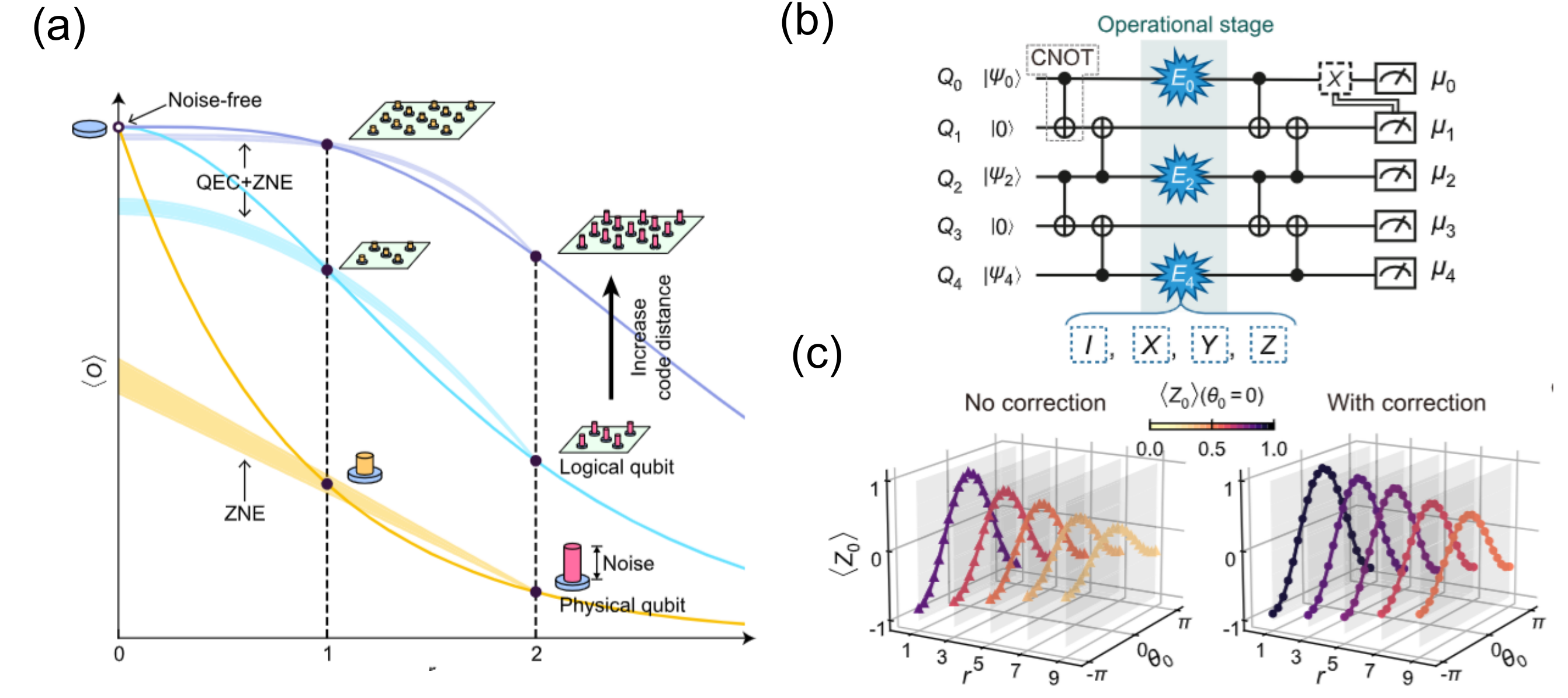}
\caption{{Zero-noise extrapolation on error-corrected and logical qubits.}
(a) Schematic of ZNE applied to physical and logical qubits. The expectation value of an observable is evaluated at several amplified noise strengths $r$, and the noise-free value is inferred by polynomial extrapolation. Error correction reduces the sensitivity to physical noise, with larger code distance producing a flatter logical-noise dependence; combining QEC with ZNE can therefore reduce both the residual bias and the extrapolation uncertainty.
(b) Parity-check circuit used to demonstrate ZNE in the presence of error correction. The data qubits are initialized in parametrized states, parity information is encoded onto syndrome qubits, and Pauli errors are controllably inserted during the operational stage. The final correction is obtained from the measured bit strings through post-selection and a numerically implemented feedback $X$ operation.
(c) Measured $\langle Z_0\rangle$ as a function of the input angle $\theta_0$ and noise-scaling factor $r$, shown without and with error correction. Error correction slows the degradation of the observable with increasing noise strength, providing the corrected data used for subsequent ZNE.
Panel (a) is adapted from Fig.~1 of Ref.~\cite{zhangDemonstratingQuantumError2026}, and panels (b,c) are adapted from Fig.~2 of the same reference.}
    \label{fig:qmqec}
\end{figure}

A complementary strategy was recently demonstrated experimentally by Zhang \textit{et al.}, who applied ZNE directly to quantum-error-correction circuits by controllably amplifying errors on the underlying physical qubits
\cite{zhangDemonstratingQuantumError2026}.
The basic idea is illustrated in Fig.~\ref{fig:qmqec}(a). For an unencoded physical qubit, the observable decays relatively rapidly as the physical-noise scaling factor $r$ increases. Error correction makes this dependence flatter, and increasing the code distance further suppresses the sensitivity of the logical observable to physical noise. ZNE then uses measurements obtained at finite $r$ to extrapolate back toward the noise-free point $r=0$. In this picture, QEC and ZNE play complementary roles: QEC reduces the physical-noise dependence before extrapolation, while ZNE further removes the residual post-correction bias.

The corresponding experimental implementation is shown in Fig.~\ref{fig:qmqec}(b). Three data qubits, $Q_0$, $Q_2$, and $Q_4$, are coupled to two syndrome qubits, $Q_1$ and $Q_3$, through parity-check CNOT operations. During the operational stage, controlled Pauli errors are inserted on the data qubits to realize different noise strengths. The parity information is then decoded and used to correct the measured data, including post-selection and a feedback $X$ correction on $Q_0$. As shown in Fig.~\ref{fig:qmqec}(c), the measured $\langle Z_0\rangle$ decays substantially more slowly with increasing $r$ after error correction than without correction, directly demonstrating that QEC reshapes the noise dependence into a form more favorable for ZNE.

More generally, for a code with effective distance $d$, faults involving fewer than $\lceil d/2\rceil$ errors can be corrected, so the leading dependence on the amplified physical noise is shifted to higher order. The fitting model used in Ref.~\cite{zhangDemonstratingQuantumError2026} is therefore
\begin{equation}
\langle O\rangle(r)
=
\langle O\rangle_{\rm em}
+
\sum_{k=\lceil d/2\rceil}^{\lceil d/2\rceil+K-1}
a_k r^k ,
\end{equation}
where $r$ is the physical-noise amplification factor. This differs from conventional NISQ ZNE, whose expansion generally begins with a linear contribution in $r$.

The same QEC--ZNE strategy was further demonstrated using repetition codes with distances $d=3,5,7$ and a distance-$3$ rotated surface code. For the $d=7$ repetition code, the residual bias was reduced to approximately $10^{-4}$ with a sampling overhead of about $5$, and the mitigation performance remained nearly unchanged as the number of parity-check rounds increased
\cite{zhangDemonstratingQuantumError2026}.

Randomization-based mitigation also extends naturally to encoded computations. In NISQ circuits, randomized compiling and Pauli twirling are used to convert coherent physical errors into more stochastic effective noise \cite{wallman2016noise}. This remains important in the early fault-tolerant regime because coherent physical errors can propagate through syndrome-extraction circuits and appear as coherent logical bias. Although repeated stabilizer measurements can partially decohere physical errors, they do not automatically guarantee that all residual logical noise is purely stochastic or Pauli-like \cite{beale2018decoheres,huang2019performance}. Randomized compiling may therefore be applied at several levels: to the physical gates used inside syndrome-extraction circuits, to encoded logical gates, or to logically equivalent implementations of the same protected operation \cite{jain2023improved}. The goal is the same as in the NISQ regime: to replace structured coherent errors by a simpler effective noise model that is easier to decode, extrapolate, or invert.

These examples show that the distinction between NISQ mitigation and early fault-tolerant mitigation is mainly a distinction of level. At the physical level, error mitigation acts on noisy physical circuits and measured observables. At the logical level, it acts on residual logical channels, encoded observables, syndrome trajectories, and decoder outputs \cite{suzuki2022universal,smith2024logical,zhangDemonstratingQuantumError2026}. The mathematical tools may look similar---extrapolation, inverse-channel reconstruction, randomization, post-selection, and calibrated inference---but the noise objects being mitigated are different. This shift is important because logical noise is not simply a smaller copy of physical noise. It can be highly structured, nonlocal, decoder-dependent, and correlated across syndrome rounds. Therefore, extending NISQ mitigation to early FTQC requires not only reusing existing protocols, but also reformulating them in terms of logical circuit structure and syndrome-resolved data.

\subsection{Outlook: A stepped progression from noisy physical circuits to logical quantum computation}

The transition from NISQ hardware to fault-tolerant quantum computing should no longer be viewed as a distant change that begins only after physical devices become nearly perfect.
Recent experiments have already demonstrated many of the ingredients needed for early logical computation, including repeated syndrome extraction, real-time decoding and feedback, logical error rates below corresponding physical baselines, below-threshold surface-code memories, fault-tolerant logical gates, and programmable logical processors
\cite{ryananderson2021realtime,krinner2022repeated,postler2022faulttolerant,paetznick2024demonstration,bluvstein2024logical,google2025quantum}.
For example, trapped-ion experiments have reported encoded two-qubit operations with logical error rates ranging from approximately the physical-level baseline to several hundred-fold below it, depending on the code and use of post-selection
\cite{paetznick2024demonstration}.
On Google's superconducting processor, a distance-seven surface-code memory used 101 physical qubits and achieved a logical error rate of approximately $0.143\%$ per error-correction cycle, with the logical error decreasing by a factor of about $2.14$ whenever the code distance was increased by two
\cite{google2025quantum}.
Neutral-atom experiments have meanwhile implemented programmable circuits with up to 48 logical qubits using arrays of up to 280 physical atoms, including circuits containing 228 logical two-qubit gates and 48 logical CCZ gates
\cite{bluvstein2024logical}.
These advances indicate that the field is entering an intermediate regime in which physical NISQ circuits, error-detected circuits, and small logical circuits coexist.
The central question is therefore how mitigation, correction, decoding, and compilation should be combined during this transition.

At the physical-circuit stage, quantum computations are still executed mainly on unencoded physical qubits.
This is the regime in which the mitigation techniques reviewed in this article apply most directly.
Hardware-aware compilation can reduce unnecessary routing and control overhead, readout mitigation can correct terminal measurement bias, and methods such as zero-noise extrapolation, probabilistic error cancellation, symmetry verification, and circuit randomization can improve estimates obtained from noisy circuits
\cite{preskill2018quantum,cai2023quantum}.
These methods can substantially improve near-term estimates of observables, but their accuracy and cost remain limited by physical error rates, circuit depth, model mismatch, calibration drift, and sampling overhead.
They therefore extend the useful reach of physical-qubit hardware without turning it into a fault-tolerant processor.

The next stage is not simply ``larger NISQ'', but error-detected computation.
Here, additional measurements are used to determine whether the computation has left an intended logical, physical, or symmetry-defined subspace.
These checks may arise from stabilizer codes and repeated syndrome measurements
\cite{ryananderson2021realtime,krinner2022repeated,zhao2022surfacecode},
parity or Pauli checks
\cite{gonzales2023paulicheck},
leakage-detection procedures
\cite{chen2016measuring},
gauge constraints
\cite{stryker2019gauss,ballini2025nonabelian,carena2024gaugeredundant},
or problem-specific symmetries
\cite{mcclean2017hybrid,mcardle2019errormitigated,kakkar2022qaoa}.
In many experiments, the resulting check record is used for post-selection, symmetry verification, or reweighting rather than for complete active correction
\cite{mcardle2019errormitigated,gonzales2023paulicheck,botelho2022midcircuit}.
This regime is important because it introduces many operational primitives required for FTQC---mid-circuit measurement, reset, real-time feedforward, syndrome extraction, and decoding---while still retaining the statistical character of NISQ mitigation
\cite{ryananderson2021realtime,postler2022faulttolerant,iqbal2024topological,koh2026readout}.
Error mitigation at this stage becomes more dependent on the measurement trajectory: estimators can be conditioned not only on the final measurement outcome, but also on whether the intermediate syndrome or check record is consistent with the intended evolution.

A further stage is small logical-qubit computation.
Here, quantum information is encoded, and repeated syndrome extraction begins to suppress physical errors dynamically.
The relevant noise object is therefore no longer only the physical noise channel acting on individual gates, but also the residual logical channel that remains after syndrome processing and decoding.
Current experiments have already shown that encoded circuits can outperform corresponding physical implementations in selected settings
\cite{paetznick2024demonstration,bluvstein2024logical,google2025quantum}.
For example, repeated error correction with trapped-ion codes has produced logical circuit errors below physical-circuit baselines, while surface-code experiments have demonstrated that logical memory improves as the code distance increases
\cite{paetznick2024demonstration,google2025quantum}.
These are major milestones, but the resulting logical qubits remain finite-distance and resource constrained.
Logical gates, logical measurements, syndrome extraction, leakage removal, decoding, and feedforward can all introduce residual errors.

Logical-level error mitigation therefore remains meaningful in this regime.
Possible approaches include extrapolating with respect to code distance or logical error rate, correcting residual logical measurement bias, tailoring coherent logical noise through randomized compiling, and post-selecting or reweighting data using syndrome histories
\cite{wahlZeroNoiseExtrapolation2023,smith2024logical,zhangDemonstratingQuantumError2026,dutkiewicz2025error}.
Importantly, this is no longer only a theoretical possibility.
Recent experiments have applied zero-noise extrapolation directly to quantum-error-correction circuits and shown that mitigation can reduce errors arising from imperfect correction operations
\cite{zhangDemonstratingQuantumError2026}.
This provides direct evidence that error mitigation and error correction can operate together rather than only in separate hardware regimes.

In the fully fault-tolerant limit, the primary source of reliability should come from quantum error correction itself.
Logical error rates must be reduced to the level required by the target computation through improved physical hardware, increasing code distance, fault-tolerant logical operations, optimized syndrome extraction, and scalable decoding.
In that regime, quantum error mitigation will no longer be the main mechanism that enables long computations.
Nevertheless, mitigation may remain useful as a supporting tool.
It can help characterize logical noise, reduce residual bias in near-threshold logical experiments, validate decoder outputs, improve estimates of selected observables, and reduce the resource requirements of carefully chosen subroutines
\cite{suzuki2022universal,smith2024logical,jeon2026qecmitigated}.
As quantum processors move from physical circuits to logical computation, quantum error mitigation will therefore not simply disappear.
It will become more structured, more syndrome-aware, and more closely integrated with   compilation, decoding, and fault-tolerant control.

\section*{Acknowledgements}  
This work is supported by the Singapore Ministry of Education Academic Research Fund Tier-I preparatory grant (WBS no. A-8002656-00-00) and  MOE's Tier-II grant (WBS no: A-8003505-00-00). In addition, it is supported by the National Research Foundation, Singapore through the National Quantum Office, hosted in A*STAR, under the Advanced Quantum Algorithms and Solutions (AQAS) Funding Initiative (S25Q9DA001). H.-Z.\ Li is supported by the China Scholarship Council Scholarship (Grant No.\ 202506890103).

\bibliography{references}

\onecolumngrid
\flushbottom
\newpage
\appendix
\setcounter{equation}{0}
\setcounter{figure}{0}
\setcounter{table}{0}
\setcounter{section}{0}
\renewcommand{\theequation}{S\arabic{equation}}
\renewcommand{\thefigure}{S\arabic{figure}}
\renewcommand{\thesection}{S\arabic{section}}
\renewcommand{\thepage}{S\arabic{page}}
\renewcommand{\thetable}{S\arabic{table}}
%


\end{document}